\documentclass[b5paper,11pt,twoside,openright]{book}

\usepackage{epsfig}
\usepackage{etex}

\usepackage{longtable}              
\usepackage{graphicx}
\usepackage[english,dutch]{babel}
\usepackage{amscd,amssymb,amsmath,amsthm}
\usepackage{bbm}
\usepackage{cite}                   
\usepackage{verbatim}               
\usepackage{multirow}               
\usepackage{fancybox}               
\usepackage{caption}[2005/05/05]
\usepackage[title,titletoc,toc,page]{appendix}
\usepackage{titlesec}

\usepackage{chngcntr}

\usepackage{slashed}

\usepackage{array}

\usepackage{enumitem}

\usepackage{latexsym}
\usepackage{color}
\usepackage{tabularx}
\usepackage{amsfonts,amsbsy}
\usepackage{comment}
\usepackage{tikz,tikz-cd}
\usepackage[font=footnotesize,labelsep=newline,labelfont=sc,justification=centering,
position=top]{caption}

\usepackage{float}

\usepackage{multirow}                     
\usepackage{float}                          
\usepackage{lscape}                         
\usepackage{bm}

\definecolor{MyDarkBlue}{rgb}{0.15,0.15,0.45}
\definecolor{MyGreen}{rgb}{0.15,0.45,0.45}
\definecolor{MyPurple}{rgb}{0.55,0.25,0.55}
\usepackage[linktocpage=true]{hyperref}
\hypersetup{colorlinks=true,citecolor=MyPurple,linkcolor=MyPurple,urlcolor=MyGreen}

\usepackage[utf8]{inputenc}
\usepackage[T1]{fontenc}
\usepackage{lmodern}

\usepackage[nottoc]{tocbibind} 

\usepackage{eso-pic}

\usepackage{pst-all}
\usepackage{pstricks-add}

\usepackage{type1cm}

\usepackage{fancyhdr}

\usepackage[includeheadfoot]{geometry}

\usepackage{bookmark}

\usepackage{etoolbox}

\newcounter{appendix}[chapter]

\makeatletter
\patchcmd{\@chapter}{\protect\numberline{\thechapter}#1}
{\@chapapp~\thechapter: #1}{}{}

\makeatother

\usepackage{braket}

\usepackage[grey]{quotchap}

\makeatletter

\renewcommand*{\chapnumfont}{%
  \usefont{T1}{\@defaultcnfont}{b}{n}\fontsize{95}{130}\selectfont
  \color{chaptergrey}%
}
\makeatother

\renewcommand{\chaptermark}[1]%
              {\markboth{#1}{}}
\renewcommand{\sectionmark}[1]%
              {\markright{\thesection\ #1}}

\newcommand{\mychapterend}{\clearpage{\pagestyle{empty}\cleardoublepage}}

\newcommand{\e}{\textrm{e}}                                     

\renewcommand{\Re}{\textrm{Re}}                                 
\renewcommand{\Im}{\textrm{Im}}                                 

\usepackage{tocloft}

\newcommand{\be}{\begin{equation}}
\newcommand{\ee}{\end{equation}}
\newcommand{\ba}{\begin{eqnarray}}
\newcommand{\ea}{\end{eqnarray}}
\def\a{\alpha}
\def\bP{\bar{\Phi}}
\def\cD{{\cal D}}

\def\Kahler{K\"{a}hler~}
\def\tb{\tilde{\beta}}
\def\tNs{\tilde{N}_\star}
\def\tN{\tilde{N}}
\def\P{\Phi}
\def\p{\partial}
\def\S{S}
\def\Sb{\bar S}
\def\ib{\bar{\imath}}
\def\jb{\bar{\jmath}}
\def\K{K{\"a}hler}

\begin{document}
\selectlanguage{english}
\counterwithout{equation}{section} 
\counterwithin{equation}{chapter}  
\counterwithout{table}{section} 
\counterwithin{table}{chapter}

\pagenumbering{alph}
\thispagestyle{empty}

\begin{center}

\vspace*{0.5cm}

\begingroup
   \fontsize{22pt}{30pt}\selectfont
   {\bf Inflation, Universality\\and Attractors}\\
\endgroup

\vspace*{4.5cm}

\begin{figure}[htb]
\begin{center}
\includegraphics[width=8cm,keepaspectratio]{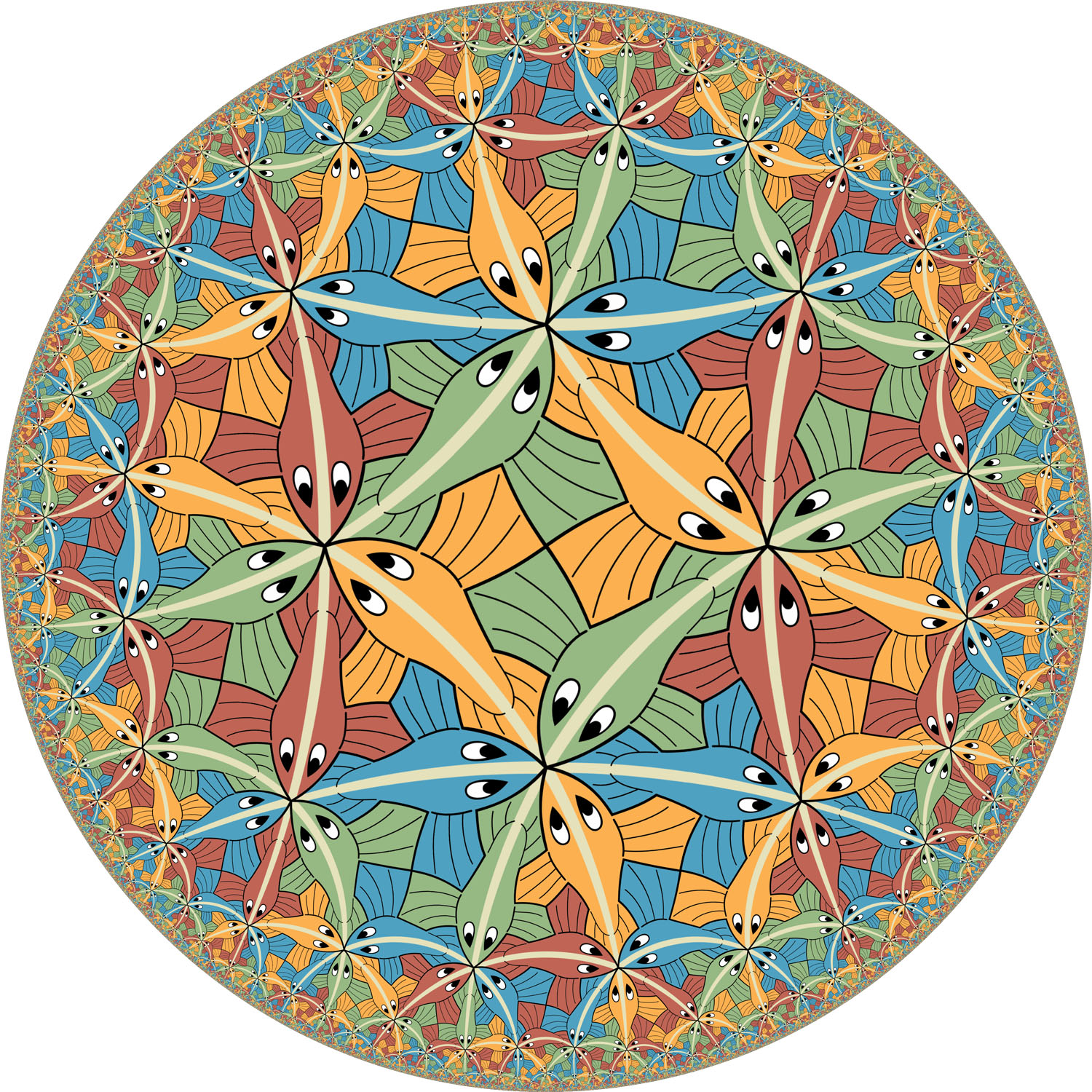}
\end{center}
\end{figure}

\vspace*{2.cm}

\begingroup
    \fontsize{13pt}{17pt}\selectfont
    \textbf{Marco Scalisi}\\
\endgroup	
\vfill

\end{center}
\clearpage

\clearpage
\thispagestyle{empty}
\null
\vfill


\vspace{.5cm}
\noindent ISBN: 978-90-367-8946-2 (printed version)

\noindent ISBN: 978-90-367-8945-5 (electronic version)

\vspace{.5cm}

\noindent The work described in this thesis was performed at the Van Swinderen \mbox{Institute} for Particle Physics and Gravity of the University of Groningen.

\vspace{.5cm}

{\small \noindent {\bf Front cover}: ``Circle Limit III'' by M. C. Escher (computer generated picture by V.~Bulatov,  \href{http://bulatov.org}{http://bulatov.org}). It represents the Poincar\'e disk model of a \mbox{2-dimensional} hyperbolic space. This type of geometry is a central topic of the present thesis, as at the core of the cosmological attractor phenomenon.

\vspace{.3cm}

\noindent {\bf Bookmark}: ``Circle Limit III'' by M. C. Escher (computer generated picture by  V.~Bulatov). It represents a 2-dimensional hyperbolic geometry in half-plane coordinates.

}
\vspace{.5cm}

\noindent Cover and  bookmark realized by Seannamon Design (Giada Maugeri).
\vspace{.1cm}

\noindent Printed by Ipskamp Printing, The Netherlands.

\vspace{.5cm}
\noindent \copyright{} 2016 Marco Scalisi

\clearpage
\thispagestyle{empty}
\includegraphics[width=0.45\textwidth]{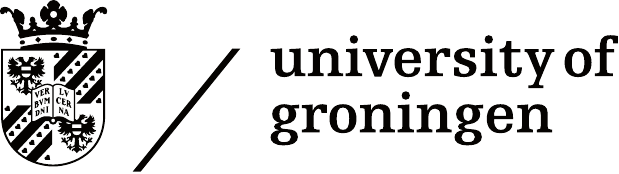}

\begin{center}
\vspace*{1cm}
\begingroup
    \fontsize{22pt}{30pt}\selectfont
   \textbf{ Inflation, Universality\\ and Attractors}\\
\endgroup

\vspace{1.8cm}

\begingroup
    \fontsize{13pt}{16pt}\selectfont
    \textbf{PhD thesis}\\
\endgroup

\vfill

\begingroup
    \fontsize{11pt}{13pt}\selectfont
 to obtain the degree of PhD at the\\
	University of Groningen\\
	on the authority of the\\	
	Rector Magnificus Prof. E.~Sterken\\
	and in accordance with\\
	the decision by the College of Deans.\\
	\bigskip
	This thesis will be defended in public on\\
	\bigskip
	Monday 13 June 2016 at 9:00 hours\\%
\endgroup

\vfill

\begingroup
    \fontsize{11pt}{12pt}\selectfont
    by
\endgroup
    
\vfill
\begingroup
    \fontsize{13pt}{17pt}\selectfont
    \textbf{Marco Scalisi}\\
\endgroup	
\vfill

\begingroup
    \fontsize{11pt}{15pt}\selectfont
    \vspace*{-0.4cm} born on 28 April 1986\\
    in Siracusa, Italy 
\endgroup
\end{center}

\clearpage
\thispagestyle{empty}

\begingroup
    \noindent 
    \fontsize{12pt}{15pt}\selectfont
    \textbf{Supervisors}
    
     \noindent Prof. D. Roest
     
    \noindent Prof. E. A. Bergshoeff

    \vspace{.7cm}
     \noindent \textbf{Assessment committee}

    \noindent Prof. A. Ach\'ucarro
    
    \noindent  Prof. A. Hebecker
        
    \noindent  Prof. R. van de Weijgaert

\endgroup

%


\clearpage
\thispagestyle{empty}

\vspace*{2.5cm}
\begin{flushright}
{\it to Anika}
\end{flushright}

\newpage\null\thispagestyle{empty}\newpage

\clearpage
\pagenumbering{Roman}
\tableofcontents
\mychapterend

\pagenumbering{arabic}
\mainmatter


\renewcommand{\chapterheadstartvskip}{\vspace*{-3.5\baselineskip}}
\chapter{Introduction}

\begingroup
\begin{flushright}
\vspace{.5cm}
\end{flushright}
\begin{quote}
{\it We present a non-technical overview of this thesis. After providing a general background and motivations for this work, we review the main subject and discuss the challenges one would like to face. Finally, we provide an outline of the following chapters.}
\end{quote}

\endgroup

\newpage

\section{Scientific background}

Nature is beautifully various. This is one of the most evident features one may notice of the objects and phenomena surrounding us. Specifically, the scales involved in the natural processes are very diverse. The size of an atom is around $10^{36}$ times smaller than the typical diameter of a galaxy. The energy stored in a usual alkaline AA battery is around 20 orders of magnitudes smaller than the total energy released by the Sun in one second. 

An interesting aspect of this diversity is that, in order to extract information on a specific physical system, we can usually focus just on the relevant scale of that process. An astrophysicist, studying the motion of a planet, does not need to know about the quantum effects governing the behavior of the single atoms composing the body. Similarly, if we want to study the properties of molecules we will not take into account the gravitational effects of those, because just too small and, hence, negligible.

Technically speaking, one uses an effective theory that must be valid up to a critical energy scale. This way to proceed has been always used in a huge variety of contexts (e.g. in hydrodynamics or thermodynamics, one does not resolve the behavior of the single atoms). However, just in the 70’s, the Nobel laureate Wilson formalized such a method \cite{Wilson:1971bg,Wilson:1971dh}. The high-energy or UV degrees of freedom can affect the effective low-energy theory but they are usually suppressed by inverse powers of the typical UV scale. The effective theory can be seen as an expansion in $1/M$, where $M$ is the typical scale of the UV theory. If we probe energies $E$ smaller than $M$, we are not able to resolve the UV degrees of freedom since they are usually suppressed by powers of $E/M$. In the limit $E\ll M$, one can usually forget of the physics happening at energy scales close to $M$.

The history of physics is full of examples, one of the most famous being the electroweak theory of the Standard Model replacing Fermi theory at energies around 100 GeV.

Something similar happens with any quantum field theory coupled to gravity, when the latter is described by General Relativity (GR). In fact, Einstein's theory of gravitation predicts its own breakdown at the {\it Planck scale}, which is equal to
\be\label{PlanckScale}
M_{Pl}\equiv \sqrt{\frac{\hbar c}{8 \pi G}}=2.4\times 10^{18}\,\text{GeV}\,,
\ee
where $c$ is the speed of light, $\hbar$ is the reduced Planck constant and $G$ is the gravitational constant.
This is the energy scale around which a quantum formulation of the gravitational interaction should come into play. A theory of {\it quantum gravity} is indeed needed whenever this force acts on tiny distances and at very high energies. Important examples of such circumstances include the very first moments of the Universe and the interior of black holes.

The last century has seen much effort in the direction of unifying gravity and quantum mechanics in a unique and solid physics framework. The approaches and research lines have been several with rather different perspectives (see  e.g. \cite{Oriti:2009zz,Kiefer:2012boa}).  Nonetheless, they have yielded very interesting and useful insights into the generic properties this ultimate theory should exhibit. The quest for a consistent theory of quantum gravity, and for an observational test of it, has become arguably the greatest theoretical challenge in fundamental physics.

\begin{figure}[htb]
\hspace{-3mm}
\begin{center}
\includegraphics[width=12.cm,keepaspectratio]{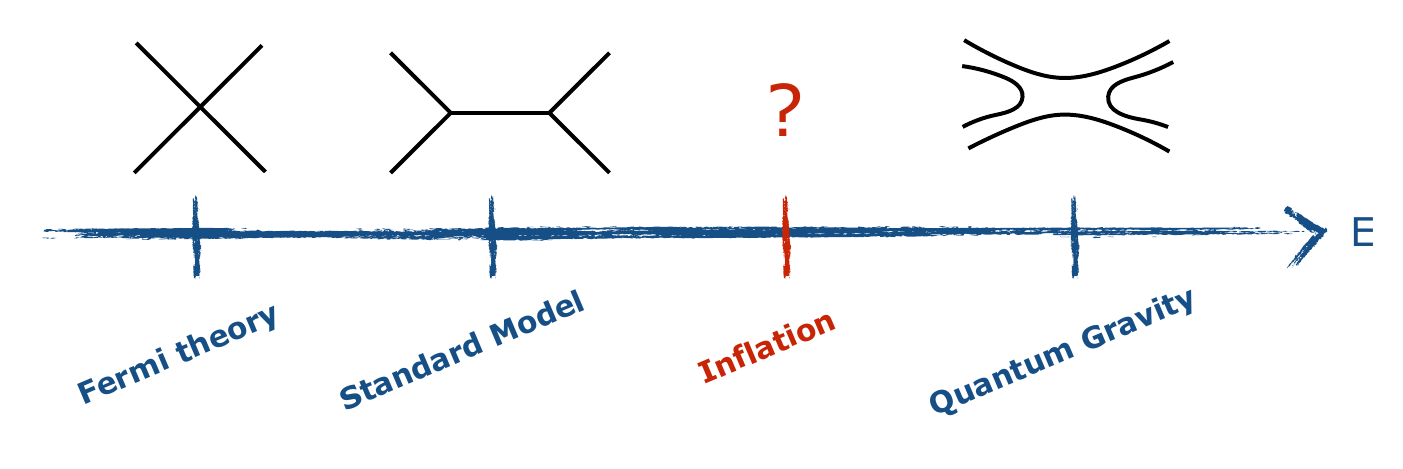}
\vspace*{0.2cm}
\caption{ \it Hierarchy of energy scales with typical Feynman diagrams.\\Inflation happens just some orders below the Planck scale.}\label{scales}
\end{center}
\vspace{-0.2cm}
\end{figure}

One of the most promising candidate scenarios has turned out to be {\it string theory} \cite{Green:1987sp,Green:1987mn,Polchinski:1998rq,Polchinski:1998rr}. This has been put forward to give a proper description of physics at really short scales and a unifying view of Nature. However, despite the advanced level of mathematical tools developed, the accumulated results have struggled to get in touch with the observations. The fundamental reason is because of the extremely high energies involved in typical quantum gravity processes. The Planck scale \eqref{PlanckScale} is indeed around 14 orders of magnitude greater than the highest energies we can currently reach on Earth at the Large Hadron Collider (LHC). Quantum gravity simply occupies the top place in the hierarchy of the energy scales present in Nature, as it is schematically shown in Fig.~\ref{scales}.

Any physical process happening at scales lower than $M_{Pl}$ is usually insensitive to quantum gravity effects, in line with the general effective behaviour described above. Nevertheless, there is an exception to this common rule: this is {\it inflation}, the primordial phase of accelerated expansion which gave the effective start to our cosmic history and which constitutes the main subject of the present thesis. This phenomenon indeed exhibits a peculiar {\it UV-sensitivity} to Planck degrees of freedom although its energy scale is expected to be some orders below. This represents one of the main motivations for this work.  We will explain some of the most basic aspects here in the following.

\section{Main subject and challenges}

In the 80’s, Guth, Linde and others \cite{Guth:1980zm,Linde:1981mu,Albrecht:1982wi} revolutionized the field of cosmology when they first proposed the inflationary paradigm as solution to longstanding problems in the standard scenario. Through inflation, the Universe naturally becomes flat, homogeneous and isotropic, just as we observe today in the distribution of radiation and matter in the sky. This has been measured and analysed with unprecedented accuracy by experiments such as Planck \cite{Adam:2015rua,Planck:2015xua,Ade:2015lrj} and the Sloan Digital Sky Survey (SDSS) \cite{Ahn:2013gms,Betoule:2014frx}. Strikingly, the observed pattern still preserves the imprints of those initial quantum perturbations set by inflation. These are nothing but the seeds which ultimately evolve into the stars and all the cosmic structures we see around us.

The inflationary phase acts as a natural amplifier connecting the world of the very small, determined by the laws of quantum mechanics, with that of the very large, obeying Einstein’s theory of gravity. It provides the sole example in physics where genuine quantum effects become that large to be visible to the naked eye. There are indeed two direct consequences of inflation:
\begin{itemize}
\item First of all, it induces perturbations on the primordial plasma of photons and baryons. The associated temperature thus presents characteristic fluctuations over the average value. The famous picture of the temperature anisotropy in the {\it cosmic microwave background} (CMB) radiation is simply the clearest evidence of this remarkable stretching of quantum effects.

\item Secondly, it produces ripples in the space-time fabric. These so-called {\it primordial gravitational waves} currently constitute the research target of several operating and planned experiments (e.g. KECK \cite{Keck}, BICEP3 \cite{Bicep3} and POLARBEAR \cite{Polar}). After hearing the echoes of two colliding black holes thanks to the LIGO collaboration \cite{Abbott:2016blz}, looking for the echoes of the big bang seems indeed to be the natural next step.

\end{itemize}

The study of the primordial Universe thus provides a unique window onto the smallest scales of Nature. Today, thanks to remarkable developments on the experimental side, we can basically probe scales orders of magnitudes above the limits of the currently operating particle colliders.  The interplay between theory and observation has become more concrete than ever.



Inflation definitively offers the best chance we have to probe regions where quantum gravity effects become manifest. The underlying reason is the following: even though the typical energy should be some orders below the Planck scale (we will explain this in detail in Ch.~\ref{chapter:Inflation}), the duration of the inflationary expansion and the way this accelerates strongly depend on gravity’s quantum mechanical features. Therefore, Planck degrees of freedom easily enter the low effective action of inflation thus generically spoiling its dynamics. In the parametrization of inflation by means of a scalar field $\phi$, this situation becomes even worse when its range exceeds Planckian values, that is, $\Delta\phi >M_{Pl}$ (we will analyse the properties of this important variable in Ch.~\ref{chapter:Universality}). This so-called UV-sensitivity thus not merely allows but requires investigating inflation within a quantum gravity framework.

String theory seems to offer the proper machinery to have best control over these UV-interactions. The grand challenge is then constructing a proper embedding of inflation into this complete framework of physics (see \cite{Baumann:2014nda} for a recent review on this topic). Achieving this would shed light on the microscopic mechanism of this primordial phase as well as providing a very important test for string theory. Nevertheless, the route towards this goal has turned out to be dotted with many obstacles and we still lack a solid theoretical underpinning for this phenomenon.

Focusing on effective limits of string theory and extracting universal properties has proved to be very successful in terms of investigation and comparison with the experimental data. Studying the generic mechanism beyond the particular details (e.g. number of fields involved, mechanism to end inflation, value of the current acceleration and others) of a large class of models becomes a powerful tool in order to better define the path towards this final understanding. This is the approach we will be taking in the rest of the thesis.

\section{Outline of the thesis}

The present work intends to build on and go beyond the ideas presented above. It aims to explore generic features of inflation which appears intimately related to fundamental aspects of quantum gravity. The outline of the thesis is as follows:

\begin{itemize}
\item Chapters \ref{chapter:standard} and \ref{chapter:Inflation} contain basic material which will set the stage for the following research investigation. This includes the standard cosmological scenario, its shortcomings and the inflationary paradigm as a solution to these. In addition, we present the latest constrains provided by Planck and discuss their implications on the dynamics of inflation. Although these topics are usually covered in most of the available cosmology books and lecture notes (see e.g. \cite{Dodelson:2003ft,Mukhanov:2005sc,Lyth:2009zz,Baumann,Calcagni} ), we present our own perspective and set the notation being used afterwards.

\item Chapter \ref{chapter:Universality} is devoted to discussing the small sensitivity CMB observations have to the entire period of the inflationary expansion. Nevertheless, we  show that it is possible to extract {\it universality properties} of the fundamental dynamics and still yield very stringent bounds on the observational predictions. Specifically, we will be interested in analysing the behaviour of the inflationary field range $\Delta\phi$, whose properties are of utmost importance for the consistency of an effective description of inflation.

\item In Chapter \ref{chapter:supergravity}, we discuss the embedding of the inflationary scenario into supergravity, seen as an effective limit of string theory. We highlight how very general properties regarding the geometry of the internal moduli space and the directions of supersymmetry breaking can highly constrain the physics of inflation. Remarkably, non-trivial hyperbolic \Kahler geometries, naturally arising in string theory, lead to the concept of {\it attractors}: the specific details of the model get washed out and the observational predictions all converge towards a single value.

\item Finally, in Chapter \ref{chapter:Landscape}, we discuss the implications of building a unified framework for inflation and dark energy, in the context of supergravity. The inclusion of a nilpotent sector will considerably simplify the underlying construction thus yielding to remarkable flexibility in terms of the cosmological predictions. The case of hyperbolic \Kahler geometry is  again special as one can show that unifying early- and late-time acceleration basically enhances the attractor property of the system.

\end{itemize}

At the beginning of every chapter, we provide an abstract which gives an overview of the content and of the main results. These are based on the works [{\sc iii}]-[{\sc x}] of the \hyperref[chapter:Publications]{List of Publications}. Throughout the following, we will work in reduced Planck mass units and then set $M_{Pl}=1$.

\renewcommand{\chapterheadstartvskip}{\vspace*{-3.5\baselineskip}}

\chapter{Standard Cosmology}
\label{chapter:standard}
\begingroup
\begin{flushright}
\vspace{.5cm}
\end{flushright}
\quote{\it In this chapter, we introduce the reader to the basic ingredients of the standard cosmological model. This perfectly describes the evolution of our Universe from few moments after the so-called ``Big Bang'' up to now. We start the discussion with Hubble's discovery of the cosmic expansion which gave the start to the development of this standard scenario. A proper embedding of these ideas within the language of general relativity is certainly needed. Then, we present the fundamental tools of relativistic cosmology, specify the geometry of our Universe, its matter-energy content and the cosmic dynamics encoded in the Friedmann equations. Finally, we discuss the main shortcomings which represent a serious threat to the physics of this model.}

\endgroup

\newpage

\section{The expanding Universe}

Cosmology is a very old subject. Throughout millennia, humans have looked up to the sky and tried to decipher our Universe. The cosmic picture has changed several times. However, something radical happened in the second quarter of the $20^{\text{th}}$ century. A simple astronomical observation has revolutionized the way we understand the Universe as a whole, and has given rise to the subsequent establishment of cosmology as a science.

\begin{figure}[htb]
\hspace{-3mm}
\begin{center}
\includegraphics[width=11cm,keepaspectratio]{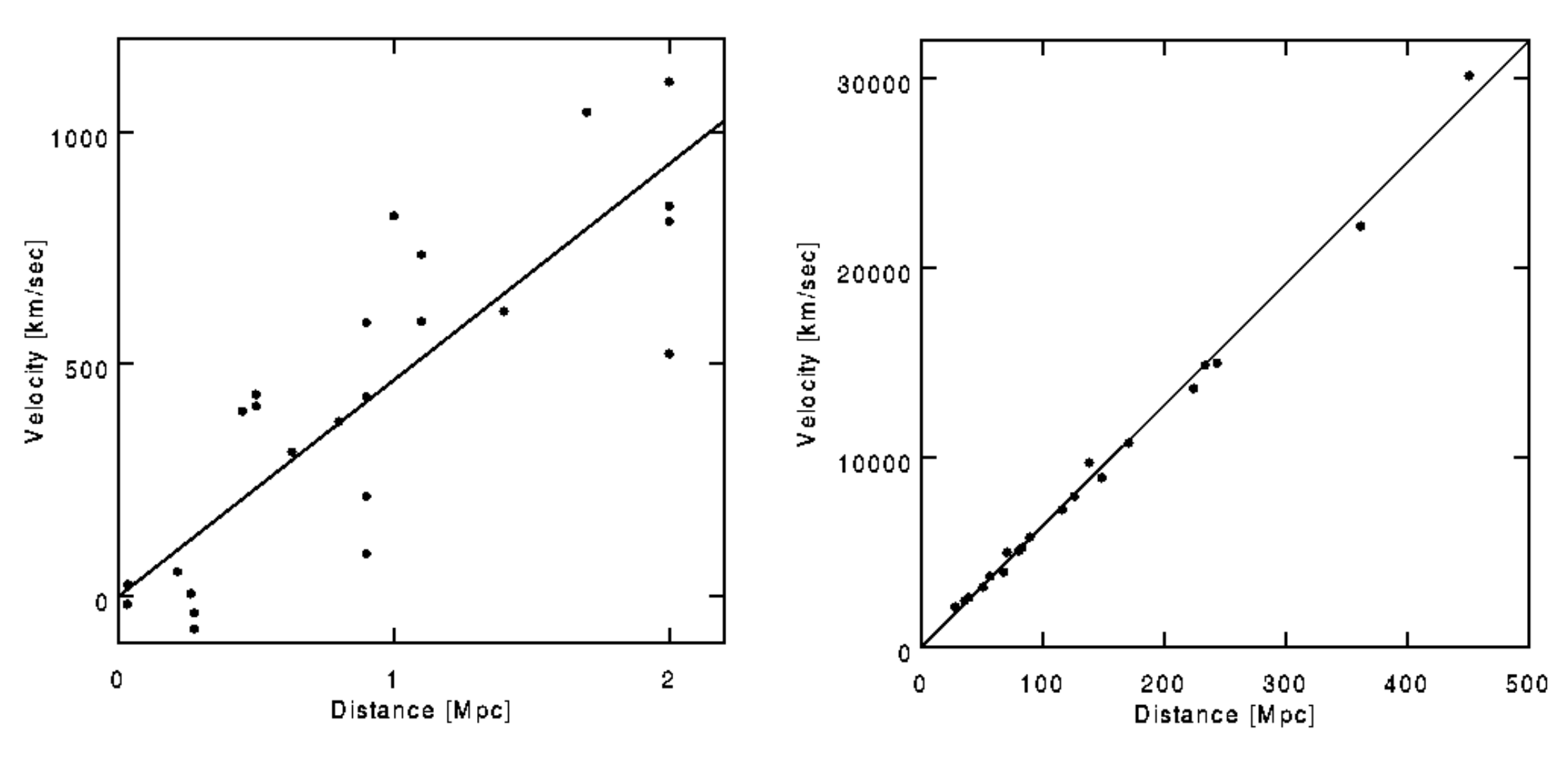}
\caption{\it Hubble diagrams (as replotted in \cite{Wright}) showing the relationship between recession velocities and distances of several astronomical objects. The left panel presents Hubble's original measurements \cite{Hubble:1929ig}. The right panel shows more recent data of very distant objects \cite{Riess:1994nx,Riess:1996pa}.}\label{Hubblediag}
\end{center}
\vspace{-0.5cm}
\end{figure}

In 1929 the astronomer Edwin Hubble observed the recession of different galaxies from us \cite{Hubble:1929ig} with velocity equal to
\be\label{Hubblelaw}
v=H(t) d\,,
\ee
with $d$ being the physical distance from our observational point and $H$ a proportionality parameter. The formula \eqref{Hubblelaw} is often refereed to as ``Hubble law'' and it has been tested along the years with increasing accuracy, as shown in Fig.~\ref{Hubblediag}.

The {\it Hubble parameter} $H$ is assumed to be dependent  just on time. The independence on the space coordinates simply reflects the very old idea that we do not occupy any special place in the Universe (this is the so-called {\it cosmological principle}); any observer would see the astronomical bodies moving away with velocity equal to \eqref{Hubblelaw}.

This empirical observation was almost immediately interpreted as first evidence that we live in a dynamical Universe whose physical size is growing with time. Every object would move away from each other as the effect of the expansion of the space itself. The latter is encoded in the evolution of so-called {\it scale factor} $a(t)$ such that the physical distance can be rewritten as
\be
d=a(t) x\,,
\ee
with $x$ being the {\it comoving distance} and decoupled from the effect of the expansion. 

\begin{figure}[htb]
\hspace{-3mm}
\begin{center}
\includegraphics[width=12cm,keepaspectratio]{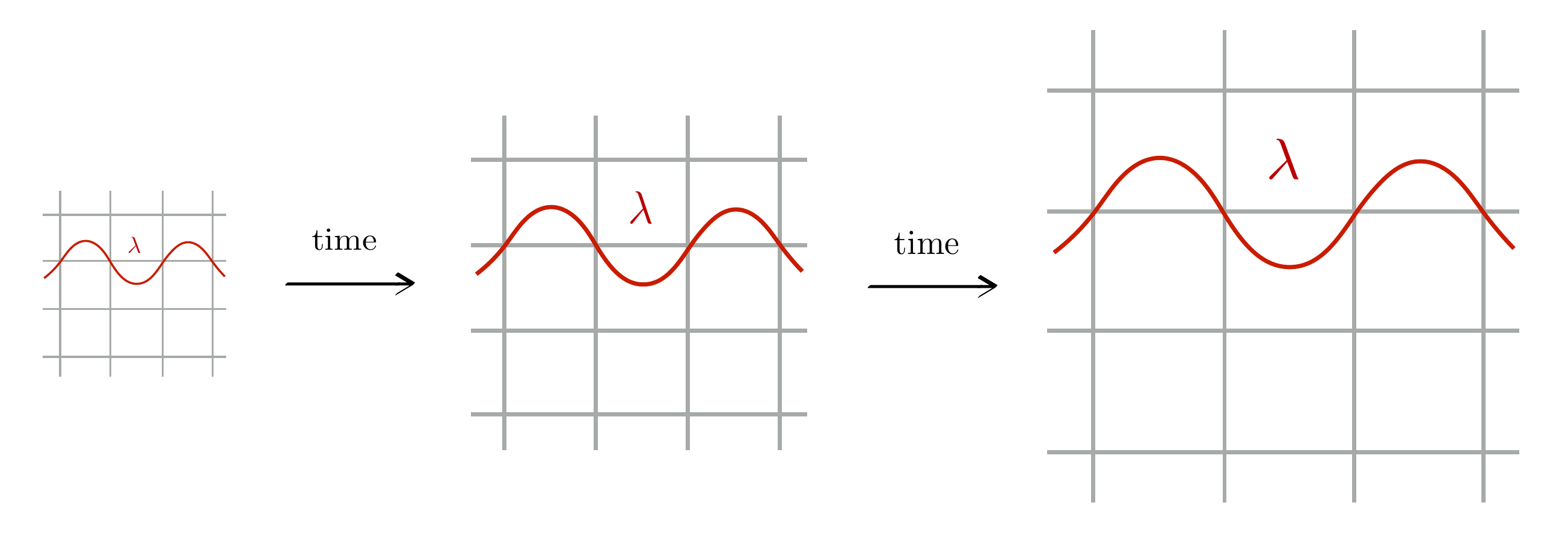}
\caption{\it The expanding Universe with a typical scale $\lambda$. The grid schematically represents comoving coordinates which do not change with time. Physical distances increase proportionally to the scale factor $a(t)$.}\label{FRWgrid}
\end{center}
\vspace{-0.5cm}
\end{figure}

In order to better visualize the situation, we could imagine the spacetime texture as an expanding grid where objects are fixed on (i.e.~at constant comoving coordinates) and still recede from each other as the effect of a growing scale factor. In a realistic description of the Universe, we need to assume also a non-zero peculiar velocity (corresponding to $\dot{x}\neq 0$, with the overdot denoting a time derivative); local gravitational inhomogeneities would indeed contrast the global expansion and, then, induce such effects. However, at very large distances, this can be neglected as the biggest contribution comes from the recessional component.  In an expanding Universe, typical scales, e.g. the wavelength $\lambda$ of a photon, will increase as $\lambda \propto a$ as the expansion proceeds. However, the comoving wavelenght $\lambda/a$ will remain constant in time, if no other external process occurs (see Fig.~\ref{FRWgrid}). 

The expansion rate of the Universe is thus described by the Hubble parameter, defined as
\be\label{HubbleDEF}
H(t)\equiv \frac{\dot{a}(t)}{a(t)}\,,
\ee
generically dependent on the particular cosmic era. Its numerical value can be simply obtained by measuring the ratio between the recessional velocities and the distances of astronomical objects such as Cepheid variables or Type IA supernovae (these provide a reliable estimate of astronomical distances). Its current value is given by 
\be \label{currentH}
H_0= 100 h\ \text{km sec$^{-1}$ Mpc$^{-1}$}
\ee
where
\be \label{currenth}
h = 0.678 \pm 0.009\ (68\% \text{CL})\,,
\ee
as given by Planck2015 \cite{Planck:2015xua}, combining information on the temperature spectrum, on the polarization spectra at low multipoles and on lensing reconstruction.

The value of the Hubble parameter can give us a quite good estimate of the age of our Universe. Specifically, in a very simplified model,  we can extract a rough value of the time passed since everything was concentrated on a point (the so-called ``Big Bang''). If we indeed assume a constant recession velocity $v$ and neglect the effect of gravity, we obtain that points, separated by a distance $d$ today, were in contact at the time $t\simeq d/v=1/H$. Plugging the current measurements \eqref{currentH} and \eqref{currenth} in, we obtain $t_0\simeq14.4$ Gyr. This value is not far from the more precise estimate given by Planck2015
\be
t_0 =13.799\pm 0.038\ \text{Gyr}\ (68\% \text{CL})\,,
\ee
which takes the different matter-energy contributions at any cosmic era into account.

\begin{figure}[htb]
\hspace{-3mm}
\begin{center}
\includegraphics[width=7cm,keepaspectratio]{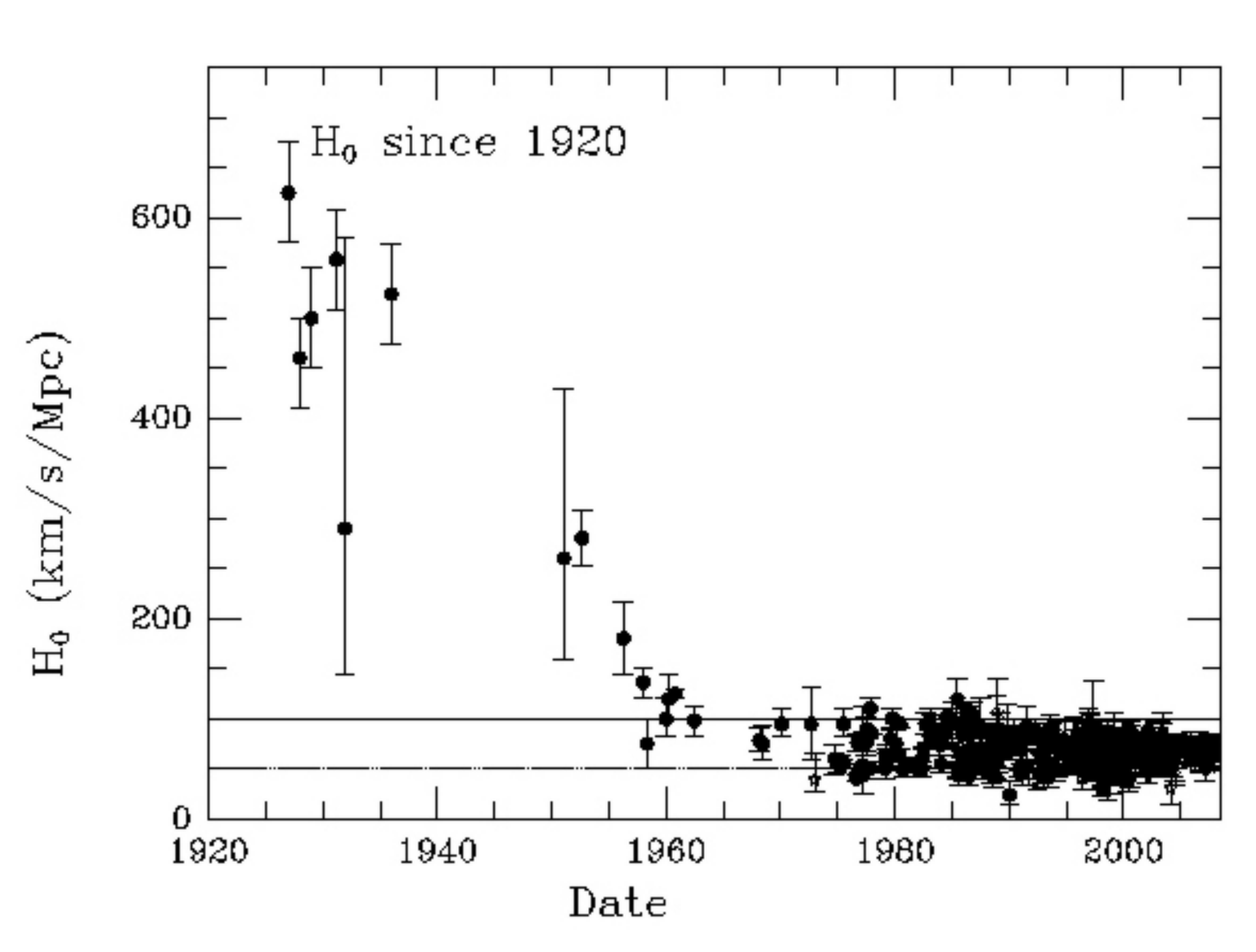}
\vspace*{0.3cm}
\caption{\it Evolution of the estimates of the current Hubble parameter $H_0$ during the last century.}\label{evolutionHubble}
\end{center}
\vspace{-0.5cm}
\end{figure}

It is interesting to notice how the value of the current Hubble parameter $H_0$ has changed during the last century (Fig.~\ref{evolutionHubble} shows this evolution) due to the increasing accuracy of our measurements. Curiously, the first estimates given by Hubble were leading to a very low estimate for the age of the Universe in contradiction  with other astronomical evidences, such as the age of the Earth inferred from radioactive decay. This embarrassing conflict contributed to initially discard the interpretation of Hubble's observation as evidence for the cosmic expansion\cite{Kirshner}.

Hubble's discovery and the consequent idea of the expanding Universe led to the development of the standard model of Big Bang cosmology, whose predictions are in excellent agreement with observations. Despite the name, the model says nothing about the Big Bang which remains a mathematical singularity as well as an unsolved physical question. On the other hand, it furnishes a clear and precise picture of the cosmic evolution from a few seconds after this mysterious start: the temperature decreases as the expansion of the Universe proceeds, light elements form during a process called Big Bang Nucleosynthesis (BBN), recombination of nuclei and electrons takes place followed by the last scattering of photons which freely reach us today as cosmic microwave background (CMB) radiation, observed in the sky at the temperature $T=2.73 \ K$.

Although the model has had many successful experimental confirmations, it contains some serious theoretical shortcomings which can be better understood once we know the geometric and dynamical properties of the Universe we  live in.

\section{Relativistic cosmology}

Prior to Hubble's discovery, Einstein had already noticed that a genuine prediction of his newly born theory of general relativity  was a non-static Universe. Puzzled by its cosmological implications, in 1917 he decided to augment his equations with a specific cosmological constant \cite{Einstein:1917ce} in order to avoid such a phenomenon. However, Hubble's observation and the consequent interpretation confirmed that we do live in a non-static Universe.

A dynamical Universe is indeed what comes naturally from Einstein theory of gravity which relates the geometry of spacetime to its matter-energy content, through the field equations 
\be\label{Einstein}
G_{\mu \nu} = T_{\mu \nu}\,.
\ee
The Einstein tensor is defined as
\begin{align}
G_{\mu \nu} \equiv R_{\mu \nu} -\frac{1}{2} g_{\mu \nu} R\,,
\end{align}
where $ R_{\mu \nu} $ and $R$ are respectively the Ricci tensor and the Ricci scalar depending on the metric $ g_{\mu \nu}$ and its derivatives. The energy-momentum tensor is defined as
\begin{align}
T_{\mu \nu}\equiv g_{\mu \nu} \mathcal{L}_m - 2 \frac{\partial\mathcal{L}_m}{\partial g^{\mu \nu}}\,,
\end{align}
where $\mathcal{L}_m$ is the matter Lagrangian. 

In the following, we discuss the implication of assuming some natural symmetries dictated by observations both for the metric and for the matter source.
 
\subsection{Homogeneity and isotropy}

\begin{figure}[t]
\vspace*{-1mm}
\begin{center}
\includegraphics[width=8cm,keepaspectratio]{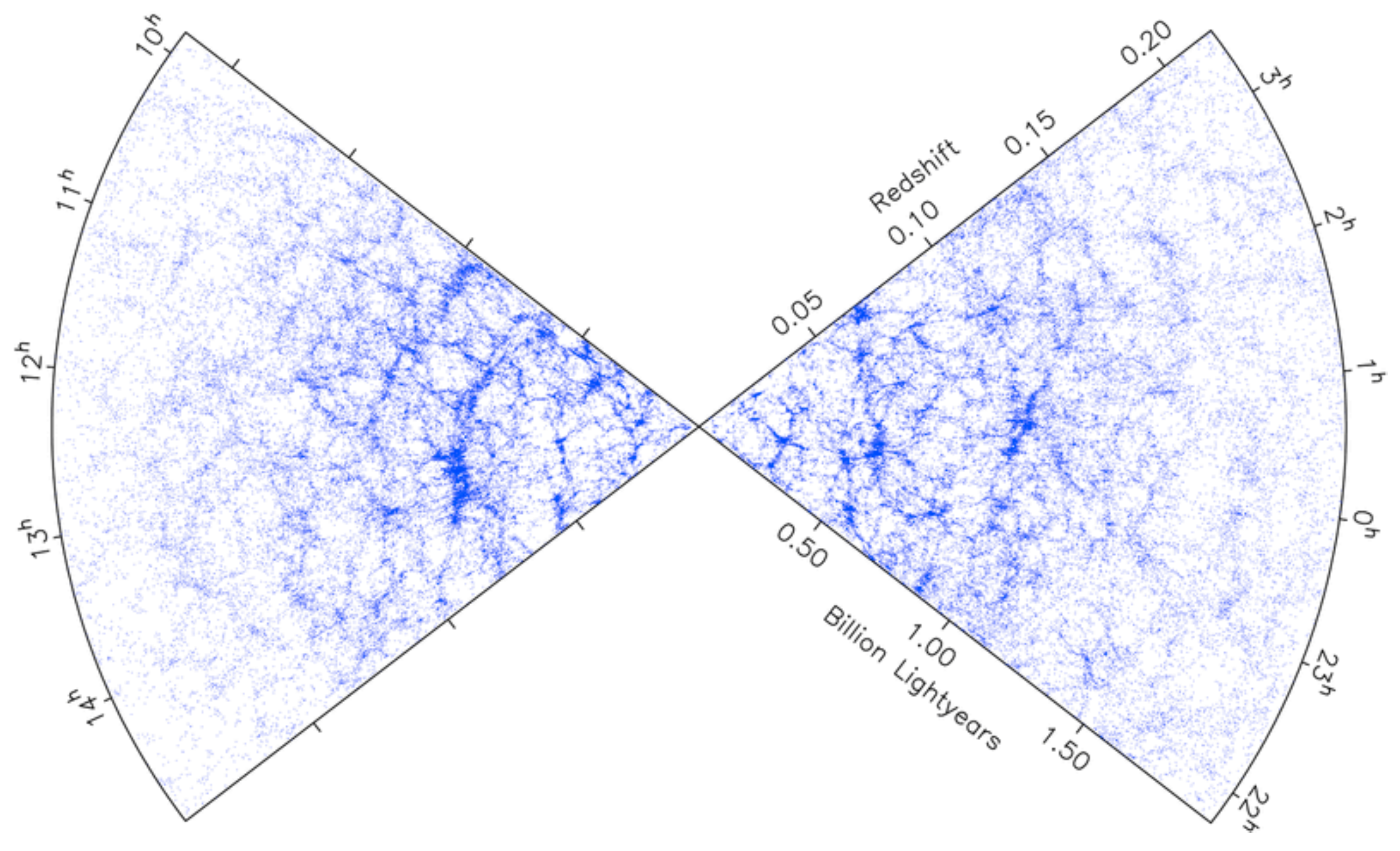}
\caption{\it 2-degree Field Galaxy Redshift Survey (2dFGRS) map \cite{Cole:2005sx}.\\The distribution of the cosmic structures is quite irregular at small distances but becomes more uniform towards the past and, then, at large scales.}\label{LSS}
\end{center}
\end{figure}

The homogeneity and isotropy of our Universe at large scales ($>100$ Mpc) has always been taken as a reasonable assumption, known as the cosmological principle. This has been mainly driven by the simple idea that we do not occupy any special place in the cosmos. In addition, it has been beautifully confirmed by several recent cosmological observations, specifically the ones concerning the distribution of galaxies at large scales, shown in Fig.~\ref{LSS}, and the all-sky map of the CMB radiation. Pushing our observations to large distances simply means looking back in time. Then, the basic picture is the one of a very isotropic and homogeneous initial state which, during its evolution, has developed the cosmic structures we observe today in the sky as the result of gravitational instabilities.

Assuming these fundamental symmetries imposes stringent constraints on the form of both sides of Eq.~\eqref{Einstein} as we will see below. 

\subsubsection{FLRW metric}

Homogeneity and isotropy highly restricts the possibilities of the spatial metric. The most general class is represented by the Friedmann-Lema\^{i}tre-Robertson-Walker (FLRW) solution which, written in terms of polar spherical coordinates $(r,\theta, \sigma)$, reads
\be\label{FRW}
ds^2=-dt^2 + a(t)^2 \left[\frac{dr^2}{1- \kappa r^2} + r^2 (d\theta ^2+ \sin^2 \theta\ d\sigma^2) \right] \,.
\ee
The scale factor $a(t)$ sets the physical distances among objects and can vary with respect to the cosmic time $t$ (the proper time as measured by a comoving observer at constant spatial coordinates) allowing, then, for an expanding Universe. The coordinates $(r,\theta, \sigma)$ reflect the symmetries assumed and are called comoving, as they are decoupled from the effect of the expansion.

The assumed symmetries still allow for a constant curvature of the \mbox{3-dimensional} spatial slices which can correspond to an open, flat or closed Universe, respectively parametrized by $\kappa=-1,0,1$. These hypersurfaces are the natural spatial slices at any constant time. The homogeneous evolution allows us indeed to use a universal clock at each point. However, being the time coordinate not a physical time, one may adopt other parametrizations. A very useful one is characterized by defining  the {\it conformal time} $\tau$, such as
\be\label{conformaltime}
d \tau \equiv \frac{d t}{a(t)}\,,
\ee
which brings Eq.~\eqref{FRW} into 
\be\label{FRWconformal}
ds^2=a(\tau)^2\left[-d\tau^2 + \frac{dr^2}{1- \kappa r^2} + r^2 (d\theta ^2+ \sin^2 \theta\ d\sigma^2) \right] \,.
\ee
It is possible to further simplify the metric by introducing a new radial coordinate $\chi$, defined as
\be
d \chi \equiv \frac{d r}{\sqrt{1- \kappa r^2}}\,,
\ee
through which we may rewrite Eq.~\eqref{FRWconformal} as
\be\label{FRWconformalnew}
ds^2=a(\tau)^2\left[-d\tau^2 + d\chi^2+ S^2_{\kappa}(\chi) (d\theta ^2+ \sin^2 \theta\ d\sigma^2) \right] \,,
\ee
where 
\be
S_{\kappa}(\chi) \equiv \begin{cases}
    \sinh\chi,& \kappa=-1\\
    \chi,              & \kappa=0\\
     \sin\chi & \kappa=1\\
\end{cases}
\ee

This particular gauge choice simplifies the description of the causal structure of the FLRW metric: the propagation of light is the same as in Minkowski space and takes place diagonally (at 45 degrees) in the $(\chi,\tau)$ plane.

\subsubsection{Perfect fluid}

It is possible to show that the stress-energy tensor $T_{\mu \nu}$, compatible with such homogeneity and isotropy, is the one of a perfect fluid (this is well explained in \cite{Baumann}), that is
\be
{T^\mu{}_\nu}= \text{diag}(-\rho, p , p, p)\,,
\ee
where $\rho$ is the energy density and $p$ the pressure as measured in the rest frame of the fluid.

\subsection{Friedmann equations and the cosmic history}

After specifying both sides of Eq.~\eqref{Einstein}, we can extract the cosmic dynamics of a FLRW universe. Due to the symmetries assumed, the independent equations turn out to be two, which are known as {\it Friedmann equations} and read
\begin{align}\label{Fried}
 H^2&=\left(\frac{\dot{a}}{a}\right)^2=\frac{\rho}{3}-\frac{\kappa}{a^2} \,,\\
\label{Fried2}H^2 + \dot{H}&= \frac{\ddot{a}}{a}=-\frac{1}{6}(\rho+3p)\,,
\end{align}
where dots denote derivatives with respect to the time $t$ and $H$ is defined by Eq.~\eqref{HubbleDEF}.

In order to extract the evolution of the scale factor $a(t)$, one must specify the type of matter and solve Eq.~\eqref{Fried} and  Eq.~\eqref{Fried2}. In fact, these two equations can be combined into the continuity equation
\be\label{continuity}
\dot{\rho} + 3H(\rho + p) =0\,,
\ee
which, alternatively, can be also derived from the condition of energy conservation $\nabla_\mu T^{\mu\nu}=0$. Depending on the relation between energy density and pressure, dictated by the equation of state parameter
\be
p= w \rho\,,
\ee
we obtain the following scaling for the energy density
\be \label{endensity}
\rho\propto a^{-3(1+w)}\,,
\ee
which, plugged back into Eq.~\eqref{Fried}, yields 
\be \label{scalefactor}
a(t) \propto \begin{cases}
    t^{\frac{2}{3(1+w)}},& w\neq-1\\
    \text{e}^{Ht},              & w=-1
\end{cases}
\ee
in the case of flat curvature ($\kappa=0$). The parameter $w$ can be assumed to be constant and depends on the specific species filling the Universe at any epoch:
\begin{itemize}
\item {\it Matter}, or any pressureless species where the kinetic energy is negligible with respect to the mass (that is {\it non-relativistic matter}, e.g. baryons or dark matter), is characterized by $w=0$. One has $\rho\propto a^{-3}$ and a Universe dominated by matter will have a scaling $a \propto t^{2/3}$. This fact is quite intuitive as this type of matter is the one we are most familiar with. The scaling can be easily understood by the following argument. Imagine a cubic portion of the expanding Universe of volume $V(t)=L^3=[a(t) X]^3$ where $L$ and $X$ are respectively the physical and the comoving length of one side of the cube. When this is filled with ordinary matter, the energy will be basically determined by the mass thus having no scaling with respect to the physical volume. The energy density being equal to $\rho=E/V$ will have the expected scaling with respect to the scale factor $a(t)$. This situation is shown in Fig.~\ref{matter}. 

\begin{minipage}{\linewidth}
\begin{figure}[H]
\vspace*{-1mm}
\begin{center}
\includegraphics[width=8.5cm,keepaspectratio]{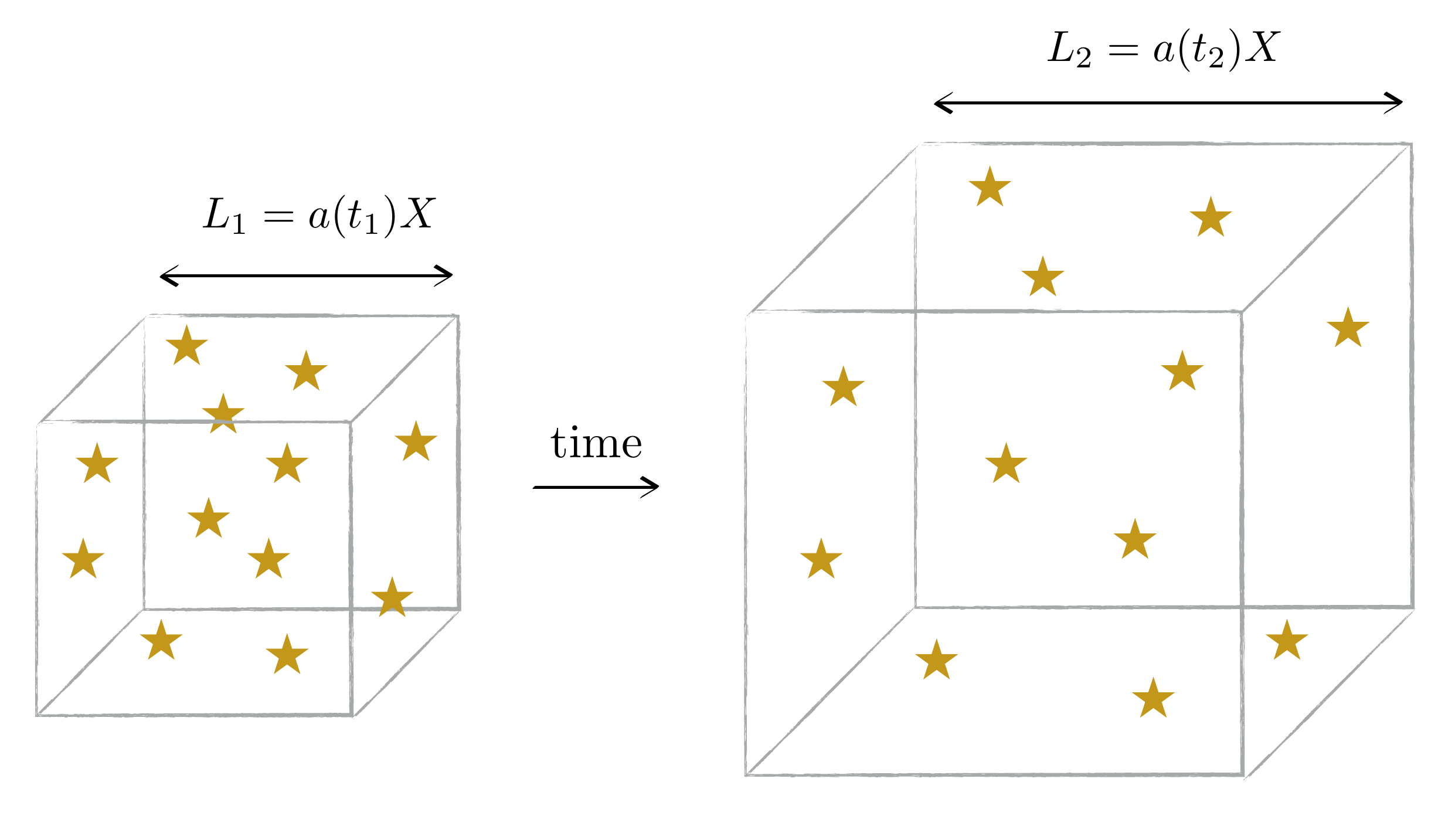}
\vspace*{.3cm}
\caption{\it Non-relativistic matter in a cubic portion of the expanding Universe. While the length of the side increases with time ($L_1>L_2$),  the matter content gets diluted and the energy density scale as $\rho\propto a^{-3}$.}\label{matter}
\end{center}
\end{figure}
\end{minipage}

\item {\it Radiation}, or any species with dominating kinetic energy (that is {\it relativistic matter}, e.g. photons or neutrinos), is characterized by $w=1/3$. The energy density scales as $\rho\propto a^{-4}$ which implies that a Universe dominated by such type of matter expands as $a \propto t^{1/2}$. One can understand the scaling of the energy density by means of an analogous picture to the one above. Imagining the cubic portion filled with photons, now the main difference is that the energy is equal to $E=2\pi/\lambda$ ($\hbar=c=1$) thus scaling as the inverse of the wavelength $\lambda$. The latter is a physical length linearly depending on the scale factor (see Fig.~\ref{radiation}) and this provides the quartic power in the scaling of $\rho$.

\begin{minipage}{\linewidth}
\begin{figure}[H]
\vspace*{-1mm}
\begin{center}
\includegraphics[width=8.5cm,keepaspectratio]{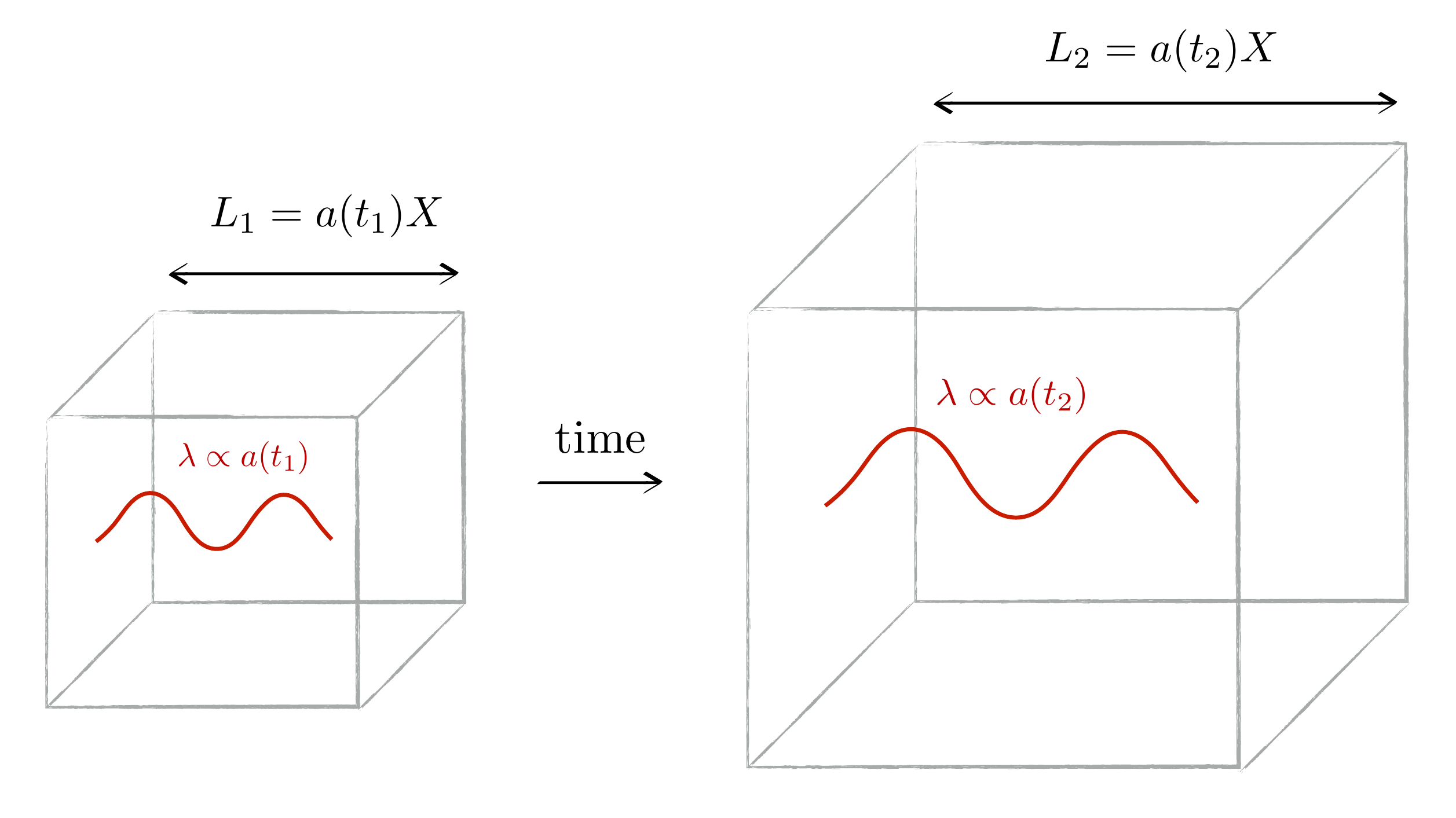}
\vspace*{.4cm}
\caption{\it Radiation in a cubic portion of the expanding Universe.  The length of the side increases with time ($L_1>L_2$) proportionally to the scale factor $a(t)$ as well as the photon wavelength $\lambda$. Consequently the energy density scale as $\rho\propto a^{-4}$.}\label{radiation}
\end{center}
\end{figure}
\end{minipage}

\item {\it Dark energy}, the mysterious component dominating the Universe nowadays, is characterized by $w=-1$ (when described by a cosmological constant) with negative pressure and constant energy density. A Universe dominated by that will expand exponentially as given by Eq.~\eqref{scalefactor}. This situation is quite counterintuitive if we want to draw an analogy with the cases above. If we indeed imagine this component filling a part of our Universe, new dark energy ``atoms'' must necessarily appear out of nothing during the expansion, in order to keep the energy density constant (this is shown in Fig.~\ref{DEscaling}) .

\begin{minipage}{\linewidth}
\begin{figure}[H]
\vspace*{-1mm}
\begin{center}
\includegraphics[width=8.5cm,keepaspectratio]{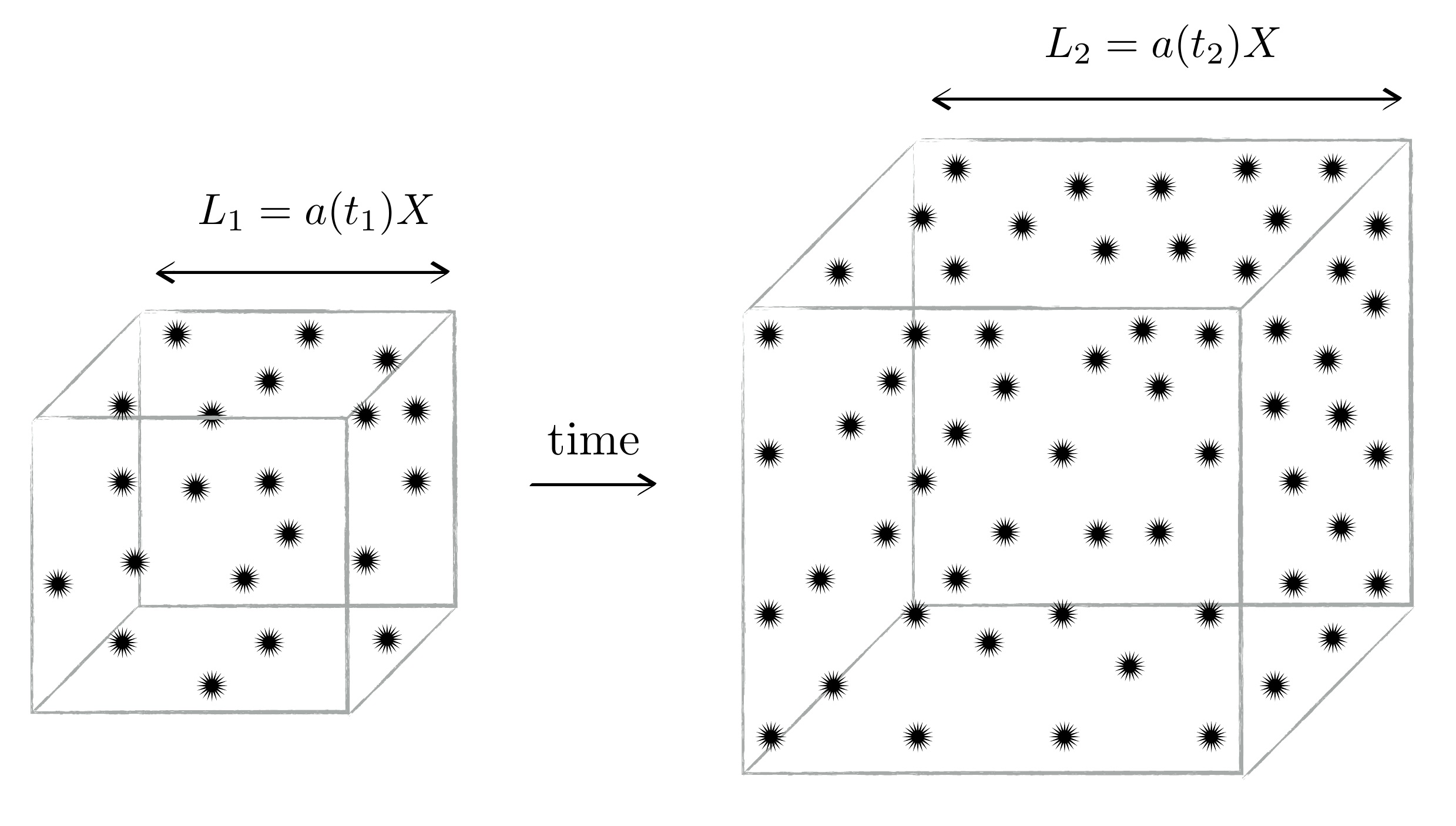}
\vspace*{.4cm}
\caption{\it Dark energy in a cubic portion of the expanding Universe.  While the length of the side increases with time ($L_1>L_2$), new dark energy ``atoms'' appear in order to mantain a constant energy density.}\label{DEscaling}
\end{center}
\end{figure}
\end{minipage}

\end{itemize}

\begin{figure}[htb]
\hspace{-3mm}
\begin{center}
\includegraphics[width=11.5cm,keepaspectratio]{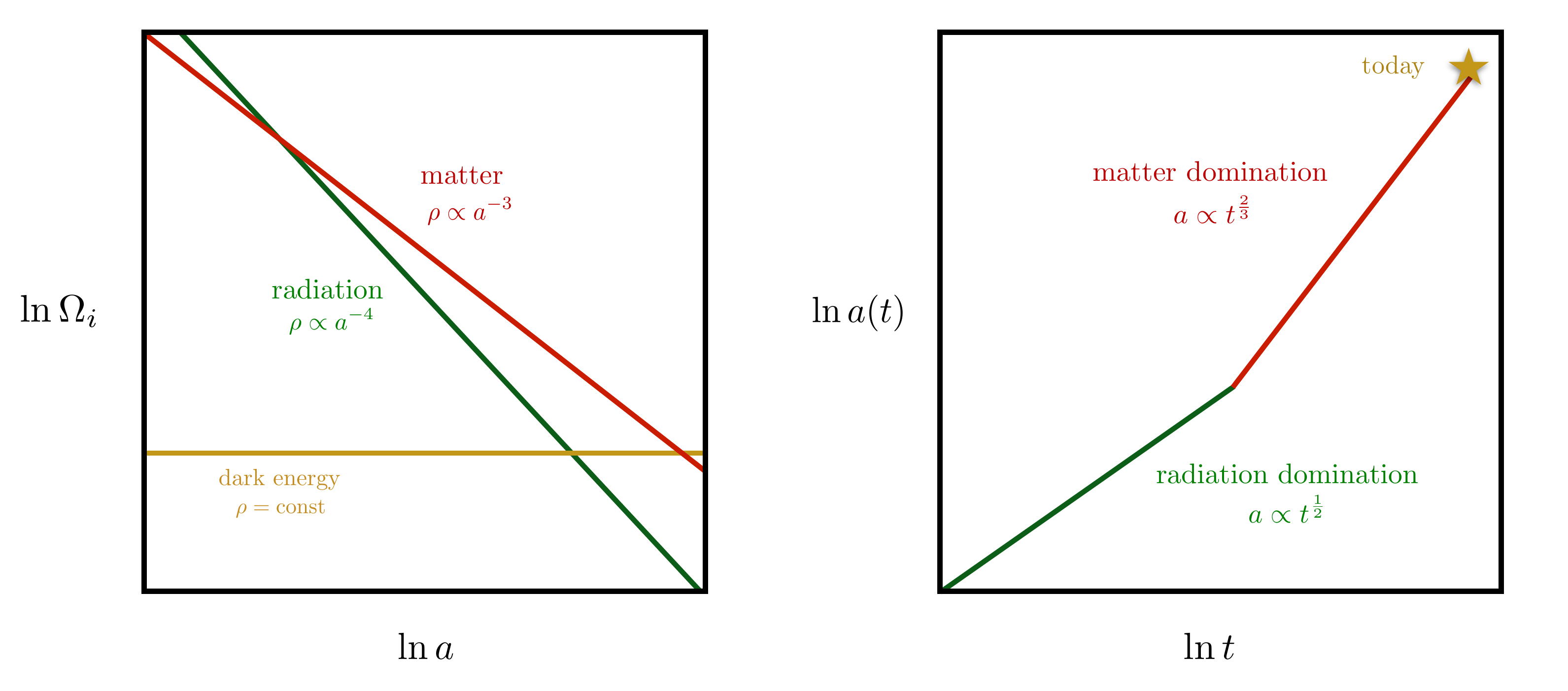}
\caption{\small \it Standard evolution of the energiy densities (left panel) and the scale factor (right panel) in logarithmic scales. According to the standard cosmological model, going back in time, the Universe becomes radiation dominated and the scale factor shrinks up to a singular point $a=0$, commonly called ``Big Bang''.}\label{history}
\end{center}
\vspace{0.cm}
\end{figure}

In standard cosmology, therefore, the history of the Universe is characterized by early times dominated by radiation, a moment of matter-radiation equality and subsequent domination of matter. Just recently we have entered an era in which dark energy constitutes most of the total energy in the Universe, at present $68.3\%$ of the entire content. This evolution is shown in Fig.~\ref{history}.

Finally, one may write the Friedmann equation in a form which is better for the discussion of the shortcomings affecting the standard cosmological model. By looking at Eq.~\eqref{Fried}, one may define, at any time, a critical energy density
\be
\rho_c\equiv 3H^2
\ee
corresponding to a perfect flat sectional curvature $\kappa=0$. After normalizing all energy densities as
\be
\Omega_i \equiv \frac{\rho_i}{\rho_c}\,,
\ee
one can rewrite Eq.~\eqref{Fried} as
\be\label{Fried1NEW}
\Omega\equiv\sum_i \Omega_i = 1 + \frac{\kappa}{(a H)^2}
\,.
\ee
where the index $i$ runs over the different matter-energy species.

\section{Big Bang shortcomings}

According to what we have presented in the previous section, the Universe, for most of its evolution, has been dominated by non-relativistic matter and radiation satisfying the strong energy condition (SEC)
\be
1+3w\geq 0\,.
\ee
Specifically, just after the Big Bang (corresponding to the singularity $a=0$), the standard picture sees the cosmos mostly filled with radiation (see Fig.~\ref{history}). In the following, we will discuss the problematic consequences of assuming these familiar matter sources up to the beginning of the cosmic history.

\subsection{Flatness problem}

\begin{figure}[htb]
\hspace{-3mm}
\begin{center}
\includegraphics[width=8.5cm,keepaspectratio]{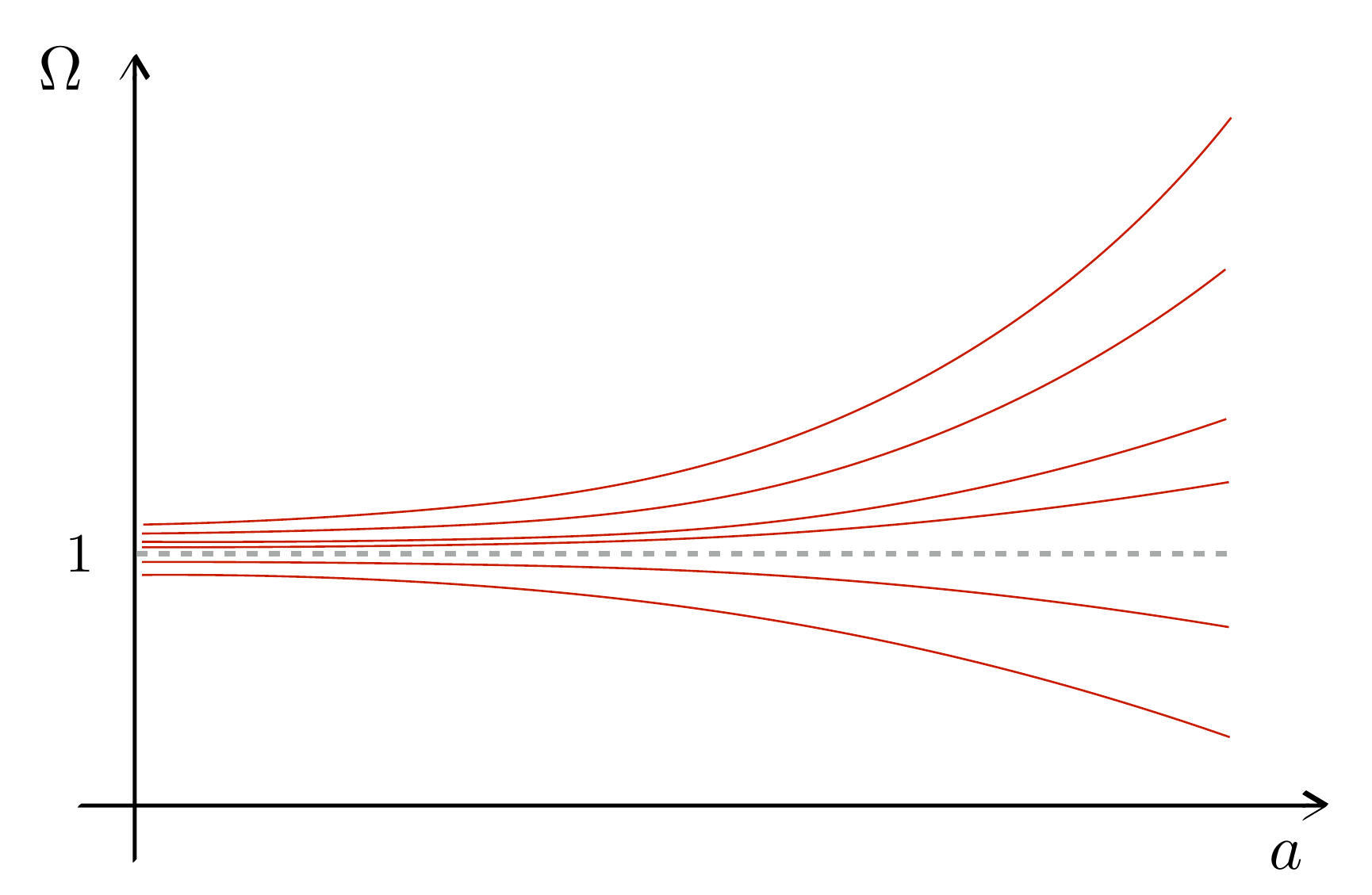}
\caption{\small \it Evolution of the total energy density in standard cosmology. The point $\Omega=1$, corresponding to flat curvature, is a repeller.}\label{unstable}
\end{center}
\vspace{0.cm}
\end{figure}

In standard cosmology, an expanding Universe is naturally driven away from flatness. This can be well understood by differentiating Eq.~\eqref{Fried1NEW}, that is
\be
\dot{\Omega}= H \Omega \left(\Omega -1\right) (1+3w)\,,
\ee
which can be rewritten as
\be
\frac{d |\Omega -1|}{d \ln a}= \Omega |\Omega -1| (1+3w)\,.
\ee
A Universe with a growing scale factor $a(t)$ that is dominated by ordinary matter (subject to the strong energy condition $1+3w\geq 0$ ) therefore has $\Omega=1$ as an unstable fixed point as displayed in Fig.~\ref{unstable}.

This is exactly what happens in the standard cosmological picture where the Universe has been dominated by such type of energy from the beginning until the present time, as shown in Fig.~\ref{history}. A Universe starting with generic initial curvature is driven away from flatness during its evolution. The same conclusion can be reached by looking at Eq.~\eqref{Fried1NEW} and noticing that, in a Universe filled with radiation or matter, the sum of the energy densities $\Omega_i$ diverges from unity as the quantity $(aH)^{-1}$ increases with time.

The surprise comes with cosmological observations that suggest that the Universe today must be flat with extreme accuracy. Specifically, the latest Planck data  combined with BAO give
\be
|\Omega-1| = 0.000 \pm 0.005\ (95\% \text{CL})\,.
\ee

This implies that, going back in time, the curvature of the Universe should have been even closer to perfect flatness: at the BBN epoch $|\Omega -1|\lesssim 10^{-16}$, at the Planck scale $|\Omega -1|\lesssim 10^{-64}$. Generally, such an incredible amount of fine-tuning for the initial conditions of the Universe makes physicists uncomfortable. A dynamical explanation of what we observe today would  be certainly more desirable.

\subsection{Horizon problem}
 
Given a space-time, the scale of causal physics is set by null geodesics, being the paths of photons. In a FLRW Universe, with flat curvature, radial null geodesics (i.e. at constant $\theta$ and $\phi$) are defined as
\be
ds^2 = - dt^2 + a(t)^2 dr^2 = 0 \quad \Rightarrow \quad dr= \pm \frac{dt}{a(t)}\equiv \pm  d \tau 
\ee 
where, in the last step, we have used the conformal time $\tau$ defined in Eq.~\eqref{conformaltime}.

If we assume the standard picture given by Fig.~\ref{history}, the Universe was dominated by ordinary matter with state parameter $w>-1/3$ for most of its evolution and, going back in time, the scale factor $a(t)$ decreases up to the singular point $a(0)=0$. In this case there is a maximum distance to which an observer, at a given time $t_0$, can see a light-signal sent at $t=0$. In comoving coordinates, this is given by the so-called  {\it comoving particle horizon}, that is
\be\label{compartHor}
r_{ph} = \int_0^{t_0} \frac{dt}{a(t)} = \int_0^{a_0} (a H)^{-1} d \ln a\,.
\ee
If the comoving distance between two particles is greater than $r_{ph}$, they could have never talked to each other. Assuming Eq.~\eqref{scalefactor} and integrating Eq.~\eqref{compartHor}, we get
\be\label{compartHor2}
r_{ph} \sim a_0^{\frac{1}{2}(1+3w)}\sim (a_0H_0)^{-1}\,.
\ee
where the index ``$0$'' means calculated at time $t=t_0$.

The quantity $(aH)^{-1}$ is called {\it comoving Hubble radius} and determines the distance over which one cannot communicate at a given time. It basically fixes the causal structure of the space-time and its time-evolution is crucial for the particle horizon in Eq.~\eqref{compartHor}. The comoving Hubble radius and the particle horizon are basically the same in standard cosmology, as one can see from Eq.~\eqref{compartHor2}. However, we will see in the next Chapter that inflation modifies this correspondence. This will make a crucial difference.

The resulting picture is that, in an expanding Universe filled with ordinary matter, the horizon grows with time as given by Eq.~\eqref{compartHor2}. This means that comoving scales entering the horizon today have been never in causal contact before. These regions should look quite different from each others as they could never exchange any information before. 

The problem arises when we look at the largest scales we can observe in the sky, such as CMB and LSS scales. Their homogeneity is not just remarkable but very curious, according to the causal structure discussed above. Specifically, the CMB radiation has the same temperature in any direction we could look although the naive horizon scales would be just around one degree in the sky. How can this so-called {\it horizon problem} be explained? In the next Chapter, we will present the inflationary paradigm as a solution to this problem.

\subsection{Additional challenges}

The standard cosmological model contains other puzzling issues weakening its robustness and internal consistency. Surprisingly these will find a solution again in the next Chapter via inflation. We list these challenging problems below.

\subsubsection{Unwanted relics}
Going back in time, near the Big Bang, the temperature increases and the energy reaches values where UV-physics completion should play an important role. According to different UV-physics scenarios, new objects might be produced and their effect would survive until the present time, following the standard cosmological evolution. Typical examples are the following:
\begin{itemize}
\item {\it Magnetic monopoles}, very massive objects with a net magnetic charge, are copiously produced in the early Universe, according to most of grand unified theories (GUT). Combining these predictions with the standard cosmological model, Preskill \cite{Preskill:1979zi} found that today its abundance would be 12 orders higher than any other particle. This would have dramatic consequences on the age of the Universe and, then, it is in clear contrast with observations \cite{Guth:2007ng}.

\item The {\it gravitino}, the supersymmetric particle of the graviton, may ruin the success of the BBN cause of its late-time decay.

\item {\it Topological defects}, such as cosmic strings or domain walls, might bring worrisome consequences for current observations.
\end{itemize}

Today we do not observe any direct effect due to the production of these objects in the early Universe. An elegant explanation is provided by inflation which simply dilutes their density to a very negligible level.

\subsubsection{Homogeneity VS Inhomogeneities}

It is very remarkable how the Universe is extremely homogeneous at large scales while being very irregular and clumpy at small scales. However, the standard picture of cosmology does not provide any compelling explanation for the origin of this fundamental difference. Which is the origin of the inhomogeneities over a very smooth background?

\clearpage
\thispagestyle{empty}

\chapter{Inflationary Cosmology}
\label{chapter:Inflation}

\begingroup
\begin{flushright}
\vspace{.5cm}
\end{flushright}
\quote{\it This chapter is devoted to the inflationary paradigm as solution to the standard cosmological puzzles discussed in the previous chapter.  We present the basic features of inflation, how this modifies the causal structure of the space-time and its implementation through a scalar field. Then, we discuss the implications of treating this scenario quantum-mechanically: the zero-point fluctuations of the inflaton field become the fundamental origin of those perturbations we can measure in the sky in the form of the CMB temperature anisotropies and primordial gravitational waves. In the end, we present the latest Planck data which provide stringent constraints on the fundamental dynamics of inflation.}

\endgroup

\newpage

\section{Inflation and the smooth background}

The shortcomings of standard cosmology concern the initial conditions of our Universe that require serious fine-tuning in order to reproduce what we observe today. The flatness problem can be solved by assuming that the initial value of the curvature was precisely flat. Similarly, in order to solve the horizon problem,  one should imagine at least $10^6$ causally disconnected spatial patches to have started their evolution exactly in the same physical conditions, in particular at the same temperature and same magnitude of perturbations. Postulating all this is possible but hardly attractive to a physicist that aims to understand the very early Universe.

In order to do better, inflation was proposed in the 1980's \cite{Guth:1980zm,Linde:1981mu,Albrecht:1982wi} to solve these problems all at once.  The fundamental idea is that the primordial Universe underwent a finite phase of quasi-exponential expansion (similar to the one we are experiencing nowadays with dark energy) which changed the causal structure and how information propagates. As a bonus, one gets a physical mechanism to explain the presence of very small inhomogeneities as quantum fluctuations in the very early Universe; ultimately, these represent the seeds for the large scale structures we observe in the sky.

\subsection{Basic idea}

Standard cosmology assumes that the early Universe was dominated by some form of energy satisfying the strong energy condition $\rho+3p\geq0$, which implies a decelerating phase of the scale factor, $\ddot{a}<0$, as dictated by Eq.~\eqref{Fried}. This is at the core of both the flatness and horizon problems.

Inflation is nothing but inverting such a behavior and postulating a phase of accelerated expansion such as
\be\label{acceleration}
\ddot{a}>0\,,
\ee
which implies that the Universe was filled with some kind of matter with negative pressure, satisfying
\be\label{negpressure}
\rho+3p<0\,.
\ee
The idea that, at very early times, neither matter nor radiation represented the dominant components of energy is not in contrast with any well-tested physical theory. In fact, the standard model of particles physics (SM) cannot be assumed to work up to the first moments after the Big Bang, when energies were several orders of magnitude higher than the domain of validity of the SM (which extends up to around one TeV). Inflation lives off the idea that something non-trivial might have happened due to high-energy physics.

\subsection{Decreasing Hubble radius}

\begin{figure}[htb]
\hspace{-3mm}
\begin{center}
\includegraphics[width=10.4cm,keepaspectratio]{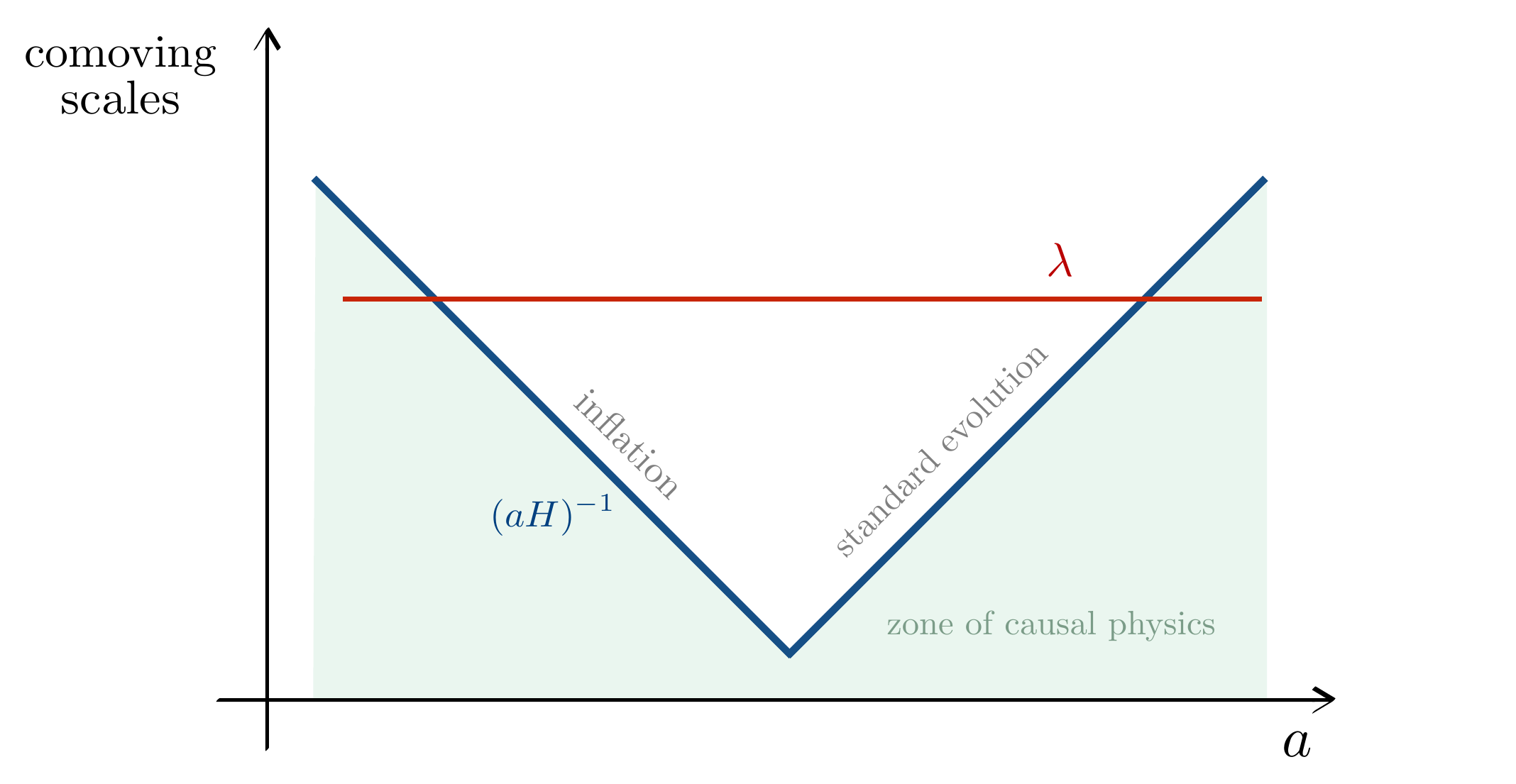}
\caption{\small \it The comoving Hubbe radius and a typical comoving scale as a function of the scale factor. Due to the anomalous scaling of the comoving Hubble radius, which does not remain constant in time as it happens for all typical scales, the zone of causal physics change with time.}\label{horizon}
\end{center}
\vspace{0.cm}
\end{figure}

Interestingly, the condition Eq.~\eqref{acceleration} turns out to be equivalent to a decreasing comoving Hubble radius
\be\label{decreasingHubbleradius}
\frac{d}{dt} (aH)^{-1}<0\,,
\ee
which gives a deeper insight into the causal structure of a Universe undergoing a phase of inflationary expansion. Typical scales, being initially inside the horizon, leaves the radius of causal contact as inflation proceeds and the Hubble radius $(aH)^{-1}$ decreases. They start reentering the horizon when inflation ends, the standard cosmological evolution progresses and $(aH)^{-1}$ increases. This situation is illustrated in Fig.~\ref{horizon}.

\subsection{Quasi-de Sitter phase and Hubble flow functions}

The rate of change of the Hubble radius with respect to the time $t$ can be expressed also as
\be
\frac{d}{dt} (aH)^{-1}= \frac{\epsilon_1 -1}{a}\,,
\ee
with
\be \label{EPSILON1}
\epsilon_1 \equiv -\frac{\dot{H}}{H^2}=\frac{d \ln H}{d N}\,,
\ee
and $N$ being the {\it number of e-folds} defined as
\be\label{efoldsN}
d N \equiv d\ln a = - H dt\,,
\ee
where we have assumed that $N$ decreases as time progresses (one can find also the opposite convention in literature). Then, inflation takes place as long as
\be
\epsilon_1 <1\,,
\ee
which means that the Hubble parameter must vary very slowly in time. The inflationary phase corresponds to a quasi-exponential expansion with $H$ almost constant for the whole duration of the process. The extreme limit $\epsilon_1=0$  is an exact de Sitter phase sourced by an pure cosmological constant and then defined by Eq.~\eqref{scalefactor}  for $w=-1$. However, we know inflation must end in order to give rise to the standard cosmological evolution. This point is defined by $\epsilon_1=1$.

In order to assure inflation to last long enough, thus solving the standard shortcomings, we define a second important parameter controlling the duration of the process (in the next subsection, we quantify the amount of inflation needed in order to solve the cosmological puzzles). This is\footnote{Unfortunately, in the scientific literature, there is no unique convention regarding the symbols identifying the parameters which control the dynamics of inflation. Different authors may assign different symbols to the same parameter. Specifically, $\epsilon_1$ and $\epsilon_2$ are often referred to as $\epsilon$ and $\eta$ in other references (e.g. in \cite{Baumann}). However, throughout the thesis, we will reserve the symbols $\epsilon$ and $\eta$ for the slow-roll parameters later introduced in Sec.~\ref{SUBSECTIONslowroll}.}
\be
\epsilon_2\equiv \frac{d \ln \epsilon_1}{d N}\,.
\ee

The condition $|\epsilon_2|<1$ basically means having a small fractional variation of $\epsilon_1$ which guarantees that inflation persists enough time. 

We can further proceed constructing the entire tower of the so-called {\it Hubble flow functions} $	\epsilon_i$ defined iteratively as \cite{Schwarz:2001vv,Schwarz:2004tz}
\be\label{HubbleFlowFunctions}
\epsilon_{i+1}\equiv \frac{d \ln |\epsilon_i|}{d N}\,,
\ee
with the first of these quantities identical to the Hubble parameter, $\epsilon_0=H$.

\subsection{Puzzles resolution and the amount of inflation} \label{amountinflation}

The horizon problem is solved if one allows for enough inflation such that also the largest scales we observe in the sky today (CMB and LSS scales) were inside the horizon at early times. Quantitatively, this means that the comoving scales of the observable Universe today $(a_0H_0)^{-1}$ must fit inside the comoving Hubble radius at the beginning of inflation  $(a_iH_i)^{-1}$, that is
\be\label{horizonsolution}
(a_iH_i)^{-1}>(a_0H_0)^{-1}\,.
\ee

The amount of inflation needed to allow for this resolution is conveniently quantified by the number of e-folds $N$ defined by Eq.~\eqref{efoldsN} and determined by the increase of the scale factor during inflation. Specifically, a total number $N\gtrsim 50-60$ suffices to explain the thermalization of the largest observational scales at present.

A rough estimate can be obtained by assuming that the Universe has been dominated mainly by radiation since the end of inflation (at that moment, the comoving Hubble radius was equal to $(a_eH_e)^{-1}$). This implies that the Hubble parameter scales as $H\propto a^{-2}$. Then we have
\be
\frac{a_0H_0}{a_eH_e} \sim \frac{a_e}{a_0} \sim \frac{T_0}{T_e} \sim 10^{-28}\,,
\ee
where we have assumed $T_0= 10^{-3}$ eV, as the CMB temperature measured today, and $T_e= 10^{15}$ GeV as the typical expected inflationary energy. Then, Eq.~\eqref{horizonsolution} becomes
\be
(a_iH_i)^{-1}>10^{28}(a_eH_e)^{-1}\,,
\ee
which means that the Hubble radius had to shrink 28 orders magnitude in order to solve the horizon problem. Since during inflation $H$ is almost constant, we have $H_i\approx H_e$ and then 
\be
\frac{a_{e}}{a_{i}}>10^{28}\,,
\ee
which, using Eq.~\eqref{efoldsN}, implies 
\be
N>64\,.
\ee

The flatness problem is overcome by means of the same mechanism. A decreasing comoving Hubble radius $(aH)^{-1}$ drives the value of the total energy density $\Omega$ to unity, providing a physical explanation for this apparently fine-tuned configuration. After inflation, the curvature will start diverging from $\Omega \approx 1$, as it happens in a Universe filled with ordinary matter. Interestingly, the same amount of inflation needed to solve the horizon problem is enough to explain the flatness we observe today. In fact, during inflation we have
\be
\Omega -1 = \frac{\kappa^2}{(aH)^2} \propto e^{-2N}\rightarrow 0\,.
\ee
The same number of e-folds quoted before would give the accuracy required for the value observed today.

\subsection{Scalar field dynamics and slow-roll inflation} \label{SUBSECTIONslowroll}

The Einstein equations tell us that inflation should be supported by some form of matter with a negative pressure, as given by Eq.~\eqref{negpressure}. However, we are still left with the issue of identifying the origin of such an incredible energy which led the scale factor to increase by an order of $10^{28}$. 

The simplest example is to imagine that (a small portion of) the primordial Universe is filled with a scalar field, often called {\it inflaton} field, minimally coupled to gravity with Lagrangian
\be
\mathcal{L}= \sqrt{-g} \left[\tfrac{1}{2}  R - \tfrac{1}{2} g^{\mu\nu}\partial_\mu \phi\ \partial_\nu \phi -V(\phi)\right]\,,
\ee
leading to the energy-momentum tensor
\be
T_{\mu\nu}= \partial_\mu\phi\ \partial_\nu\phi - g_{\mu\nu}\left[\tfrac{1}{2} \partial^\sigma\phi\ \partial_\sigma\phi + V(\phi)\right]\,.
\ee
In the case of a homogeneous scalar field $\phi(t)$ filling a patch of the Universe with flat FLRW metric \eqref{FRW}, the energy density and pressure turn out to be simply
\begin{align}
\rho&\equiv T_{00}=\tfrac{1}{2} \dot{\phi}^2 + V(\phi)\,, \quad
p \equiv T_{ii}= \tfrac{1}{2} \dot{\phi}^2 - V(\phi)\,.
\end{align}
The dynamics and interaction of the spacetime metric and scalar field  is described by the two equations
\begin{align}
&H^2= \frac{1}{3} \left[\frac{\dot{\phi}^2}{2} + V(\phi) \right]\,, \quad
\ddot{\phi}+3H\dot{\phi}+V'=0\,,\label{EOMScalar}
\end{align}
where primes denote derivatives with respect to $\phi$. The first is simply the Friedmann equation \eqref{Fried}, with $\kappa=0$. The second is the equation of motion for the scalar field which is  derived by varying its action. It describes a particle rolling down along its potential and subject to a friction due to the expansion term $3H\dot{\phi}$. The second Friedmann equation \eqref{Fried2} simply becomes
\be\label{Fried2Scalar}
\dot{H}=-\frac{{\dot{\phi}}^2}{2}\,,
\ee
which implies that the time evolution of the Hubble parameter depends on the kinetic energy of the field. Alternatively, it is possible to obtain the second equation of \eqref{EOMScalar} by taking the time derivative of the first equation and combining this with Eq.~\eqref{Fried2Scalar}.

\begin{figure}[htb]
\hspace{-3mm}
\begin{center}
\includegraphics[width=8.5cm,keepaspectratio]{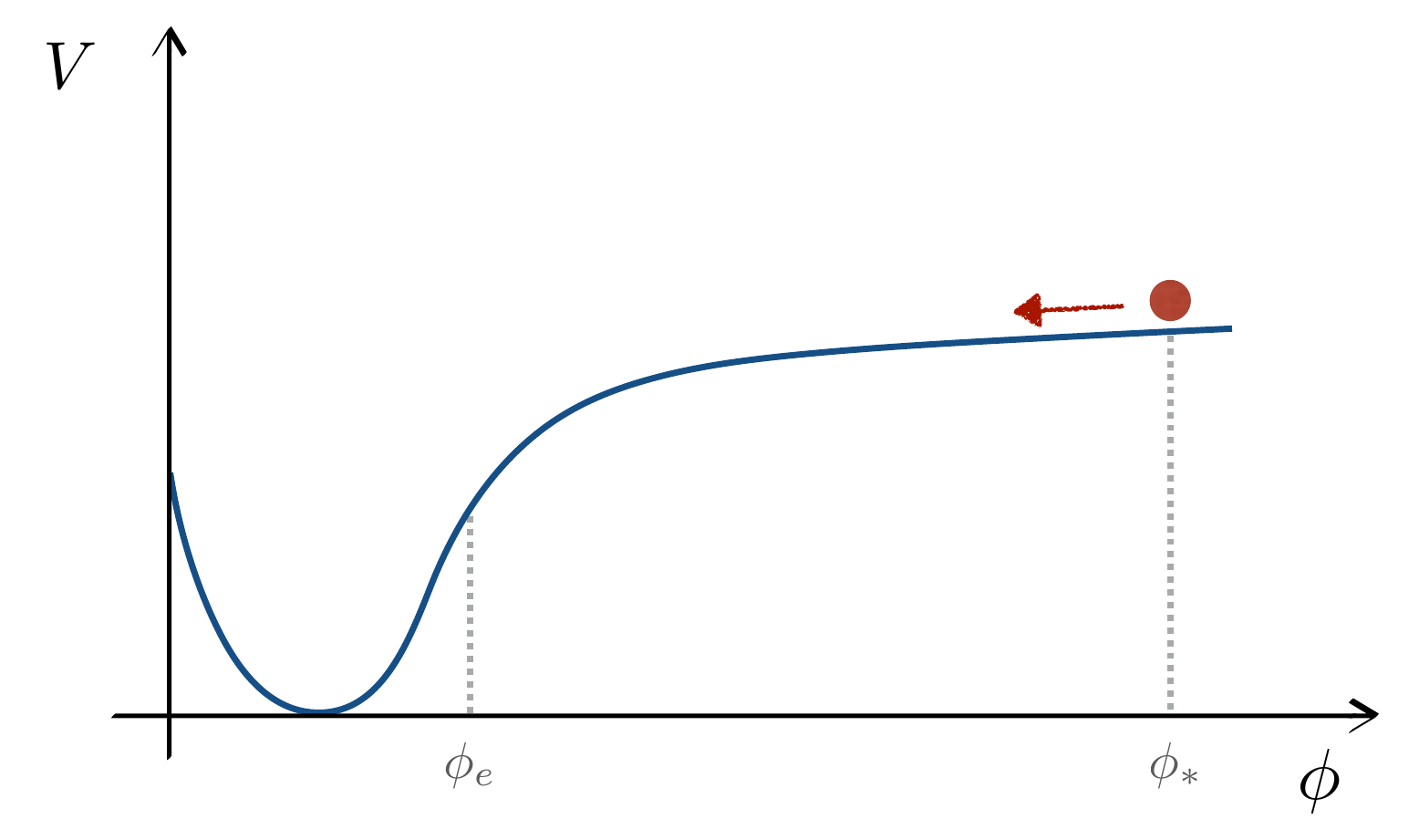}
\caption{\small \it Cartoon picture of a typical inflationary potential. The scalar field slowly rolls down along the  shape driving the quasi-exponential expansion. Inflation ends at $\phi_e$ and starts at $\phi_*$, at least around 60 e-foldings before the end.}\label{SRpot}
\end{center}
\vspace{0.cm}
\end{figure}

This region of the Universe inflates if the state parameter $w=p/\rho<-1/3$, which is easily realizable if the potential energy dominates over the kinetic energy, that is
\be\label{SlowRoll}
V(\phi)\gg\dot{\phi}^2\,.
\ee
The regime described by Eq.~\eqref{SlowRoll} is said {\it slow-roll inflation} as the field will evolve really slowly with respect to the quasi-exponential growth of the scale factor.  Further, in order to have an inflationary period lasting long enough, one must ensure a small acceleration of the field and therefore impose 
\be\label{smallacceleration}
|\ddot{\phi}|\ll |3H\dot{\phi}| \,.
\ee
Intuitively, such a scenario is possible any time that the shape of the potential is sufficiently flat (in some measure) as it is shown in the cartoon of Fig.~\ref{SRpot}. 

 Within the slow-roll regime, the dynamical equations \eqref{EOMScalar} become
\begin{align}
H^2&\approx  \frac{V(\phi)}{3}\approx \text{constant}\,, \quad
\dot{\phi}\approx -\frac{V'}{3H}\,.\label{SRdotPhi}
\end{align}
Given a scalar field with its potential $V(\phi)$, one can verify whether such scenario is suitable for inflation or not by calculating the so-called {\it slow-roll parameters}, defined as
\be\label{epseta}
\epsilon\equiv \frac{1}{2}\left(\frac{V'}{V}\right)^2\,,\qquad\eta\equiv \frac{V''}{V}\,,\
\ee
and check that
\be\label{SRconditions}
\{\epsilon,|\eta|\}\ll 1\,,
\ee
which is equivalent to Eq.~\eqref{SlowRoll} and Eq.~\eqref{smallacceleration}.

Strictly within the slow-roll approximation, the slow-roll parameters are related to the Hubble flow functions through the following 
   \be\label{flowSR}
  \epsilon_0 = (V/3)^{1/2} \,, \quad \epsilon_1 = \epsilon \,, \quad \epsilon_2 = - 4 \epsilon + 2 \eta \,.
 \ee

Eventually, inflation must end and give way to the standard cosmological evolution (with an increasing Hubble radius and ordinary matter domination). This happens when the conditions \eqref{SRconditions} are violated: the trajectory becomes first too steep and the inflaton eventually falls into a local minimum. The oscillations around the vacuum convert the inflationary energy into ordinary particles, within a process called {\it reheating} (see \cite{Allahverdi:2010xz} for a review on this topic and \cite{Cook:2015vqa} for a recent work).


\section{Inflation and the background perturbations} \label{FLUCTUATIONS}

\subsection{The inhomogeneous Universe}

The inflationary paradigm elegantly solves the standard cosmological puzzles, providing a natural explanation for the homogeneity and isotropy at large distances. However, at scales smaller than 100 Mpc, we do observe structures in form of galaxies, stars and so on. The standard cosmological theory allows us to accurately trace the evolution of such structures back in time. We are able to identify their origin in the gravitational instability of small density perturbations of a primordial plasma made up of photons and baryons, which have evolved into the large-scale structures of the present Universe. 

\begin{figure}[htb]
\vspace{-3mm}
\begin{center}
\includegraphics[width=8.4cm,keepaspectratio]{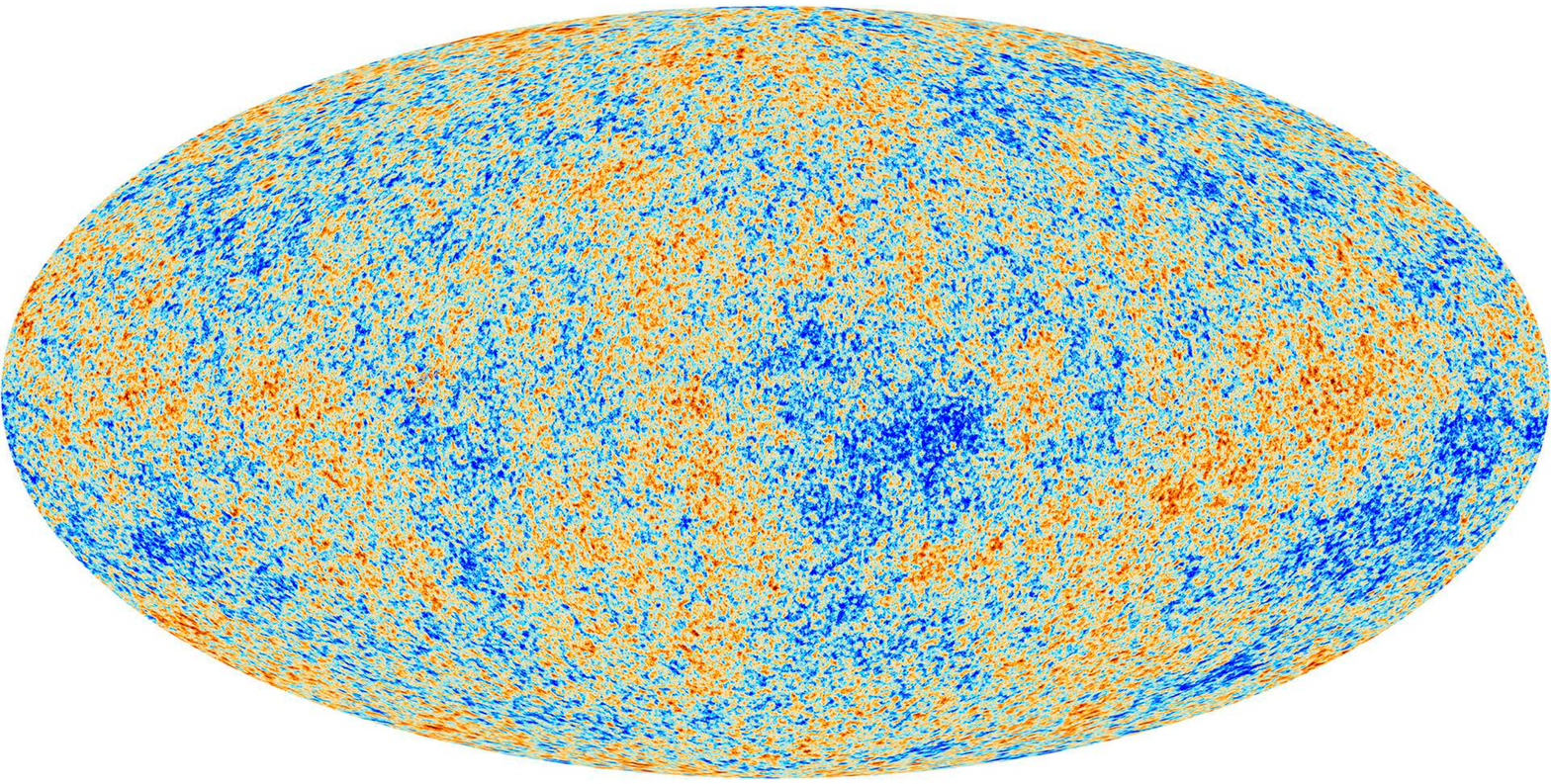}
\vspace*{-0.3cm}
\caption{ \it The fluctuations of 1 part in $10^5$ around the average temperature of $T= 2.73$ of the CMB.}\label{CMB}
\end{center}
\vspace{-0.5cm}
\end{figure}

This idea of structure formation is confirmed by the oldest snapshot we have of our Universe: the cosmic microwave background (CMB). It was produced at the time when electrons and nuclei have just recombined, around 300.000 years after the Big Bang, leaving the CMB photons to freely stream. The tiny temperature fluctuations of order $\delta T / T \sim 10^{-5}$, indicated in Fig.~\ref{CMB}, reflect the presence of regions with slightly different densities; the wavelength of the photons is red-shifted or blue-shifted depending on the value of the local density. Indeed the properties of the CMB can be time-evolved into a forecast for the Universe that has an excellent match with our observed one.

Despite the stunning success of the theory of structure formation, we are left with some puzzling questions: {\it what set those initial density perturbations? Which is their fundamental origin? Why were they there at all? }

Surprisingly, inflation suggests a possible answer that is in excellent agreement with observations, thus definitively establishing itself as the leading paradigm for the understanding of the early Universe physics. This answer stems from adding quantum mechanics to the fundamental inflationary dynamics. The scalar field implementation provides once more a very useful stage in order to discuss such a physics. In fact, quantum fluctuations $\delta\phi$ are  unavoidable in the homogeneous background represented by $\phi(t)$. These source metric perturbations via the Einstein equations and vice versa according to the following scheme
\be\label{decomposition1}
\phi(t,{\bold x}) = \phi(t) + \delta\phi(t,{\bold x}) \quad\Leftrightarrow\quad g_{\mu\nu}(t,{\bold x}) = g_{\mu\nu}(t) + \delta g_{\mu\nu}(t,{\bold x})\,,
\ee
where $g_{\mu\nu}(t)$ is simply the unperturbed FLRW metric, as given by Eq.~\eqref{FRW}. Due to the symmetries and gauge invariance of the coupled system, the resulting physical perturbations reduce to a scalar and a tensor one (vector perturbations decay during the quasi-exponential expansion). Intuitively, quantum fluctuations excite all the light particles, in the minimal scenario being the inflaton and the graviton. The scalar perturbations couple to the energy density and eventually lead to the inhomogeneities and anisotropies observed in the CMB. The tensor perturbations are often referred to as primordial gravitational waves. They do not couple to the density but induce polarization in the CMB spectrum \cite{Seljak:1996ti,Kamionkowski:1996zd,Seljak:1996gy,Zaldarriaga:1996xe,Kamionkowski:1996ks,Hu:1997hv}. This is considered to be a unique signature of inflation and many current experiments are searching for it in the sky.
 
A detailed treatment of the cosmological perturbations theory goes beyond the aim of the present thesis. The interested reader might consult the references \cite{Mukhanov:1990me,Brandenberger:2003vk,Dodelson:2003ft,Mukhanov:2005sc,Baumann:2009ds}. In the following, we would like just to sketch the main consequences of a consistent quantum formulation of the inflationary paradigm. In order to simplify the discussion, we will firstly discuss the pure de Sitter and massless case. In the next Sec.~\ref{OBSERV}, we will focus on the proper inflationary analysis, regarded as a small deviation from the case studied here, and eventually extrapolate the significant observational parameters.
 
 \subsection{Quantum scalar fluctuations during inflation} \label{InflatonFluctuations}
 
Scalar fluctuations can be fully attributed to the quantum nature of the inflaton field living in an unperturbed FLRW background. This corresponds to a specific gauge (usually called {\it spatially flat slicing}) where metric perturbations are set equal to zero. It is a perfectly consistent choice in order to discuss the relevant physics and show how scalar fluctuations behave in an inflationary background metric. The decreasing Hubble radius $(aH)^{-1}$ will play again a crucial role, as we will see.

Let us consider the inflaton field $\phi(t,{\bold x})$ with a small spatial dependence as given by Eq.~\eqref{decomposition1}. The corresponding equation of motion is
\be\label{EOMspatial}
\ddot{\phi}+3H\dot{\phi}-\frac{\nabla^2}{a^2}\phi+V'=0\,,
\ee
which differs from the homogeneous equation \eqref{EOMScalar} of the background field $\phi(t)$ for the third extra term. We can Fourier-expand the fluctuations such as
\be\label{FourierDec}
\delta\phi(t,{\bold x}) =  \int \frac{d^3 {\bold k}}{(2\pi)^{3/2}} \delta\phi_{k}(t) e^{i {\bold k}\cdot{\bold x}}\,,
\ee
with ${\bold x}$ and ${\bold k}$ being respectively the comoving coordinates and momenta. Note that the Fourier modes $\delta\phi_{k}$ depend just on the modulo $k=|{\bold k}|$ because of the isotropy of the background metric. Then, we can perturb at first order Eq.~\eqref{EOMspatial}, plug the decomposition \eqref{FourierDec} in and get
\be\label{EQpert}
\delta\ddot{\phi_k}+3H\delta\dot{\phi_k} + \frac{k^2}{a^2}\delta\phi_k =0\,,
\ee
where we have neglected the additional term $V''\delta\phi_k$ due to the slow-roll conditions Eq.~\eqref{SRconditions} during inflation. Eq.~\eqref{EQpert} can be rewritten in a simpler form, without the Hubble friction term, once we introduce the variable
\be\label{variablev}
v_k\equiv a \delta\phi_k\,,
\ee
and switch to the conformal time $\tau$.  This was defined by Eq.~\eqref{conformaltime} and it is naturally related to the comoving Hubble radius as
\be\label{tauDS}
\tau=-\frac{1}{aH}\,,
\ee
during a perfect exponential expansion with $H$ constant. Then, the dynamics of the scalar perturbations can be described simply by the equation of a collection of independent harmonic oscillators
\be\label{EQmodes}
\frac{d^2}{d\tau^2} v_k +\omega^2_k(\tau) v_k=0\,,
\ee
with time-dependent frequencies
\be\label{freq}
\omega^2_k(\tau) = k^2 -\frac{2}{\tau^2}=  k^2 -2 (aH)^2\,.
\ee

The quantization of the physical system now becomes very easy and one proceeds as in the case of the simple harmonic oscillator, following the canonical procedure. In particular, the modes $v_k$ become nothing but the coefficients of the decomposition of the quantum operator
\be
\hat{v}(\tau,{\bold k}) =  v_k (\tau) \hat{a}_{\bold k}+v^\ast_k (\tau) \hat{a}^\dagger_{\bold k}\,,
\ee
where the creation and annihilation operators satisfy the canonical commutation relation
\be
\left[\hat{a}_{\bold k},\hat{a}^\dagger_{\bold k'}\right]= \delta^3\left({\bold k}-{\bold k}'\right)\,.
\ee

The quantum zero-point fluctuations are given by
\be\label{zerofluc}
\Braket{0|\hat{v}^\dagger(\tau,{\bold k})  \hat{v}(\tau,{\bold k'})|0 } =|v_k (\tau)|^2\delta^3\left({\bold k}-{\bold k}'\right)
\ee
where the vacuum is defined by $\hat{a}_{\bold k}\Ket{0}=0$ for any ${\bold k}$. Therefore, computing the quantum perturbations of the inflaton field reduces to solving the classical equation \eqref{EQmodes} and, then, extracting the time dependence of the Fourier modes $v_k(\tau)$.

The physics of the mode functions $v_k$, during inflation, is non-trivial and crucially depends on the fact that the comoving Hubble radius shrinks with time. In fact, fluctuations are produced on every scale $\lambda$ and therefore with any momentum $k$. While initially being inside the horizon, they leave the zone of causal physics at one point of the accelerated expansion, as schematically shown in Fig.~\ref{horizon}. 

One can prove that an exact solution of Eq.~\eqref{EQmodes} is
\be\label{solutionvk}
v_k(\tau) = \alpha \frac{e^{-i k \tau}}{\sqrt{2k}} \left(1-\frac{i}{k\tau}\right) + \beta \frac{e^{i k \tau}}{\sqrt{2k}} \left(1+\frac{i}{k\tau}\right)\,,
\ee
where $\alpha$ and $\beta$ are some free parameters to be set by means of the initial conditions. These are defined at very early times, when the relevant scales were still inside the horizon. In the {\it sub-horizon limit} ($k\ll aH$), that is when $k|\tau|\rightarrow\infty$,  the frequencies \eqref{freq} become time-independent and Eq.~\eqref{EQmodes} reduces to
\be
\frac{d^2}{d\tau^2} v_k +k^2 v_k=0\,,
\ee
basically the one of a simple harmonic oscillator. We can exploit this fact in order to get the correct normalized solution
\be
\lim_{k|\tau|\rightarrow\infty} v_k = \frac{e^{-i k \tau}}{\sqrt{2k}}\,,
\ee
which comes from the requirement of a unique vacuum (so-called {\it Bunch-Davies} vacuum) being the ground state of energy. This sets $\alpha=1$ and $\beta=0$ in Eq.~\eqref{solutionvk}, thus yielding the definitive expression for the Fourier modes
\be\label{CompleteSolution}
v_k(\tau) = \frac{e^{-i k \tau}}{\sqrt{2k}} \left(1-\frac{i}{k\tau}\right)\,.
\ee

Once we have the complete solution Eq.~\eqref{CompleteSolution}, we are particularly interested in studying when the modes leave the horizon. We would like indeed to understand how they behave after inflation and affect late time physics. How can quantum fluctuations, produced during inflation, source density perturbation at CMB decoupling? These events are separated by a huge amount of time where physics is very uncertain. Fortunately, something special happens as we explain below.

The {\it super-horizon limit} ($k\gg aH$), that is when $k|\tau|\rightarrow 0$, corresponds to the solution
\be\label{superhor}
\lim_{k|\tau|\rightarrow 0} v_k = -\frac{i}{\sqrt{2}k^{3/2}\tau}\,.
\ee
Since the conformal time is related to the scale factor by Eq.\eqref{conformaltime}, the latter represents a growing mode $v_k\propto\ a$, in de Sitter background. Switching to the physical scalar perturbations by means of Eq.~\eqref{variablev}, one obtains that the amplitude $\delta \phi_k$ remains constant as long as the Hubble radius is smaller than their typical length.  Modes freeze outside the horizon and this is a crucial result in order to connect the physics of the early Universe to the time when the density perturbations are created. It is a great bonus we get from inflation as we do not need to worry about the time evolution of such fluctuations for a very substantial part of the cosmic evolution.

Now we can return to Eq.~\eqref{zerofluc} and properly evaluate the dimensionless  {\it power spectrum} $\Delta_v^2$ of the quantum fluctuations $v_k$, defined as 
\be\label{powerv}
\Braket{0|\hat{v}^\dagger(\tau,{\bold k})  \hat{v}(\tau,{\bold k'})|0 } \equiv \frac{2 \pi^2}{k^3} \Delta_v^2(k)\ \delta^3\left({\bold k}-{\bold k}'\right)\,.
\ee
Then, the power spectrum of the fluctuations after horizon crossing is
\be
\lim_{k|\tau|\rightarrow 0} \Delta_v^2(k) = \frac{k^3}{2 \pi^2} |v_k|^2 =  \left(\frac{aH}{2 \pi}\right)^2\,, 
\ee
where we have used Eq.~\eqref{zerofluc} in the first step while Eq.~\eqref{superhor} and Eq.~\eqref{tauDS} in the last. Therefore, the power spectrum of the physical fluctuations of the inflaton field on super-horizon scales is
\be\label{powerinfl}
\Delta_{\delta\phi}^2(k) =  \left(\frac{H}{2 \pi}\right)^2\,, 
\ee
which is scale-invariant as no $k$-dependence enters the expression above. Note that this result was first derived in \cite{Bunch:1978yq}, in a perfect de Sitter approximation, before inflation was proposed. A proper inflationary analysis would bring corrections of order $\mathcal{O}(\epsilon,\eta)$. This is shown below in Sec.~\ref{OBSERV}.

 \subsection{Classical curvature and density perturbations}

In the previous section, we have learned that quantum fluctuations, produced during inflation, stop oscillating once they are stretched to super-horizon scales. Their amplitude freezes at some nonzero value, with scale invariant power spectrum given by Eq.~\eqref{powerinfl}. This situation lasts for a very long period until the point when the modes re-enter the horizon, during the standard cosmological evolution, as schematically shown in Fig.~\ref{horizon}. At horizon re-entry, the amplitude of the modes starts oscillating again inducing the density perturbations. However, the energy density directly interacts with the gravitational potential. Therefore, {\it how do quantum fluctuations of the inflaton affect the metric curvature and ultimately become density perturbations?} Here, we present a very simple and heuristic derivation, mainly based on the {\it time-delay formalism} developed in \cite{Guth:1982ec}.

The presence of quantum fluctuations $\delta\phi(t,{\bold x})$ over the smooth background $\phi(t)$ translates into local differences $\delta N$ of the duration of the inflationary expansion, directly related to curvature perturbations $\zeta$. In fact, not every point in space will end inflation at the same time thus leading to local variations of the scale factor $a$. Then,  fluctuations $\delta\phi$ induce curvature perturbations equal to
\be
\zeta= \delta N = 	H \frac{\delta \phi}{\dot{\phi}}=\frac{\delta a }{a}\,.
\ee
The corresponding dimensionless power spectrum is
\be
\Delta^2_\zeta (k)=\frac{H^2}{\dot{\phi}^2} \Delta^2_{\delta \phi} (k)=\frac{H^4}{4\pi^2\dot{\phi}^2}\,,
\ee
which, during slow-roll, reads
\be\label{PowerZeta}
\Delta^2_\zeta = \frac{1}{12 \pi^2} \frac{V^3}{V'^2} = \frac{1}{24 \pi^2} \frac{V}{\epsilon}\,,
\ee
where we have used Eq.~\eqref{SRdotPhi} in the first equality and Eq.~\eqref{epseta} in the second one.

Once inflation ends and the standard cosmological history begins, the energy density will evolve as $\rho=3H^2$ and, then, decrease as given by Eq.~\eqref{endensity} (the evolution is shown in Fig.~\ref{history}). Local delays of the expansion lead to local differences in the density, schematically being $\delta N \sim \delta \rho/\rho$. The amplitude of the density fluctuations will be directly related to the amplitude of the curvature perturbations with power spectrum Eq.~\eqref{PowerZeta}.

 \subsection{Primordial gravitational waves}

Primordial quantum fluctuations excite also the graviton, corresponding to tensor perturbations $\delta h$ of the metric. These have two independent and gauge-invariant degrees of freedom, associated to the polarization components of gravitational waves (usually denoted by $h_+$ and $h_\times$). One can prove that the Fourier modes of these functions satisfy an equation analogous to Eq.~\eqref{EQpert}. Therefore, one may proceed identically to what done in Sec.~\ref{InflatonFluctuations}. The dimensionless power spectrum turns out to be
\be\label{powerinfl-gr}
\Delta_{h}^2(k) = 2\times 4\times  \left(\frac{H}{2 \pi}\right)^2\,, 
\ee
where the factor $2$ is due to the two polarizations and the factor 4 is related to different normalization.

\section{Inflation and observations}\label{OBSERV}

The last 50 years have seen extraordinary success in the development of observational techniques and in the experimental confirmation of our cosmological theories.  The discovery of the CMB in 1965 \cite{Penzias:1965wn} gave the start to a new scientific era where our most speculative ideas have found empirical verification. Analyzing this primordial light has become our fundamental tool for the investigation of the very early Universe physics.

The CMB is essentially the farthest point we can push our observations to. It is nothing but an almost isotropic 2D surface surrounding us and beyond which nothing can directly reach our telescopes. One can draw an analogy to  the surface of the Sun: the inner dense plasma does not allow any light to freely stream outwards and the analysis of the last scattering photons (around 8 minutes old) becomes essential in order to probe the internal structure. In fact, the homogeneity and isotropy of the CMB  together with its tiny and characteristic temperature anisotropy (see Fig.~\ref{CMB}) naturally led us to consider inflation as what lies beyond that last scattering surface, around 13.8 billions years old.

\begin{figure}[htb]
\vspace{3mm}
\begin{center}
\includegraphics[width=12.5cm,keepaspectratio]{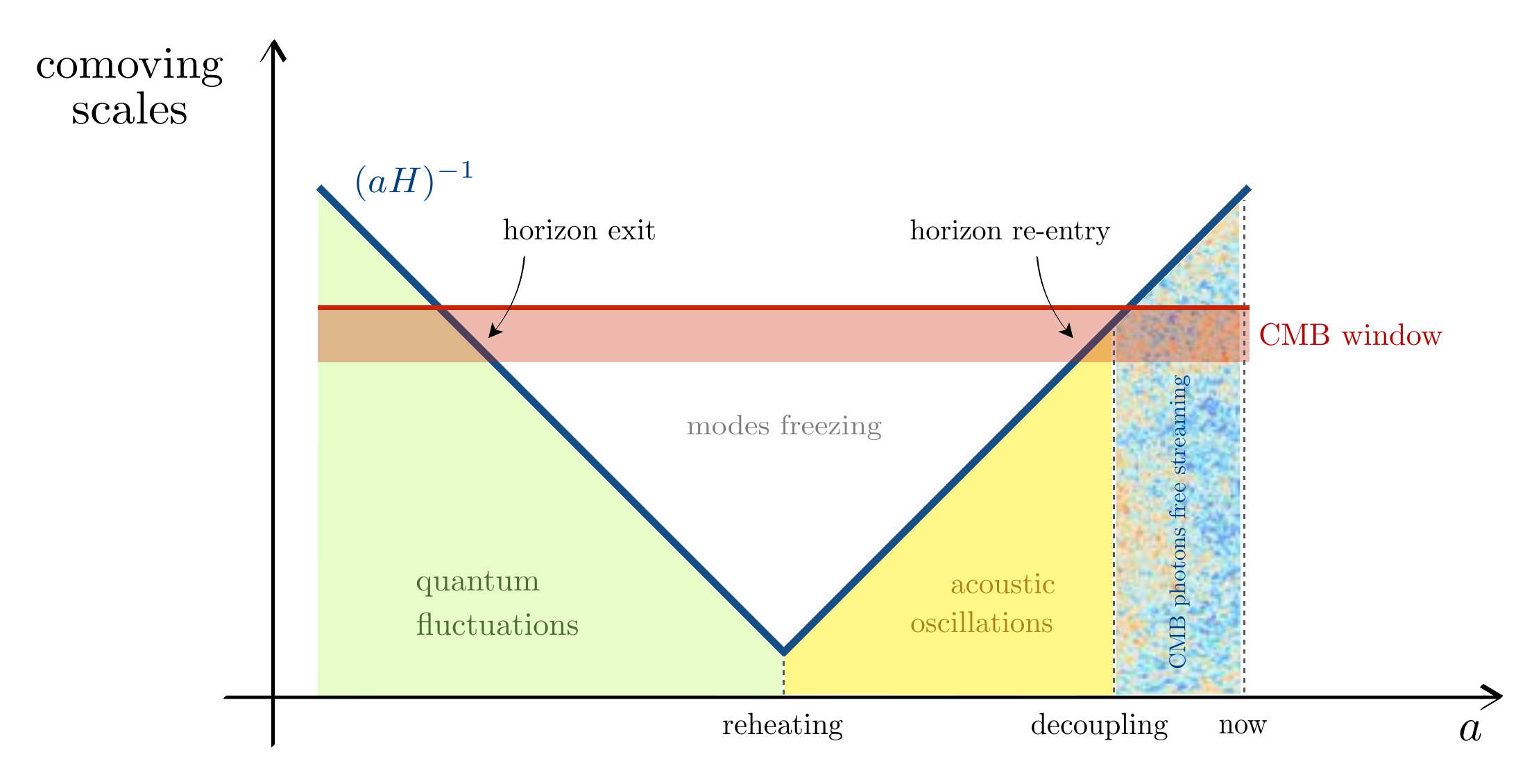}
\vspace*{0.3cm}
\caption{\it Quantum fluctuations produced during inflation (green area)  freeze at  the horizon exit. They reenter the horizon after reheating thus sourcing acoustic oscillations of the plasma (yellow part). At decoupling time, the CMB photons freely stream towards us who measure their power spectrum just in the small red window.}\label{evolution}
\end{center}
\vspace{-0.5cm}
\end{figure}

Via CMB measurements, we are able to probe the inflationary era and set stringent constraints on the fundamental dynamical mechanism. In the language of the scalar field implementation, we can use observational inputs to impose restrictions on the form of the scalar potential $V(\phi)$. The reason why we are able to have access to such a primordial era is closely connected to the mechanism outlined in the previous section:  fluctuations produced during inflation freeze outside the horizon thus providing a link between two very separated moments in time. This situation is depicted in Fig.~\ref{evolution}.

\subsection{CMB power spectrum and inflationary observables} 	\label{CMBInflation}

\begin{figure}[htb]
\vspace{3mm}
\begin{center}
\includegraphics[width=9cm,keepaspectratio]{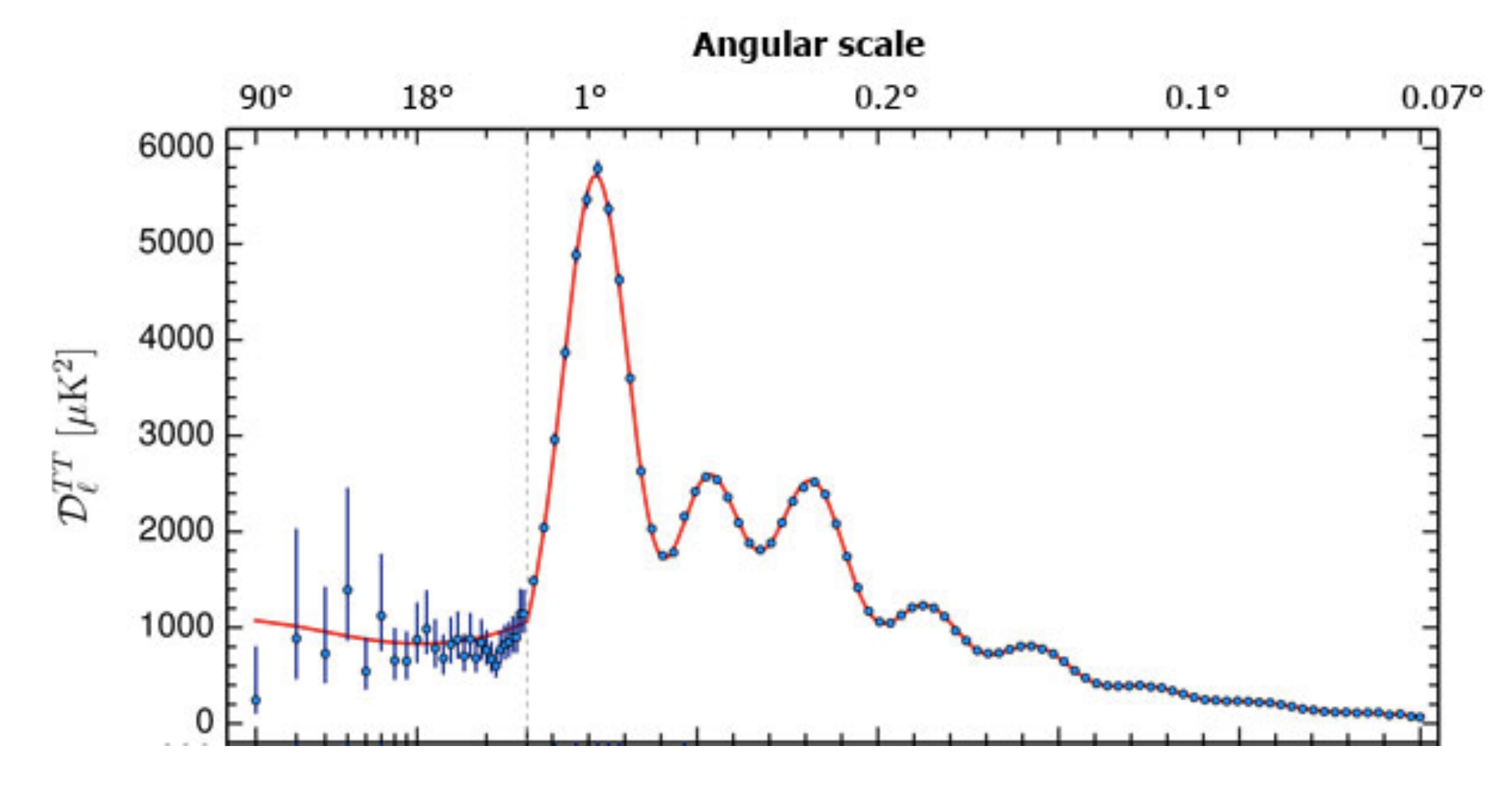}
\vspace*{0.3cm}
\caption{\it Power spectrum of the CMB temperature anisotropy as measured by Planck 2015 \cite{Planck:2015xua,Ade:2015lrj}.}\label{power}
\end{center}
\vspace{0.cm}
\end{figure}

The power spectrum of the temperature fluctuations in the CMB contains valuable information on the dynamics of inflation. The characteristic shape is simply dictated by the two-point correlation function of the inflaton fluctuations calculated in Sec.~\ref{FLUCTUATIONS}. A proper investigation of the CMB physics is required in order to understand the functional form. However, this goes beyond the scope of the present work (see e.g.\cite{Dodelson:2003ft,Hu:2008hd} for a detailed treatment). In practice, it is the so-called {\it transfer function} which relates the two power spectra: it contains all the information regarding the evolution of the initial fluctuations from the moment when they re-enter the horizon to the time of photon-decoupling (yellow part in Fig.~\ref{evolution}) and, subsequently, their projection in the sky as we observe them today. The final result is the solid line of Fig.~\ref{power} with the peculiar Doppler peaks originated from the acoustic oscillations of the baryon-photon plasma. The first peak corresponds to a mode that had just time to compress once before decoupling. The other peaks underwent more oscillations and, on small scales, are damped. The high suppression of  the power spectrum, at  small angular scales, reflects why we are able to probe just a small window of the inflationary era. In terms of the number of e-folds this corresponds to about $\Delta N\approx 7$. On the contrary, scales to the left of the first peak show no oscillations as they were superhorizon at the time of decoupling, and hence have not experienced any oscillations.

In Sec.~\ref{FLUCTUATIONS}, we have derived the power spectrum of perturbations in a perfect de Sitter ($H\approx \text{const}$) and massless ( $V''\approx 0$) approximation. However, an appropriate inflationary analysis would bring some corrections (order slow-roll) and hence a small $k$-dependence. This is because, during inflation, the energy scale (set by $H$) will slightly change together with time and the inflaton mass is non-zero, although being very small (order $\eta$). In order to parametrize the deviation from scale-invariance, we introduce the {\it spectral indexes} $n_s$ and $n_t$ defined by
\be
n_s -1 \equiv \frac{d \ln \Delta_\zeta^2}{d \ln k}\,, \qquad n_t \equiv \frac{d \ln \Delta_h^2}{d \ln k}\,,
\ee
respectively for scalar and tensor perturbations. In terms of the slow-roll parameters, they read
\be\label{nsslowroll}
n_s -1 = 2\eta - 6\epsilon\,, \qquad n_t = -2\epsilon\,.
\ee

Furthermore, since observations probe just a limited range of $k$, we can express the deviation from scale-invariance by means of the power laws
\be
\Delta^2_\xi (k) = \Delta^2_\zeta (k_0) \left(\frac{k}{k_0}\right)^{n_s-1}\,,\qquad \Delta^2_h (k) = \Delta^2_h (k_0) \left(\frac{k}{k_0}\right)^{n_t}\,,
\ee
where $k_0$ is a normalization point called {\it pivot scale}. Note that we have only included the first coefficients of scale-dependence; higher-order effects lead to a scale dependence of these coefficients themselves (referred to as running). Finally, the {\it tensor-to-scalar ratio} is defined by
\be\label{rslowroll}
r\equiv \frac{\Delta^2_h (k_0) }{\Delta^2_\zeta (k_0) } = 16 \epsilon\,,
\ee
and indicates the suppression of the power of tensor with respect to scalar modes.

\subsection{Planck data}

The Planck satellite \cite{Planck:2015xua,Ade:2015lrj} has mapped the Universe with unprecedented accuracy. In this way, it has set stringent constraints on the parameters related to the inflationary dynamics. First of all, at $k_0 = 0.05\ \text{Mpc}^{-1}$, the experimental value for the scalar amplitude (first detected by COBE \cite{COBE}) is
\be\label{deltazeta}
\Delta^2_\zeta (k_0) = (2.14 \pm 0.10)\times 10^{-9}\,.
\ee

\begin{figure}[htb]
\vspace{-3mm}
\begin{center}
\includegraphics[width=10cm,keepaspectratio]{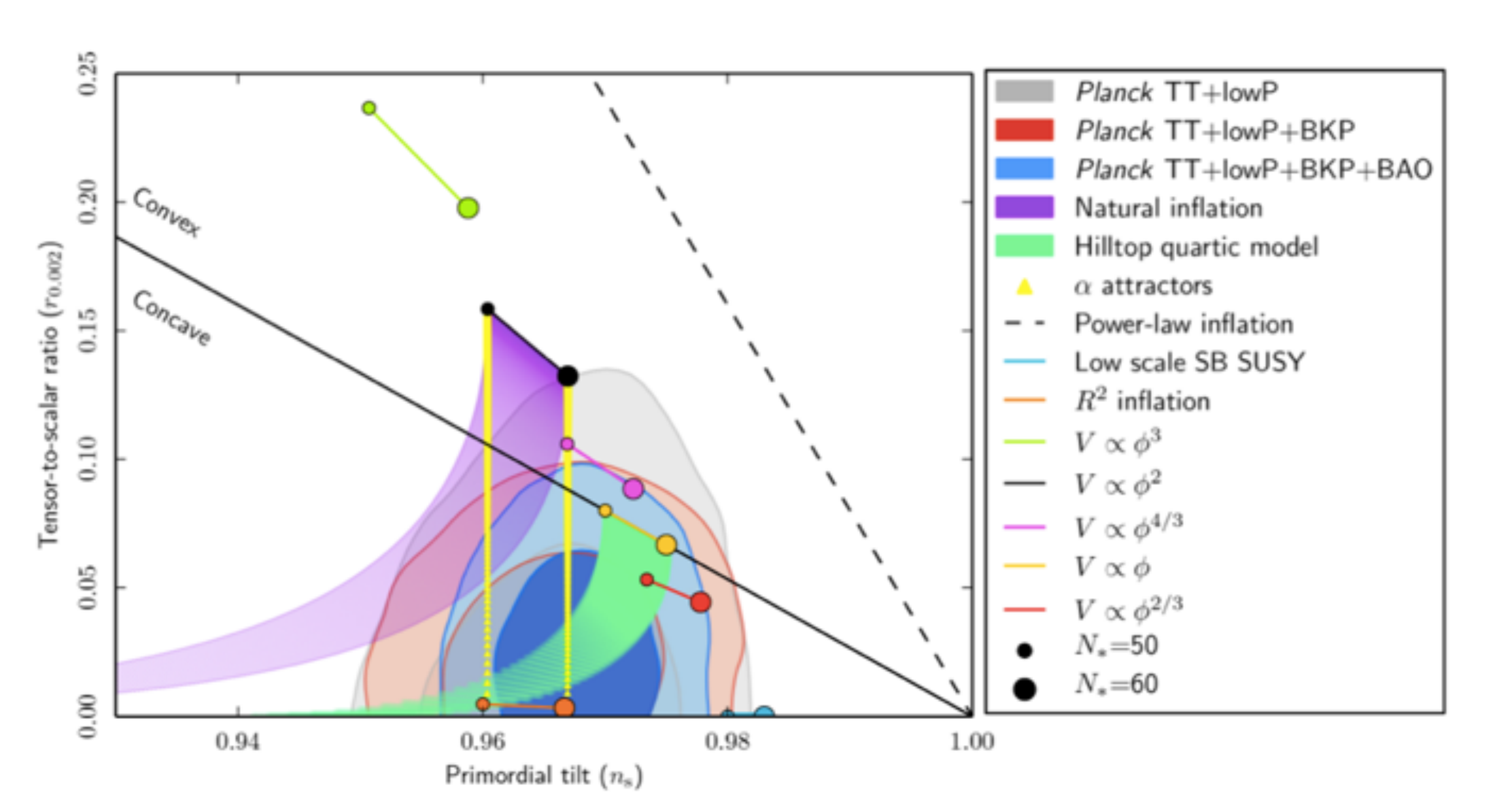}
\vspace*{0.3cm}
\caption{\it \small Planck 2015 results \cite{Planck:2015xua,Ade:2015lrj} for the spectral index and tensor-to-scalar ratio with the predictions of different inflationary models superimposed.}\label{nsr}
\end{center}
\vspace{0.cm}
\end{figure}

Secondly, the deviation from perfect scale-invariance has been definitively confirmed and the scalar spectral index $n_s$ has been measured to be \cite{Planck:2015xua,Ade:2015lrj,Ade:2015tva}
 \be\label{ns}
  n_s = 0.968\pm 0.006\ (68\% \text{CL})\,.
 \ee

\begin{figure}[htb]
\vspace{-3mm}
\begin{center}
\includegraphics[width=7cm,keepaspectratio]{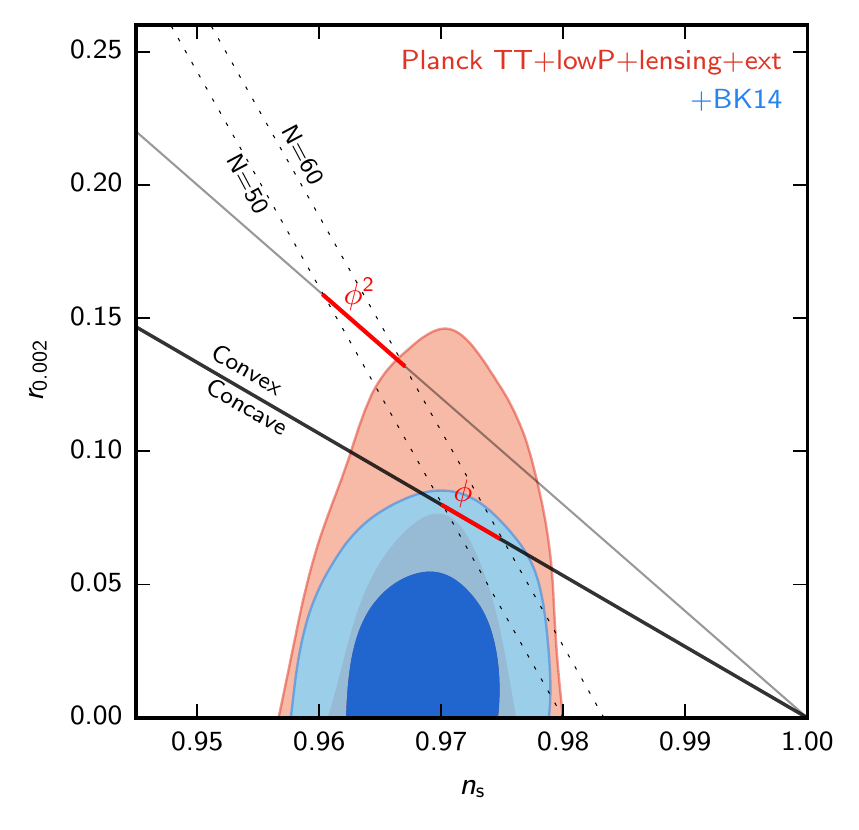}
\vspace*{0.3cm}
\caption{\it \small Improved observational constraints for the spectral index and tensor-to-scalar ratio, after including the 95 GHz data from Keck array \cite{Array:2015xqh}. The quadratic model of inflation is basically ruled out.}\label{nsrimproved}
\end{center}
\vspace{0.cm}
\end{figure}

On the other hand, the value of the tensor-to-scalar ratio has been observationally bounded to be \cite{Ade:2015tva,Array:2015xqh}
\be\label{r}
r<0.07\quad (95\% \text{CL})\,.
\ee
These can be read in Fig.~\ref{nsr}, where the predictions of different models of inflation are superimposed\footnote{Given a potential $V(\phi)$, one can calculate the observational predictions $n_s$ and $r$ by means of the formulas \eqref{nsslowroll}, \eqref{rslowroll} and \eqref{epseta}.} and in Fig.~\ref{nsrimproved}, where the constraints have been improved once including the 95 GHz data from Keck array \cite{Array:2015xqh}.

Finally, one can get an upper bound on the value of the inflationary energy at horizon crossing. This is given by
\be
V^{1/4}\simeq 1.93 \times 10^{16} \left(\frac{r}{0.12}\right)^{1/4} \ \text{GeV}\,,
\ee
which is obtained by combining Eq.~\eqref{PowerZeta} and Eq.~\eqref{deltazeta}. This value indicates that inflation should happen at very high energies (i.e. around $10^{15}-10^{16}$ GeV), below the Planck scale though. The quartic root dependence on $r$ implies that even a very small value of the tensor-to-scalar ratio (e.g. $10^{-3}$ or $10^{-5}$) does not correspond to a relevant decrease of energy. 

\clearpage
\thispagestyle{empty}


\chapter{Universality, Observations and the Inflaton Range}
\label{chapter:Universality}

\begingroup
\begin{flushright}
\vspace{.5cm}
\end{flushright}
\quote{\it The small window we can have access to via CMB measurements does not allow us to probe the full inflationary trajectory. This implies that disparate models of inflation can yield the same observational predictions and be organized in terms of universality classes, as long as they agree on the CMB window.  We present a description of the inflationary background dynamics fully in terms of the number of e-folds $N$. This becomes a very useful language in order to describe universality properties of inflation. Then, we examine the properties of the inflaton range, a variable depending on the whole trajectory. We show its degeneracy in the super-Planckian regime; namely, its value is not uniquely determined by the inflationary observables. On the other hand, we provide strong evidence for its universality properties when the value is sub-Planckian. Both the tensor-to-scalar ratio and the spectral tilt are essential for the field range. Remarkably, this results into strengthening the usual Lyth bound by two orders of magnitudes.
The novel results of this Chapter are based on the publications {\normalfont[\hyperref[chapter:Publications]{{\sc iv}}]} and {\normalfont[\hyperref[chapter:Publications]{{\sc v}}]}.}

\endgroup

\newpage

\section{The $N$-formalism of inflation}

In the previous chapters, we have expressed the time evolution of the system mainly in terms of the cosmic time $t$ or the conformal time $\tau$. However, it is possible to make other choices which may appear more natural depending on the specific context. Eventually they may lead to some simplifications in the description of the physics.

A famous example is provided by the {\it Hamilton-Jacobi formalism} \cite{Salopek:1990jq,Muslimov:1990be}. This turns out to be a very natural choice in a Universe dominated by a scalar field. The fundamental idea is that the field $\phi$ itself plays the role of an internal clock. This is possible as long as $\phi$ evolves monotonically in time, that is $\dot{\phi}\geq0$. Then, the relevant cosmological quantities can be regarded as a function of $\phi$ rather than $t$. This formalism is very useful for describing the background dynamics of inflation. However, it proves not to be suited for reheating where the field oscillates around the minimum and $\dot{\phi}$ changes sign.

Another very natural choice for describing the time evolution of an inflationary Universe is the number of e-folds $N$. This variable is directly related to the scale factor by means of Eq.~\eqref{efoldsN}, thus representing a concrete measure of time. In an expanding Universe, observers can simply use the physical growth of space as a universal clock.

The inflationary phase can then be specified fully in terms of $N$ \cite{Garcia-Bellido:2014gna}. This has the direct advantage to go beyond an explicit description of the microscopic mechanism generating the accelerated expansion and the deviation from a scale invariant spectrum.

The Hubble flow functions, defined by Eq.~\eqref{HubbleFlowFunctions}, provide the ideal tool to extract the relevant information from a description in terms of $N$ \cite{Schwarz:2001vv,Schwarz:2004tz,Martin:2013tda,Garcia-Bellido:2014eva}. The inflationary observables, in particular the spectral index and the tensor-to-scalar ratio, can then be compactly expressed as

\be\label{inflaparam}
n_s =1+\epsilon_2 -2\,\epsilon_1,    \qquad  r=16\,\epsilon_1 \,.
\ee
In order to connect these to CMB observations, one needs to evaluate these quantities at horizon crossing, denoted by $N_\star$.

It is easy to check that it is possible to recover the expressions for $n_s$ Eq.~\eqref{nsslowroll} and $r$ Eq.~\eqref{rslowroll}, given in the previous chapter in terms of the slow-roll parameters $\epsilon$ and $\eta$, simply by substituting Eq.~\eqref{flowSR} into Eq.~\eqref{inflaparam}.

The link between the formulation in terms of $\phi$ and the one in terms of $N$ is provided by the important relation
\be\label{dphi1}
\frac{d \phi}{dN} = \sqrt{2 \epsilon_1} \,,
\ee 
which is easily obtained by combining Eq.~\eqref{EPSILON1}, Eq.~\eqref{efoldsN} and Eq~\eqref{Fried2Scalar}. Note, however, that this last expression is exact and it is valid beyond the slow-roll approximation.

This last equation can be interpreted as a background field redefinition from $\phi$, with canonical kinetic terms, to the field $N$ with Lagrangian
 \begin{align}
 \mathcal{L}  = \sqrt{-g}\,[ \tfrac12 R -  \epsilon_1 (N) (\partial N)^2 - V(N) ] \,,
 \end{align}
 Generically, the functional form of the potential $V$ will be very different when expressed in terms of $\phi$ or in terms of $N$.

\section{Universality classes at large $N$}

In Sec.~\ref{OBSERV}, we have seen that the window we can probe by means of CMB observations corresponds to a small portion of the whole inflationary period. This is mainly due to the suppression of the CMB power spectrum at small angular scales, as it is shown in Fig.~\ref{power}. This sensitive region amounts to $\Delta N \simeq 7$ and it is located at present around 60 e-folds before the end of inflation (we derived this number in Sec.~\ref{amountinflation} in order to account for the homogeneity and isotropy of the CMB at its largest scale). This is indeed the point when the modes relevant for the CMB power spectrum left the region of causal physics. Its position is determined by $N$ being equal to the number of e-folds between the points $N_\star$ of horizon crossing and $N_e$ where inflation ends, that is 
\be\label{eqN}
N = N_\star - N_e\,.
\ee

\begin{figure}[htb]
\vspace{-3mm}
\begin{center}
\includegraphics[width=9.5cm,keepaspectratio]{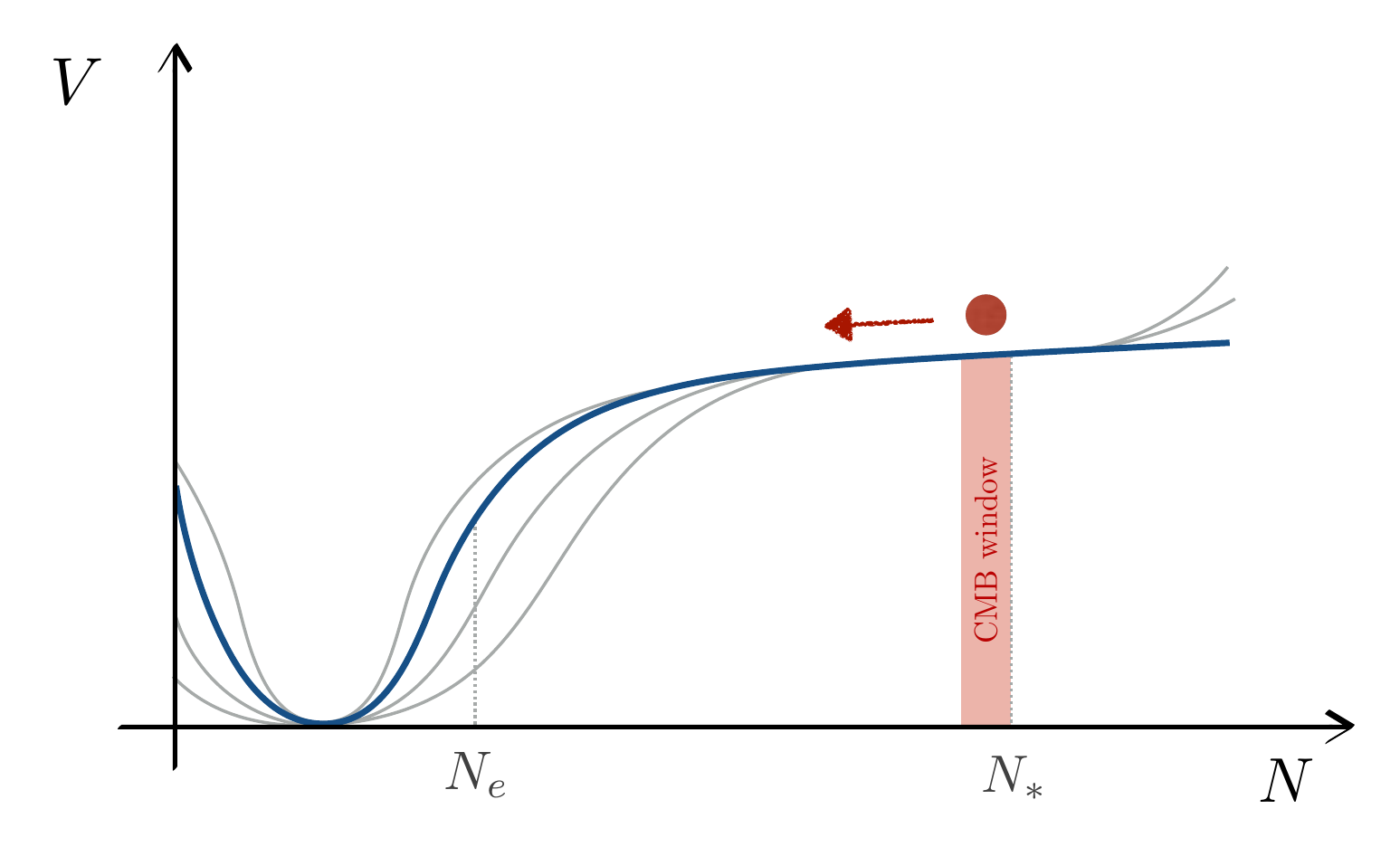}
\vspace*{0.3cm}
\caption{\it Cartoon of a typical inflationary scalar potential (blue line) with different deviations (grey lines). The details of the models are different but they agree on the CMB window thus yielding identical observational predictions.}\label{Nlarge}
\end{center}
\vspace{0.cm}
\end{figure}

Then, the measured values of the cosmological parameters Eq.~\eqref{ns} and Eq.~\eqref{r} constrain the form of the scalar potential just on a limited part. The practical situation is that several scenarios can give rise to the same predictions despite the details of the specific model. This situation is visually explained in Fig.~\ref{Nlarge}.

A lower limit on $N$ can be set by the temperature of reheating \cite{Dai:2014jja} (see also \cite{Allahverdi:2010xz, Martin:2010kz, Adshead:2010mc,Mielczarek:2010ag,Dodelson:2003vq,Easther:2011yq}). On the other hand, there is no compelling reason to assume the number $N$, quantifying the amount of exponential expansion of the Universe, has an upper bound; in fact, it seems natural that inflation extends a long way further into the past than the portion we can observe (see \cite{Remmen:2014mia} for a study on this topic). 

The above argument seems to suggest $1/N$ as a natural small parameter to expand our cosmological variables \cite{Mukhanov:2013tua,Roest:2013fha,Garcia-Bellido:2014gna} (see also \cite{Boyanovsky:2005pw}). This approach is also motivated by the percentage-level deviation of the Planck reported value for the spectral index \eqref{ns} from unity which can be interpreted as
\be \label{nsN}
n_s= 1 -\frac{2}{N}\,.
\ee

This argument naturally leads to assume the function $\epsilon_1$ (or equivalently the first slow roll parameter $\epsilon$) scaling as
\be\label{epsilonN}
\epsilon_1 = \frac{\beta}{N^p}\,,
\ee 
where $\beta$ and $p$ are constant and we have neglected higher-order terms in $1/N$ as not relevant for observations. This simple assumption leads to the so-called {\it perturbative} class defined by
\begin{align} \label{puniv}
   r =  \frac{16\beta}{N^p} \,, \qquad
   n_s=\begin{cases}
1 - \frac{2\beta+1}{N} \,, \quad &p=1\,, \\
1 - \frac{p}{N} \,, \quad &p>1\,, 
 \end{cases}
\end{align}
where we have discarded the case $p<1$ as generically not compatible with the current cosmological data. Eq.~\eqref{puniv} identifies the families of universality classes which any specific scenario belongs to, for fixed values of $\beta$ and $p$.

Most of the inflationary models in literature have an equation of state parameter scaling as a power of $1/N$, thus falling into the perturbative class. These includes the chaotic monomial inflation scenarios, the Starobinsky model, hilltop models and many others. It is possible to consider also other functional forms for $\epsilon_1$, as it was investigated in \cite{Garcia-Bellido:2014gna} (see also \cite{Binetruy:2014zya} for a related analysis with a different approach). However, the cosmological predictions of these classes are generically more in tension with the observational data. Further, the number of well motivated models of inflation falling into these other classes is more restricted.

The analysis at large-$N$ proves to be a powerful tool in order to organize different inflationary models just in terms of their cosmological predictions. Physically different scenarios may predict the same values of $n_s$ and $r$ in the leading approximation in $1/N$. The details of any specific model, encoded in the subleading terms (higher powers of $1/N$), are washed out and not relevant for the observational predictions. Examples of these circumstances are listed in \cite{Roest:2013fha,Garcia-Bellido:2014gna}.

\begin{figure}[h]
\vspace{3mm}
\begin{center}
\includegraphics[width=8cm,keepaspectratio]{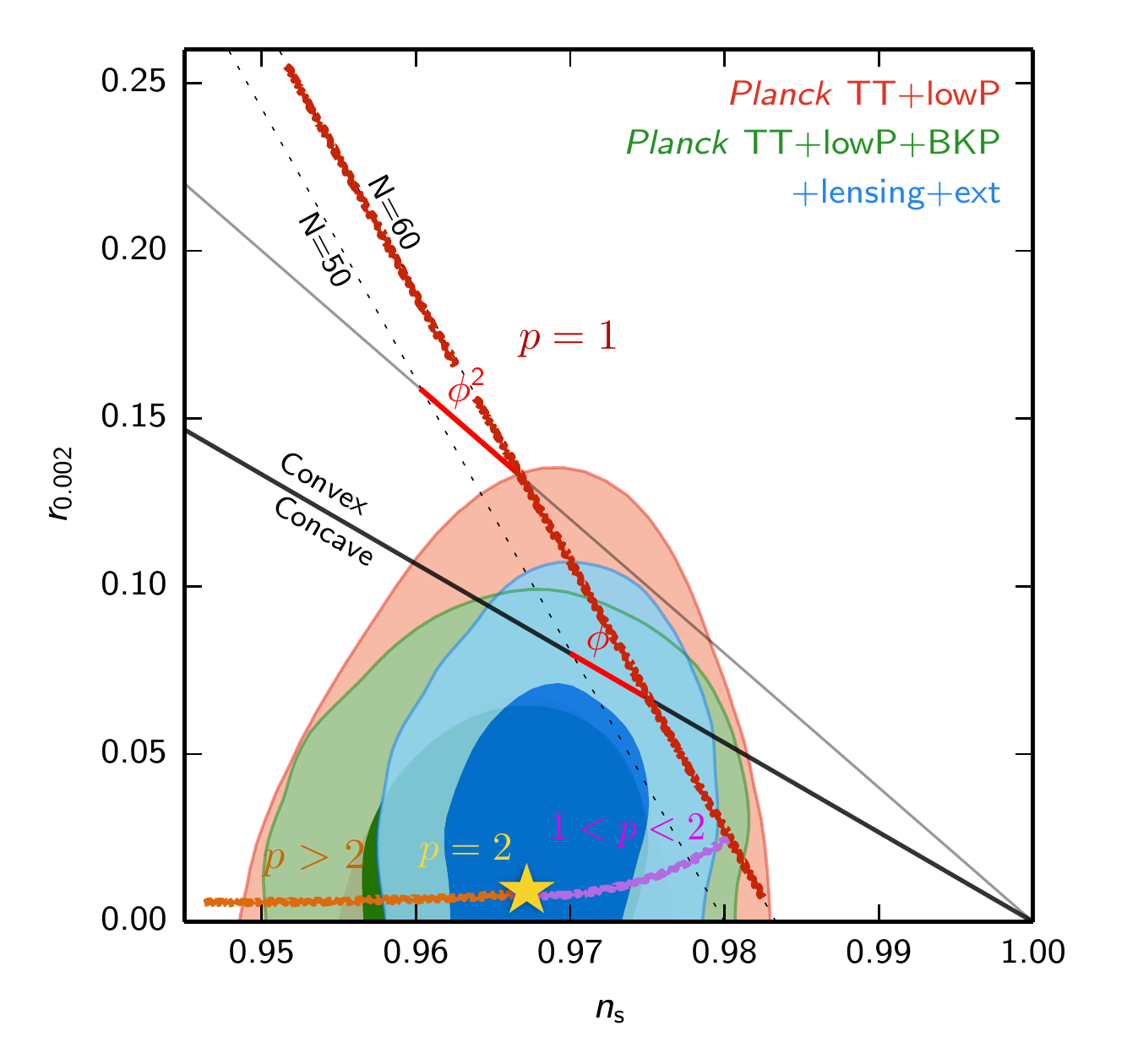}
\vspace*{0.3cm}
\caption{\it Predictions of the inflationary scenarios with equation of state parameter given by Eq.~\eqref{epsilonN} superimposed over the Planck data. Given the favored value of the spectral index Eq.~\eqref{nsN}, one has generically a forbidden region for value of the tensor-to-scalar ratio $r$. }\label{nsrN}
\end{center}
\vspace{0.cm}
\end{figure}

Now, we want to show that the simple assumption of a scalar spectral tilt scaling as $1/N$ can exclude a consistent region of the $(n_s, r)$ plane and yield definite predictions for our cosmological variables \cite{Roest:2013fha,Creminelli:2014nqa}. The allowed regions are shown in Fig.~\ref{nsrN}. In particular, given the best fit value for $n_s$ and the strict bound on $r$, we will generically expect a very low value for the tensor-to -scalar ratio, probably order $10^{-3}$. 

Finally, in a pure large-$N$ description, one can identify the benchmark potentials for this Ansatz. Let us recall the relation between the Hubble parameter $H$ and $\epsilon_1$ given by Eq.~\eqref{EPSILON1}.  Within the slow-roll approximation, employing $H^2=V/3$, one can integrate this equation and obtain an expression for the potential in terms of $N$ which reads
\begin{align} \label{VN}
V(N)=\begin{cases}
V_0\,N^{2\beta} \,, \quad &p=1\,, \\
V_0\left[1 - \frac{2\beta}{(p-1) N^{p-1}}\right] \,, \quad &p>1\,,
 \end{cases}
 \end{align}
where $V_0$ is an integration constant related to the energy scale of inflation.  By means of Eq.~\eqref{dphi1} and Eq.~\eqref{epsilonN}, one  gets the asymptotic form of $V$ in terms of the canonical scalar field $\phi$, that is
\ba \label{VP}
V(\phi)=\begin{cases}
V_0\,\phi^n \,, \quad &p=1\,, \\
V_0\left[1-\exp\left(- \phi/\mu\right)\right] \,, \quad &p=2\,, \\
V_0\left[1-\left(\phi/\mu\right)^{n}\right] \,, \quad &p>1\,, p\neq2\,,
 \end{cases}
 \ea
where $\mu$ and $n$ are related to $\beta$ and $p$ as dictated by \eqref{dphi1}. In particular, for $p>1$ and $p\neq2$, the power  $n$ is related to $p$ through the following equation
\be\label{np}
n=\frac{2(1-p)}{2-p}\,,
\ee 
where $p<2$ or $p>2$  determine respectively the negative or positive sign of $n$. The inverse relation $p=p(n)$ turns out to be of the same form. 

In the large-$N$ limit, any model belonging to these universality classes will have a potential asymptotically approaching well-known scenarios such as chaotic monomial inflation ($p=1$), inverse-hilltop models ($1<p<2$), Starobinsky-like inflation ($p=2$) and hilltop potentials ($p>2$). As already explained, the reason for such simplicity is that, in this limit, we are probing just a limited part of the inflationary trajectory,  close to  horizon crossing. Peculiarities among different models appear  when we go away from this region. In general, the situation near the end-point of inflation will be very different from one model to another, even though they belong to the same universality class.

\section{The inflaton range and observations}\label{inflatonrangeobservations}

In this Section, we intend to examine features of a variable crucial for the construction of inflationary models at high energies, namely the inflaton field range $\Delta\phi$. A crucial distinction is indeed between small- and large-field models, defined by sub- and super-Planckian field ranges. Generic quantum corrections to a tree-level scalar potential come in higher powers of $\phi$, and hence large-field models are particularly sensitive to these. This puts the consistency of an effective field theory description \cite{Cheung:2007st,Weinberg:2008hq,Senatore:2010wk} of such models into doubt. A key question in theoretical cosmology is therefore whether $\Delta\phi$ exceeds the Planck length or not.

\begin{figure}[htb]
\vspace{-3mm}
\begin{center}
\includegraphics[width=9.5cm,keepaspectratio]{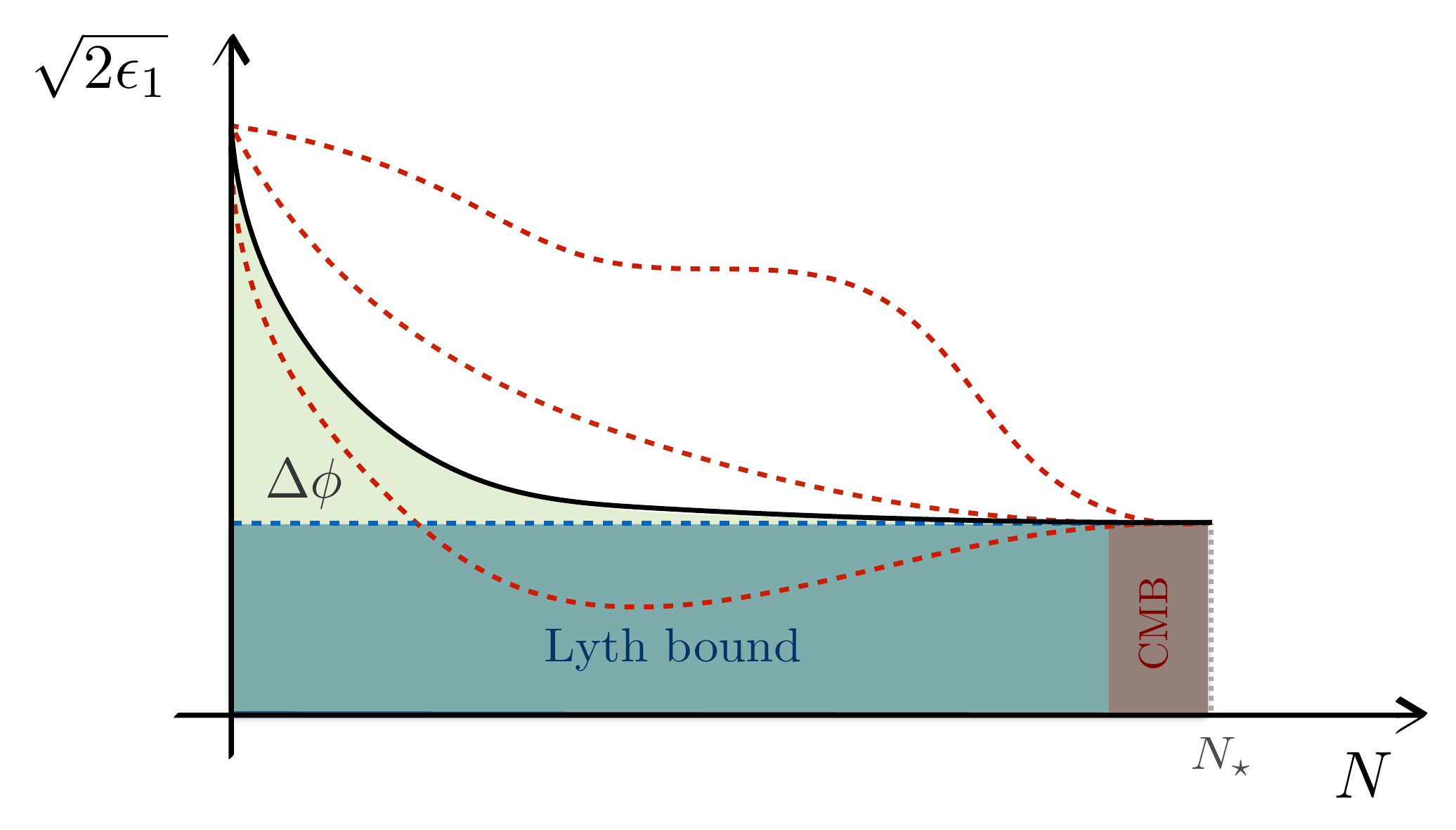}
\vspace*{0.3cm}
\caption{\it Identical observational predictions can correspond to very different inflaton excursions, which are given by the area underneath the curves. Although the solid black line and its variations (red dashed lines) agree in the CMB window, they have very different integrals. The blue rectangular area depicts the Lyth bound.}\label{DeltaPhiPicture}
\end{center}
\vspace{0.cm}
\end{figure}

Knowledge of the evolution of $\epsilon_1(N)$, during all e-foldings $N$ of the inflationary period, determine the field range by means of Eq.~\eqref{dphi1}. Therefore, it is the area underneath the curve $\sqrt{2 \epsilon_1(N)}$ which determines the excursion of the scalar field $\phi$ during the expansion. However, cosmological observations allow us to constrain just a small part. The situation resembles what already seen in the previous section: generally it is not possible to uniquely connect CMB data with a precise value of the inflaton excursion. This is depicted in Fig.~\ref{DeltaPhiPicture}.

However, a first estimate of $\Delta \phi$ can be obtained by the assumption that $\epsilon_1(N)$ is constant throughout inflation. This is referred as the {\it Lyth bound}\footnote{To be more precise, Lyth's analysis concerns just the small window accessible via CMB observations. One must note that in 1997, at the time of publication of his paper, experiments could probe just around $\Delta N\simeq 4$. This certainly leads to a milder bound than Eq.~\eqref{Lbound}. On the other hand, this result is always valid within the slow-roll approximation and does not make any assumption on the form of $\epsilon_1(N)$.} \cite{Lyth:1996im} and leads to \cite{Boubekeur:2005zm,Boubekeur:2012xn}:
\be \label{Lbound}
\Delta\phi \sim \left(\frac{r}{0.002}\right)^{1/2} \left(\frac{N_\star}{60}\right) \,,
\ee
where we have set the number of e-folds at horizon exit $N_\star$ equal to $60$ (other values allow for a similar analysis). Therefore, a sub-Planckian excursion for the inflaton field requires a very small value of $r\lesssim 2 \cdot 10^{-3}$. For monotonically increasing $\epsilon_1(N)$, this represents a lower bound. The blue rectangular area in Fig.~\ref{DeltaPhiPicture} provides a visual representation of this bound.

In the following, we start  the discussion with some specific examples where a point in the $(n_s, r)$ plane does correspond to a wide spectrum of values for $\Delta\phi$. We consider chaotic inflation models with monomial potentials as the benchmark scenarios to show such a {\it degeneracy} of the inflaton excursion. Intriguingly, we find that the field range cannot exceed an upper-bound due to the slow-roll conditions.

On the other hand, in the sub-Planckian regime and for a range of universality classes, we prove that it is possible to precisely connect observations to a unique value of $\Delta\phi$. Information on both the tensor-to-scalar ratio and the spectral tilt uniquely determines the value of $\Delta\phi$. This remarkable {\it universality} of the inflaton range will lead to a stronger bound than the usual estimate given by Eq.~\eqref{Lbound}.

\section{Degeneracy of the inflaton range}\label{SecDegeneracy}

We now discuss the field range in different classes of models. In particular, we are interested in exploring the correspondence between a specific point in the $(n_s,r)$ plane and the values of $\Delta \phi$. We will prove that it is possible to have exactly the same cosmological predictions, in terms of the scalar tilt and the amount of gravitational waves, while the field excursion may vary over several orders of magnitude.

For simplicity we will consider the monomial inflation scenarios as benchmark models for our study. However, note that other models can be straightforwardly studied following the same reasoning.

In the following, we analyze three classes of inflationary models with a specific dependence on $N$ for the Hubble flow parameters. Such classes, discussed at length in \cite{Garcia-Bellido:2014gna}, reproduce the large-$N$ behavior of most of the inflationary models available in the literature.

As a first case, we  discuss the so-called {\it perturbative} class, characterized by a leading term in $\epsilon_1$ scaling as $1/N^p$, with $p$ being a constant positive coefficient. Then, we analyze models where {\it logarithmic} terms, such as $\ln^q(N+1)$,  appear in the leading part of $\epsilon_1$. In a  third class of models,  we consider the parameter $\epsilon_1$ having a {\it non-perturbative} form, in the limit at large-$N$, of the type $\epsilon_1 \sim \exp(-cN)$. We will consider the possibility of letting the total number of e-folds $N$ vary over a certain interval which is related to reheating details of the specific model. Interestingly, we find an {\it upper bound} on $\Delta\phi$ and the total number of e-folds which sets connections among the three classes of models considered.

As final part of our analysis, we focus on the logarithmic class and we explore the possibility of playing  with the power coefficient $q$, while keeping $N$ fixed. This is an alternative way to get the same predictions of quadratic inflation, while having quite different values for the inflaton range. We will consider the possibility of going beyond single-field and/or slow-roll inflation and getting a sub-Planckian $\Delta \phi$.

Throughout this section, we assume that the inflationary parameters $\epsilon_1$ and $\epsilon_2$ of each class are exact over the whole inflationary trajectory, as it happens for chaotic scenarios. In several cases, this may be a very good approximation and may capture most of the essential properties of the models falling into the specific universality classes. Anyhow, we will take advantage of a formulation purely in terms of $N$ and extract the information we are interested in, without referring to the particular form of the scalar potential $V(\phi)$. In fact, for any specific parametrization of each class, the latter may be very complicated when expressed in terms of the canonical scalar field $\phi$.

In what follows, the benchmark will be the value of $\Delta \phi$ for chaotic models corresponding to a quasi exponential expansion of $N=60$. This sets
\be \label{Nstpert}
N_\star= N_e + 60\,,
\ee
 as corresponding to horizon exit. Moreover, all symbols with a tilde  will be reserved for the classes being examined, while the benchmark models will have no tilde.  

\subsection{Chaotic inflation as benchmark}

Chaotic scenarios are usually characterized by monomial potentials when expressed in terms of the canonical scalar field $\phi$. Further, they naturally lead to a large value of $r$ together with a super-Planckian excursion of the inflaton field.

In a large-$N$ description, the first three Hubble flow functions turn out to be
\be\label{chaopar}
\epsilon_0= h N^\beta\,,    \qquad \epsilon_1= \frac{\beta}{N}\,,    \qquad  \epsilon_2=-\frac{1}{N}\,,
\ee
where $h$ is an integration constant and $\beta$ is related to the specific universality class.

The description in terms of $N$ is exact for these models (there are no subleading corrections) and hence captures all of their fundamental features. However, even if there would be subleading corrections, e.g.~at the level $1/N^2$, observables calculated at horizon exit, such as $n_s$ and $r$, will be observationally insensitive to these (as they are too much suppressed for $N \gtrsim 50$). Therefore, these are universal predictions of entire classes of models that agree in the large-$N$ limit. 

The same universality holds for the inflaton range. In the case of chaotic models with parameters \eqref{chaopar}, the inflaton excursion $\Delta \phi$ will be basically determined just by the leading term in $N$ \cite{Garcia-Bellido:2014wfa} through Eq.~\eqref{dphi1}.

As these models receive most of their e-foldings at large-$N$, one can safely assume that restricting to the leading term of $\epsilon_1$ is a very good approximation over the relevant part of the inflationary trajectory. The expression for the inflaton field range will therefore read
\be\label{DPchaotic}
 {\Delta \phi}_c = 2 \sqrt{2\beta} \left( N_\star ^{1/2}-N_e^{1/2} \right)\,,
 \ee
where the subscript $c$ is added in order to refer more easily to the benchmark field excursion of monomial models throughout the following part of the thesis. Further, $N_e=\beta$, when assuming that inflation ends at $\epsilon_1=1$, and $N_\star$ is found through Eq.~\eqref{eqN}. 

With the above relations, potentials of the type $V (\phi)= \lambda_n \phi^n$ will keep monomial form even when formulated in terms of $N$, namely $V(N)=h^2 N^{2\beta}$, and vice versa. The relation between the two power coefficients reads
\be
\beta =\frac{n}{4}\,,
\ee
and can be found by using Eq.~\eqref{dphi1}. As an explicit example, a quadratic potential  corresponds to $\beta=1/2$ and an inflationary period of $N=60$ leads to $\Delta \phi  \simeq 14.14$. Of course, this is identical to the value of $\Delta \phi$ calculated through the scalar potential $V$, within the slow-roll paradigm.

\subsection{Perturbative class} \label{PClass}

We start considering the possible degeneracies within the perturbative class of models. In this case, the relevant Hubble flow parameters for determining the observational data  have the following $N$-dependence:
\be\label{pertpar}
\epsilon_1= \frac{\tb}{N^p}\,,    \qquad  \epsilon_2=-\frac{p}{N}\,.
\ee
The case discussed above is easily recovered for $p=1$ and $\tb=\beta$.

\begin{figure}[htb]
\hspace{-3mm}
\begin{center}
\includegraphics[width=8.5cm]{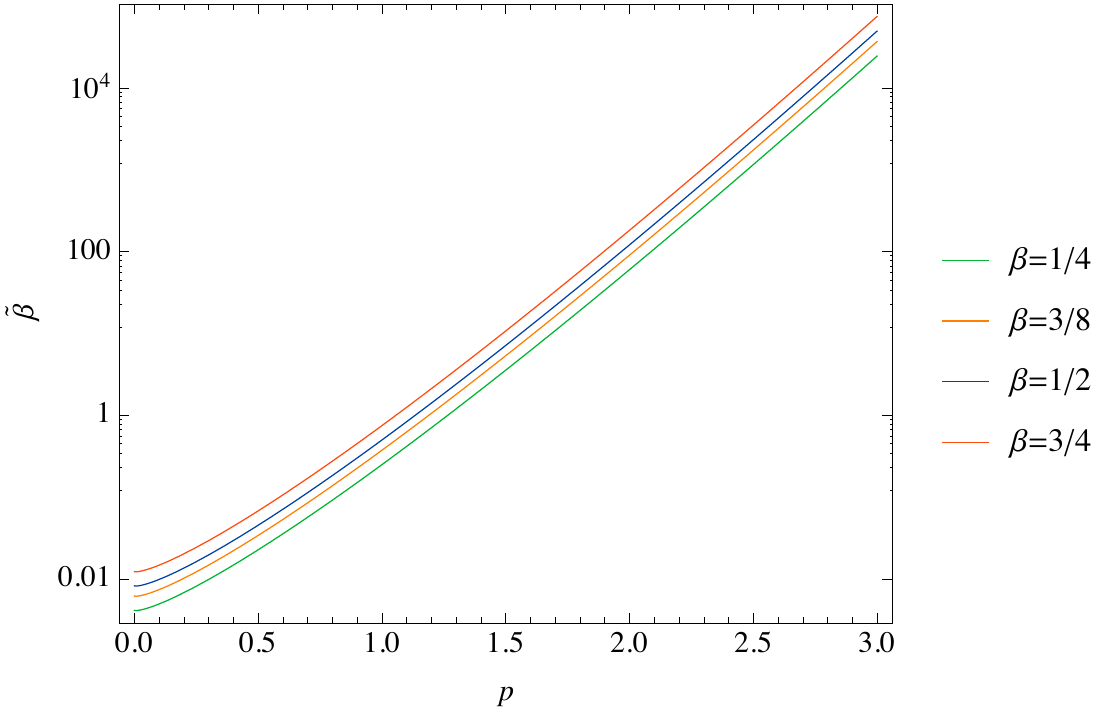}
\vspace*{0.4cm}
\caption{\it Behavior of $\tb$ as function of $p$ in a log-plot. It generally blows up for $p\gtrsim2$, where the  perturbative model should be highly fine-tuned in order to reproduce the same $(n_s,r)$ of chaotic scenarios. The four lines correspond to the same observational predictions of models with potential of the type $V = \lambda_n \phi^n$, with $n$ respectively equal to $1$, $3/2$, $2$ and $3$. }\label{FIGbeta}
\end{center}
\end{figure}

We would like to reproduce the same $n_s$ and $r$ of the benchmark chaotic model through a generic pertubative model with $p\neq1$. This translates into equating both $\epsilon_1$ and $\epsilon_2$ of \eqref{chaopar} to the functions \eqref{pertpar} at horizon exit, respectively at $N_\star$ and $\tNs$. As result, we have the following relations:
\ba\label{btper}
\tb &=& \beta\ \frac{\tNs ^p }{N_\star} \,, \qquad
p =  \frac{\tNs}{N_\star}\,.
\ea 
This allows  to express $\tb$ as
\be\label{beta}
\tb = \beta\ p^p N_\star^{p-1} \,,
\ee
where $N_\star$ is given by \eqref{Nstpert}. Eq. \eqref{beta} gives us an estimate of how fine-tuned the model is in order to reproduce the same predictions of the chaotic models. Curiously, for any  $\beta=\mathcal{O}(1)$ (corresponding to different chaotic models), the corresponding perturbative model will start to be severely fine-tuned in the region $p>2$, as is shown in Fig.~\ref{FIGbeta}.

\vspace*{-0.5mm}

\begin{figure}[htb]
\hspace{-3mm}
\begin{center}
\includegraphics[width=8.5cm]{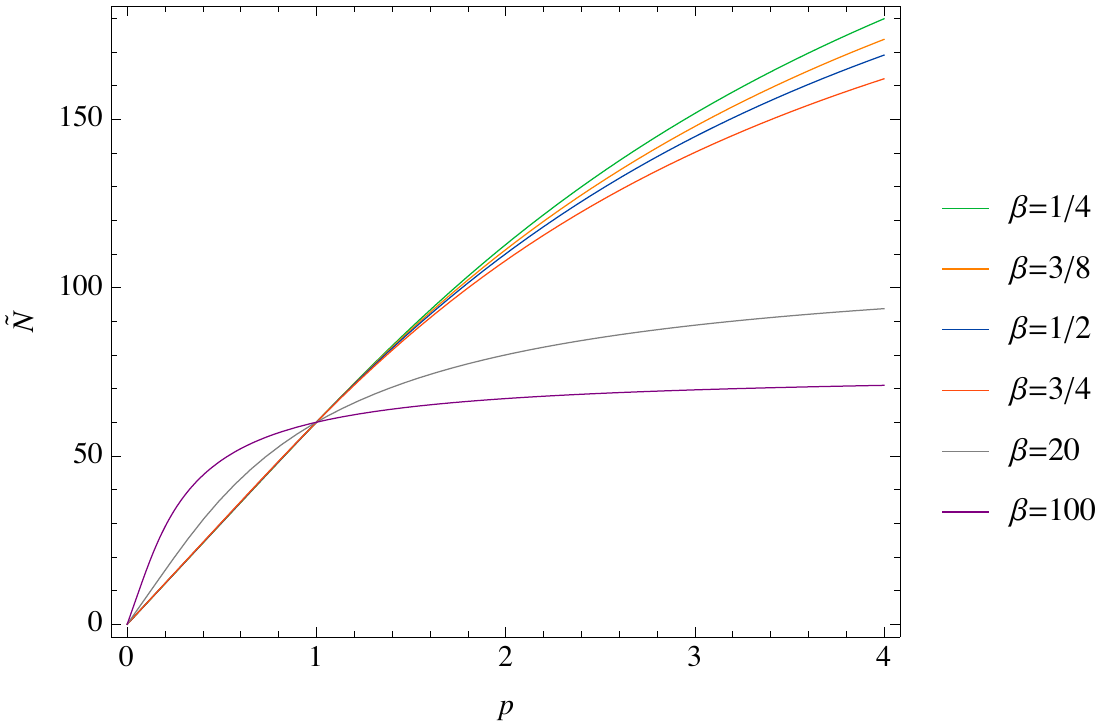}
\vspace*{0.4cm}
\caption{\it The total number of e-folds $\tN$ as function of $p$ for models belonging to the perturbative class. Lines follow a linear relation for low values of $p$ and $\beta=\mathcal{O}(1)$. For larger values of $\beta$, lines have different behaviors, as shown by the grey and purple lines. In this case, the unique intersection with the value $p=1$ becomes evident.} \label{FIGnpert}
\end{center}
\end{figure}

Demanding that inflation ends at $\epsilon_1=1$ turns into 
\be\label{Neper}
\tilde{N}_e = \tNs - \tilde{N} = \tb ^{1/p} \,,
\ee
where the total number of e-foldings $\tilde{N}$ in principle could span a range of different values related to reheating properties of the model. Using \eqref{btper}, Eq.~\eqref{Neper} gives us the functional form of the total number of e-folds $\tilde{N}$ as a function of $p$, for any $\beta$, that is
\be\label{Nper}
\tilde{N} = p \,N_\star \left[1-  \left(\frac{\beta}{N_\star}\right)^{1/p}\right]\,.
\ee
A period of inflation $\tilde{N}=60$ necessarily corresponds to $p=1$, which is the benchmark of our analysis. For any other value of $\tilde{N}$, there exist several possibilities with $p\neq1$, reproducing exactly the same predictions of chaotic models, while having a viable mechanism to end inflation ($\epsilon_1=1$). Nevertheless, in the region $p<2$ and for $\beta=\mathcal{O}(1)$, solutions in $p$ are highly close together and they follow a linear relation, as shown in Fig.~\ref{FIGnpert}.

\begin{figure}[htb]
\hspace{-3mm}
\begin{center}
\includegraphics[width=7.6cm]{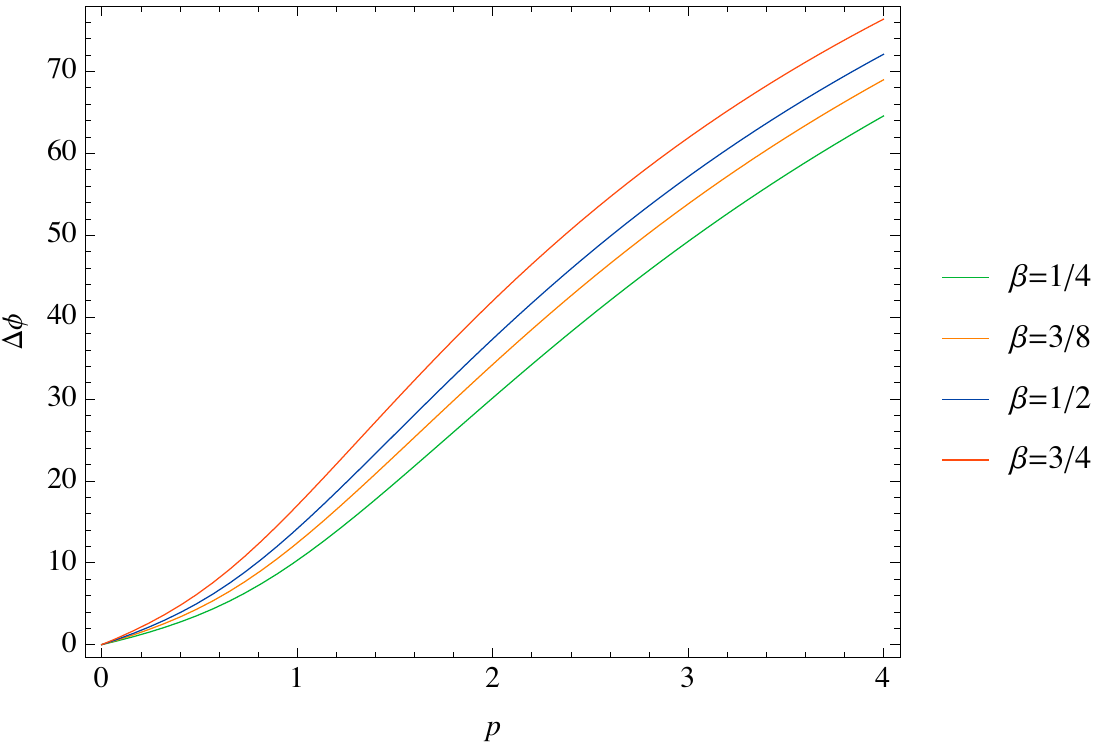}
\includegraphics[width=7.6cm]{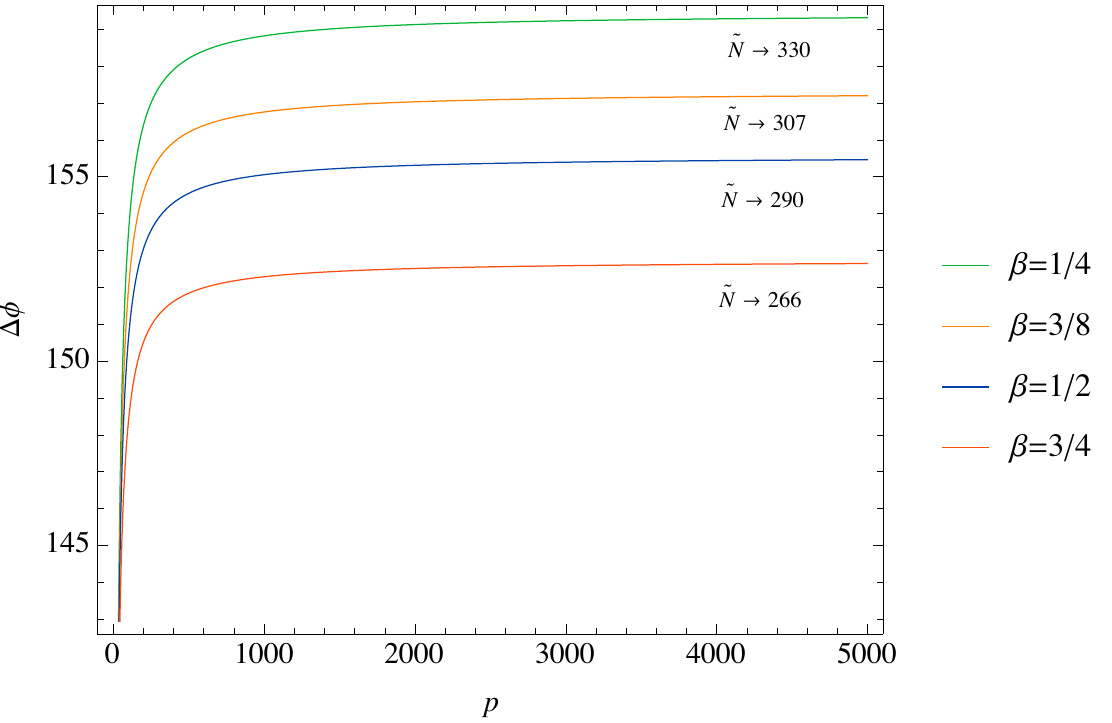}
\vspace*{0.4cm}
\caption{\it The inflaton range $\Delta\phi$ as function of $p$ in two different limits for the pertubative class of models. For low values of $p$ (no fine-tuning), $\Delta\phi$ may vary over a range related to the total number of e-folds $\tN$. In the large $p$ region, the upper bounds on $\Delta\phi$  becomes evident.} \label{FIGdeltapert}
\end{center}
\end{figure} 

The inflaton range can be easily computed by integrating Eq.~\eqref{dphi1} and one obtains
\be
\Delta \phi = \frac{2\sqrt{2\tilde\beta}}{2-p}  \left(\tilde N_\star^{1-\frac{p}{2}} -\tilde N_e^{1-\frac{p}{2}}\right)\,.
\ee
The latter formula can be written as function of $p$, for any value of $\beta$. By substituting \eqref{btper}, one gets
\be\label{DPpert}
\Delta \phi = 2\sqrt{2\beta} \frac{p}{2-p}  \left[ N_\star^{\frac{1}{2}}- \beta^{\frac{2-p}{2p}}N_\star^{\frac{p-1}{p}}\right]\,.
\ee
Figure \ref{FIGdeltapert} shows the main results on the inflaton range, given by \eqref{DPpert}, for models belonging to the perturbative class. At this point, we identify the two regions and get the following conclusions:
 \begin{itemize}
 \item For small values of $p$, the inflaton excursion $\Delta\phi$ is a continuously increasing function. It has a typical dependence $p/(2-p)$ in the region $p\lesssim1$, where the first term of \eqref{DPpert} dominates over the second one; it has a mild transition for $1\lesssim p \lesssim2$, while it starts to show a really different behavior in the region\footnote{The value $p=2$ is special as the two contributions of Eq.~\eqref{DPpert} become the same while the factor $p/(2-p)$ blows up.} $p>2$. The field range covers a wide spectrum of values depending on the total number of e-folds $\tN$ of this perturbative class. As a consequence, it can be quite different from the corresponding chaotic one, which is given by $p=1$. In particular, we can reproduce the same values $(n_s,r)$ of a quadratic potential with $N = 60$ and still have a $\Delta\phi$ running from  5 to 32, in Planck units, corresponding to $\tN$ approximately between 30 and 100. Note that $p$, as well as $\Delta\phi$, cannot be arbitrarily small as we need a minimum amount of exponential expansion, quantified by $\tN$.
 
 \item For large values of $p$, the inflaton range approaches a constant value, setting an {\it upper bound} on $\Delta\phi$ for each specific value of $\beta$.  This can be  seen explicitly by taking the limit of  \eqref{DPpert} for $p\to\infty$, this becomes:
  \be\label{dfbound}
 \Delta \phi \to 2\sqrt{2}N_\star^{1/2}\left[N_\star^{1/2} -\sqrt{\beta}\right]\,.
\ee
 This corresponds to an upper bound also on $\tN$, as can be seen again by taking the limit for $p\to \infty$ of equation \eqref{Nper}, which gives:
 \be\label{Nbound}
 \tilde N \to  N_\star \ln \frac{N_\star}{\beta} \,.
 \ee
 This limit  cannot be appreciated in Fig.~\ref{FIGnpert}, given the reported limited range of $p$. Plugging the values of the parameters for quadratic inflation into \eqref{dfbound}  and \eqref{Nbound}, one gets the approximate bounds\footnote{Such large values of $\tilde N$ are not necessarily realistic (see e.g.~the discussion in \cite{Liddle:2003as} for an upper estimate); nevertheless, it is interesting to study the behaviour of the field range for such models.}
 \be
\Delta\phi\to 155.56\,, \qquad \tN\to 290\,.\label{boundquadr}
\ee
Curiously, the hierarchy of ranges is inverted with respect to the one present at small $p$, as it is clear by comparing the two pictures of Fig.~\ref{FIGdeltapert}: at higher values of the tensor-to-scalar ratio $r$, we have smaller ranges. 
 
 We will see that the bounds for $\Delta\phi$ and $\tN$ found here are recovered in the next two cases we consider in  Sec.~\ref{Log} and \ref{NP}, within the analysis of the logarithmic and non-perturbative classes of models. 
\end{itemize}

\subsection{Logarithmic class}\label{Log}

As a second case, we consider models with a first subleading correction to the Hubble flow parameters. While still neglecting higher order $1/N$ terms, one can imagine including a logarithmic dependence on $N$ such as
\be \label{logep1}
\begin{aligned}
\epsilon_1 &= \frac{\tb}{N^p\ln^q(N+1)}\,,\\
\epsilon_2 &= -\frac{p}{N} -\frac{q}{(N+1)\ln(N+1)} \,.
\end{aligned}
\ee
Inflationary models having similar dependence can be found e.g. in \cite{Roest:2013fha}.

As in the previous case, in order to mimic the observational predictions of chaotic models in terms of ($n_s,r$), we equate ($\epsilon_1$, $\epsilon_2$) of \eqref{chaopar} to \eqref{logep1} at horizon exit, respectively at $N_\star$ and $\tNs$. As result, we obtain
\ba\label{deg2}
&&\tb = \beta\ \frac{\tNs^p\ \ln^q(\tNs +1)}{N_\star} \\
&&  q= \left(\frac{1}{N_\star}-\frac{p}{\tilde N_\star}\right)(\tilde N_\star + 1) \ln(\tilde N_\star +1)\label{deg3}
\ea
where  $\tNs = \tilde N_e + \tN$ and $\tilde N_e$ is determined by the condition $\epsilon_1=1$:
\be\label{Nf2}
\frac{\tilde \beta}{\tilde N_e^p \ln^q(\tilde N_e +1)} =1\,.
\ee

We follow the same approach as in the perturbative case and allow $\tilde N$  to vary as function of $p$, while fixing $q$. The range of the inflaton $\Delta \phi$ can be determined by integrating \eqref{dphi1} as before. However, we have to rely on numerics as obtaining an analytic expression both for $\tN$ and $\Delta\phi$ turns out to be not as trivial as in the previous case. For this reason, we restrict our analysis just to the benchmark of a quadratic potential, namely just to $\beta=1/2$.

\begin{figure}[htb]
\hspace{-3mm}
\begin{center}
\includegraphics[width=8.5cm]{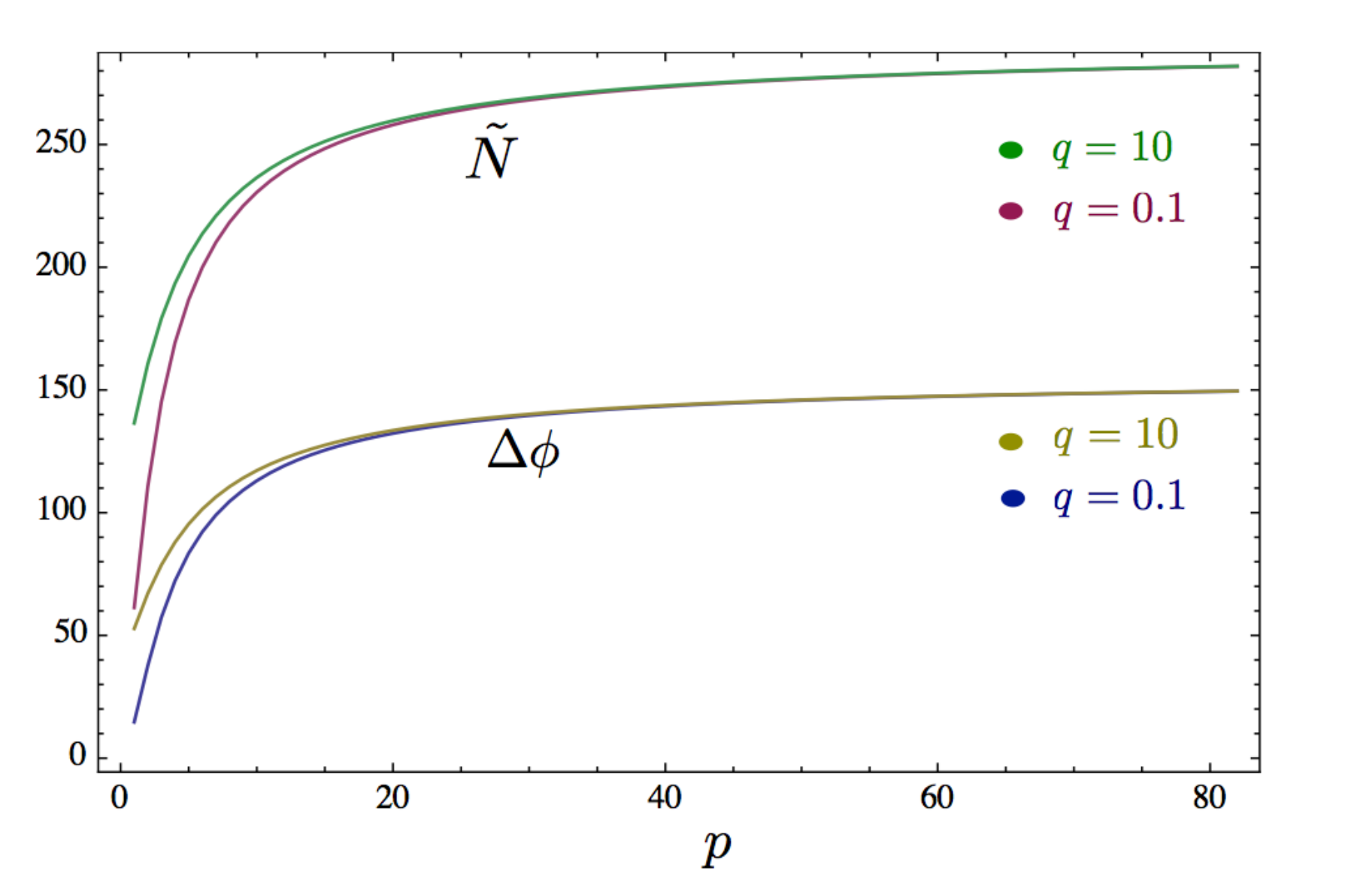}
\vspace*{0.4cm}
\caption{\it Field range $\Delta\phi$ and total number of e-folds $\tilde N$ as functions of $p$ in the logarithmic class, for two fixed values of $q$. The lines correspond to the same predictions of quadratic inflation ($\beta=1/2$). The bounds on $\Delta \phi$ and $\tN$ can be appreciated at large values of $p$.}
\label{FRLogdNJoined}
\end{center}
\end{figure}

The results for the field range and the total number of e-folds are summarized in Fig.~\ref{FRLogdNJoined}, for two different values of $q$. As we can see, for large values of $p$, we recover exactly the same bounds \eqref{boundquadr} found within the analysis of the perturbative class. This is a remarkable result, though it may be understood from the large $p$ behaviour of $\epsilon_1$. In this limit, $\tN$ also increases and hence subleading terms, in $\epsilon_n$ for $n \geq 2$, will be increasingly irrelevant. The two lines in Fig.~\ref{FRLogdNJoined}, corresponding to different values of $q$, do differ for smaller values of $p$. However, they show identical behavior when $p$ increases, which correspond to a large-$\tN$ limit.

\subsection{Non-perturbative class}\label{NP}

As a third class, we consider models with Hubble flow functions such as
\be\label{npertpar}
\epsilon_1= \ e^{-2cN}\,,    \qquad  \epsilon_2=-2 c\,,
\ee
where $c$ is a constant. Note that we are not including any coefficient for $\epsilon_1$ as this can be set equal to one by a shift in $N$.

We proceed as in the previous cases by equating ($\epsilon_1$, $\epsilon_2$) of \eqref{chaopar} to the functions \eqref{npertpar} at horizon exit, in order to reproduce the same observational predictions of chaotic inflation models. We get the following relations:
\ba\label{btnper}
\tNs &=& \frac{1}{2c} \ln\frac{N_\star}{\beta} \,, \qquad
c =  \frac{1}{2N_\star}\,. 
\ea 
Moreover, imposing that inflation ends at $\epsilon_1=1$ translates into
\be\label{Nenper}
\tilde{N}_e = \tNs - \tilde{N} = 0 \,,
\ee
which can be manipulated, using \eqref{btnper}, in order to get the following condition on the total number of e-foldings:
\be \label{TNnper}
\tN = N_\star \ln\frac{N_\star}{\beta}\,,
\ee
expressed just in terms of parameters of the benchmark models, where $N_\star$ is given by \eqref{Nstpert}. Eq.~\eqref{TNnper} fixes uniquely the total amount of exponential expansion required to give the same $(n_s,r)$ of the chaotic scenarios, with parameter $\beta$, and to end inflation via the condition $\epsilon_1=1$. Note that this coincides exactly with the large-$p$ limit of the perturbative case, namely Eq.~\eqref{Nbound}. 

\begin{figure}[htb]
\begin{center}
\includegraphics[width=7.6cm]{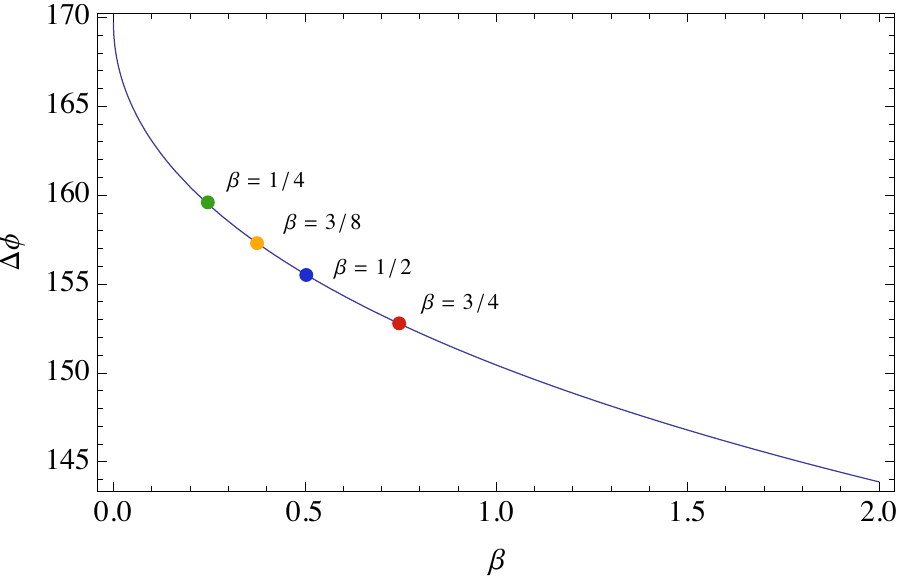}
\includegraphics[width=7.6cm]{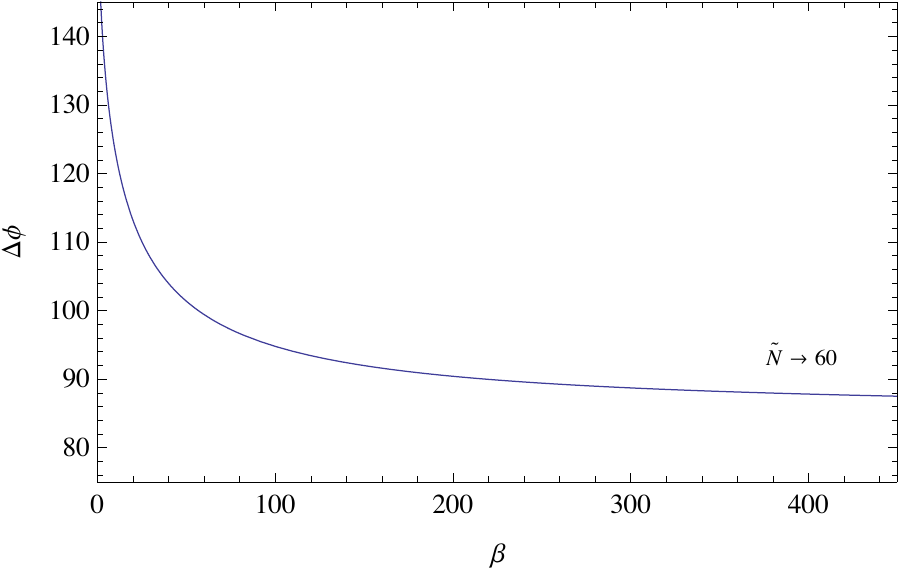}
\vspace*{0.4cm}
\caption{\it The inflaton range $\Delta\phi$ as function of $\beta$ in two different limits for the non-perturbative class of models. For low values of $\beta$ (physical values for the tensor-to-scalar ratio $r$), $\Delta\phi$ is a decreasing function. The four coloured points correspond to the upper-bounds already found in sec.~\ref{PClass}  and \ref{Log}. In the large-$\beta$ limit, the inflaton range cannot arbitrarily decrease and approaches a lower-limit.} \label{FIGDeltaPhiNP}
\end{center}
\end{figure}

The inflaton range is given by integrating Eq.~\eqref{dphi1} between $\tN_e$ and $\tNs$:
\be\label{DPnper1}
\Delta \phi = \frac{\sqrt{2}}{c}  \left( 1-e^{-c\tNs}\right)\,.
\ee
The latter can be written just in terms of the benchmark parameters by using \eqref{btnper} and it reads
\be\label{DPnper}
\Delta \phi = 2\sqrt{2} \left(N_\star-\sqrt{\beta N_\star}\right)\,,
\ee
which yields the field range in terms of $\beta$. Note that this again coincides exactly with the large-$p$ limit of the field range in the perturbative case, that is \eqref{dfbound}. Fig.~\ref{FIGDeltaPhiNP} shows such functional dependence; the negative slope of the curve makes explicit the inversion of hierarchy of field ranges with respect to the one which naively one would expect. In fact, lower values of $r$ (lower values of $\beta$) will correspond to larger $\Delta\phi$. This is exactly the same finding for the upper bounds in the perturbative class of models. Such behavior becomes explicit once we express Eq.~\eqref{DPnper} in terms of the typical inflaton range ${\Delta\phi}_c$ for the chaotic models, given by Eq.~\eqref{DPchaotic}. The relation turns out to be:
\be
\Delta \phi = 2\sqrt{2} N - {\Delta\phi}_c \,,
\ee
where $N$ is the total number of e-folds for the benchmark chaotic models and, throughout our study, it is fixed to be equal to $60$. However, it is not possible to arbitrarily decrease $\Delta\phi$ even going to really large values of $\beta$. In fact, by taking the limit for\footnote{Such a limit is anyway not physical as it would correspond to an infinitely large amount of primordial gravitational waves.} $\beta\rightarrow\infty$ of \eqref{DPnper}, we obtain
\be
\Delta\phi \rightarrow \sqrt{2} N\,,
\ee
as can be seen in the second plot of Fig.~\ref{FIGDeltaPhiNP}, where $N=60$. This corresponds to a lower-bound on $\tN$ which, in the same limit, approaches the benchmark number of e-folds $N$, as it is clear by taking the limit of \eqref{TNnper}. The field range of $\sqrt{2} N$ can then be understood from an $\epsilon_1$ parameter that is approximately equal to one during almost the entire inflationary period.

Within the non-perturbative models, it is then possible to mimic chaotic scenarios in terms of their cosmological observables $n_s$ and $r$. Nevertheless, both the total number of e-foldings $\tN$ and the field excursion $\Delta\phi$ are uniquely determined once we choose the power coefficient of the chaotic scenario, namely once we fix $\beta$. Curiously, the resulting values perfectly correspond to the upper-limits we found in the previous sections. In the specific example of quadratic inflation, that is for $\beta=1/2$, one obtains again $\Delta\phi\approx 155.56$ and $\tN\approx290$, as  expected from the discussion in sec.~\ref{PClass}  and \ref{Log}. 

Note, however, that this limit appears only in the large-$N$ limit of the non-perturbative class. Specific models of this class are discussed in \cite{Garcia-Bellido:2014gna}. An example is natural inflation \cite{Freese:1990rb}, which has specific subleading corrections in addition to \eqref{npertpar}. In the limit of a large periodicity, corresponding to small $c$, this model asymptotes to quadratic inflation and therefore has the same field range as this benchmark model. The origin of this difference with \eqref{DPnper} lies in the subleading corrections, that exactly become increasingly important when $c$ is small (the effective expansion parameter being $1/cN$). In other models, like hybrid inflation \cite{Linde:1993cn}, which end by the action of a transverse symmetry breaking field, the excursion can be even smaller, and still satisfy the observational constraints. We will discuss a similar phenomenon in the next subsection.

\subsection{Sub-Planckian field ranges}

Now we take a  different approach within the logarithmic class of models, in order to illustrate the possibility of obtaining smaller field ranges as compared to the benchmark model of quadratic inflation. The idea is to reproduce the same observational predictions in terms of $(n_s,r)$ by fixing $\tilde N$ (in what follows, we assume $\tilde N=60$) and letting $q$ vary as a function of $p$ through the relation \eqref{deg3}.
  
Once again, the inflationary field range $\Delta \phi$ can be determined by integrating \eqref{dphi1} numerically. We find a striking difference between values of $p$ that are larger or smaller than around $1.1$.

We find that setting the end of inflation by $\epsilon_1=1$ turns out to be possible only for $p$ not exceeding a value around $1.1$. For $p>1.1$ the function $\epsilon_1(N)$, given by \eqref{logep1}, never reaches the unity and, then, a viable inflationary scenario has to be ended through some other mechanism.

\begin{figure}[thb]
\hspace{-3mm}
\begin{center}
\includegraphics[width=8.5cm]{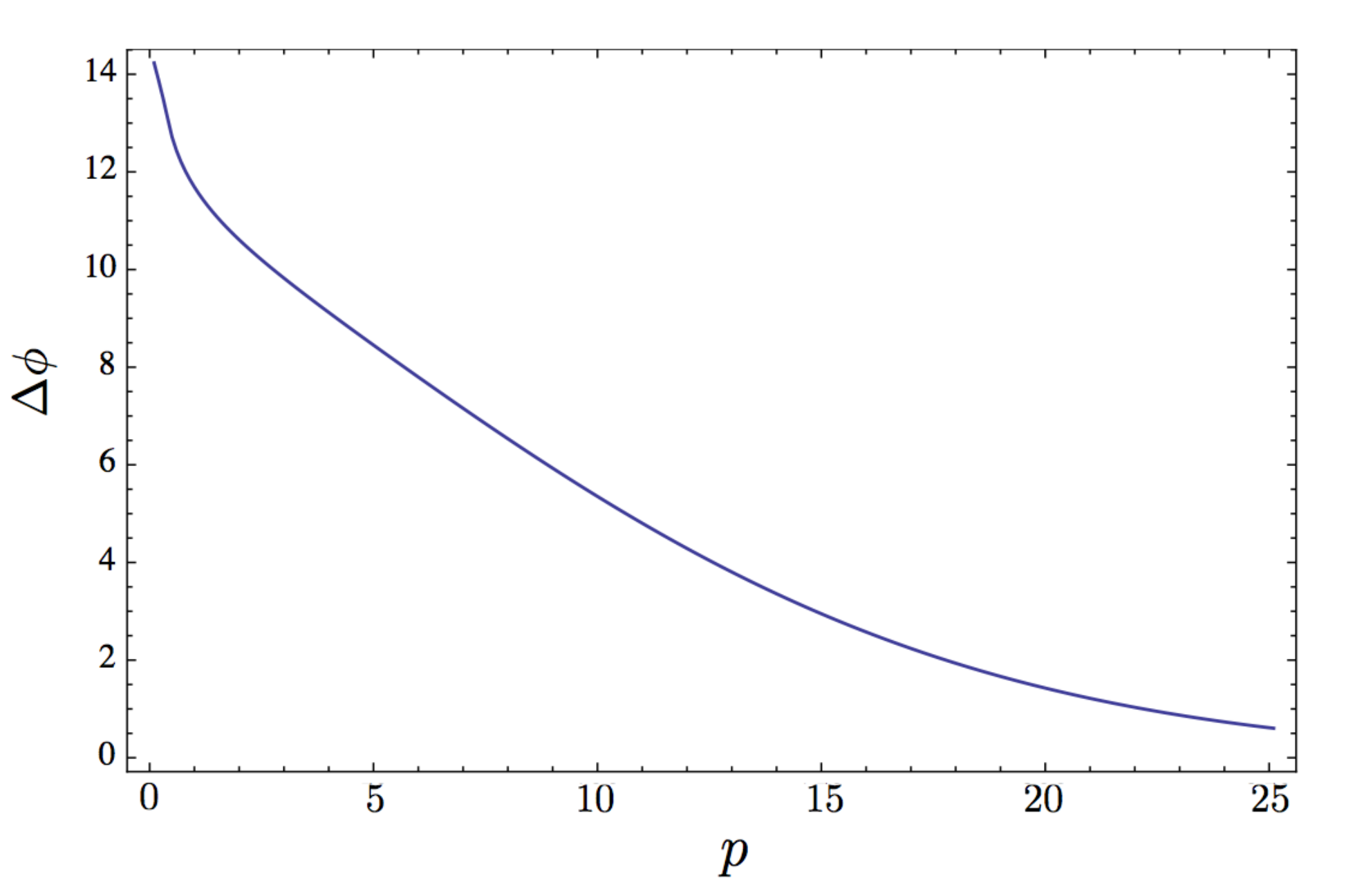}
\vspace*{0.4cm}
\caption{\it Slow-roll field range $\Delta\phi$ as function of $p$. The end of slow-roll inflation is set through the condition $\epsilon_2=1$. Sub-Planckian field ranges can be obtained if inflation ends through a second field or some other mechanism.} \label{FRLogepsJoined}
\end{center}
\vspace{-0.3cm}
\end{figure}

On the other hand, one can still set the end of slow-roll inflation via the condition $\epsilon_2=1$. In this case, the field range is a decreasing function of $p$, as showed in Fig.~\ref{FRLogepsJoined}, and the values of $\Delta\phi$ correspond to the distance which the canonical field $\phi$ travels within the slow-roll approximation. For sufficiently large $p$, such excursion becomes even sub-Planckian. However, note that these models generically would not correspond to slow-roll inflation throughout the whole period of exponential expansion and they would need to end inflation e.g. via a second field, or some other mechanism.

\subsection{Discussion}
\label{sec:disc}

In this Section, we have investigated the implications of the CMB data for the inflationary field range. More precisely, we have tried to answer to what extent one can infer $\Delta \phi$ from a measurement of $(n_s, r)$. We have analyzed this question by comparing three different classes of models -- perturbative, logarithmic and non-perturbative -- to the benchmark models of chaotic inflation, with particular attention to the quadratic scenario. 

Surprisingly, we have found that the field range can vary an order of magnitude; while the quadratic model implies $\Delta \phi \approx 14$ in Planck units, the non-perturbative class gives the same observables while $\Delta \phi$ is a factor 11 larger. Moreover, we have identified a continuous degeneracy in the other classes: different one-parameter families of models yield identical $(n_s,r)$ while $\Delta \phi$ spans over a quite large range. Remarkably, $\Delta \phi$ can be increased by exactly the same factor by varying this parameter in both the perturbative and the logarithmic class. Therefore, this constitutes an upper bound for these classes of models.

\begin{figure}[htb]
\hspace{-3mm}
\begin{center}
\includegraphics[width=8cm]{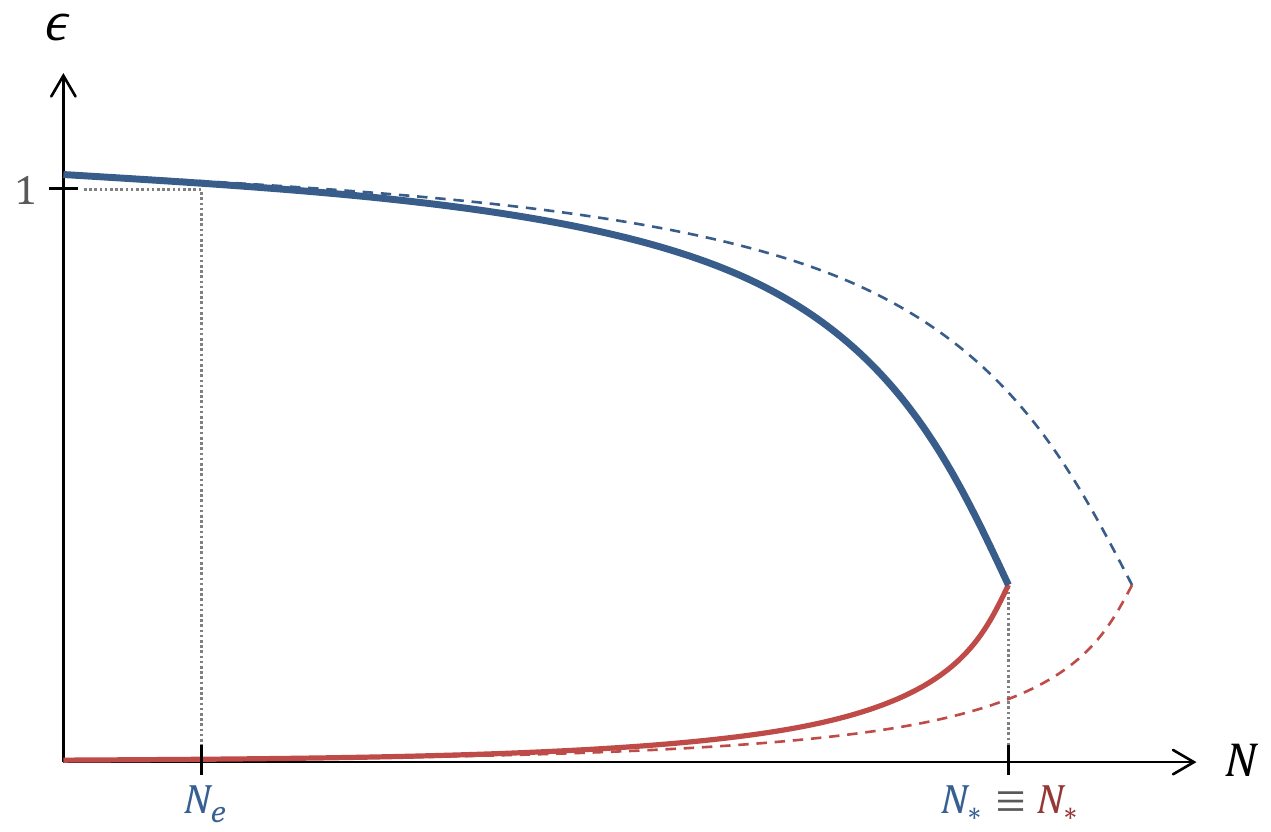}
\vspace*{0.2cm}
\caption{\it Two possible scenarios for the $N$-dependence of $\epsilon$: the solid upper line aims to maximize the field range while the lower minimizes it. The dashed lines correspond to the same scenarios where one increases $N$.} \label{epsilon-N}
\end{center}
\vspace{-0.3cm}
\end{figure}

It might be surprising that there is an upper limit on the field range. After all, we are allowing in principle for an infinite number of e-foldings, hence one would expect it to be possible to hover just below $\epsilon_1 = 1$ for an infinitely long period in terms of $N$; such a scenario is illustrated by the upper line in Fig.~\ref{epsilon-N}. This period would contribute an infinitely large field range $\Delta \phi$ as well. This raises the question: why do we not find such infinitely large field ranges? We suspect that the answer lies in the Hubble flow equations for the slow-roll parameters. For slow-roll inflation, in the approximation where we are only keeping the lowest two slow-roll parameters, these can be written as (where $\epsilon = \epsilon_1$ and $\eta = 2\epsilon_1 + 1/2\ \epsilon_2$)
\begin{align}
\frac{d \epsilon}{dN} = 2\epsilon ( \eta - 2\epsilon) \,, \qquad \frac{d \eta}{d N} = \epsilon (\eta - 3 \epsilon) \,. \label{slowroll}
\end{align}
Note that one cannot have both right-hand sides vanishing at the same time when $\epsilon \neq 0$; therefore it is impossible to keep $\epsilon$ constant over a large range of e-foldings. As a consequence, there is a limit on the number of e-foldings between horizon exit and the end of inflation, for a generic slow-roll model. This is a consequence of the generic lower limit on $d \epsilon / dN$, and translates into a limit on the field range during this period.

Nevertheless, the above discussion constitutes only a generic argument; in fact,  specific and non-generic inflationary models could have yet larger field ranges. Examples are in fact provided by models in the perturbative and the logarithmic classes, with parameter $p<0$. In these models the field range can be arbitrarily large. However, these models are contrary to the large-$N$ approach that we have taken in this section, where the inflationary period approaches a De Sitter phase as $N$ becomes infinite. For $p$ negative it turns out that one has a cut-off on the number of e-folds preceding the moment of horizon exit. Therefore these do not extend infinitely into the past, approaching a De Sitter phase. In this way it turns out to be possible to evade the generic argument for the upper limit based on \eqref{slowroll} above.

From the perspective of UV-sensitivity, yet more interesting is the question how small $\Delta \phi$ can be, and in particular whether it can reach sub-Planckian values. This point has been discussed in some detail recently in literature. In order to minimise the field range, one would like to have the area under the curve $\epsilon(N)$ in Fig.~\ref{epsilon-N} as small as possible; this case is illustrated by the lower line. Starting at horizon exit, one would therefore need to suppress $\epsilon$ as fast as possible \cite{Hotchkiss:2011gz,German:2014qza, Gao:2014pca, Bramante:2014rva}. In \cite{Antusch:2014cpa}, however, it was pointed out that this is impossible in the slow-roll approximation, exactly due to Eq.~\eqref{slowroll}; as the right hand sides are bilinear in percent-level slow-roll parameters, these can only vary rather slowly as a function of $N$. This upper bound on the change of $\epsilon$ implies a lower bound on the field range. Amusingly, this is the exact opposite reasoning which led to the large field range discussion above.

The issue of getting a smaller $\Delta \phi$ with respect to the benchmark of the quadratic model
has been investigated explicitly in the different classes. In the single-field slow-roll approximation, we have found that sub-Planckian field ranges do not seem to be possible, in agreement with the recent bound \cite{Antusch:2014cpa}: we could only reduce $\Delta \phi$ by a factor of three, down to $\Delta \phi\approx5$ in Planck units. However, these classes of models allow for a much stronger reduction of the inflationary field range, provided one allows for an alternative end of inflation (a related interesting analysis in the context of hybrid natural inflation was done in \cite{Hebecker:2013zda}). In particular, by imposing the condition $\epsilon_2 = 1$, we have found sub-Planckian inflationary trajectories that satisfy all slow-roll single-field requirements. Nevertheless, within these models, the parameter $\epsilon_1$ never reaches the unity and the inflationary expansion needs to be stopped by some other mechanism. Such models could be viable when performing a full fast-roll analysis, or when embedded e.g.~in a multi-field model. Note that this type of multi-field is markedly different from those studied in Ref.~\cite{McDonald:2014oza}; in contrast to that reference, our entire inflationary trajectory is purely single-field, and we only appeal to the second field for a waterfall transition to end inflation.

\section{Universality of the inflaton range}\label{SecUniversalityRange}

In the previous Section, we have shown concrete examples where a value of $n_s$ and $r$ does not correspond to a specific estimate for the inflationary field range $\Delta\phi$. This is indeed what one would expect generically from a variable depending on the entire inflationary trajectory.

Nevertheless, it is possible to identify different regions where the field range does exhibit a universal behavior. We have proved this remarkable fact in the publication \cite{Garcia-Bellido:2014wfa} and we will present again the main results below.

In the following, we will restrict our analysis to the perturbative class of models, characterized by an equation of state parameter given by Eq.~\eqref{epsilonN}. We will not assume this expression to be exact but allow for subleading contributions which generically may play an important role towards the end of inflation.

\subsection{Universality at large $N$}

In order to get the expression for $\Delta\phi$, one must integrate Eq.~\eqref{dphi1} along the entire inflationary trajectory. By considering a large-$N$ behavior such as that in Eq.~\eqref{epsilonN}, for $p\neq2$, we obtain
\be \label{range}
\Delta\phi= \frac{2\sqrt{2\beta}}{2-p} \,N^{1-\frac{p}{2}}-\phi_e\,,
\ee
where $\phi_e$ is a constant piece related to the value of the inflaton when inflation ends. Then, we run into two possible  situations, depending on whether $p$ is smaller or larger than $2$.

In the first case, for $p<2$, the inflaton range $\Delta \phi$  is  proportional to a positive power of $N$. In the large-$N$ limit, the constant part $\phi_e$ is subleading and one can  argue that, within any universality class, the magnitude of the field excursion will be model-independent and therefore universal. Furthermore, given that $\Delta \phi$ keeps increasing together with $N$, one can correctly refer to such scenarios as genuine large field models. 

In the second case, for $p>2$, the $N$-dependent term of \eqref{range} is subleading with respect to the constant term $\phi_e$, in the large-$N$ limit. The value of $\Delta \phi$ is therefore determined by the point where inflation stops and generically not universal: for instance, $\Delta\phi$ can already obtain  a super-Planckian contribution during the last e-fold \cite{Easther:2006qu}. This model-dependent piece is generically sub-dominant for models with $p<2$ while it represents the main contribution when $p>2$.  

Finally, the remaining possibility is $p=2$ where the functional form of the field range reads
\be \label{range2}
\Delta\phi= \sqrt{2\beta}\, \ln N-\phi_e\,.
\ee

The log-dependence leads to a situation where $\Delta \phi$ mildly increases together with $N$. The special role of this point, corresponding to Starobinky-like scenarios, has been recently highlighted in the context of the inflationary attractors \cite{Kallosh:2013tua, Kallosh:2013yoa,Galante:2014ifa,Roest:2015qya,Scalisi:2015qga} as well as non-compact symmetry breaking \cite{Burgess:2014tja}. Moreover, a change of behavior around the point $p=2$ was noticed also in the analysis on the degeneracy of the inflaton range done in \cite{Garcia-Bellido:2014eva} and presented in the previous Section.  Here we stress its peculiarity  also as  marking the separation between a region of authentic large field models ($p<2$), whose $\Delta \phi$ exhibits universality features, and a region ($p>2$) where models can have the same $r$ and $n_s$ at leading order (and, thus, belonging to the same universality class) but still very different field ranges. 
\\

\subsection{Universality at small $\mu$}

The results presented above are obtained in a pure large-$N$ expansion, that is, in the limit $N\rightarrow \infty$. However, physical values usually amount to an exponential expansion of around 50 to 60 e-foldings preceding the end of inflation. Although the latter is a big number, the universal regime can be easily affected by tuning specific parameters of the models.

For large enough values of $N$, any model, characterized by an equation of state parameter such as Eq.~\eqref{epsilonN}, will be represented by a potential, which is parameterized as a small deviation from the benchmarks potentials \eqref{VP}. Specifically, for $p>1$ and $p\neq 2$, the generic form of $V$ will include higher order corrections and read
\be\label{Vdev}
V(\phi) = V_0 \left[1-\left(\frac{\phi}{\mu}\right)^{n} + \sum^{\pm\infty}_{q=n\pm 1} c_q \left(\frac{\phi}{\mu}\right)^q\right]\,,
\ee 
where $n$ is related to $p$ through Eq.~\eqref{np} and the plus or minus sign depends respectively on $p>2$ or $p<2$. Then, the coefficients $c_q$ parameterize the deviation  from hilltop or inverse-hilltop models respectively.

Now we show that, at small $\mu$ and for finite values of $N$, we recover universality: in addition to the cosmological observables $n_s$ and $r$, the inflaton excursion will be model-independent. Interestingly, this is exactly the regime we will consider  to derive  the {\it field range bound} in the next Section.

The spectral index $n_s$ and tensor-to-scalar ratio $r$ will be generically insensitive to higher order terms in the expansion \eqref{Vdev} as they are calculated at horizon exit. In fact, the inflationary regime is restricted to the region $\phi<\mu$, for hilltop models ($p>2$), and $\phi>\mu$, for inverse hilltop potentials ($1<p<2$); therefore, the farther one is located from the end-point of inflation the more one can ignore higher order corrections in the scalar potential. Then, the large-$N$ regime provides an accurate estimate of such observables which, at small $\mu$, read
\ba \label{nsrHT}
n_s&=& 1-\frac{p}{N} \,,\quad 
r = 2^{5-2p} \frac{(p-2)^{2p-2}}{(p-1)^{p-2}}\frac{\mu^{2p-2}}{N^p}\,.
\ea
The coefficients $c_q$ will appear only in subleading terms in $N$. The family of models represented by Eq.~\eqref{Vdev} will have identical behavior in the small-$\mu$ limit and for large enough values of $N$. Conversely, this is generically not the case for large values of $\mu$; in such a limit, the end-point of inflation is pushed  towards the region where the coefficients $c_q$ play an important role and dissimilarities become important; consequently, going 50-60 e-foldings back, even the point at horizon crossing will start to be sensitive to $c_q$ corrections. For large values of $\mu$, the large-$N$ expansion is not well defined and scenarios belonging to the same universality class at small $\mu$, may give quite different predictions in terms of $n_s$ and $r$. 

In the limit of large $N$ and small $\mu$, the field range turns out to be
\be\label{deltaVdev}
\begin{aligned}
\Delta\phi= & \left[\frac{2-p}{\sqrt{2}(1-p)}\right]^{1-\frac{2}{p}}\mu^{2-\frac{2}{p}} - \frac{(\frac{p}{2}-1)^{p-2}}{(p-1)^{\frac{p}{2}-1}}\, \mu^{p-1} N^{1-\frac{p}{2}}\,,
\end{aligned}
\ee
where the first term is clearly related to the end-point of inflation while the second one is the $N$-dependent term. For the  reasons given above, $c_q$ corrections will not enter the $N$-dependent part which gives the main contribution to the field range for $1<p<2$ while it is subleading for $p>2$. Things are different when calculating the end-point $\phi_e$;  this piece is sensitive to higher-order corrections in $\mu$. As soon as $\mu$ increases, this point is pushed away towards a region where differences among the models begin to appear. If, for simplicity, we focus on the case $n=3$ (examples belonging to this universality class are hilltop inflation and the models referred to as RIPI and MSSMI in \cite{Martin:2013tda}) and consider terms up to fifth order in the expansion \eqref{Vdev}, the end-point reads
\be\label{phie3}
\phi_e= \sqrt{\frac{\sqrt{2}}{3}}\, \mu^{3/2}+ \frac{2\sqrt{2}}{9}\,c_4 \mu^2 + \frac{5(4\,c_4^2 + 3 \,c_5)}{27\cdot 2^{1/4} \sqrt{3}}\mu^{5/2}\,.
\ee 
Crucially, the coefficients $c_q$ appear just with higher powers of $\mu$; this holds even for other values of $n$ (both positive and negative) as well as the special point $p=2$. This implies that one obtains universal predictions in the small-$\mu$ limit, not just in terms of $n_s$ and $r$, but also in terms of $\Delta\phi$, whose form approaches Eq.~\eqref{deltaVdev}.
\\

\section{The Lyth bound with a tilt}

In Sec.~\ref{inflatonrangeobservations}, we have seen that the Lyth bound provides an optimal estimate of the field range, given a measurement of $r$ which is simply related to $\epsilon_1$ through Eq.~\eqref{inflaparam}. However, starting from the same value of $\epsilon_1$ at horizon crossing, one can imagine different behaviors $\epsilon_1(N)$ that give rise to either smaller\footnote{The Lyth bound can also be evaded using multiple scalar \cite{McDonald:2014oza} or vector fields \cite{Maleknejad:2011jw}. An extension   to fast roll can be found in \cite{Baumann:2011ws}.} \cite{BenDayan:2009kv, Hotchkiss:2011gz, Antusch:2014cpa} or larger areas \cite{Garcia-Bellido:2014eva}. This situation is shown in Fig.~\ref{DeltaPhiPicture}.

\begin{figure}[htb]
\vspace{1mm}
\begin{center}
\includegraphics[width=9.5cm]{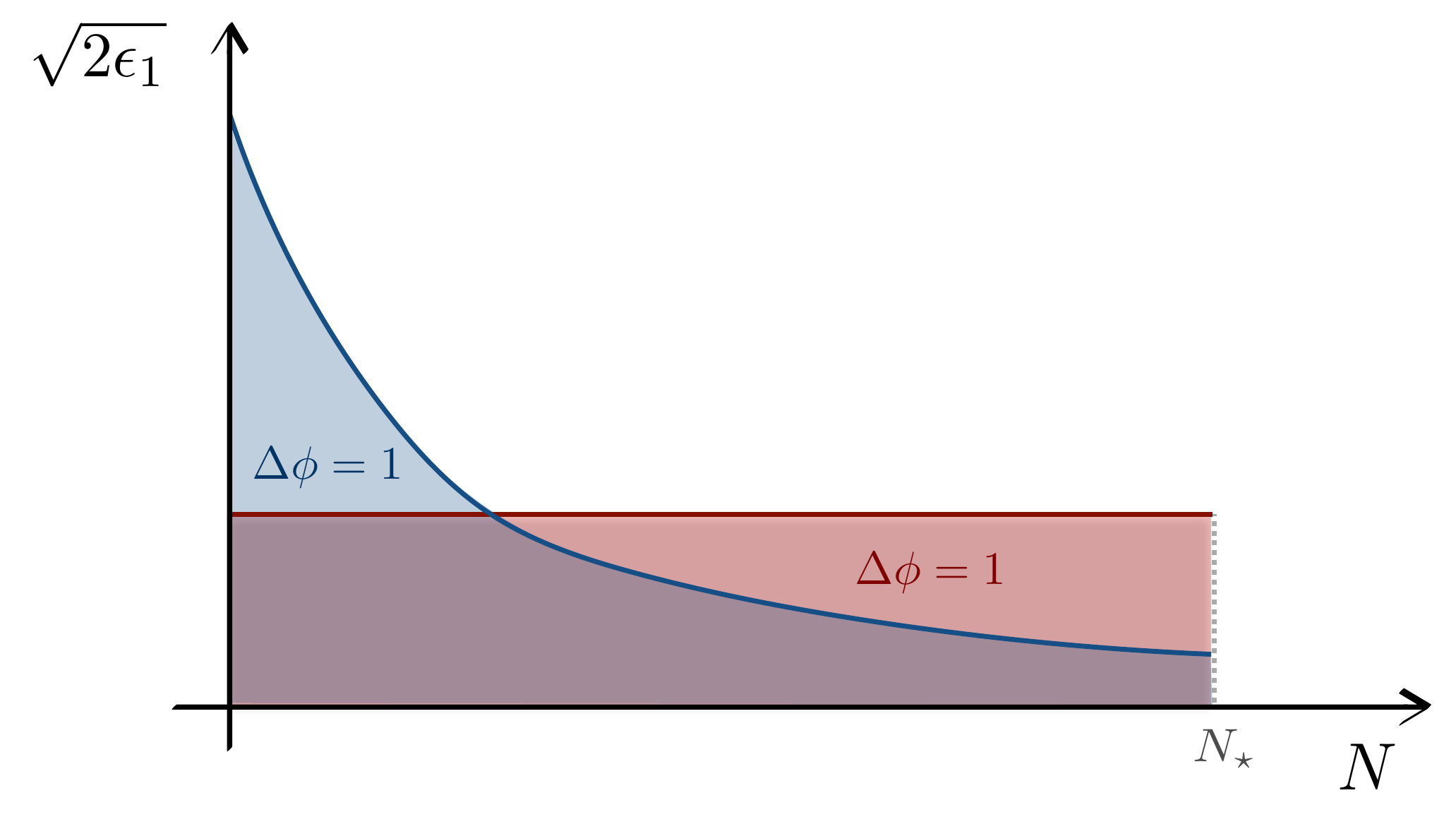}
\vspace{1mm}
\caption{\it Two curves indicating $\sqrt{2 \epsilon_1}$ with identical areas $\Delta \phi=1$. The flat curve depicts the Lyth bound, while the tilted curve indicates the improvement when taking the spectral index into account.}\label{LythTilt}
\end{center}
\vspace{-0.5cm}
\end{figure}

We would like to show that this estimate becomes stronger when one takes the additional information of the spectral index into account. In particular, given the redshifted value \eqref{ns} and assuming $r$ to be small, the dependence $r=16\epsilon_1(N)$ is tilted upwards at horizon crossing\footnote{Note that our approach differs from \cite{German:2014qza,Bramante:2014rva}, which also include the spectral tilt in their expressions: while these references derive a {\it minimal} value for $\Delta \phi$, we aim to provide a {\it generic} estimate by making use of its universal properties.}. The natural history therefore leads to a larger area than that of the corresponding rectangle. As a consequence, the requirement  $\Delta \phi=1$ implies a lower value of $r$, as illustrated by the blue line in Fig.~\ref{LythTilt}. This is our main message: by including constraints on $n_s$ one can strengthen considerably  the Lyth bound. Thanks to the results on the universality of $\Delta\phi$ in the sub-Planckian regime\footnote{Strictly speaking, this is true for values $\Delta\phi \lesssim 10^{-2}$, which define more accurately  small field inflation. In this  region $\mu<1$ and thus sub-leading corrections are suppressed, strengthening the results on universality.},  we will show that the reported value \eqref{ns} leads to $r \lesssim 2 \cdot 10^{-5}$ for sub-Planckian field ranges. This constitutes a bound which is two orders of magnitude stronger than the usual estimate as given by Eq.~\eqref{Lbound}.

\subsection{Strengthening the Lyth bound}

We now use the results derived in the previous Section in order to revisit the discussion on small- and large-field excursions and derive a stronger field range bound than the usual estimate Eq.~\eqref{Lbound}.

The findings on the universality of the field range translate into the possibility of inferring an accurate estimate of $\Delta\phi$ given a point in the $(n_s,r)$ plane. This is certainly true in the small-$\mu$ limit where $\Delta\phi$ is given by Eq.~\eqref{deltaVdev}. One can properly argue that sub-Planckian field ranges will be model-independent and uniquely determined by a measurement of the cosmological observables. The situation changes when $\mu$ increases; already for $\mu \gtrsim \mathcal{O}(1)$, in the region $p>2$ (corresponding to $n_s\lesssim0.96$), universality breaks down (as can be seen from Eq.~\eqref{phie3} where each contribution is order one); differently, for $p<2$, universality can hold even for some orders of magnitude larger than the reduced Planck mass $M_P=1$, thanks to the dominant $N$-dependent term as set by Eq.~\eqref{range}.

\begin{figure}[htb]
\begin{center}
\includegraphics[width=8cm]{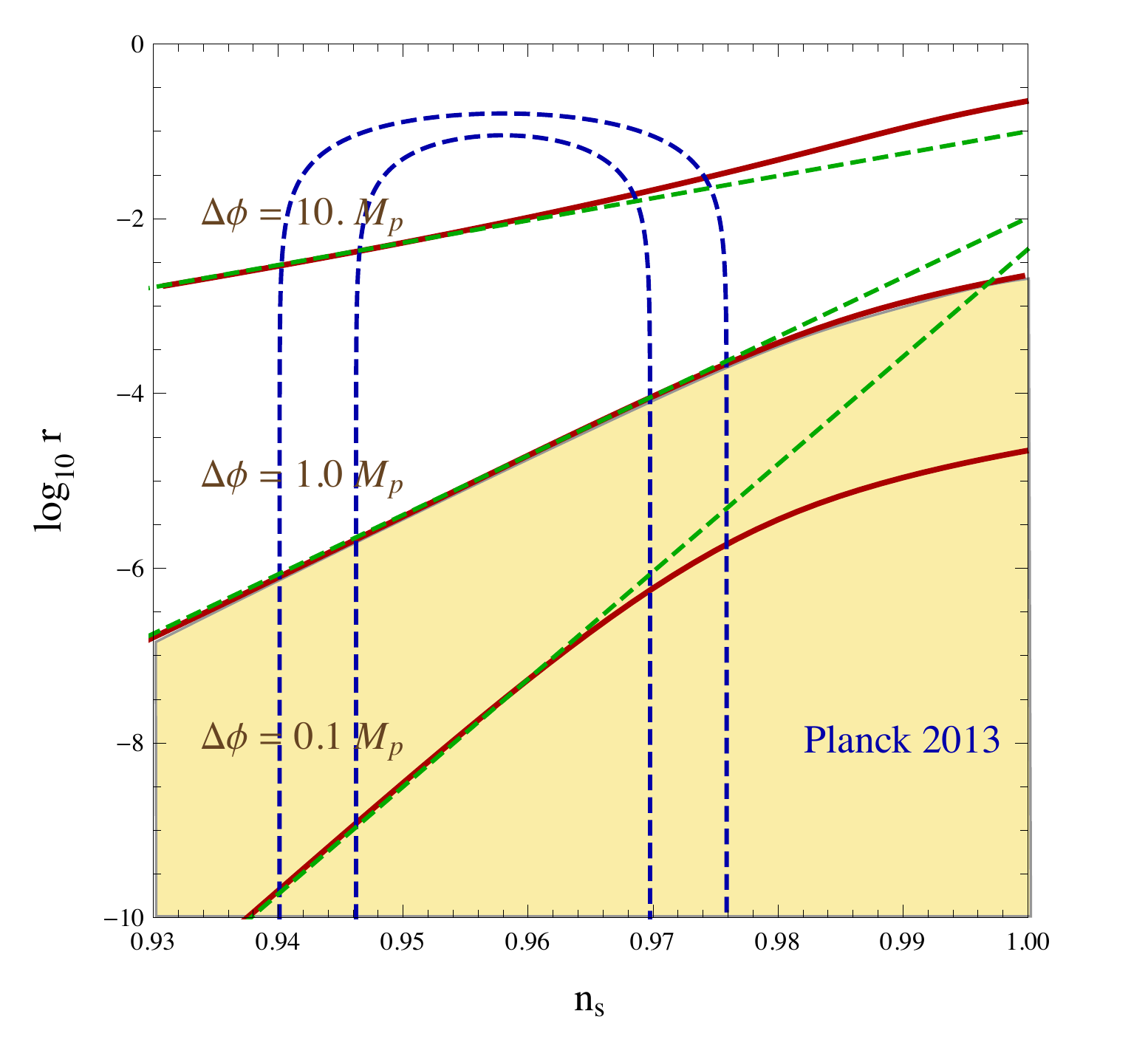}
\vspace{2mm}
\caption{\it Field ranges corresponding to $\Delta \phi = (0.1,1,10)$ in the plane ($n_s,~\log_{10}{(r)}$). The green straight dashed lines represent the asymptotic behaviour for large $p$. The yellow area corresponds to sub-Planckian values of the field excursion and, then, to the universality region.} \label{fig.LythBound}
\end{center}
\vspace{-0.5cm}
\end{figure}

Then, if we plot lines of constant $\Delta\phi$ in a  $(n_s,r)$ plane, the one corresponding to unity $\Delta\phi=1$ will be a good estimate of the border above which universality breaks down, regardless the value of $n_s$. This will be taken as the new, stronger bound. As can be seen from Fig.~\ref{fig.LythBound}, the line is tilted as it is a function also of the spectral index $n_s$. Interestingly, for $n_s=1$ it approaches the value of the original Lyth bound, which is a constant value not depending on the tilt. On the other hand, in the Planck-range, an excellent fit is provided by the following expressions, corresponding to the (green) dashed straight lines in Fig.~\ref{fig.LythBound},

\be\label{improved-Lyth}
\begin{aligned}
   \log_{10}r  & = & -1.0 + 25.5\,(n_s -1)\,, \hspace{9mm} \Delta\phi = 10 \,,\\
   \log_{10}r  & = & -2.0 + 68.0\,(n_s -1)\,, \hspace{8mm} \Delta\phi = 1.0 \,,\\
   \log_{10}r  & = & -2.35 + 123\,(n_s -1)\,, \hspace{8mm} \Delta\phi = 0.1 \,. 
 \end{aligned}
\ee

The range of values of ($n_s,\,r$) consistent within those of Planck reduces the values of $\Delta\phi$ during inflation by at least an order of magnitude. For the central value $n_s\simeq 0.96$, imposing that $\Delta\phi \leq 1$ leads to the bound $r\lesssim 2\cdot10^{-5}$, which is two orders of magnitude below the usual Lyth bound.

\begin{figure}[htb]
\vspace*{5mm}
\begin{center}
\includegraphics[width=6.8cm]{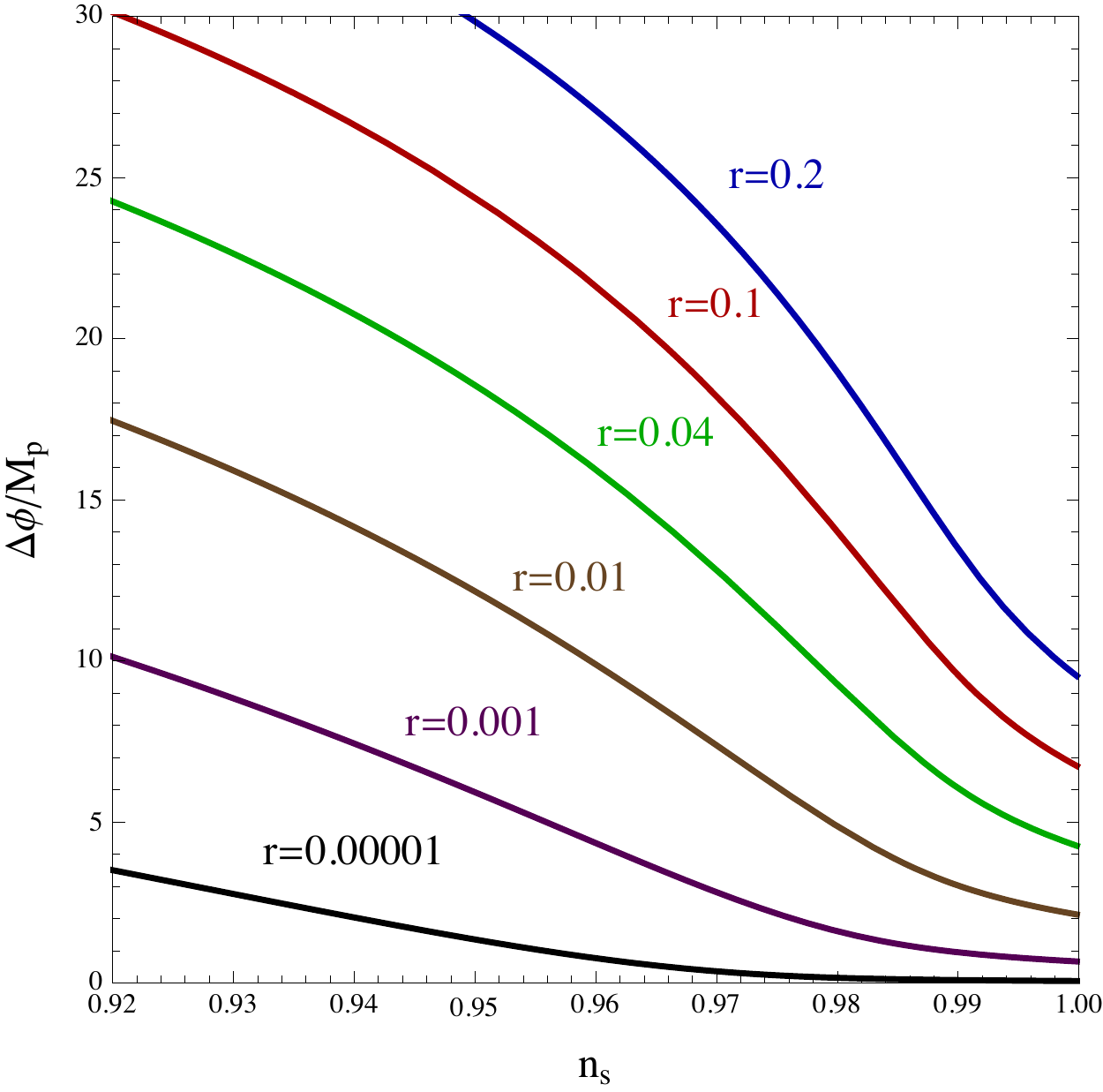}
\vspace{2mm}
\caption{\it The range of field values corresponding to $r = 0.2,\,0.1,\,004,\,0.01,\,0.001,\,0.00001$ in the plane ($n_s,~\Delta\phi$).}\label{fig.dphins}
\end{center}
\end{figure}

On the other hand, if we impose that the ratio $r$ be bigger than a certain value, then we find a lower bound on $\Delta\phi$. Fig.~\ref{fig.dphins} shows the field range as a function of the scalar spectral index for different values of the ratio $r$. Again, in the range consistent with Planck, the field range is always super-Planckian, for all values of the ratio $r \gtrsim 2 \cdot 10^{-5}$. This conclusion can only be avoided by going to unrealistically large spectral indices $n_s$ close to $1$. 

Similarly to the original Lyth bound, the relations \eqref{improved-Lyth} provide generic estimates of the field range, which could be avoided only by a very specific (non-generic) behavior of $\epsilon_1(N)$. However the existence of such counterexamples is of limited importance: one would like to understand when large field inflation is expected given a measurement of $r$ even if there might be fine-tuned models which give smaller field ranges for this value of~$r$.

Given the central value for $n_s$ from Planck, our results imply that super-Planckian field ranges require a tensor-to-scalar ratio that exceeds $2 \cdot 10^{-5}$. Planned future CMB experiments, such as COrE \cite{Bouchet:2011ck, Core} and PRISM \cite{Andre:2013afa,Andre:2013nfa, Prism}, might bring the sensitivity down to $10^{-4}$. In contrast to what one would conclude from the original Lyth bound, our results  imply that a small detectable $r$ still corresponds to super-Planckian field ranges.  

\clearpage
\thispagestyle{empty}


\chapter{Inflation and Attractors in Supergravity}
\label{chapter:supergravity}

\begingroup
\begin{flushright}
\vspace{.5cm}
\end{flushright}
\quote{\it In this chapter, we discuss the problem of realizing a consistent inflationary scenario within a supergravity framework. We discuss its relations to string theory and present its basic properties. Then, we discuss the most common and challenging obstacles, and its possible solutions, to a successful realization of inflation in supergravity. These include the well known $\eta$-problem and the dynamical restrictions arising from the interplay between the inflaton and supersymmetry breaking sectors. Remarkable simplifications arise in the case of complete orthogonality of these two sectors. An arbitrary inflaton potential can indeed be obtained when the internal \Kahler manifold is flat. On the other hand, assuming a hyperbolic geometry has dire implications for inflation: the \Kahler curvature controls the amount of primordial gravitational waves and the value of the scalar tilt tends automatically to the “sweet spot” of Planck, no matter the details of the superpotential. The non-trivial \Kahler geometry basically induces an attractor for observations. Finally, we present a novel supergravity construction, dubbed $\a$-scale model, which turns out to be at the origin of the attractor mechanism. This will allow us to construct the first single superfield formulation of $\a$-attractors and ultimately shed light on the connection between flat and curved internal space.
The novel results of this Chapter are based on the publications {\normalfont[\hyperref[chapter:Publications]{{\sc iii}}]}, and {\normalfont[\hyperref[chapter:Publications]{{\sc viii}}]}.}

\endgroup

\newpage

\section{Inflation in supergravity}

In the Introduction of this thesis, we have already discussed the importance of embedding the inflationary paradigm into a complete high-energy physics scenario. The UV sensitivity of inflation makes indeed its implementation into a concrete framework of quantum gravity a primary challenge to face.

String theory \cite{Green:1987sp,Green:1987mn,Polchinski:1998rq,Polchinski:1998rr} offers a great arena where to construct our cosmological models of the early Universe. Its control over Planckian degrees of freedom seems indeed to provide a robust environment where to investigate inflation. However, properly realizing inflation within a concrete stringy scenario has turned out to be quite challenging. First of all, in order to bring this complex framework in contact with reality, one must find a suitable mechanism to reduce the number of spacetime dimensions from ten to four. This procedure is named {\it compactification} \cite{Grana:2005jc,Douglas:2006es} of the six extra dimensions and it has become a central research topic in theoretical physics. In addition, a successful compactification usually leads to the appearance of many light moduli, namely scalar fields with no mass. These must be {\it stabilized} by means of specific mechanisms which produce the appropriate potential constraining their dynamics (famous examples are the so-called GKP \cite{Giddings:2001yu} and KKLT \cite{Kachru:2003aw} mechanisms). Finally, in order to have best control on the string inflationary models, one must ensure that a desirable hierarchy of scales holds. Specifically, one would like to have 
\be\label{hierarchyscales}
H<M_{KK}<M_s<M_{Pl}\,,
\ee
where $H$ denotes the Hubble scale during inflation, $M_{KK}$ denotes the compactification Kaluza-Klein scale below which one may consider an effective 4-dimensional description of physics, $M_s$ is the scale at which one resolves the string structure and $M_{Pl}$ is the Planck scale.

The route towards a complete embedding of inflation in string theory is still in progress (see \cite{Baumann:2009ni,Silverstein:2013wua,Baumann:2014nda,Westphal:2014ana,Cicoli:2016ygh} for some reviews on this topic). Along the way, it has produced very interesting results (see e.g. \cite{Dvali:1998pa,Burgess:2001fx,Kachru:2003sx,Kallosh:2007ig,Cicoli:2008gp,Hebecker:2011hk,Hebecker:2012aw,Burgess:2013sla,Hebecker:2014eua}) which have shed light on the basic properties a consistent cosmological scenario should have. Questions about the fundamental behavior of the inflaton field can be often translated into questions about the geometry of the internal manifold. Understanding how the physics of the many moduli naturally arising in string theory can (or cannot) be decoupled from the inflaton dynamics becomes of utmost importance in this context (much effort in this direction has been made by works such as \cite{Achucarro:2008sy,Achucarro:2008fk,Achucarro:2010jv,Achucarro:2010da,Achucarro:2012sm,Achucarro:2012yr,Achucarro:2012fd,Achucarro:2013cva,Achucarro:2014msa} and \cite{Buchmuller:2014pla,Buchmuller:2015oma,Dudas:2015lga}).

The usual strategy is investigating inflation within an effective {\it supergravity} (SUGRA) description (i.e. at energies lower than $M_s$), whose consistency is automatically satisfied if one assumes the hierarchy of scales Eq.~\eqref{hierarchyscales}. This is the limit where the strings can be simply approximated by point particles and one can use a convenient field theory description. In fact, supergravity \cite{Nath:1975nj,Freedman:1976xh} (see \cite{Freedman:2012zz} for a comprehensive book on this topic), as a local extension of supersymmetry (a perfect correspondence between bosons and fermions \cite{Gervais:1971ji,Volkov:1973ix,Ramond:1971gb,Wess:1974tw}), was discovered independently from string theory. It was then realized that some supergravity incarnations could arise as effective limits of this complete theory of Nature.

However, string theory compactifications usually yields severe constraints on the possible internal geometries. This implies that not every supergravity model can be regarded as an effective limit coming from string theory. One can easily encounter the risk of studying a supergravity cosmological construction which has no correspondent in the UV limit.

In this Chapter, we intend to face this last issue and prove that investigating inflation within a pure supergravity context can still yield crucial insights into the general structure of an effective UV description. In fact, the sole ingredient of supersymmetry (SUSY) can yield very strong constraints on the inflationary dynamics. In addition, the form of the scalar potential and the stability of the scalar manifold can be studied in full generality. We will show the generic properties a supergravity model should have in order to reproduce successful inflationary scenarios compatible with the current observational data. We will see that it is possible to draw very general conclusions, independently from the specific details of the model at hand. The generality of these results will assume particular relevance in the light of building a consistent inflationary model within string theory.

In the following, we will start our discussion by presenting the main properties and most common problems when one tries to embed inflation in $\mathcal{N}=1$ supergravity\footnote{$\mathcal{N}$ defines the number of supersymmetry transformations of the theory. The higher the number of supersymmetries is,  the more constrained are the field content and its dynamics. If we require the absence of particles with spin higher than 2, $\mathcal{N}=8$ is the maximum possible value as there are no more than eight half-steps between spin -2 and 2.}, without referring to any specific model. We will present the general form of the inflaton Lagrangian in supergravity. Both the scalar potential and the kinetic terms will be functions of fundamental geometric properties of the internal manifold where the fields are defined. We will then discuss a generic problem threatening the flatness of the inflaton potential (namely the smallness of the $\eta$ parameter) in supergravity and its possible ways-out. We will then examine the connections between spontaneous SUSY breaking and the inflationary dynamics. This will have direct consequences on the possibility of yielding inflation by means of just a single superfield. Finally, we will discuss to what extent one can regard a supergravity model as an effective limit coming from string theory.

After these first general considerations, we will devote the following sections of this Chapter to some concrete models of inflation. Specifically, we will show how the geometry of the internal field space can have dire implications on the final result of inflation in terms of $n_s$ and $r$. In Sec.~\ref{SECflat} and Sec.~\ref{SECHyp}, we will see the main differences between the cases of flat and hyperbolic geometry. The latter leads naturally to the concept of {\it cosmological attractors} providing universal observational predictions. We will devote the last section to a novel supergravity construction, dubbed {\it $\alpha$-scale supergravity}, which proves to be at the origin of the attractor mechanism. This will also shed light on the link between the two benchmark geometries considered before.

\subsection{Basics of 4D $\mathcal{N}=1$ supergravity}

The field content of a four-dimensional $\mathcal{N}=1$ SUGRA theory is given by the graviton $g_{\mu\nu}$, the gravitino $\psi_\mu$ coupled to an arbitrary number $n$ of chiral supermultiplets\footnote{We are considering the case of vector supermultiplets playing a subdominant role, that is, the effective description is the so-called F-term supergravity. When gauge interactions become relevant, they induce an additional so-called D-term piece in the potential $V$.}. Each of these contains a chiral spin-$1/2$ field and a complex scalar field. In the following, we discuss just the bosonic sector as the most relevant for the next sections. We will return on the fermionic sector in the next Chapter.

Scalar fields are ubiquitous in supergravity theories. Specifically, they always come in pairs (being complex scalars) and their number is not constrained. This provides a very flexible and natural setup to embed the physics of the inflaton field, as it was introduced in Sec.~\ref{SUBSECTIONslowroll}.

The dynamics of the complex scalar fields $\Phi_i$ (with $i=1, ... ,n$) is fully determined by two functions:

\begin{itemize}
\item The {\it K\"ahler potential} $K(\Phi_i,\bar{\Phi}_{\ib})$, being a hermitian function of the fields $\Phi_i$ and their complex conjugates $\bar{\Phi}_{\ib}$. 

\item The {\it superpotential} $W(\Phi_i)$, being a holomorphic function of the fields $\Phi_i$.
\end{itemize}

The Lagrangian of the scalar fields turns out to be\footnote{In this thesis, we do not intend to provide the detailed derivation of the supergravity action but just the relevant formulas for a proper discussion of inflation in this framework.  We refer the interested reader to the seminal papers \cite{Cremmer:1979up,Cremmer:1982en}.}
\be\label{LagrangianFSUGRA}
\mathcal{L} = -K_{i\jb}\ \partial_\mu \Phi^{i} \partial^\mu \Phi^{\jb} - V\,,
\ee
where $K_{i\jb}\equiv \partial_{\Phi_i}\partial_{\bar{\Phi}_{\jb}} K$ is the metric of the {\it \Kahler manifold} spanned by the fields $\Phi_i$. Then, the geometry of this internal space is fundamentally related to the kinetic terms of the scalars. In the case of one superfield with canonical kinetic terms, a very natural choice is $K=\Phi \bar{\Phi}$ which corresponds to a flat \Kahler geometry ($K_{\Phi\bP}=1$). However, other forms of $K$ are allowed in supergravity. Normally, string theory compactifications yield severe constraints on the geometric properties of this internal space.

The form of the F-term scalar potential is
\be \label{scpot}
V= e^K \left(K^{i\jb}\ \cD_i W\cD_{\jb} \bar{W}-3|W|^2\right),
\ee
where $K^{i\jb}$ is the inverse matrix of the \Kahler metric  $K_{i\jb}$ and
\be \label{Fterms}
\cD_i W \equiv \p_iW+K_i W\,,
\ee
is the \Kahler covariant derivative, referred to as F-term and being the order of parameter for spontaneous SUSY breaking. The potential \eqref{scpot} is always given by two opposing contributions where the negative definite term sets the AdS scale. Developing a generic mechanism which yields a positive potential $V>0$ is of primary interest for cosmological applications. We will return to this issue in Sec.~\ref{SgoldstinoOthogonal} and discuss its connection with the SUSY breaking directions.

The squared mass matrix of the scalar fields is given by
\begin{align}
m^2= 
 \begin{pmatrix}
  K^{i \bar{k}}\ \cD_{\bar{k}} \partial_j V &   K^{i \bar{k}}\ \cD_{\bar{k}}\partial_{\jb} V \\
   K^{\ib k}\ \cD_{k} \partial_{j} V&  K^{\ib k}\ \cD_{k} \partial_{\jb} V
 \end{pmatrix}\,,
\end{align}
where the \Kahler covariant derivative $\cD$ acts on $\partial V$ as in Eq.~\eqref{Fterms} for $W$.

The physics described by Eq.~\eqref{LagrangianFSUGRA} is invariant under a \Kahler transformation, namely
\begin{align}\label{Ktransformation}
K &\rightarrow K + \lambda + \bar{\lambda}\,,\\
W 	&\rightarrow e^{- \lambda} W\,,
\end{align}
where $\lambda$ is a holomorphic function of the fields. 

In order to implement single-field inflation in this framework, firstly we need to identify one of the real degrees of freedom with the inflaton field. In the simplest case of employing just one single superfield $\Phi$, one must choose an appropriate decomposition and assure stability of the other direction. A common choice is
\be\label{decomposition}
\Phi = \frac{\phi + i \chi}{\sqrt{2}}\,,
\ee
where the real direction may be identified with the inflationary trajectory while the orthogonal direction must be stabilized by means of an appropriate mechanism. Alternatively, one may allow for a multi-field dynamics which appears to be quite natural from a UV perspective, given the abundance of scalar fields in these scenarios. These additional degrees of freedom may be light thus participating in the inflationary process (see \cite{Wands:2007bd} for a review on this topic). Conversely, they  might be heavy and still produce observational features, as it was shown in the series of works \cite{Achucarro:2008sy,Achucarro:2008fk,Achucarro:2010jv,Achucarro:2010da,Achucarro:2012sm,Achucarro:2012yr,Achucarro:2012fd,Achucarro:2013cva,Achucarro:2014msa}.

\subsection{The $\eta$-problem in supergravity}

A generic problem arising in supergravity constructions describing inflation is the so-called $\eta$-problem: due to the specific form of the scalar potential of this class of theories, as given by Eq~\eqref{scpot}, the second slow-roll parameter $\eta$ generically receives contributions of order one \cite{Copeland:1994vg,Roest:2013aoa}. This can be immediately seen for the choice of a canonical \Kahler potential $K=\Phi\bP$. Expanding the overall exponential term of the scalar potential, we have
\be
V\approx(1+|\Phi|^2+...)\ V_0(\Phi)\,,
\ee
where the factor $V_0$ is determined by the superpotential. Then, the second slow-roll parameter obtains contributions such as
\be
\eta \approx 1 + \frac{V_0''}{V_0}+...\,,
\ee
which is order one for generic choices of $W$.

The exponential term in Eq.~\eqref{scpot} plays a very dangerous role and generically spoils the flatness of the inflaton potential. The problem becomes even more severe for super-Planckian values of the inflaton field.

In order to realise slow-roll inflation, one must therefore either resort to an undesired amount of fine-tuning to cancel the order one terms, or eliminate these contributions altogether by means of a symmetry. The latter is termed natural inflation \cite{Freese:1990rb}. It was first employed in a supergravity context to realise chaotic inflation \cite{Kawasaki:2000yn}. Instead of the canonical \Kahler, the authors opted for 
\begin{align}
  K = - \tfrac12 \left(\Phi - \bar \Phi\right)^2 \,. \label{shift-K}
\end{align} 
Due to the absence of the real part of the superfield $\Phi$, the \Kahler potential has a shift symmetry $\Phi \rightarrow \Phi + a$ with $a \in \mathbb{R}$.  This symmetry is the key to avoid the $\eta$-problem; relatedly, it prevents the inflaton potential from blowing up for large values of the inflaton field Re$(\Phi)$. It is evident that the dangerous term $e^K$ keeps increasing exponentially in the direction Im$(\P)$ while it remains constant  in Re$(\P)$. The shift symmetry is broken only by $W$, thus generating the inflaton potential.

However, the above discussion can also be misleading as it fails to take into account the following subtlety. The careful reader may have noticed that the two quoted \Kahler potentials are in fact related by a \Kahler transformation, and hence are physically equivalent. Yet more strikingly, the \Kahler potential can even be brought to the form 
 \begin{align}
  K  = \tfrac12 \left(\Phi + \bar \Phi\right)^2 \,,
\end{align}
by means of an additional \Kahler transformation.

The three forms of $K$ suggest symmetry protection for either none\footnote{In fact, the canonical $K$ does not depend on the complex phase of the field $\Phi$; however, in the origin of the orthogonal, radial direction, the phase is not a physical field.}, the real or the imaginary components, respectively. How does it come about that one \Kahler potential suffers from the $\eta$-problem, while physically equivalent potentials avoid it by means of a shift symmetry, which however protects different components? The answer to this apparent conundrum is that the $\eta$-problem is not only a statement about the \Kahler potential, but also about the `naturalness' of the superpotential. For {\it generic} choices of the superpotential, one needs a shift symmetry in $K$ to keep $\eta$ small; without that shift symmetry in the \Kahler potential, one needs a carefully picked $W$ to compensate for the order-one contribution to $\eta$. These two situations can be related by a \Kahler transformations and are exactly the options alluded to  above, i.e.~fine-tuning or symmetry. Thus the form of the \Kahler potential is not the only ingredient in evading the $\eta$-problem; also the generic or fine-tuned form of the superpotential comes into play.

\subsection{sGoldstino directions and inflation}\label{SgoldstinoOthogonal}

A successful implementation of the inflationary paradigm in supergravity requires a mechanism that assures a positive definite potential during the whole cosmological evolution. Achieving this is not always trivial due to the delicate balance between the two contributions in Eq.~\eqref{scpot}. Specifically, during inflation supersymmetry must be broken as we need $\cD_i W\neq 0$. This fact has dire consequences for inflation.

First of all, spontaneous breaking of supersymmetry at the inflationary scale naturally implies a second quasi-light field around the Hubble scale\footnote{This is termed quasi-single field inflation in \cite{Chen:2009we}, where the inflationary consequences of an additional scalar field in the complementary series of De Sitter's unitary irreps with $0 < m^2 \leq 9/4$ were investigated.}.  Any scalars other than the inflaton - of which there is always at least one, given that scalars come in pairs in SUSY models - therefore naturally acquire Hubble-scale masses. A concrete demonstration of this phenomenon can be found in \cite{Baumann:2011nk}.

Secondly, one can prove that the inflationary dynamics is highly constrained by the direction of SUSY breaking. This is defined by $\cD_i W$ in the scalar manifold and it is usually referred to as {\it sGoldstino directions}. The latter is a pair of scalar fields that is singled out by the spontaneous breaking of supersymmetry.

It is worth elaborating on the latter point as it has an interesting group-theoretical underpinning. As emphasised in the effective field theory approach to inflation \cite{Cheung:2007st}, the inflaton can be seen as the Goldstone boson arising from the spontaneous breaking of time translation invariance: this symmetry is necessarily broken during inflation (as exemplified by the value of the spectral tilt Eq.~\eqref{ns}) giving rise to a Goldstone boson, which can be seen as either a scalar field or an additional helicity-0 component in the metric. This is analogous to the additional degrees of freedom of the $W^\pm$ and $Z^0$ vector bosons as arising in the Higgs mechanism. A slightly different reasoning applies to spontaneous breaking of SUSY. In this case, the Goldstone modes are a pair of spin-1/2 fermions, whose supersymmetric partners are spin-0 fields. These are referred to as Goldstini and sGoldstini fields, respectively. Their emergence is completely analogous to the Higgs boson itself in the spontaneous breaking of gauge symmetry: the Higgs is the gauge partner of the aforementioned Goldstone bosons. Thus, there are interesting similarities and differences between these interpretations of the inflaton and the sGoldstini scalar fields: both arise as a consequence from the spontaneous breaking of a local symmetry (i.e.~time translational invariance and SUSY), in which they are Goldstone modes or the partners thereof.

Studying the trajectories of the sGoldstini fields turns out to be very important for the dynamics of the whole system. Indeed, these scalars generically correspond to unstable directions on the scalar manifold, signaling instabilities \cite{GomezReino:2006dk, GomezReino:2006wv, GomezReino:2006wv}. In addition, depending on the angle between the inflaton and sGoldstini directions one can draw very general conclusions about the inflationary dynamics \cite{Covi:2008cn,Borghese:2012yu}. Specifically, one can show that, in the case these directions have a non-negligible overlap and the gravitino mass is orders below the inflationary scale, single-field, slow-roll and small field inflation cannot be realize in supergravity \cite{Borghese:2012yu}. This follows from a general inequality that we will refer to as the {\it geometric bound}, as it involves the curvature of the \Kahler manifold spanned by the scalars.

There are two extreme cases one may consider. We list them below.

\subsubsection{sGoldstino inflation}\label{subsecsgold}
When the directions of the inflaton and of the sGoldstino coincide, we refer to this scenario as {\it sGoldstino inflation} (this framework has been investigated by several studies such as  \cite{Goncharov:1983mw,Goncharov:1985yu,AlvarezGaume:2010rt,AlvarezGaume:2011xv,Achucarro:2012hg,Ketov:2014qha,Ketov:2014hya,Linde:2014ela,Linde:2014hfa,Roest:2015qya,Linde:2015uga,Terada:2015sna,Ketov:2015tpa}). In the most economical scenario, this is the situation when just a single superfield is involved in the supergravity construction. In this case, the inflaton plays a double role: it drives the quasi-exponential expansion and it breaks supersymmetry at the same time. However, one cannot evade the geometric bound of \cite{Borghese:2012yu} and, then, one needs to take a number of facts into account in order to realize a successful cosmological scenario. In addition, obtaining an arbitrary inflationary potential becomes very challenging.

Here, we present three possible settings in terms of their internal \Kahler space:

\begin{itemize}

\item A first natural choice for $K$ is given by the shift-symmetric function \eqref{shift-K} corresponding to a flat \Kahler geometry. This allows for a truncation to only the imaginary part of $\Phi$ provided one takes
 \begin{align}
  W = f(\Phi) \,, \label{real-W}
 \end{align}
where the function $f$ is a real holomorphic function of $\Phi$; in other words, when expanded in terms of its holomorphic argument, all coefficients are required to be real. The mass spectrum for this model reads
\be
 \begin{aligned}
  m^2_{{\rm Re}(\Phi)} & = - 6 f'^2 - 6 f f'' + 2 f''^2 + 2 f' f''' \,,  \\ 
  m^2_{{\rm Im}(\Phi)} & = 4 V + 4 f^2 + 2 f'^2 -2 f f'' + 2 f''^2 -2 f' f''' \,.
 \end{aligned}
 \ee
when evaluated at $\Phi = \bar \Phi$ and where primes denote derivatives with respect to the variables the function depends on. In this set-up, the imaginary part of the superfield $\Phi$ will generically give rise to a Hubble-scale field, which is stabilised at zero. This is exactly as expected, as the \Kahler potential contributes order one to $\eta$, and the contribution from $W$ will generically be much smaller (think e.g.~a sum of exponential - the shift symmetry of the imaginary part is mildly broken, leading to a small contribution). Hence the total $\eta$ will be order one, allowing stabilisation of this field. In contrast, the real part Re$(\P)= \phi$ will be light and play the role of the inflaton.

The resulting scalar potential, along $\Phi=\bP$, reads
 \begin{align}\label{potsingleflat}
  V = f'(\Phi)^2 - 3 f(\Phi)^2 \,.
 \end{align}
This model thus allows for a truncation to a single field. However, the form of this potential is clearly not the most general due to the appearance of both the function $f$ and its derivatives. For large field inflation with a single monomial dominating the superpotential at large field values, the negative-definite contribution dominates the scalar potential. Thus, it is impossible to realise in particular chaotic inflation with a monomial in this way (two explicit examples with polynomials of fourth order as superpotentials are given in \cite{Achucarro:2012hg}).

\item A second natural possibility is given by taking a logarithmic rather than polynomial \Kahler potential. The choice
 \begin{align}\label{hyperbKahlerpotential}
   K = - 3\alpha \ln \left(\Phi + \bar \Phi\right) \,,
 \end{align}
leads to a hyperbolic \Kahler geometry, a manifold $SU(1,1) / U(1)$, whose curvature is parameterised by $\alpha$. Thus, it is well motivated from a supergravity point of view as well as from string theory, as we will discuss below. In addition it eliminates the dangerous exponential terms arising from the overall \Kahler exponential in the scalar potential. Therefore, there is no longer a compelling reason to identify the imaginary part of $\Phi$, which now enjoys the shift symmetry of $K$, with the inflaton. This is good news, as it is generically {\it inconsistent} to set the real part of $\Phi$ equal to zero in order to obtain a single field model. The fact that this is consistent in the model with the shift symmetric \Kahler potential is a consequence of the square in \eqref{shift-K}. As the logarithmic \Kahler potential no longer has this feature, the only consistent truncation is to the real part. For this one needs to take the same requirement on the superpotential \eqref{real-W} being a real function of $\Phi$.

This model, with the same $W$ as in Eq.~\eqref{real-W} and along $\Phi=\bP$, leads to the scalar potential:
 \begin{align}\label{potsinglelog}
V=  8^{-\alpha } \Phi^{-3 \alpha } \left[\frac{\left(3 \alpha  f(\Phi)-2 \Phi f'(\Phi)\right)^2}{3 \alpha }-3 f(\Phi)^2\right]
 \end{align}
The choice $\alpha = 1$ is special due to the no-scale structure \cite{Cremmer:1983bf,Ellis:1983sf,Lahanas:1986uc}, where the negative definite term is exactly cancelled. However, the functional form of Eq.~\eqref{potsinglelog} appears to be even more complicated than the flat correspondent Eq.~\eqref{potsingleflat}.

\item A final possibility was recently proposed by Ketov and Terada in \cite{Ketov:2014qha,Ketov:2014hya} (see also \cite{Linde:2014ela,Terada:2015sna,Ketov:2015tpa} for subsequent developments) where the authors considered a logarithmic \Kahler potential of the form
\be\label{KK}
 K= -3\ln \left[ 1 + \frac{\Phi + \bar{\Phi} + \zeta \left( \Phi + \bar{\Phi}\right)^4}{\sqrt{3}}\right]\,.
\ee
In this setup, the role of the inflaton is played by the Im$(\Phi)=\chi$. The term with constant parameter $\zeta$ serves to stabilize the field Re$(\Phi)=\phi$ during inflation at  $\phi \approx 0$. The main idea is that by making $\zeta$ sufficiently large one can make the field component $\phi$ heavy and constrained to a very small range of its values, 
$\phi \ll 1$, so it plays almost no role during inflation with the inflaton field $\chi \gg 1$. 
For superpotentials
\be
W = {1\over \sqrt 2} f(-\sqrt 2 i\Phi) \,,
\ee
where $f$ is a real function of its argument, the potential along the inflaton direction $\phi \ll 1$ becomes
\be\label{quadr}
V \approx \bigl[f'(\chi)\bigr]^{2} \ .
\ee
For example, for $W = {1\over 2}m\Phi^{2}$ one recovers the simplest chaotic inflation potential $V = {m^{2}\over 2}\chi^{2}$ along the direction $\phi = 0$. A numerical investigation of this scenario in \cite{Ketov:2014hya} confirms that for sufficiently large $\zeta$, the field $\phi$ practically vanishes during the main part of inflation. Its evolution begins only at the very end of inflation, so the cosmological predictions almost exactly coincide with the predictions of the quadratic scenario. At the end of inflation, the field rolls down towards its supersymmetric Minkowski vacuum at $\Phi =0$, where $V =0$, $W =0$, and supersymmetry is restored.

However, this scenario does not lead to a pure single-field truncation  and a two fields dynamics generically appears near the minimum\footnote{In order to truncate consistently the orthogonal direction to the inflaton, Re$\Phi$, one has to ensure that its equation of motion is satisfied. It can be checked that this receives contributions from the Christoffel symbols, which do not allow us to decouple the field. However, the extra factors obtained are proportional to the slow-roll parameters and, therefore, the inflationary trajectory occurs approximately along the imaginary part of $\Phi$. Just after inflation, we can notice a shift of the inflaton from the initial straight initial direction.}. We will return to this model in the next Chapter.

\end{itemize}

In conclusion, obtaining a general mechanism which assures always successful inflation, within a single superfield context, seems to be a rather non-trivial and challenging task. However, later in Sec.~\ref{SECalphascale}, we will clarify the fundamental steps to follow in order to achieve a consistent inflationary scenario with a model involving simply one superfield. We will do this in a separate Section given the relevance of the proposed original recipe. The basic mechanism indeed involves a novel supergravity construction (named $\alpha$-scale model) firstly discovered in \cite{Roest:2015qya}. Surprisingly, it sheds light on the deep connection between the flat \Kahler \eqref{shift-K} and the hyperbolic one \eqref{hyperbKahlerpotential}.

\subsubsection{Orthogonal inflation} \label{orthogonalinflation}
	
When the directions of the inflaton and of the sGoldstino are orthogonal to each other, we refer to this scenario as {\it orthogonal inflation}. This particular framework necessarily involves at least two superfields: $\Phi$, responsible for inflation, and the complex scalar $S$ breaking supersymmetry. In addition, it provides a unique way to evade the geometric bound of \cite{Borghese:2012yu} thus allowing for remarkable flexibility.

The benchmark model is characterized by a superpotential linear in $S$ such as
\be\label{superpotorthogonal}
W= S f(\Phi)\,,
\ee
with $f$ being again a real holomorphic function. Then, after choosing a suitable \Kahler potential which allows for a consistent truncation along the direction $S=0$ (this is usually assured if $K$ is invariant under $S\rightarrow -S$ and, then, e.g. depending on $S\Sb$), the scalar potential assumes the form 
\be\label{scalarpotentialorthogonal}
V= e^K K^{S\Sb} |f(\Phi)|^2\,.
\ee
The latter is always a positive function as the negative definite contributions of Eq.~\eqref{scpot} disappears at $S=0$. Along the same trajectory, the superpotential is indeed identically zero and the F-terms are
\be
\cD_\Phi W=0\,, \qquad \cD_S W=f\,.
\ee
The latter fact sheds light on the peculiar role of the complex scalar $S$ in this construction: this field belongs to the sGoldstino supermultiplet and inflation happens in the  orthogonal direction to the one defined by the sGoldstino along which supersymmetry is broken.

Note that the potential Eq.~\eqref{scalarpotentialorthogonal}  still depends on the two real degrees of freedom of the complex field $\Phi$ and, in order to have single-field inflation, one must truncate along a specific direction and assure stabilization of the trajectory. However, both the consistency of the final truncation and stabilization issues strictly depend on the specific form of the \Kahler potential. The simple framework, with canonical shift-symmetric \Kahler in the inflaton sector coupled to $S$, was first proposed in \cite{Kawasaki:2000yn} and further developed in \cite{Kallosh:2010ug,Kallosh:2010xz}. On the other hand, examples of working models of orthogonal inflation with logarithmic \Kahler of the form \eqref{hyperbKahlerpotential} in the inflaton sector (that is, at $S=0$) already appeared in \cite{Cecotti:1987sa,Kallosh:2013lkr,Ellis:2013nxa}. However, a full general analysis with an arbitrary superpotential, such as the one of Eq.~\eqref{superpotorthogonal}, was first performed in \cite{Roest:2013aoa} and, then, in the context of the $\alpha$-attractors model in \cite{Kallosh:2013yoa,Kallosh:2014rga,Carrasco:2015rva,Scalisi:2015qga,Carrasco:2015pla} (see also \cite{Lahanas:2015jwa} for related analysis).

To conclude, a model with the inflaton orthogonal to the sGoldistini fields always allows for remarkable flexibility in terms of the scalar potential.  We will discuss the two important scenarios of orthogonal inflation with flat and hyperbolic \Kahler geometry respectively in Sec.~\ref{SECflat} and in Sec.~\ref{SECHyp}.

\subsection{Towards an embedding in string theory}

Here we would like to discuss to what extent successful supergravity models of inflation can be implemented in string theory. Which are the typical form of $K$ and $W$ following from a string-theoretic configuration\footnote{When restricting to the fields $\S$ and $\Phi$, which will generically be a subset of all fields in string-theoretic scenarios, we are assuming that this is a consistent procedure and will not address the subtleties of such truncations as pointed out in e.g.~\cite{Hardeman:2010fh}.}? Specifically, is it possible to realize orthogonal inflation by means of sectors naturally arising in string theory?

Let us start by discussing {\it open string fields} as candidates for inflation, the most famous case being D-brane inflation \cite{Dvali:1998pa,Burgess:2001fx,Kachru:2003sx}, where the position of a D-brane in the internal compact dimensions plays the role of the inflaton. However, other open strings, such as more generic matter fields, can also be considered as inflaton candidates. Matter fields (including open string moduli) in  string theory obtain a K\"ahler potential of the form\footnote{In this subsection the fields $\Phi$ and $S$ do not necessarily denote the inflaton and the sGoldstino; instead, their role should be clear from the context.}
\ba\label{KMatter}
&& K= \alpha \Phi \bar \Phi   \qquad {\rm or } \qquad K=  \alpha \left( \Phi - \bar \Phi \right)^2  \,.  
\ea
Here we have assumed that any closed string moduli have been stabilised and their vevs are taken into account in the constant $\alpha$. Hence matter fields can satisfy all symmetry requirements for a shift-symmetric or simply a canonical \Kahler potential such as Eq.~\eqref{KMatter}. Therefore, from the point of view of the \Kahler potential, the matter sector alone can provide both sGoldstino and inflaton candidates. Examples of matter fields with a shift symmetry have been discussed in the context of D-brane inflation with D3/D7 in \cite{Kallosh:2007ig} and in the context of fluxbrane inflation with D7/D7 in \cite{Hebecker:2012aw}. These K\"ahler potentials can also arise for some matter fields in heterotic theory \cite{LopesCardoso:1994is,Antoniadis:1994hg}. Moreover, the superpotential for matter fields generically turns out to be of the form\footnote{The somewhat unconventional $i$ arises in the superpotential as a consequence of the choice of \Kahler potential \eqref{KMatter} depending on $\Phi - \bar \Phi$ rather than $\Phi + \bar \Phi$, as often considered in the string theory literature.}
\be\label{WM}
W = \beta \sum_n^N \prod_{\alpha_n} i \Phi_{\alpha_n} \,,
\ee
where again, we take into account a likely dependence on any closed string moduli vevs into the constant $\beta$.
From the structure above we have the following properties:
\begin{itemize}
\item We generically expect to get the sum over several couplings for all the fields involved, including the sGoldstino, which is in contrast to the linear structure of \eqref{superpotorthogonal}. 
\item On the other hand, there is a simple case which can fit completely. If one is allowed to truncate the superpotential to only a single term in the sum over $n$ in $W$ above, then it is always possible to  add a phase such that the superpotential has the form $W= S f(\Phi_i)$ with $f$ real.
\end{itemize}
In conclusion, having matter fields alone in a configuration where some of these have a shift symmetry, it is possible, restricting to a single term in $W$, to identify the relevant sectors needed for inflation. The detailed implementation of this model will be discussed in Sec.~\ref{SECflat}. Extra sectors in the configuration can be added to $W$ so long a separation is possible, for example as it happened in \cite{Kallosh:2011qk}.

The other possibility to consider is geometric \textit{closed string moduli} in string theory. Generically these fields are the so-called dilaton $S$, the complex structure moduli $U$ and the K\"ahler moduli $T$. These have a well known K\"ahler potential, which takes the form such as Eq.~\eqref{hyperbKahlerpotential} where we denote $\Phi = \{ S, T, U \}$. Such fields do enjoy the shift symmetry (which in this case we take in the imaginary part) of $\Phi$ but not the $\mathbb{Z}_2$ symmetry $\Phi \rightarrow - \Phi$. Therefore, there is no field which can be identified with the sGoldstino direction, as it was employed for orthogonal inflation (we remind the reader that the function $K$ must allow for consistent truncation along $S=0$). Turning to the superpotential:
\begin{itemize}
\item  If we consider the shift symmetry to be broken only by {\em non-perturbative} effects, the superpotential turns out to be a function of $\Phi$. For example, this is the case of $W \propto e^{-a \Phi}$ for $T$ in type IIB compactifications with fluxes considered widely in the literature.

\item On the other hand, if the shift symmetry is broken at tree, {\em perturbative} level, then $W$ is a function of $i\Phi$. This is the case  of the tree-level superpotentials for the $S T U $-moduli generated via bulk, geometric and non-geometric fluxes.
\end{itemize}

In conclusion the closed string sector provides promising inflationary directions; however, the lack of $\mathbb{Z}_2$-symmetric \Kahler potentials prevents the implementation of orthogonal inflation.

From the discussion above, an interesting hybrid emerges naturally: the case when both matter and closed string moduli play a role. The identifications of the fields with the relevant sectors of SUSY breaking and inflation are clear: a K\"ahler modulus is identified with the inflaton sector, while a matter field is identified with the sGoldstino sector. In this case generically  we expect  the K\"ahler potential to be of the form:
 \be
  K = - 3\alpha  \ln \left( \Phi + \bar \Phi - S \bar S\right) \,,\quad {\rm or } \quad   K = - 3\alpha  \ln \left( \Phi + \bar \Phi\right) - S \bar{S} \,, \label{log-K}
 \ee
where $\Phi$ is identified with a closed string modulus, for example the K\"ahler modulus, and $S$ is identified with some matter field, for example a brane position. From a string theory perspective, natural values of curvature parameter $\alpha$  are of order one (relevant examples are $1, 2/3$ and $1/3$). We will discuss the details of such a model in Sec.~\ref{SECHyp}.


\section{Flat K\"ahler geometry and arbitrary inflation}\label{SECflat}

In this section, we will discuss a number of two-superfield models of inflation with the common feature of the orthogonality between the sGoldstino directions and the inflaton. We have indeed already seen in Sec.~\ref{orthogonalinflation} that this construction allows for remarkable flexibility. We will show the details below. In addition, we focus on flat \Kahler geometry for the inflaton field $\Phi$ such that the metric takes the form
\be\label{flatspaceK}
ds^2= d\Phi\ d\bP\,.
\ee
A useful parametrization for $K$ which avoids the $\eta$-problem and allows for the orthogonal separation between the inflaton and the sGoldstino was firstly introduced in \cite{Kawasaki:2000yn}. Successfully realizing chaotic  inflation in supergravity was the original motivation of  this investigation. This pioneering model consists of
 \begin{align}
  K = - \tfrac12 \left(\Phi - \bar \Phi \right)^2 + S \bar{S} \,, \qquad W = M S \Phi \,, \label{KYY}
 \end{align}
in terms of a real constant $M$. Inflation can be chosen to take place along $\Phi - \bar \Phi = S = 0$, while the remaining degree of freedom Re$(\P)=\phi$ has a quadratic scalar potential:
  \begin{align}
  V = M^2 \phi^2 \,.
 \end{align}
However, in this case the three truncated fields are not yet stabilised: while the imaginary part of $\Phi$ has a Hubble-scale mass in compliance with the $\eta$-problem, this is not the case for the $S$-field. To this end one can add a higher-order term to the \Kahler potential,
 \begin{align} \label{Kflatstabilized}
  K = - \tfrac12 \left(\Phi - \bar \Phi\right)^2 + S \bar{S} + \zeta  \left( S \bar{S}\right)^2 \,,
 \end{align}
which parametrises its curvature. The ensuing mass eigenvalues are
\begin{align}
 m^2_{{\rm Im}(\Phi)} = V + M^2 \,, \quad m^2_{S} = \zeta V + M^2 \,. 
\end{align}
For coefficients $\zeta $ of order one this will indeed allow both Im($\Phi)$ and $S$ to be stabilised.

Subsequently, a new development has build on this model to generate other inflationary potentials in a similar manner, see e.g.~\cite{Kallosh:2010ug}. This has culminated in a model by Kallosh, Linde and Rube (KLR) \cite{Kallosh:2010xz}, which consists of a prescription of how to build a class of supergravity models allowing for a completely arbitrary inflaton potential $V(\phi)$. Similar to the previous model, it consists of two complex scalar fields $\P$ and $\S$. The role of both fields will be identical as before; the real part Re$(\P) = \phi$ will be the inflaton field while Im$(\P)$ and $S$ will be essential in order to stabilise the inflationary trajectory, along which such fields will vanish. However, the \Kahler and superpotential are generalised to the following:
\be \label{KLR}
  K = K \left((\P - \bar \P)^2, \S \Sb, \S^2, \Sb^2\right), \quad W=S f(\P)\,,
\ee
where, similarly to the previous cases, $f(\P)$ is an arbitrary but real holomorphic function of the variable $\P$.

The \Kahler potential can be an arbitrary function of the arguments as indicated, and, as a consequence, it is separately invariant under the following transformations:
\ba
S \rightarrow -S \,, \quad \Phi\rightarrow - {\Phi}, \quad \Phi\rightarrow \Phi +a, \qquad a \in \mathbb{R} \,. \label{sym}
\ea
A first natural choice for $K$ is Eq.~\eqref{Kflatstabilized}. On the other hand, the prescription Eq.~\eqref{KLR} allows for more complicated functional forms, provided they satisfy Eq.~\eqref{sym}. Some examples with a logarithmic $K$ are given in \cite{Kallosh:2010xz,Kallosh:2013lkr}. Notice that the \Kahler curvature is still flat along the inflaton trajectory Im$\Phi=0$, although the 2-dimensional internal space may not satisfy Eq.~\eqref{flatspaceK}.

Amongst the main novelties of such a model is that a completely general inflationary potential can be generated from a supergravity model. Moreover, given $K$ and $W$,  one does not need to perform long calculations without knowing whether the final form of the potential will be actually suitable for inflation or not. Within this model, the form of the inflaton potential will always be 
\be \label{infpot}
V(\phi)=  f(\phi)^2,
\ee
which is a completely general positive function of $\phi$. This functional freedom is guaranteed by the symmetries of the K\"ahler potential $K$ and by the linearity of $W$ in $S$. 

In the above derivation we have set the three fields that are not protected by the shift symmetry, i.e.~$S$ and $\Phi - \bar \Phi$, equal to zero. The consistency of this truncation can be seen from the full mass matrix, which gives rise to the following eigenvalues:
\begin{align}
 m^2_{{\rm Im}(\Phi)} & =  f^2 \left(1 - K_{\Phi \bar \Phi S \bar S} - \tfrac12 \partial^2_\Phi \ln(f) \right) \,, \notag \\
 m^2_{S} & =  - \left( K_{S \bar S S \bar S} \pm \left|K_{SSS \Sb} - K_{SS}\right|\right) f^2 + (\partial_\Phi f)^2 \,. 
\end{align}
Thus, for suitably chosen \Kahler manifolds with the right sectional curvature, the mass of these components is indeed Hubble-scale and hence they are stabilised at their origin.

\section{Hyperbolic K\"ahler geometry and attractors}\label{SECHyp}

In this Section, we would like to turn to the other maximally symmetric possibility for the \Kahler geometry of the inflaton field. This is the hyperbolic space of the Poincar\'e disc or half-plane. The metric of the {\it unit disc} reads
\be
ds^2 = 3\alpha \frac{d\Psi\ d\bar{\Psi}}{\left(1-\Psi\bar{\Psi}\right)^2}\,,
\ee
defined for $\Psi\bar{\Psi}<1$.  The usual \Kahler potential associated with this space is 
\be
K=-3\a \ln\left(1-\Psi\bar{\Psi}\right)\,.
\ee
Its curvature is given by
\be\label{curvatureKhyp}
R_K= - K_{\Psi\bar{\Psi}}^{-1} \partial_\Psi \partial_{\bar{\Psi}} \ln K_{\Psi \bar{\Psi}}= -\frac{2}{3\alpha}\,,
\ee
that is constant and negative as expected (this space having hyperbolic geometry) and depending just on the parameter $\alpha$.

It is possible to go to the {\it half plane} representation of this geometry\footnote{The difference between the unit disk representation and the one in terms of the half plane coordinates can be visually appreciated, respectively, in the picture of the {\it front cover} and the {\it bookmark} of this thesis.} by means of a change of variables such as
\be
\Phi= \frac{1+\Psi}{1-\Psi}\,,
\ee
which leads to the metric
\be
ds^2 = 3\alpha \frac{d\Phi\ d\bP}{\left(\Phi+\bP\right)^2}\,,
\ee
defined for $\Phi+\bP>0$. The corresponding \Kahler potential for this space is the one already given in Eq.~\eqref{hyperbKahlerpotential}. The curvature, being an invariant, remains the same as in Eq.~\eqref{curvatureKhyp}. An interesting analysis regarding the properties of such a geometry, its possible representations and connections to physics is performed in \cite{Kallosh:2015zsa}.

Throughout the following, we will mainly make use of the half-plane coordinates $\Phi$ as a matter of convenience. We will present supergravity models of inflation that admit consistent truncation at $\Phi=\bP$. Along this line, the relation between the geometric field and the canonical normalized field $\varphi$ is
\be
\Phi=\bP= e^{-\sqrt{\frac{2}{3\alpha}}\varphi}\,.
\ee
This relation simply reflects the dramatic effects of the non-trivial geometry of the hyperbolic \Kahler manifold. This indeed induces a boundary in moduli space (located at $\Phi=0$) where the theory attains a conformal \cite{Kallosh:2013hoa} or a scale symmetry  \cite{Ozkan:2015kma} (depending on the parameter $\alpha$).  Inflation takes place as the inflaton moves away from this boundary, leading to universal cosmological predictions in terms of $n_s$ and $r$. In canonical coordinates, the boundary is indeed pushed at $\varphi=\infty$. Then, any generic expansion around this boundary often corresponds to a scalar potential which is an exponential fall-off from de Sitter such as
\be\label{expfalloff}
V=V_0 + V_1\ e^{-\sqrt{\frac{2}{3\alpha}}\varphi}+ \ldots\,,
\ee
where the dots represents subleading terms, irrelevant for values of $\a$ of order one. This expansion automatically yields universal values for the spectral index and tensor to scalar ratio, which read
\be\label{nsrattractors}
n_s=1-\frac{2}{N}\,, \qquad r=\frac{12 \alpha}{N^2}\,,
\ee
at large values of the number of e-folds $N$. The predictions \eqref{nsrattractors} provide an excellent fit with the latest Planck data and, for specific values of $\alpha$, simply correspond to the ones of the Starobinky model \cite{Starobinsky:1980te} (together with its supergravity implementations\cite{Cecotti:1987sa,Ellis:2013xoa,Kallosh:2013lkr,Buchmuller:2013zfa,Farakos:2013cqa,Ellis:2013nxa}) and  Higgs inflation \cite{Bezrukov:2007ep}.

\begin{figure}[htb]
\vspace{3mm}
\begin{center}
\includegraphics[width=7cm,keepaspectratio]{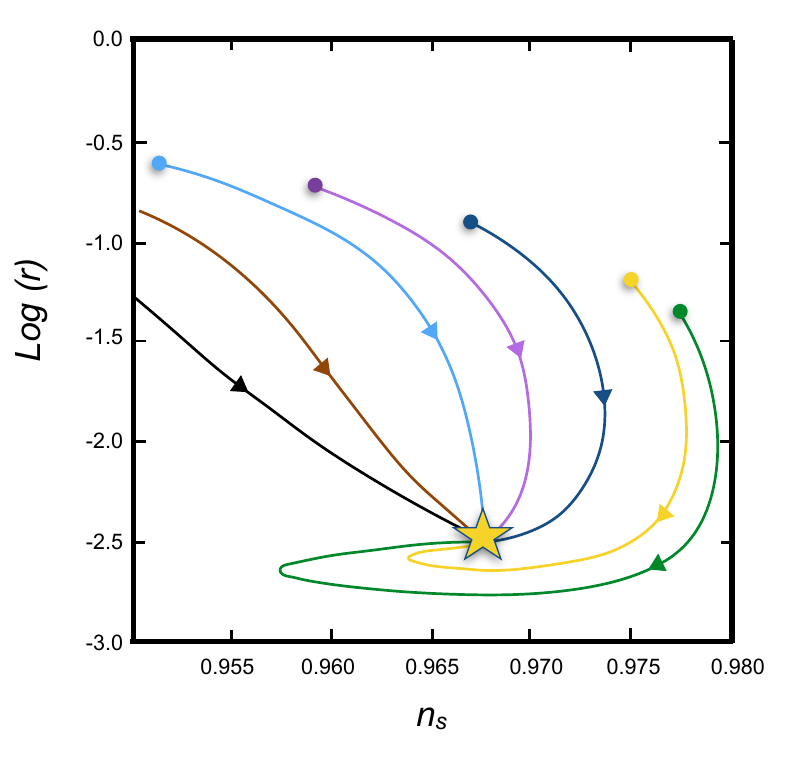}
\vspace*{0.2cm}
\caption{\it Hyperbolic \Kahler geometries as well as non-minimal couplings to gravity or non-canonical kinetic terms generically induce an attractor for observations. The specific details of the model (in this case, we show monomial chaotic models of inflation) get washed out and the observational predictions all converge to the  “attractor point” denoted by the yellow star. This evolution is normally regulated by a specific parameter of the model, in the supergravity case being the value of the \Kahler curvature. }\label{attractorFig}
\end{center}
\vspace{-0.5cm}
\end{figure}

Remarkably, the \Kahler curvature \eqref{curvatureKhyp} plays a fundamental role in this construction and, essentially, becomes a measurement of the amount of primordial gravitational waves we are currently looking for in the sky \cite{Ferrara:2013rsa,Kallosh:2013yoa,Kallosh:2014rga} (the previous works \cite{Ellis:2013nxa,Roest:2013aoa} already pointed out how sensitive $r$ is to the curvature $R_K$ and, then, to value of $\a$). As $\alpha$ varies from infinity (i.e. flat curvature) to order one or smaller, the inflationary predictions go from completely arbitrary (in the flat case) to the very specific values above\footnote{Similar attractor behaviour has been noticed also in the context of models of inflation with non-minimal coupling to gravity \cite{Kallosh:2013tua} (see also \cite{Mosk:2014cba}). Interestingly, one can find a common origin for the attractor phenomenon observed in both contexts. This is due to a pole of order two in the kinetic term of the inflaton field \cite{Galante:2014ifa} (see also \cite{Broy:2015qna,Terada:2016nqg})}. Turning on the curvature therefore ``pulls'' all inflationary models into the Planck dome in the $(n_s,r)$ plane, similarly to what is shown in Fig.~\ref{attractorFig}.

The phenomenon described above, which intimately relates geometric properties of the hyperbolic \Kahler manifold to universal observational predictions, was first discovered in \cite{Kallosh:2013hoa,Kallosh:2013yoa,Kallosh:2014rga} and it is referred to as {\it $\a$-attactors}. However, some working examples were already found in previous investigations such as \cite{Cecotti:1987sa,Kallosh:2013lkr,Ellis:2013nxa,Roest:2013aoa}, where simple and natural choices of the superpotential (such as monomial or polynomial forms)  lead to an inflationary regime such as the one of Eq.~\eqref{expfalloff}. The common feature of all these supergravity constructions is the complete orthogonality between the sGoldstino $S$ and the inflaton $\Phi$. This is indeed very useful in order to kill the negative term in the scalar potential Eq.~\eqref{scpot}. In the context of orthogonal inflation, the first analysis with varying \Kahler curvature and general superpotential $W$ was performed in \cite{Roest:2013aoa}.

It is important  to note that the final result may be different depending on how the sGoldstino enters the \Kahler potential. The field $S$ may have a simple canonical \Kahler or appear inside the argument of the logarithm. We have already outlined these two common choices in Eq.~\eqref{log-K} and we will present the details in the following subsections.

\subsection{Model with $ K = - 3\alpha  \ln ( \Phi + \bar \Phi - S \bar S)$}

Here, we follow the analysis done in \cite{Roest:2013aoa}. This supergravity construction is characterized by the following choices for the \Kahler potential and superpotential:
\begin{align}\label{KWSinside}
   K = - 3\alpha  \ln 	\left( \Phi + \bar \Phi - S \bar S\right), \quad W = S f(\Phi) \,,
\end{align}
where $f$ is still a real holomorphic function of its argument and the corresponding \Kahler manifold is $SU(2,1) / U(2)$. This model allows for a consistent truncation to the inflationary trajectory at Im$(\Phi) = S = 0$\footnote{Note that this is valid more in general for any \Kahler potentials of the form $K = K ( \Phi + \bar \Phi, S \bar S, S^2, \bar S^2)$}. However, this does not imply that this truncation is also stable. For this, one needs to consider the mass spectrum of such fields. In order for effective single-field behaviour, these will have to be super-Hubble. We will later check, in explicit examples, to what extent this condition can be met. Moreover, in the case of Eq.~\eqref{KWSinside}, the shift symmetry is enhanced to the three-dimensional Heisenberg group\footnote{The symmetry of the Heisenberg invariant \Kahler potential has also been employed to solve the $\eta$-problem in \cite{Antusch:2008pn,Antusch:2009ty}. However, that scenario differs in an important way from the present: in that case, the field $S$ is identified as the inflaton, whereas a third superfield is added to play the role of the sGoldstino. The inflationary predictions of that set-up are thus unrelated to ours.}, which acts in the following way
 \begin{align}
  \Phi \rightarrow \Phi + i a + \bar b S + \tfrac12 |b|^2 \,, \quad S \rightarrow S + b  \,, \quad a \in \mathbb{R}\,, b \in \mathbb{C} \,. \label{Heisenberg}
 \end{align}

The scalar potential as function of the fields $\Phi$ and $S$ reads
\ba\label{VHeis}
V&=& \frac{|S|^2}{3\a}\left(  X^{1-3\a}|S|^2 + X^{2-3\a} \right)\left| f' + \frac{ 3\a f }{X}\right|^2  + \frac{X^{1-3\a}}{3\a}|f|^2\left(1+ \frac{3\a |S|^2}{X}\right)^2 \nonumber \\
&& 
+\frac{X^{1-3\a}}{3\a}|S|^2\left(1+\frac{3\a |S|^2}{X}\right)\left[  \bar f \left(f' + \frac{3\a f}{X} \right) +   f \left(\bar f' + \frac{ 3\a \bar f }{X}\right)  \right]\nonumber \\
&&  
 - 3\frac{|S|^2|f|^2}{X^{3\a}}\,,
\ea
where, for convenience of notation, we have defined a compact variable $X$ such as
\be
X\equiv\Phi+\bP-S\bar{S}\,.
\ee
Thus at $S=0$ the potential is simply
\be
V= \frac{X^{1-3\alpha}|f|^2}{3\a}.
\ee
In terms of real and imaginary parts for $\Phi$ and the field $S$, the masses of the fields read:
\begin{align}
m_{{\rm Re}(\Phi)}^2 & = \frac{2}{3\a} \left[X^{1-3\a}\left(3\a -2+\frac{1}{3\a}\right) f^2 +  X^{2-3\a}\left(\frac{1}{\a} -2\right)f f'+ \frac{X^{3-3\a}}{6\a} (f f'' + f'^2)\right], \notag \\
m_{{\rm Im}(\Phi)}^2 & = \frac{2}{3\a} \left[X^{1-3\a}\left(1-\frac{1}{3\a}\right) f^2 -  \frac{X^{2-3\a}}{3\a} f f'  + \frac{X^{3-3\a}}{6\a} (f'^2-f f'')\right], \notag \\
m_S^2 & = \frac{X^{1-3\a}}{3\a} \left(3\a -2-\frac{1}{3\a}\right) f^2 + \frac{X^{2-3\a}}{3\a} \left(\frac{2}{3\a}-2\right) f'f + \frac{X^{3-3\a}}{9\a^2} f'^2. \label{spectrum}
\end{align}
Clearly, for a successful single-field inflationary model, the first of these has to be light whereas the latter three degrees of freedom need to be stabilised, either around or above the Hubble scale.

\subsubsection{Example with $\a=1$}
A first interesting example of the model above is given by the Cecotti model of inflation\footnote{
A similar set-up with an identical \Kahler potential \eqref{cecottiKW} was also used to embed the Starobinsky model in supergravity \cite{Ellis:2013xoa}. However, in that case the inflaton was identified with one of the directions of $S$, while the $\Phi$ field was argued to be stabilised by other means.}. This is found for $\a=1$ and the following potentials:
 \begin{align}
  K = - 3 \log \left( \Phi + \bar \Phi - S \bar S\right) \,, \quad
  W = 3 M S (\Phi -1) \,. \label{cecottiKW}
 \end{align}
In terms of a canonically normalised scalar field $\varphi$, this yields the scalar potential
 \begin{align}
  V = \tfrac34 M^2 \left( 1 -  e^{ - \sqrt{2} \varphi / \sqrt{3}} \right)^2 \,.
 \end{align}
Inflation takes place at large $\varphi$. In this limit the three masses become
 \begin{align}
  m^2 = \{ 0, 4 H^2, -2 H^2 \} \,,
 \end{align}
where $H^2 = V/3$. This has been demonstrated to be equivalent to Starobinsky's $R + R^2$ model of inflation \cite{Starobinsky:1980te}. The relations of this model to superconformal supergravity have been discussed in \cite{Kallosh:2013lkr}. In this reference it was also been pointed out that the $S$ field is not stable with this \Kahler choice;  to this end one could add a stabilising term $ \beta (S \bar S)^2 / (\Phi + \bar\Phi)$ to the argument of the logarithm leading to
 \begin{align}
  m_S^2 = (-2 + 4 \beta)H^2\,,
 \end{align}
which is finite along the whole inflationary trajectory and positive for an appropriate choice of $\beta$. Instead, we find that the imaginary part of $\Phi$ is stable with the present \Kahler potential and hence poses no problems for inflation.

\subsubsection{Example with $\a=1/3$}

A generalisation of previous example arises when one allows for a different curvature of the \Kahler manifold, i.e.~including the parameter $\alpha$. For simplicity we will keep the same superpotential. Following the same line of reasoning, one ends up with a scalar potential for a canonically normalised scalar field $\varphi$ that reads
 \begin{align}
   V = \frac{2^{1 - 3\alpha} (3M)^2}{\alpha} \Big[ e^{  (3 - 3\a) \varphi / \sqrt{6 \a}} - e^{(1 - 3\a ) \varphi / \sqrt{6 \a}} \Big]^2 \,.
 \end{align}
For generic values of $\alpha$ this will lead to an exponential potential for large $|\varphi|$. There are only two exceptions to this behaviour: the first is for $\alpha = 1$ discussed above, while the second is for $\a = 1/3$. Interestingly, this value is also consistent with string theory and leads to inflation for large and negative $\varphi$. The scalar potential becomes
 \begin{align}
  V = 9 M^2 \left( e^{\sqrt{2} \varphi} - 1\right)^2 \,,
 \end{align}
and thus can be obtained from the Starobinsky potential by a sign flip and stretching in the $\varphi$ direction. Nevertheless, in this case one needs to stabilise even along the $Im\Phi$ direction as the three masses \eqref{spectrum} asymptote to
 \begin{align}
  m^2 = \{ 0,  0, -6 H^2 \} \,. 
 \end{align}
Having a viable single-field scenario translates into adding a term $-\gamma S\bar S (\Phi-\bar\Phi)^2 / (\Phi + \bar\Phi)^2$, together with the same term stabilising $S$ in the case $\a=1$, to the argument of the logarithm. With these choices, the mass spectrum turns to be finite along the inflaton direction and takes the following values:
\begin{align}
  m^2 = \{ 0,  12\gamma H^2 , (-6+12\beta) H^2 \} \,. 
 \end{align}
Interestingly, a value of $\beta>1/2$ leads to positive mass of the field $S$, independently of the parameter $\a$. Moreover, this model leads to the following spectral index and tensor-to-scalar ratio for different numbers of e-foldings:
 \begin{align}
  & N = 50: \qquad n_s = 0.961 \,, \quad r = 0.0015 \,, \notag \\
  & N = 60: \qquad n_s = 0.967 \,, \quad r = 0.0011 \,.
 \end{align}
Comparable to Starobinsky's, these are also comfortably consistent with the Planck results.

\subsubsection{Original version of $\alpha$-attractors}

From the previous example, it is clear that one cannot allow for arbitrary values of the parameter $\a$ by keeping the simple superpotential as in Eq.~\eqref{cecottiKW}. Dangerous exponential terms may indeed ruin the flatness of the scalar potential and spoil inflation. The solution to this problem was found by Kallosh, Linde and Roest in the context of superconformal $\a$-attractors \cite{Kallosh:2013yoa}. It consists in allowing for an $\a$-dependence of the superpotential such as
\be\label{Woriginalattractors}
W= S \Phi^{(3\a -1)/2} f(\Phi)\,,
\ee
while still keeping the same $K$ as in Eq.~\eqref{cecottiKW}. Then, any generic expansion of $f$ around $\Phi=0$ translates into a scalar potential being an exponential fall-off from a de Sitter plateau in terms of $\varphi$ such as Eq.~\eqref{expfalloff}, typical of $\a$-attractors.

The original version of $\a$-attractors \cite{Kallosh:2013yoa} was formulated in terms of disk coordinates $\Psi$ and stability was proved for any value of the curvature. Later, it was shown in \cite{Cecotti:2014ipa,Kallosh:2014rga} that the same physics can be described by means of half-plane coordinates $\Phi$.

\subsection{Model with $ K = - 3\alpha  \ln ( \Phi + \bar \Phi) - S \bar S$}

A remarkable simplification arises when the field $S$ enters the \Kahler potential as a simple canonical sector. The framework defined by
\begin{align}\label{KWSoutside}
   K = - 3\alpha  \ln 	\left( \Phi + \bar \Phi \right) - S \bar S, \quad W = S \Phi^{3\a /2}  f(\Phi) \,.
\end{align}
yields indeed always a scalar potential which attains a plateau at infinite values of the canonical inflaton $\varphi$ and has exponential drop-off at finite values. This happens at $\Phi=\bP$ and for any generic expansion of the  function $f$ in the superpotential. The curious power of $\Phi$ in Eq.~\eqref{Woriginalattractors} becomes an overall factor which can be gauged away by means of a \Kahler transformation thus yielding a \Kahler potential which is invariant under a shift of the canonical inflaton \cite{Carrasco:2015uma}. Interestingly this case, where the \Kahler \eqref{KWSoutside} parametrizes a manifold $SU(2,1)/U(1)\times U(1)$, leads generically to an improved stability of the system \cite{Covi:2008ea} (see also Ch.~\ref{chapter:Landscape}). It was first analyzed in full generality by \cite{Carrasco:2015rva,Scalisi:2015qga,Carrasco:2015pla} (the case $\a=1$ was previously investigated by \cite{Lahanas:2015jwa}). The field $S$ has canonical kinetic terms and its directions may be stabilized by means of higher order terms in the \Kahler potential. Alternatively it may be identified as a nilpotent superfield. We will discuss this latter case in the next Chapter.


\section{$\alpha$-Scale supergravity and attractors}\label{SECalphascale}

In the previous sections, we have studied the stringent implications of the geometric properties of the internal \Kahler manifold on the physics of inflation.  Specifically, we have shown how a non-trivial hyperbolic geometry yields dire observational consequences and forces the cosmological parameters $n_s$ and $r$ to converge towards the universal values \eqref{nsrattractors}. However, we must note that all the models presented above (including the original formulation of $\a$-attractors \cite{Kallosh:2013yoa,Kallosh:2014rga}) employ the orthogonal separation between the inflationary and the supersymmetry breaking directions. Then, one would like to answer a very natural question: is the attractor phenomenon independent from the field responsible for supersymmetry breaking? Answering this question is certainly very important in order to prove the generality of the attractor mechanism, independently of the other fields involved. This is a fundamental step towards a proper string theory realization where many moduli appear naturally.

In this following we present evidence for the universality of $\a$-attractors: while the previous models contain two chiral supermultiplets and employ a separation between the inflaton and the sGoldstino, we demonstrate that the same phenomenon can be achieved in a model containing just one superfield. The economical framework of realizing inflation in single-superfield models has been discussed in \cite{Goncharov:1985yu,Goncharov:1983mw,Achucarro:2012hg,Ketov:2014qha,Ketov:2014hya,Linde:2014ela,Linde:2014hfa,Kallosh:2015lwa}, but these do not include a variable \Kahler geometry and hence lack the parameter $\alpha$.

Our construction also highlights a novel approach to Minkowski and de Sitter model building.  Whereas the classic no-scale supergravity \cite{Cremmer:1983bf,Ellis:1983sf,Lahanas:1986uc} yields two flat Minkowski directions, one of these can be lifted to a stable direction by deforming the \Kahler curvature and allowing for a more general monomial dependence of the superpotential. Interestingly, a combination of these structures leads to a De Sitter plateau. This turns out to be stable only for such $\alpha$-deformed supergravities with a smaller \Kahler curvature than the one corresponding to a combination of no-scale constructions. Remarkably, generic deformations of these De Sitter plateaus lead to inflationary regimes with prediction \eqref{nsrattractors}. This sheds light on the fundamental origin of the attractor phenomenon. 

Finally, analogous results emerge in the singular limit $\alpha\rightarrow\infty$ where the \Kahler geometry becomes flat. However, in this case the natural ingredients providing Minkowski or dS solutions and inflationary deformations will be exponentials, peculiar to this geometry.

\subsection{No-scale supergravity and de Sitter}

\begin{figure}[t!]
\hspace{-3mm}
\begin{center}
\includegraphics[width=8cm,keepaspectratio]{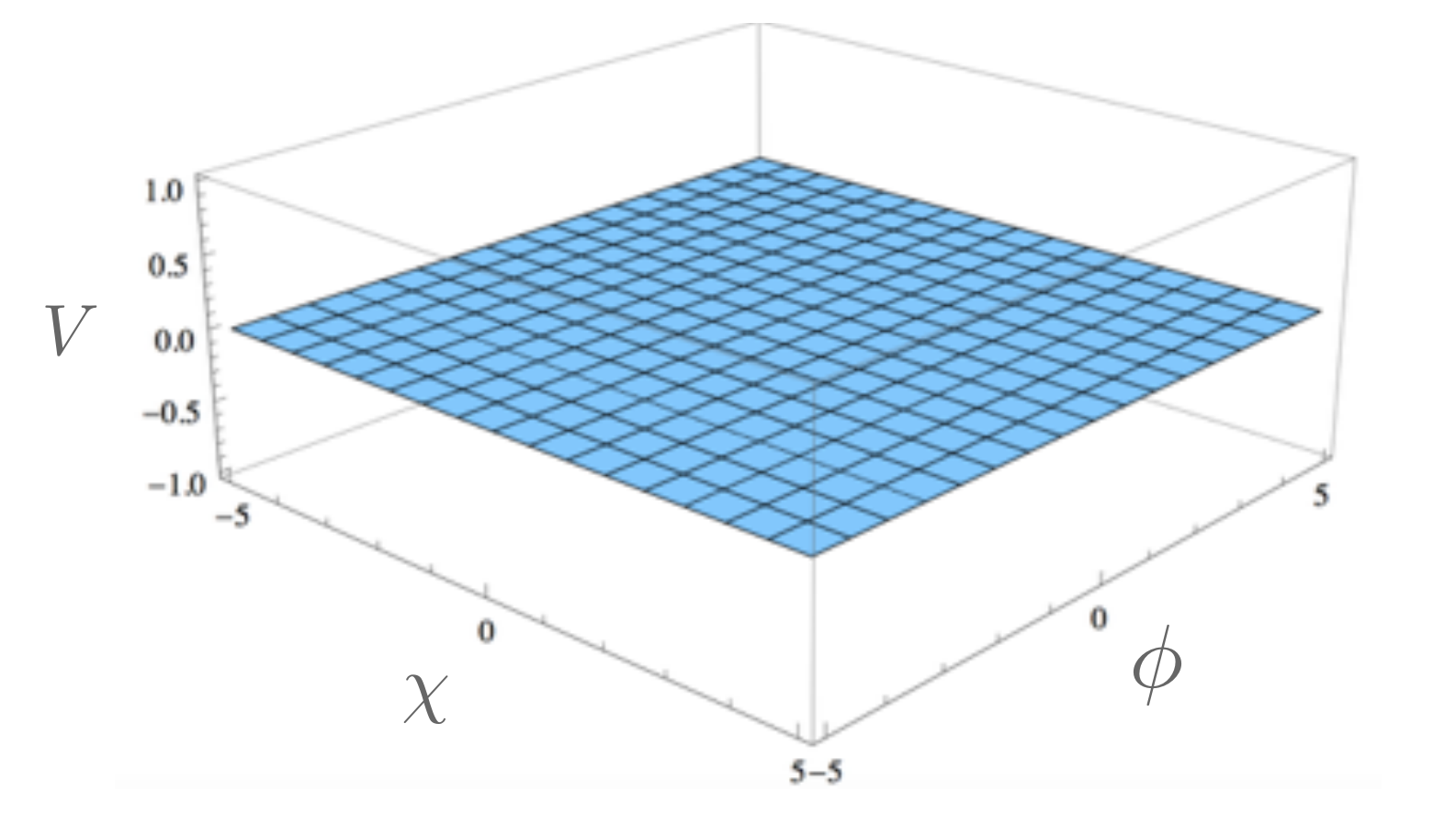}
\vspace*{0.5cm}
\caption{\it The scalar potential of the no-scale model being a flat Minkowski plane.}\label{Mplane}
\end{center}
\vspace{-0.5cm}
\end{figure}

Our starting point will be the no-scale structure for a supergravity with a single chiral superfield. In this case the \Kahler potential reads
\be
K=-3 \ln\left(\Phi+\bar\Phi\right)\,,
\ee
describing a manifold SU(1,1)/U(1) and invariant under a shift of  Im($\Phi$), while the superpotential is independent of the superfield and hence constant. This model is characterized by a Minkowski vacuum in any point in field space, as it is shown in Fig.~\ref{Mplane}. The negative definite contribution to the scalar potential, proportional to the square of the superpotential, is cancelled by the positive definite term, proportional to the square of the order parameter of supersymmetry breaking 
 \begin{align}
   D_\Phi W = \partial_\Phi W + K_\Phi W \,.
 \end{align}
Note that only the latter term of this contribution is non-vanishing due to the constancy of the superpotential. The resulting no-scale model necessarily has a flat direction along the imaginary part of $\Phi$, as this does not appear in either $K$ or $W$.

By a field redefinition, one can bring this simple no-scale model to a different form. In particular, in order to leave the \Kahler potential invariant, one can combine an inversion of the holomorphic field $\Phi$ with a specific \Kahler transformation, defined as in Eq.~\eqref{Ktransformation}, with $\lambda = -3\ln\Phi$ in this case. Under these transformations, a constant superpotential becomes cubic instead. While this model has the same scalar potential and is therefore also of the no-scale type, it receives contributions from both terms in $D_\Phi W$.

\begin{figure}[htb]
\hspace{-3mm}
\begin{center}
\includegraphics[width=8cm,keepaspectratio]{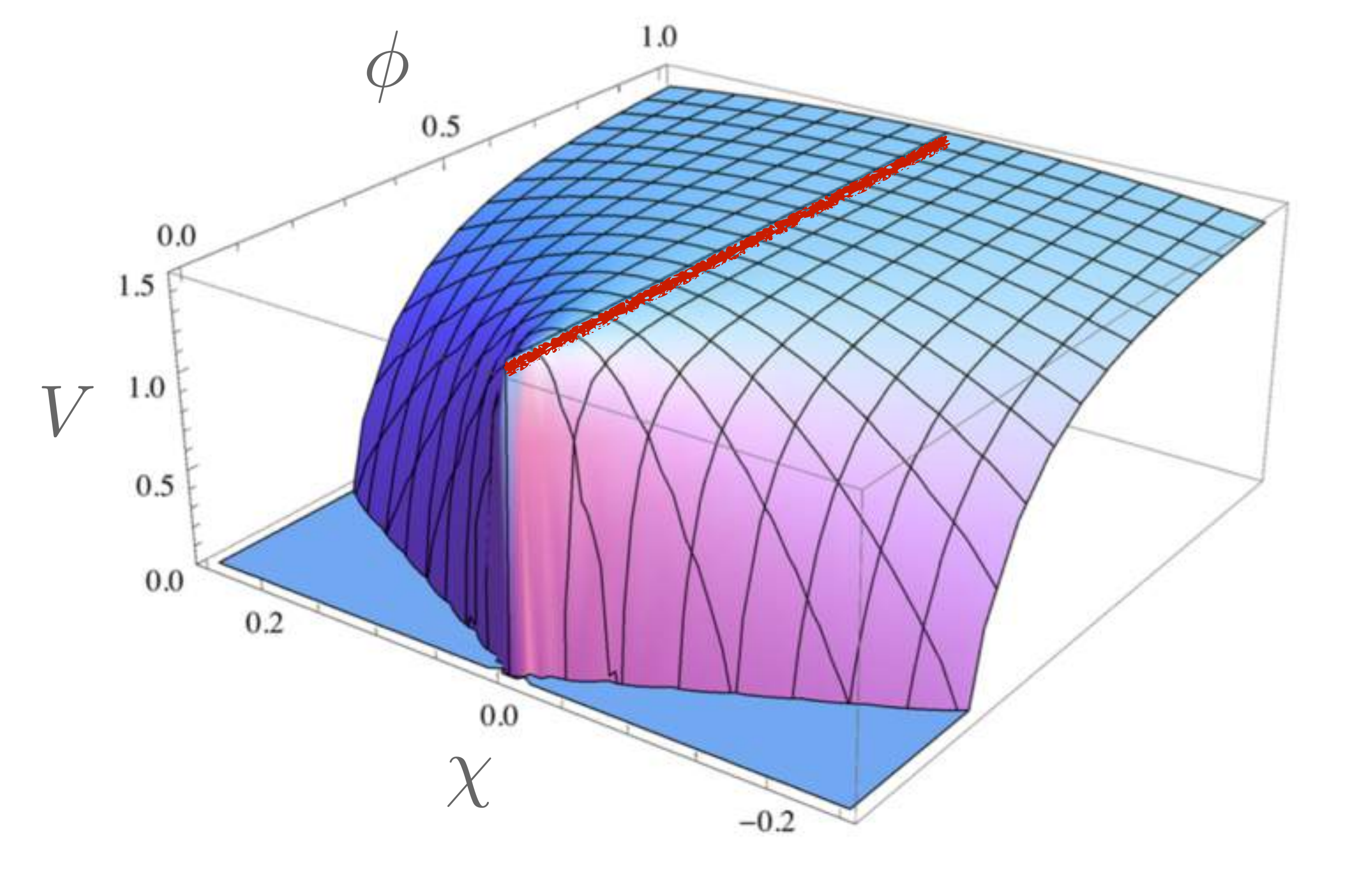}
\vspace*{0.5cm}
\caption{\it Scalar potential of the model defined by Eq.~\eqref{WNSdeSitter}. The interference of the constant and cubic superpotentials leads to an unstable de Sitter phase.}\label{NSdS}
\end{center}
\vspace{-0.5cm}
\end{figure}

Remarkably, one can combine the constant and cubic superpotentials to move away from no-scale models and generate a non-vanishing cosmological constant. In particular, the superpotential
  \begin{align} \label{WNSdeSitter}
    W = 1 - \Phi^3 \,,
\end{align}
 leads to a scalar potential with a flat direction along $\Phi=~\bar{\Phi}$, where the original Minkowski vacuum is shifted to $V =  \tfrac{3}{2}$ (while the combination with opposite sign leads to AdS), as one can see in Fig.~\ref{NSdS}. However, this De Sitter solution turns out to be unstable: the mass of the imaginary direction is given by $m_{\Im\Phi}^2 = - 2$.

\subsection{$\alpha$-Scale supergravity and stable de Sitter}

In order to improve on the previous instability, we will consider the logarithmic \Kahler potential of the form\footnote{We already presented this form of $K$ in Eq.~\eqref{hyperbKahlerpotential}, in the context of sGoldstino inflation. We explicitly show this again for a matter of convenience.}
\be
K=-3\alpha\ln\left(\Phi+\bar\Phi\right)\,. \label{Kahler}
\ee
This still parametrizes a symmetric geometry $SU(1,1)/U(1)$, whose curvature is given by Eq.~\eqref{curvatureKhyp}.

\begin{figure}[htb]
\hspace{-3mm}
\begin{center}
\includegraphics[width=7cm,keepaspectratio]{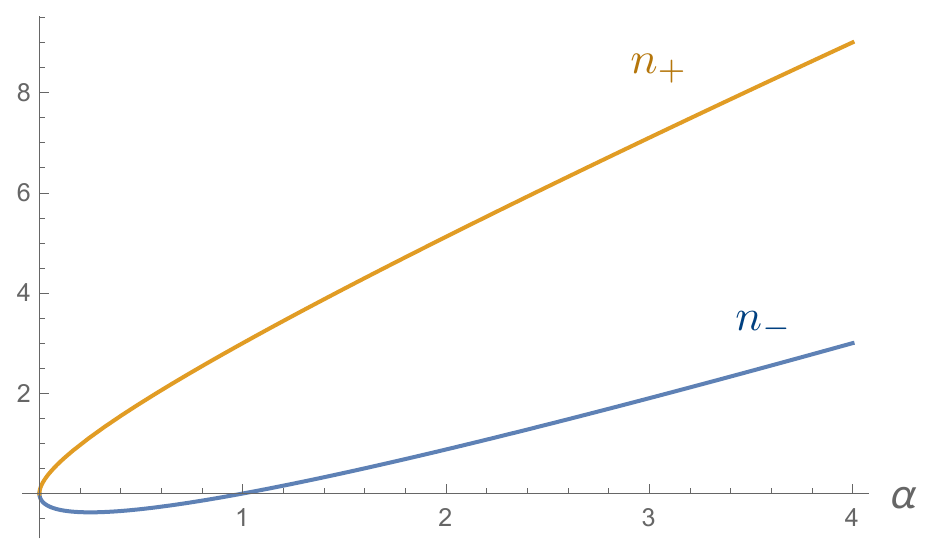}
\vspace*{0.5cm}
\caption{\it The monomial powers $n_\pm$ for the $\alpha$-scale models as function of $\alpha$.}\label{nsol}
\end{center}
\vspace{-0.5cm}
\end{figure}

A  single monomial superpotential $W=\Phi^n$ will give a scalar potential equal to
\be
V=\frac{8^{-\alpha } \left[(2 n-3 \alpha )^2-9 \alpha \right] }{3 \alpha }\Phi ^{2 n-3 \alpha }\,,
\ee
along the real direction $\Phi=\bar\Phi$. Note that a constant potential corresponds to $2n=3 \alpha$ which, for any value of $\alpha$, leads always to AdS \cite{Ellis:2013nxa}. In contrast, a vanishing scalar potential corresponds to one of the following solutions
\be\label{npm}
n_\pm = \frac{3}{2} \left(\alpha \pm \sqrt{\alpha}\right)\,,
\ee
displayed in Fig.~\ref{nsol}. These are the counterparts of the constant and cubic superpotentials of the previous section, corresponding to $\alpha =1$. We will refer to the above model as {\it $\alpha$-scale supergravities} for the following reason. 

Similarly to the standard no-scale model, the real part of $\Phi$ has flat direction. On the other hand, the mass of the imaginary part gets a dependence on the field and, along $\Im\Phi=0$, reads
\be
m_{\Im\Phi}^2 =\frac{2^{2-3 \alpha } (\alpha -1) }{\alpha }\e^{\mp \sqrt{6}\varphi}\,,
\ee
where the sign of the power depends on the choice of one of the solutions \eqref{npm}. This result assures stability of the Minkowski vacuum for $\alpha\geq1$ \cite{GomezReino}.

Following the previous construction, one obtains a de Sitter plateau along the real direction by considering a pair of monomials,
\be\label{Wpol}
 W = \Phi^{n_-} - \Phi^{n_+}\,, \quad V= 3\cdot2^{2-3 \alpha } \,.
\ee
While this generically leads to terms with irrational powers, these are integers when $\alpha$ is a perfect square. Moreover, the more general choice with $9\alpha$ a perfect square yields still integer powers multiplied by an overall phase which can be gauged away by means of a \Kahler transformation.

\begin{figure}[htb]
\hspace{-3mm}
\begin{center}
\includegraphics[width=8cm,keepaspectratio]{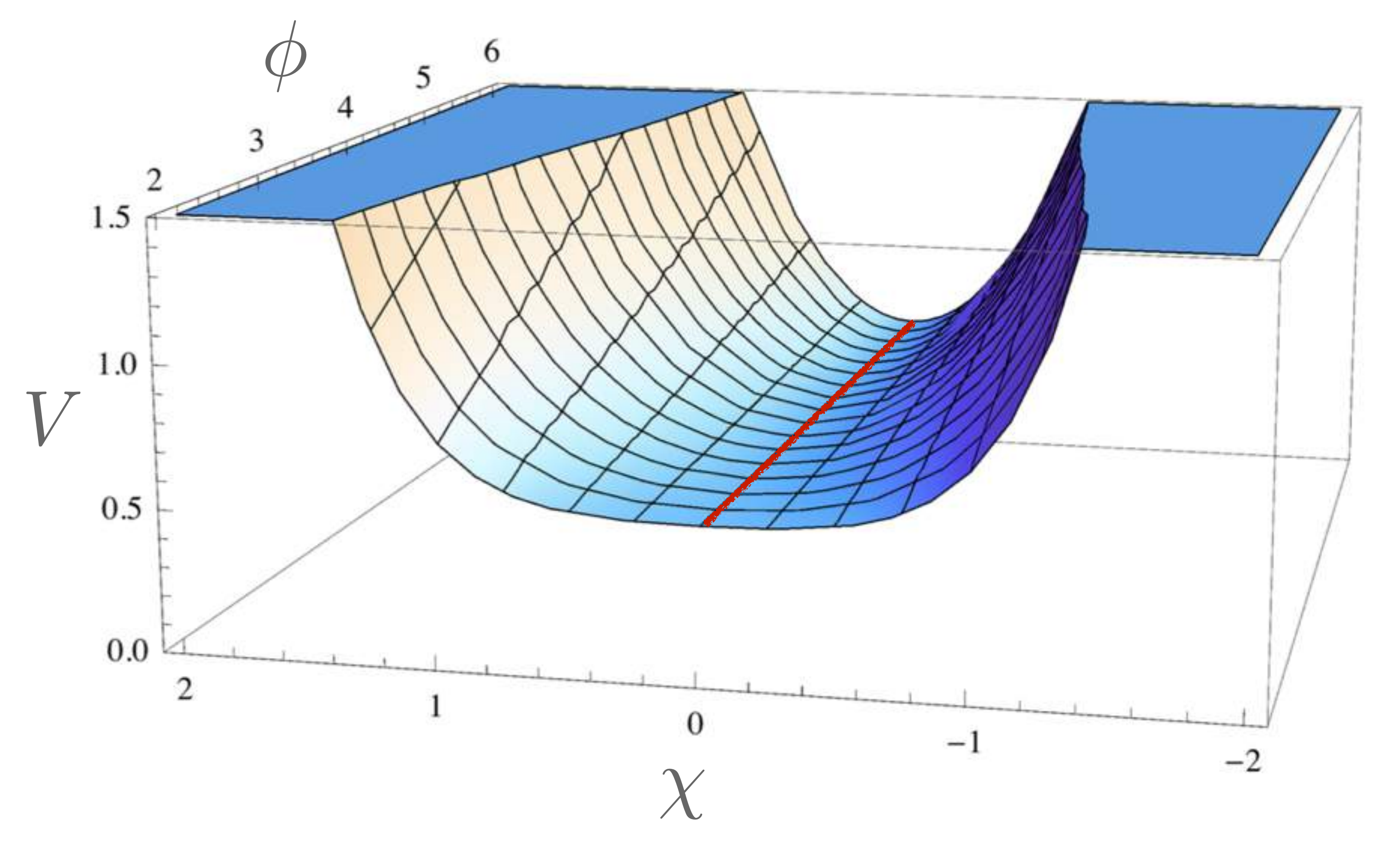}
\vspace*{0.5cm}
\caption{\it Scalar potential of the $\a$-scale model (with $\a=4$) with stable de Sitter along Im$\Phi=0$. }\label{ASdS}
\end{center}
\vspace{-0.5cm}
\end{figure}

Remarkably, unlike the standard case $\alpha=1$,  the mass of $\Im \Phi$ gets also some field dependent contributions and, along the real axis, reads
\be \label{dS-mass}
m_{\Im\Phi}^2 = -\frac{4 V}{3 \alpha} \left [ 1 - (\alpha -1) \sinh^2\left(\sqrt{\frac{3}{2}}\varphi\right)\right]\,.
\ee
Such a solution for the mass of the imaginary component allows to identify regions of stable de Sitter vacua, as one can appreciate in Fig.~\ref{ASdS}. In particular, for $\alpha>1$, the field dependent terms dominate in the limit of large $|\varphi|$, leading to a positive mass for the imaginary component. Just a small region around the symmetric point ($\varphi=0$) leads to an instability, which disappears in the limit $\alpha\rightarrow\infty$.

\subsection{Single superfield $\alpha$-attractors}

Once we know how to construct de Sitter in this context, we can add corrections to the superpotential \eqref{Wpol} in order to reproduce a consistent inflationary dynamics. Deviations from the positive plateau are given by higher powers $n$ of $\Phi$ ($n_- < n_+ < n$). In full generality, one can consider a deformation of the form
\be\label{Wgen}
W = \Phi^{n_-} - \Phi^{n_+} F(\Phi)\,,
\ee
with $F$ being a general function with an expansion $F(\Phi)=\sum_n c_n \Phi^n$. The corresponding scalar potential, in terms of the geometric field $\Phi$, reads
\be \label{potential}
V=\frac{2^{2-3 \alpha }  \left(\Phi F'(\Phi)+3 \sqrt{\alpha } F(\Phi)\right) \left( \Phi^{3 \sqrt{\alpha }+1} F'(\Phi)+3 \sqrt{\alpha }\right)}{3 \alpha }\,,
\ee
along the real axis, where primes denote derivatives with respect to $\Phi$. In the inflationary regime, close to $\Phi = 0$, only the first non-constant term is relevant: the scalar potential approximates an exponential fall-off from a de Sitter plateau such as Eq.~\eqref{expfalloff} at large values of the canonical field $\varphi$.

The inflationary scenario emerging from this construction is therefore the one typical of the $\alpha$-attractors: the \Kahler geometry, described by eq.~\eqref{Kahler}, determines unequivocally the observational predictions which, on the other hand, will be insensitive to specific changes in the superpotential.  Moreover, the predicted values for the spectral tilt and tensor-to-scalar ratio are \eqref{nsrattractors}, in the limit of large number of e-foldings $N$.

\begin{figure}[t!]
\hspace{-3mm}
\begin{center}
\includegraphics[width=7.5cm]{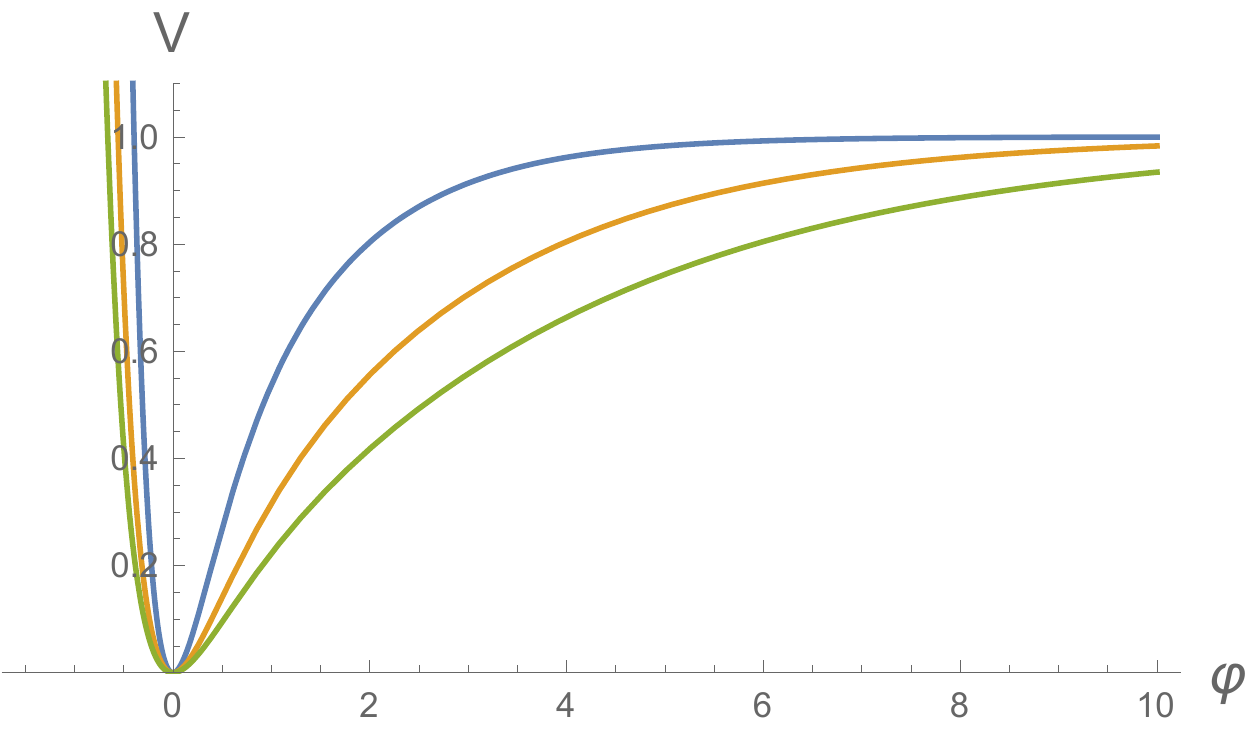}
\includegraphics[width=7.5cm]{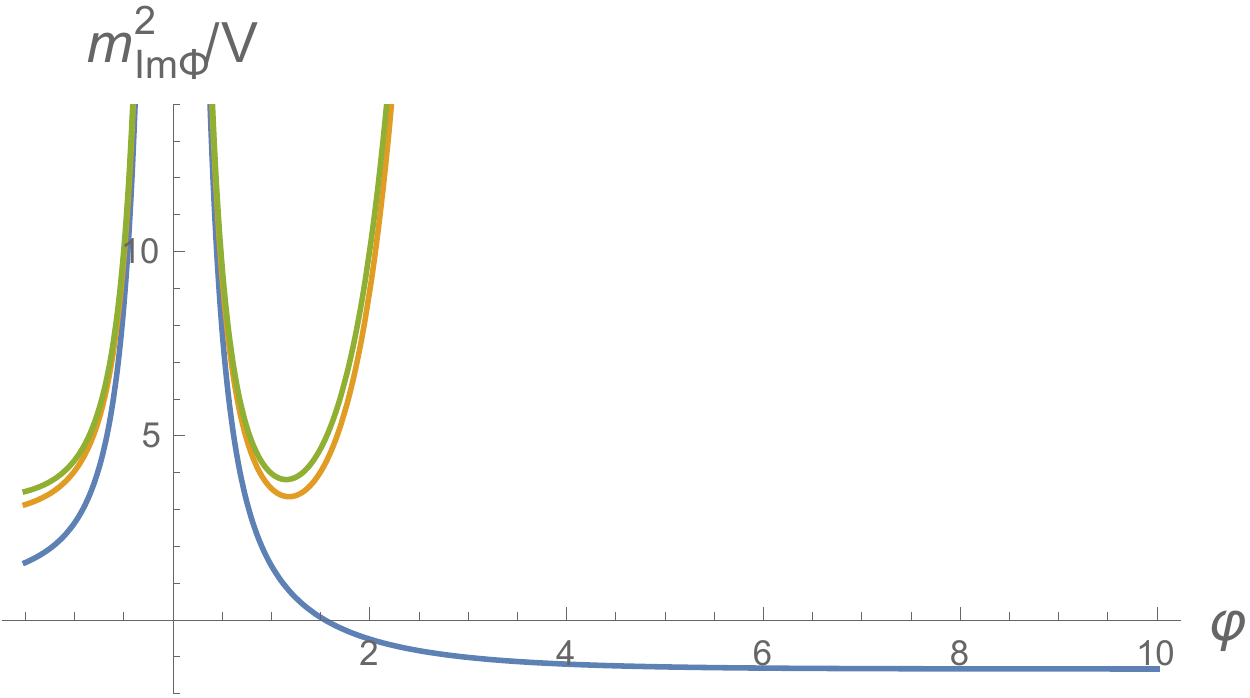}
\vspace*{0.5cm}
\caption{\it Scalar potential and imaginary mass of the model defined by Eq.~\eqref{linear} in terms of $\varphi$ for $\alpha=\{1,4,9\}$. The blue line represents the instability occurring at $\alpha=1$.}\label{Vminimal}
\end{center}
\vspace{-0.75cm}
\end{figure}

\begin{figure}[t!]
\hspace{-3mm}
\begin{center}
\includegraphics[width=8.5cm]{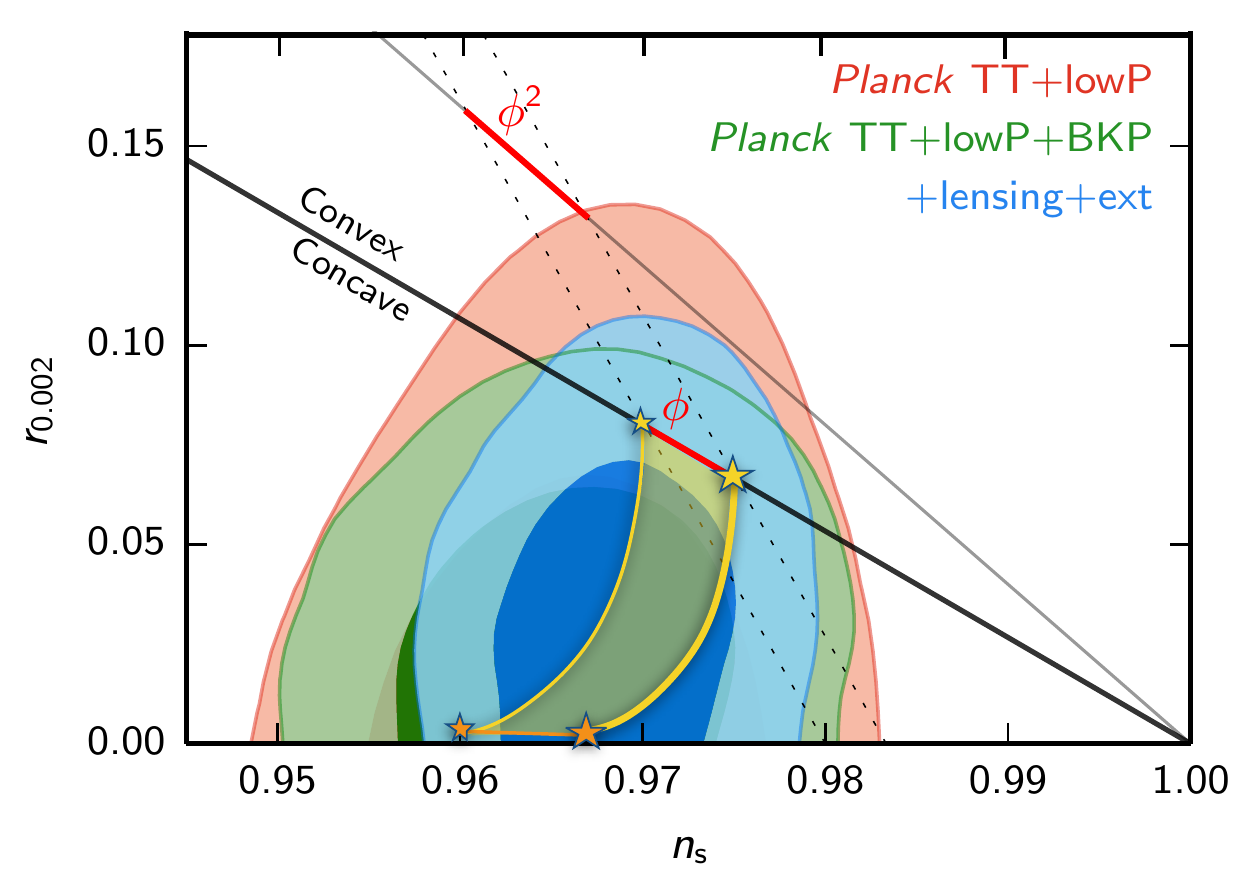}
\vspace*{0.3cm}
\caption{\it The $(n_s ,r)$ predictions for $N=50$ and $N=60$ of the model Eq.~\eqref{linear} superimposed on the Planck constraints. The predictions interpolate between \eqref{ns} and \eqref{r} for small and order-one $\alpha$ and those of linear inflation for large $\alpha$.}\label{Predictions}
\end{center}
\vspace{-0.75cm}
\end{figure}

To demonstrate the stability and vacuum structure with an explicit example, we take
 \begin{align} \label{linear}
  F(x) = 1 + 3 \sqrt{\alpha} - 3 \sqrt{\alpha} x \,.
 \end{align}
These coefficients have been chosen to have a quadratic expansion around the Minkowski minimum at $\Phi = 1$. Both the scalar potential along the real axis as well as the mass of the imaginary direction is shown in Fig.~\ref{Vminimal} for different values of $\alpha$. This model is fully stable for $\alpha>1$ while the \Kahler curvature leads to an instability along the imaginary direction when  $\alpha\leq1$. Finally, its observational predictions superimposed on the confidence levels released by Planck2015 \cite{Planck:2015xua} are given in Fig.~\ref{Predictions}. These interpolate between the $\alpha$-attractor values \eqref{ns} and \eqref{r}, and those of a linear scalar potential.

The above approach leads to a supersymmetric Minkowski minimum. Uplifting this vacuum by means of supersymmetry breaking  to include a non-zero cosmological constant  is strongly constrained \cite{Kallosh:2014oja}: generically this cannot be done with a small deformation and, within one single superfield, leads to an undesirable large gravitino mass \cite{Linde:2014ela}. An additional nilpotent sector can elegantly solve the issue of the separation of the physical scales \cite{Kallosh:2014via,Dall'Agata:2014oka,Kallosh:2014hxa,Kallosh:2015lwa,Linde:2015uga, Scalisi:2015qga,Carrasco:2015pla}. Nevertheless, as the two sectors prove to be independent from each other and play distinct roles \cite{Scalisi:2015qga}, it remains fundamental to construct a consistent inflationary dynamics in a single superfield context.

\subsection{Flat \Kahler limit}

In the singular limit $\alpha \rightarrow \infty$ the \Kahler geometry becomes flat. One could wonder whether there is a similar $\alpha$-scale model as well as de Sitter uplift in this limit. Indeed this is the case: upon a field redefinition $\Phi \rightarrow \exp( 2 \Phi / \sqrt{3 \alpha})$ and a \Kahler transformation with $\lambda = \tfrac{3}{2}\alpha\ln2 + \sqrt{3\alpha}\Phi$, the \Kahler potential \eqref{Kahler} yields
\be
K=-\tfrac{1}{2}\left(\Phi-\bar\Phi\right)^2,
\ee
in the singular limit $\alpha \rightarrow \infty$. Note that $K$ has become shift-symmetric in the inflaton field $\Re\Phi$ \cite{Kawasaki:2000yn}. This naturally provides a solution to the so-called $\eta$-problem \cite{Copeland:1994vg}, whereas, for finite values of $\alpha$, the latter is mitigated by the logarithmic form \eqref{Kahler} \cite{Roest:2013aoa}.

Under the same operations, the monomial superpotential turns into (modulo a constant, overall rescaling)
\be
W= \e^{\pm\sqrt{3} \Phi}\,.
\ee
One can check that this leads to a vanishing scalar potential along the line $\Phi = \bar \Phi$. Moreover, a linear combination of the two exponentials, such as  $W=\sinh(\sqrt{3} \Phi)$, leads to a constant and positive value of $V$ (while a cosh, instead, leads to AdS). The mass of the orthogonal imaginary component of $\Phi$ is equal to the $\alpha \rightarrow \infty$ limit of \eqref{dS-mass}.

The above construction can be perturbed to have deviations from de Sitter and produce a consistent inflationary dynamics. A first guess could be to include the same deformation in the polynomial \eqref{Wgen} and take the $\alpha \rightarrow \infty$ limit. However, in this case the field dependence of this function is washed out: for finite values of the constants $c_n$, the resulting superpotential reads 
 \begin{align}
  W = \e^{\sqrt{3} \Phi} - \e^{- \sqrt{3} \Phi} F(1) \,,
 \end{align}
 leading to a constant scalar potential.
 
A more natural possibility, given the exponential ingredients of the above superpotential, would be to take
 \be\label{Wdev}
   W= \e^{\sqrt{3} \Phi} - \e^{-\sqrt{3} \Phi} F \left(\e^{- 2 \Phi / \sqrt{3 \alpha'}}\right) \,,
 \ee
where we have parametrized the additional dependence in terms of a new parameter $\alpha'$. Remarkably, when truncating to the real axis, the scalar potential arising from this supergravity model with a flat \Kahler geometry is identical to \eqref{potential} of the supergravity model with a curved \Kahler geometry, provided one identifies $\alpha = \alpha'$. The specific choice \eqref{linear} for $F$ in this case leads to the identical predictions of Fig.~3; however, interestingly, this model proves to be stable for any positive value of $\alpha$.
 
It therefore turns out to be possible to represent the same single-field inflationary potential by means of curved or flat \Kahler geometry. Only the former has the attractive interpretation of the robustness of $\alpha$-attractors arising from a non-trivial \Kahler geometry; the same dynamics arises in the flat case by the peculiar non-polynomial form of $W$. 

The first example of such a model \cite{Goncharov:1985yu,Goncharov:1983mw,Linde:2014hfa} fits perfectly into the recipe given above: it has a superpotential 
 \begin{align}
  W = \sinh(\sqrt{3} \Phi)\tanh(\sqrt{3} \Phi)\,, 
  \end{align}
corresponding to the choice  $F(x) = (3- x)/(1+x)$ for the case of $\alpha' = 1/9$. The same inflationary potential can also be embedded in a logarithmic \Kahler structure \cite{Kallosh:2015lwa}.

\subsection{Discussion}\label{disc}

In this Section we have outlined a strikingly simple route to construct single superfield models with stable de Sitter solutions. Generic deformations of these models yield an inflationary trajectory fully consistent with Planck. The key quantity in this set of models, similar to the original $\alpha$-attractors, is the curvature of the \Kahler manifold \eqref{r}. This quantity determines both the (in)stability of such constructions as well as the inflationary predictions of the deformed models. 

Remarkably, this provides a realization of $\alpha$-attractors employing a single superfield, in constrast to the two-field model of \cite{Kallosh:2013yoa,Kallosh:2014rga}.
This suggests that the phenomenon of \Kahler curvature leading to the inflationary predictions \eqref{ns} and \eqref{r} is universal, and applies to a much larger set of \Kahler geometries than $SU(1,n) / U(n)$ with $n=1,2$. 

Given the prominence of no-scale models in the literature, it would be interesting to study other possible applications of the $\alpha$-generalization reviewed in this Section and originally proposed in \cite{Roest:2015qya}. An example could be the no-scale inflationary constructions of \cite{Ellis:2013xoa,Ellis:2013nxa,Ellis:2014gxa,Ellis:2014opa,Lahanas:2015jwa}. Moreover, while we have focused on single superfield models, it is straightforward to generalize this construction to multi-fields:
 \begin{align}
  K = \sum_i - 3 \alpha_i \log(\Phi_i + \bar \Phi_i) \,,  \quad
   W = \prod_i \Phi_i^{n_i} \,, \label{multi}
 \end{align}
where we have suppressed other fields with a different dependence. The condition for Minkowski is
 \begin{align}
 \sum_i \frac{(2n_i - 3 \alpha_i)^2}{3\alpha_i} = 3 \,.
  \end{align}
Remarkably, also in the multi-field case, the interference of superpotential terms with flat Minkowski vacua leads to a de Sitter phase, proving the generality of such a feature. It would be very interesting to investigate the stability and inflationary aspects of such constructions.
  
Finally, our construction invites investigations of string theory scenarios leading to \eqref{multi}. Many moduli contribute with a factor $\alpha_i = 1/3$ to the \Kahler potential, while flux compactifications yield polynomial contributions to the superpotential. It would be of utmost interest to realize this in a concrete setting.

\clearpage
\thispagestyle{empty}


\chapter{Inflation and de Sitter Landscape}
\label{chapter:Landscape}

\begingroup
\begin{flushright}
\vspace{.5cm}
\end{flushright}
\begin{quote}
{\it In this chapter, we discuss the possibility to construct a consistent and unified framework for inflation, dark energy and supersymmetry breaking. This approach is motivated by the idea that a vast landscape of string vacua may provide a possible explanation for the value of the current acceleration in our Universe. We employ an effective supergravity description and  investigate the restrictions and main properties coming from the interplay between the inflationary and the supersymmetry breaking sectors. Specifically, we show that the physics of a single-superfield scenario is highly constrained due to a specific no-go theorem regarding the uplifting of a SUSY Minkowski vacuum. On the other hand, the addition of a nilpotent sector yields remarkable simplifications and allows for controllable level of dark energy and supersymmetry breaking. We study this powerful framework both in the context of flat \Kahler geometry and in the case of $\a$-attractors. Interestingly, in the latter case, we prove that the attractor nature of the theory is enhanced when combining the inflationary sector with the field responsible for uplfting: cosmological attractors are very stable with respect to any possible value of the cosmological constant and, remarkably, to any generic coupling of the two sectors.
The novel results of this Chapter are based on the publications {\normalfont[\hyperref[chapter:Publications]{{\sc vi}}]}, {\normalfont[\hyperref[chapter:Publications]{{\sc vii}}]}  and {\normalfont[\hyperref[chapter:Publications]{{\sc ix}}]}.}
\end{quote}
\endgroup

\newpage

\section{Introduction and outline}

\begin{figure}[htb]
\hspace{-3mm}
\begin{center}
\includegraphics[width=9.cm]{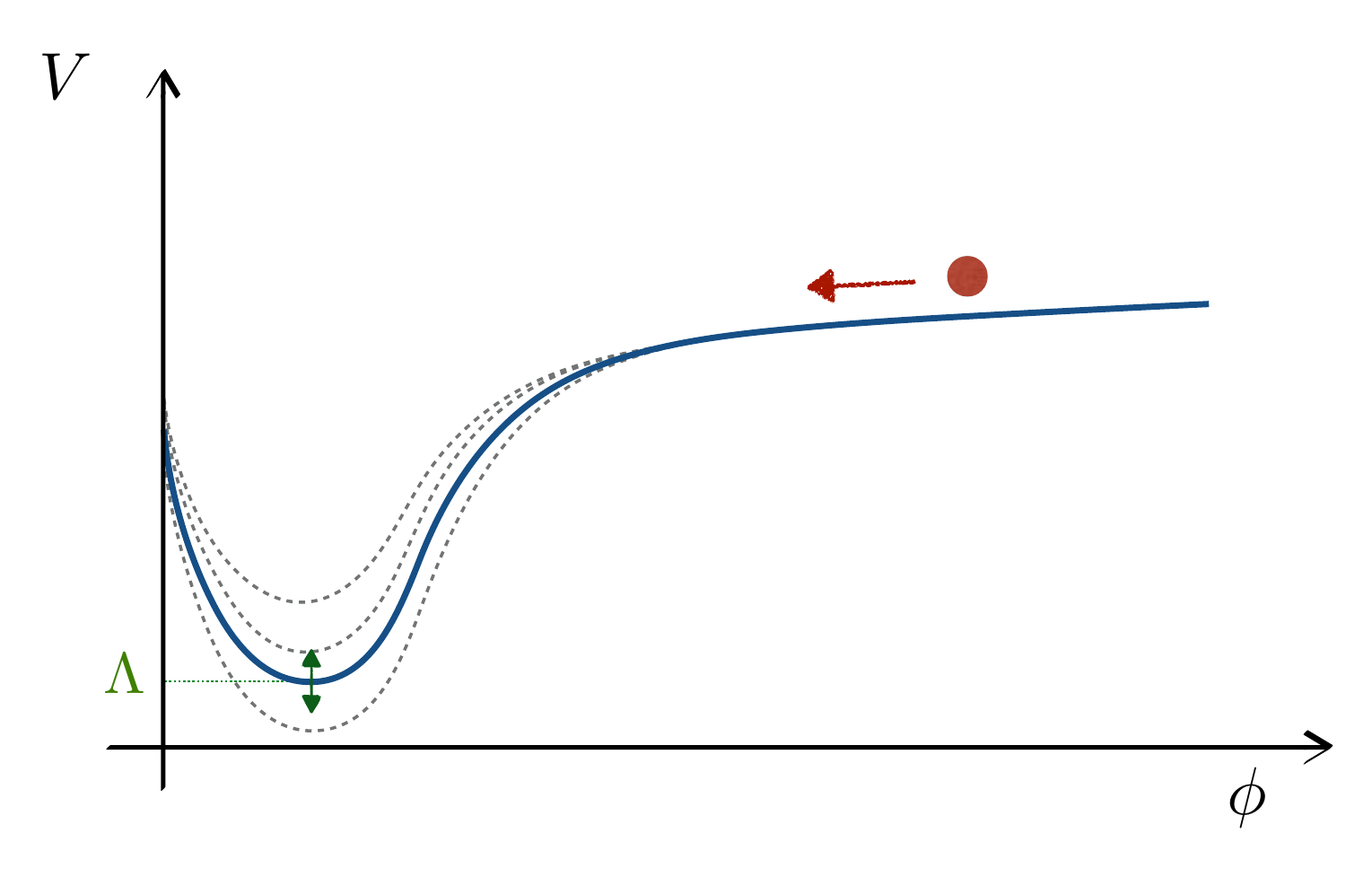}
\vspace*{0.1cm}
\caption{\it Cartoon picture of the simplest possible scenario where a single scalar field is responsible both for inflation and current acceleration of the Universe. The amount of dark energy can be controlled, following the string landscape scenario.}\label{upliftCARTOON}
\end{center}
\vspace{-0.75cm}
\end{figure}

Observational evidence \cite{Riess:1998cb,Perlmutter:1998np,Hinshaw:2012aka,Planck:2015xua,Ade:2015lrj}  seems to point at acceleration as a fundamental ingredient of our Universe. Primordial inflation is the leading paradigm to account for the origin of the anisotropies in the CMB radiation and, then, the formation of large scale structures (as we reviewed in Ch.~\ref{chapter:standard} and Ch.~\ref{chapter:Inflation} of this thesis). These are currently observed to experience a mysterious accelerating phase, whose source has been generically called dark energy. Although the origin of both early- and late-time acceleration still represents a great theoretical puzzle, the simple assumption that the potential energy of a scalar field may serve as fundamental source has turned out to be successful in terms of investigation, extraction of predictions and agreement with the present observational data (see Ch.~\ref{chapter:Inflation}). In the simplest scenario, a scalar field slowly rolls down along its potential, driving inflation, and eventually sits in a minimum with a small positive cosmological constant of the order $\Lambda\sim10^{-120}$,  as displayed in Fig.~\eqref{upliftCARTOON}.

The embedding into high-energy physics frameworks, such as supergravity or string theory, seems to be natural. On the one hand, the high energy-scale of inflation would require UV-physics control (see Ch.~\ref{chapter:supergravity} for supergravity embeddings of the inflationary paradigm). On the other hand, the anthropic argument in a landscape of many string vacua \cite{Linde:1986fd,Weinberg:1987dv, Bousso:2000xa,Kachru:2003aw,Douglas:2003um,Susskind:2003kw} would provide a possible explanation of the smallness of the current cosmological constant.

In an effective unified framework for inflation and dark energy, the concrete implementation of the idea of a de Sitter landscape would provide an enormous number of possibilities for the minimum of the scalar potential where the field eventually sits after driving inflation. Quantum corrections or interactions with other particles may certainly lead to some additional contributions to the value of the potential at the minimum. However, this should not affect the existence of a landscape of dS vacua and any possible correction to the cosmological constant (CC) would be easily faced, within a scenario with controllable level of dark energy. Therefore, we aim to construct a supergravity framework suitable for inflation with exit into de Sitter space with all possible values of the cosmological constant (see Fig.~\eqref{upliftCARTOON}).

Our starting point will be the models of inflation discussed in Ch.~\ref{chapter:supergravity}. A common property of these scenarios is that supersymmetry is restored at the minimum $V=0$ after inflation ends. Then, uplifting the SUSY Minkowski vacuum seems to be the next natural step in order to consider the current acceleration. However, it has been pointed out that obtaining a de Sitter vacuum from a SUSY one is subject to a number of restrictions encoded in a recent no-go theorem \cite{Kallosh:2014oja}  which make a unified picture of inflation and dark energy very challenging to achieve, especially when using just one chiral superfield \cite{Linde:2014ela}. Specifically, this generically yields a large Gravitino mass which is undesirable from a phenomenological point of view.  We discuss the case of one single superfield in detail in Sec.~\ref{sectionLRS}, in the context of the model proposed by Ketov and Terada in \cite{Ketov:2014qha,Ketov:2014hya}  (we have already reviewed this framework in the previous chapter). 

A way to overcome the issue of uplifting a SUSY Minkoswki minimum and still having controllable level of SUSY breaking is to employ a nilpotent superfield $S$ \cite{Volkov:1972jx,Volkov:1973ix,Rocek:1978nb,Ivanov:1978mx,Lindstrom:1979kq,Casalbuoni:1988xh,Komargodski:2009rz} (we review the important properties of this construction in Sec.~\ref{nilpotentSEC}).  In fact, the nilpotent field seems to be naturally related to de Sitter vacua when coupled to supergravity \cite{Dudas:2015eha,Bergshoeff:2015tra,Hasegawa:2015bza,Ferrara:2015gta,Kuzenko:2015yxa,Kallosh:2015tea} (see \cite{Schillo:2015ssx} for an interesting review on this topic) and it has been used in order to construct  inflationary models with de Sitter exit and controllable level of SUSY breaking at the minumum \cite{Antoniadis:2014oya,Ferrara:2014kva,KLnil,Kallosh:2014wsa,Dall'Agata:2014oka,Kallosh:2014hxa,Kallosh:2015lwa}. The two sectors appearing in these constructions have independent roles: the $\Phi$-sector contains the scalar which evolves and dynamically determines inflation and dark energy while the field $S$ is responsible for the landscape of vacua. However, in general, the inflationary regime is really sensitive to the coupling between the two sectors and to the value of the uplifting. One needs to make specific choices for the superpotential. We show the details and the limitation of this construction in Sec.~\ref{sectionKLS}.

Finally, in Sec.~\eqref{sectionScalisi}, we present special stability of $\alpha$-attractors when combined with a nilpotent sector. We prove that their inflationary predictions are extremely stable with respect to any possible value of the cosmological constant and to any generic coupling between $\Phi$ and $S$, exhibiting attractor structure also in the uplifting sector. These scenarios simply emerges as the most generic expansion of the superpotential.

\section{Single superfield inflation and dark energy}\label{sectionLRS}

In this Section, we intend to investigate the consequences of uplifting a SUSY Minkoswki vacuum in a supergravity framework consisting of just one superfield. Specifically, we consider the class of inflationary theory proposed by Ketov and Terada (KT) \cite{Ketov:2014qha,Ketov:2014hya}. Following \cite{Ketov:2014hya}, one may consider a logarithmic \K\, potential of the form\footnote{We already presented this \Kahler potential in the context of sGoldstino inflation in Ch.~\ref{chapter:supergravity}. We explicitly show this again for a matter of convenience.}
\be\label{KK}
 K= -3\ln \left[ 1 + \frac{\Phi + \bar{\Phi} + \zeta \left( \Phi + \bar{\Phi}\right)^4}{\sqrt{3}}\right]\,.
\ee
Notice that, within this model, the inflaton field is played by the Im$\Phi=\chi$, unlike the other supergravity constructions considered in Ch.~\ref{chapter:supergravity}. The quartic term in the argument of the logarithm is introduced in order  to stabilize the field $\chi$ during inflation at  $\phi \approx 0$.

As already explained in Sec.~\ref{subsecsgold}, this supergravity scenario allows to produce an almost arbitrary inflaton potential when $\phi\ll1$. After inflation, the field rolls down towards a Minkowski minimum placed at $\Phi=0$  where supersymmetry is unbroken.

This situation is valid for a large variety of superpotentials $W(\Phi)$, but not for all of them. In particular, we will show that one can have a consistent inflationary scenario in the theory with the simplest superpotential $W = c\Phi + d$, but both fields $\phi$ and $\chi$ evolve and play an important role. At the end of inflation, the field may roll to a Minkowski vacuum with $V = 0$ or to a dS vacuum with a tiny cosmological constant $\Lambda \sim 10^{{-120}}$. This is an encouraging result, since a complete cosmological model must include both the stage of inflation and the present stage of acceleration of the universe, and our simple model with a linear potential successfully achieves it. However, this success comes at a price: in this model, supersymmetry after inflation is strongly broken and the gravitino mass is  $2\times  10^{13}$ GeV, which is much greater than the often assumed TeV mass range. 
\pagebreak

In view of this result, one may wonder what will happen if one adds a tiny correction term $c\Phi +d$ to the benchmark superpotentials of the inflationary models described in \cite{Ketov:2014hya} with supersymmetric Minkowski vacua. Naively, one could expect that, by a proper choice of small complex numbers $c$ and $d$, one can easily interpolate between the AdS, Minkowski and dS minima. In particular, one could think that for small enough values of these parameters, one can conveniently fine-tune the value of the vacuum energy, uplifting the original supersymmetric minimum to the desirable dS vacuum energy with $\Lambda \sim 10^{{-120}}$.

However, the actual situation is very different. We will show that adding a small term $c\Phi +d$ always shifts the original Minkowski minimum down to AdS, which does not correctly describe our world. Moreover, unless the parameters $c$ and $d$ are exponentially small, the negative cosmological constant in the AdS minimum leads to a rapid collapse of the universe. For example, adding a tiny constant $d \sim 10^{{-54}}$ leads to a collapse within a time scale much shorter than its present age.  Thus, the cosmological predictions of the models of \cite{Ketov:2014hya} with one chiral superfield and a supersymmetric Minkowski vacuum are incredibly unstable with respect to even very tiny changes of the superpotential. Of course one could forbid such terms as $c\Phi +d$ by some symmetry requirements, but this would not address the necessity to uplift the Minkowski vacuum to $\Lambda \sim 10^{{-120}}$.

While we will illustrate this surprising result using KT models as an example, the final conclusion is very general and valid for a much broader class of theories with a supersymmetric Minkowski vacuum; see a discussion of a related issue in \cite{Kallosh:2014via}. We will show that this result is a consequence of the no-go theorem of \cite{Kallosh:2014oja} (see also \cite{GomezReino:2006dk, Hardeman:2010fh}), which is valid for arbitrary \K\ potentials and superpotentials and also applies in the presence of multiple superfields:
\begin{quote}
\it One cannot deform a stable supersymmetric Minkowski vacuum with a positive definite mass matrix to a non-supersymmetric de Sitter vacuum by an infinitesimal change of the \K\, potential  and superpotential.
\end{quote}
This no-go theorem can be understood from the role of the sGoldstino field, the scalar superpartner of the would-be Goldstino spin-1/2 field (as also emphasized in \cite{Covi:2008ea,Covi:2008cn,Achucarro:2012hg}). Since the mass of the sGoldstino is  set by the order parameter of supersymmetry breaking, it must vanish in the limit where supersymmetry is restored. The only SUSY Minkowski vacua that are continuously connected to a branch of non-supersymmetric extrema therefore necessarily have a flat direction to start with: this is the scalar field that will play the role of the sGoldstino after spontaneous SUSY breaking. A corollary of this theorem is that one cannot obtain a dS vacuum from a stable SUSY Minkowski vacuum by a small deformation. As we will see, this is exactly what forbids a small positive CC after an infinitesimal change of the KT starting point.

As often happens, the no-go theorem does not mean that uplifting of the supersymmetric Minkowski minimum to a dS minimum is impossible. In order to achieve that, the modification of the superpotential should be substantial. We will show how one can do it, thus giving a detailed illustration of how this no-go theorem works and how one can overcome its conclusions by changing the parameters of the correction term $c\Phi +d$ beyond certain critical values. For example, one can take $d =0$ and slowly increase $c$. For small values of $c$, the absolute minimum of the potential corresponds to a supersymmetric AdS vacuum. When the parameter $c$ reaches a certain critical value, the minimum of the potential ceases to be supersymmetric, but it is still AdS. With a further increase of $c$, the minimum is uplifted and becomes a non-supersymmetric dS vacuum state. Once again, we will find that the modification of the superpotential required for the tiny uplifting of the vacuum energy  by $\Lambda \sim 10^{{-120}}$ leads to a strong supersymmetry breaking, with the gravitino mass many orders of magnitude greater than what is usually expected in supergravity phenomenology.

This problem can be solved by introducing additional chiral superfields responsible for uplifting and supersymmetry breaking. However,  this may require an investigation of inflationary evolution of multiple scalar fields, unless the additional fields are strongly stabilized \cite{Dudas:2012wi} or belong to nilpotent chiral multiplets \cite{Ferrara:2014kva,Kallosh:2014via,Dall'Agata:2014oka,Kallosh:2014hxa,Linde:2014hfa}.

\subsection{Inflation and uplifting with a linear superpotential}\label{Wlin}

\begin{figure}[htb]
\begin{center}
\includegraphics[width=7cm]{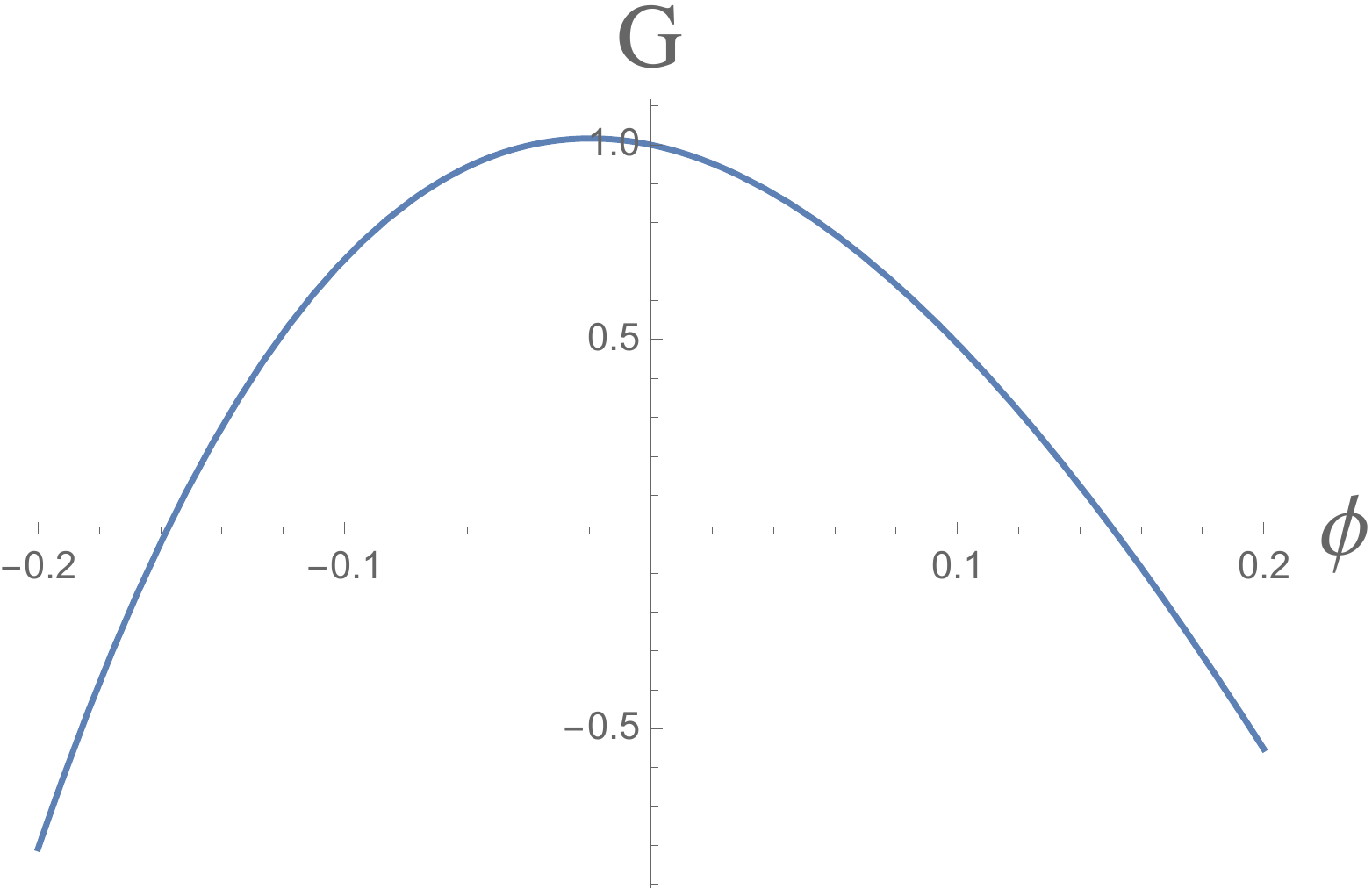}
\vspace*{3mm}
\caption{\it The coefficient in front of the kinetic term for the field $\Phi$ as a function of $\phi$ for  $\zeta =1$.}\label{1}
\end{center}
\vspace{-0.5cm}
\end{figure}

To understand the basic features of the theories with the \K\ potential \eqref{KK}, it is instructive to calculate the coefficient $G(\phi,\chi)$ in front of the kinetic term of the field $\Phi$. For an arbitrary choice of the superpotential, this coefficient is given by
\be
G(\phi,\chi)= {3 (1 + 32 \zeta^2 \phi^6 - 8 \zeta \phi^2 (3 \sqrt{3} + \sqrt{2} \phi))\over(\sqrt 3 + 
     \sqrt{2} \phi + 4  \zeta \phi^4)^2} \ .
\ee
This function does not depend on $\chi$. For small $\phi$ the fields are canonically normalized. $G(\phi,\chi)$ is positive at small $\phi$, while it vanishes and becomes negative for larger values of $|\phi|$ (provided $\zeta > 0$). Thus the kinetic term is positive definite only in a certain range of its values, depending on the constant $\zeta$. In this Section, we will usually take $\zeta =1$, to simplify the comparison with  \cite{Ketov:2014hya}, see Fig. \ref{1}.

It is equally important that the expression for the potential $V$ in this theory, for any superpotential, contains the  coefficient $1 + 32 \zeta^2 \phi^6 - 8 \zeta \phi^2 (3 \sqrt{3} + \sqrt{2} \phi)$ in the denominator, so it becomes infinitely large exactly at the boundaries of the moduli space where the kinetic term vanishes (for $\zeta=1$, the boundaries are located at $\phi\approx \pm 0.15$). For large $\zeta$, the domain where $G$ is positive definite becomes more and more narrow, which is why the field $\phi$ becomes confined in a narrow interval, whereas the field $\chi$ is free to move and play the role of the inflaton field. This is very similar to the mechanism of realization of chaotic inflation proposed earlier in a different context in Section 4 of \cite{Kallosh:2014qta}.

\begin{figure}[htb]
\begin{center}
\includegraphics[width=8cm]{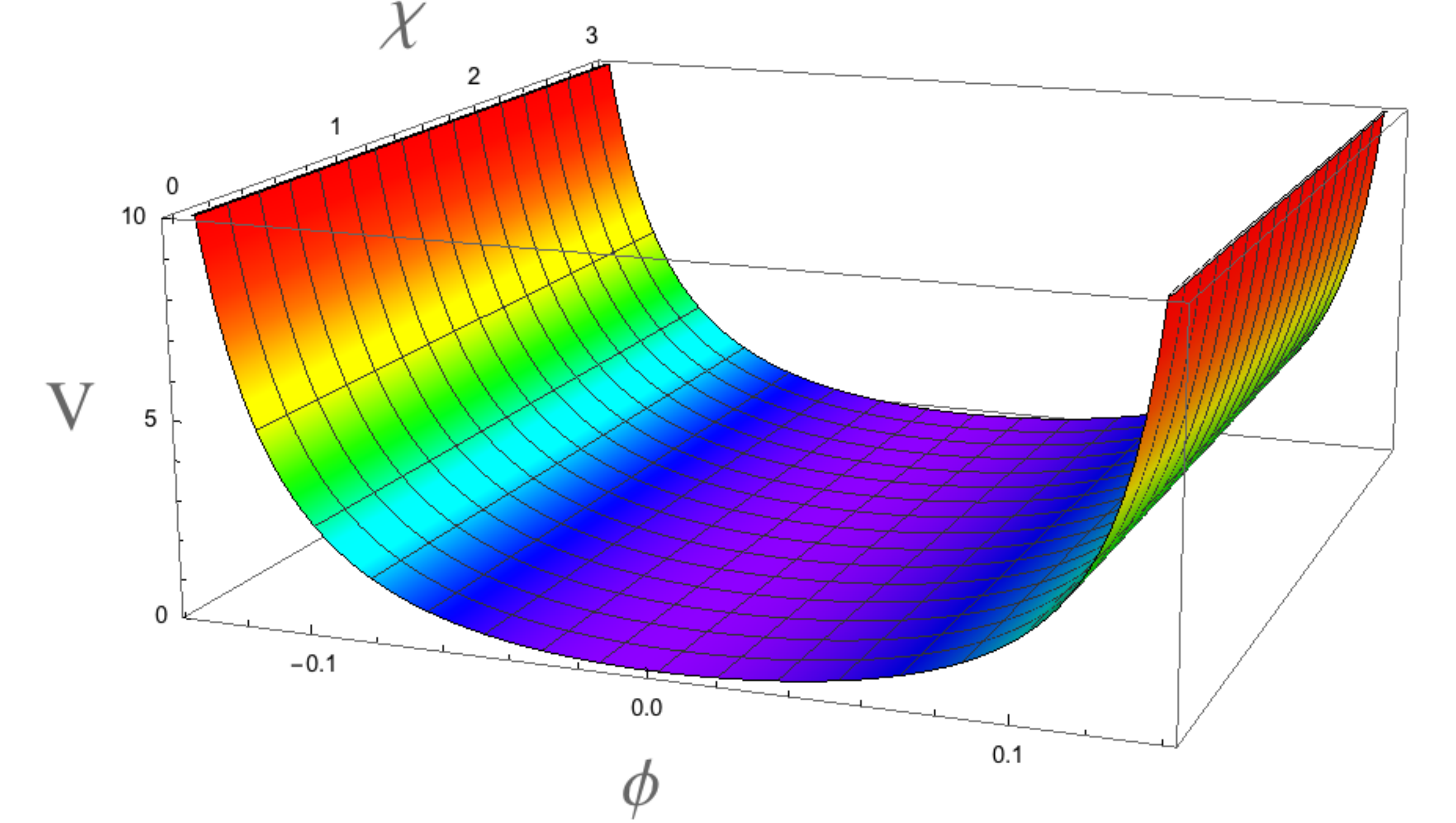}
\vspace*{3mm}
\caption{\it The scalar potential in the theory with a constant superpotential $W = m$. For $\zeta = 1$, it blows up at $\phi \approx 0.15$, and it does not depend on the field $\chi$, forming a narrow Minkowski valley surrounded by infinitely steep walls.}\label{2}
\end{center}
\vspace{-0.5cm}
\end{figure}

We will study inflation in this class of theories by giving some examples, starting from the simplest ones. The simplest superpotential to consider is a  constant one, $W = m$. In this case, the potential does not depend on the field $\chi$. It blows up, as it should, at sufficiently large $\phi$, and it vanishes at $\phi = 0$, see Fig. \ref{2}. This potential does not describe inflation.

\begin{figure}[htb]
\begin{center}
\includegraphics[width=7cm]{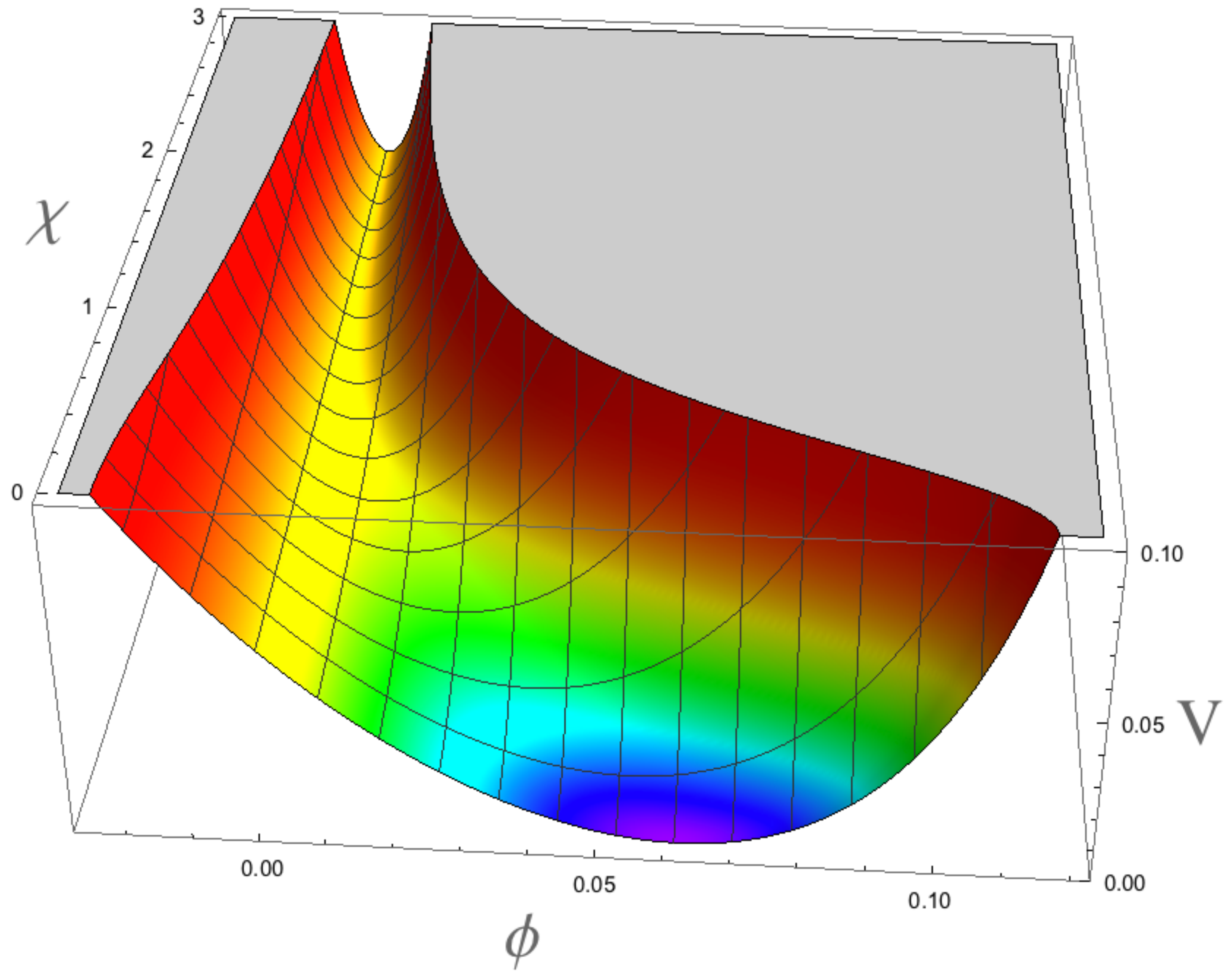}
\vspace*{4mm}
\caption{\it The scalar potential in the theory with  $W = m\, (c\Phi +1)$, for $\zeta = 1$. For $c \approx 3.671$, it has a dS valley, and a near-Minkowski minimum at $\chi =0$, $\phi \approx 0.06$. Inflation happens when the field slowly moves along the nearly flat valley and then rolls down towards the minimum of the potential. It is a two-field inflation, which cannot be properly studied by assuming that $\phi = 0$ during the process.}\label{3}
\end{center}
\vspace{-0.5cm}
\end{figure}

As a next step, we will consider a superpotential with a linear term
\be\label{linearWKT}
W = m\, (c\Phi +1) \ .
\ee
In this case, just as in the case considered above, the potential has an exactly flat direction at $\phi = 0$, but now the potential at  $\phi = 0$ is equal to
\be
V(\phi = 0,\chi) = m^{2}c\, (c-2\sqrt 3) \ .
\ee
Thus for $c < 2\sqrt 3$ it is an AdS valley, but for $c > 2\sqrt 3$ it is a dS valley. But this does not tell us the whole story. At large $\chi$, the minimum of the potential in the $\phi$ direction is approximately at $\phi = 0$, but at smaller $\chi$, the minimum shifts towards positive $\phi$. For\footnote{An understanding of this value of $c$ and its role in terms of (non-)supersymmetric branches is given in appendix A of \cite{Linde:2014ela}.} $c \approx 3.671$, the potential has a global non-SUSY Minkowski minimum with $V =0$ at $\chi =0$ and $\phi \approx 0.06$. By a minuscule change of $c$ one can easily adjust the potential to have the desirable value $\Lambda \sim 10^{{-120}}$ at the minimum. This requires fine-tuning, but it should not be a major problem in the string landscape scenario. The full potential is shown in Fig.~\ref{3}. In general, one would expect higher-order corrections which might slightly perturb the potential; however, we focus on the effect of the lower-order terms.

\begin{figure}[htb]
\begin{center}
\includegraphics[width=6.cm]{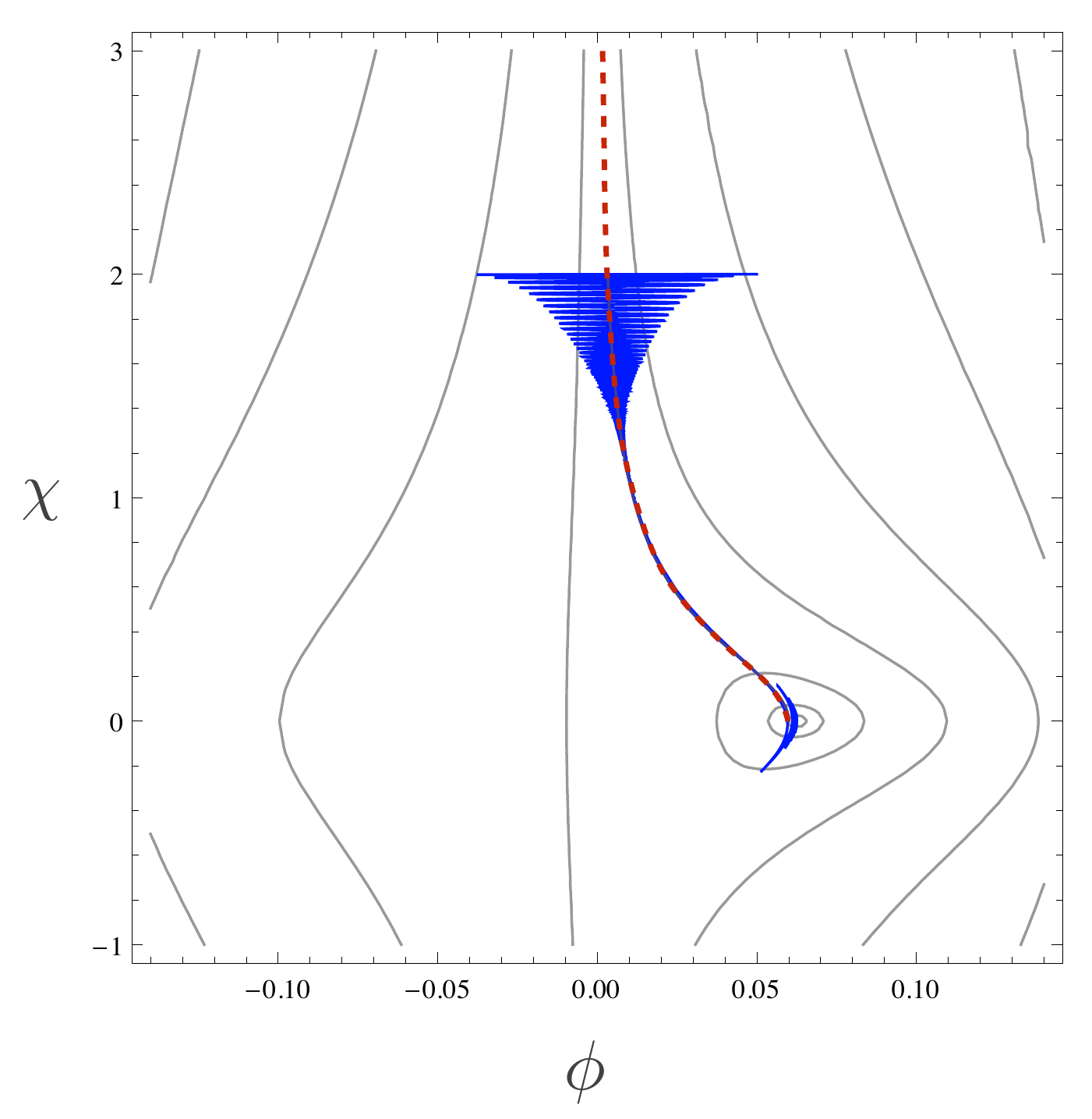}
\vspace*{4mm}
\caption{\it The dynamical evolution of the inflaton field (blue line) in the model with $W=m(c\Phi+1)$, for $\zeta=1$. The adiabatic approximation of the effective potential (dashed red line) and the contour plot of $V(\phi,\chi)$ in logarithmic scale are shown as superimposed. There is an initial stage of oscillations before the field approaches the inflationary attractor, as well as the final stage of post-inflationary oscillations. However, during inflation, which happens between these two oscillatory stages, the field accurately follows the position of the adiabatically changing minimum of the potential $V(\phi(\chi),\chi)$. }\label{Ev}
\end{center}
\vspace{-0.5cm}
\end{figure}

Inflation in this models happens when the field slowly moves along the nearly flat valley and then rolls down towards the minimum of the potential. It is a two-field dynamics, which cannot be properly studied by assuming that $\phi = 0$ during the process, as proposed in  \cite{Ketov:2014qha,Ketov:2014hya}. Indeed, the potential along the direction $\phi = 0$ is exactly constant, so the field would not even move if we assumed that during its motion. However, because of the large curvature of the potential in the $\phi$ direction, during inflation this field rapidly reaches an inflationary  attractor trajectory and then adiabatically follows the position of the  minimum of the  potential $V(\phi,\chi)$ for any given value of the field $\chi(t)$. This can be confirmed by a numerical investigation of the combined evolution of the two fields  whose dynamics is shown in Fig.~\ref{Ev}. 

Then, the adiabatic approximation of the effective potential driving inflation reads 
\be
V(\phi(\chi), \chi) = m^{2}c\, (c-2\sqrt 3) -\frac{2m^2(c-\sqrt{3})^2}{27\sqrt{3}\chi^2}\,,
\ee
neglecting higher order terms which play no role in the inflationary plateau. The effective fall-off of $1/\chi^2$ is responsible for determining the main properties of a fully acceptable inflationary scenario.

This investigation shows that this simplest model leads to a desirable amplitude of inflationary perturbations for $m \sim 7.75 \times 10^{-6}$, in Planck units. The inflationary parameters $n_{s}$ and $r$ in this model are given by (at leading order in $1/N$)
 \begin{align}
   n_{s} = 1 - \frac{3}{2N} \,, \quad r = { 2(c-\sqrt3)\over \sqrt{26 c (\sqrt{3} c -6)} \, N^{3/2}} \,.
  \end{align}
Numerically, we find $n_s \approx 0.975$ and $r \approx 0.0014$ for $N=60$, in excellent agreement with the leading $1/N$ approximation. We checked that the values of $n_{s}$ remains approximately the same in a broad range of $\zeta$, from $\zeta = 0.1$ to $\zeta = 10$. The value of the parameter $r$ slightly changes but remains in the $10^{{-3}}$ range. As of now, all of these outcomes are in good agreement with the data provided by Planck.

However, this simplest inflationary model has a property which is shared by all other models of this class to be discussed in this Section: supersymmetry is strongly broken in the minimum of the potential. In particular, for $\zeta = 1$, the superpotential at the minimum is given by $W\approx 9 \times 10^{-6}$, and the gravitino mass is $m_{3/2} \sim 8.34 \times 10^{-6}$, in Planck units, i.e.   $m_{3/2} \sim 2 \times 10^{13}$ GeV. This is many orders of magnitude higher than the gravitino mass postulated in many phenomenological models based on supergravity.

Of course, supersymmetry may indeed be broken at a very high scale, but nevertheless this observation is somewhat worrisome. One could expect that this is a consequence of the simplicity of the model that we decided to study, but we will see that this result is quite generic.

\subsection{Inflation and uplifting with a quadratic superpotential}\label{secquad}

As a second example, we will discuss the next simplest model, defined by 
\be\label{Wquad}
W=\tfrac{1}{2} m \Phi^2\,.
\ee
This case was one of the focuses of \cite{Ketov:2014hya} and gives rise to a quadratic inflationary potential. As we will demonstrate, perturbing such a superpotential by means of a linear and constant term, leads to general properties which are shared by the class discussed in the previous section.

\begin{figure}[htb]
\begin{center}
\includegraphics[width=6cm]{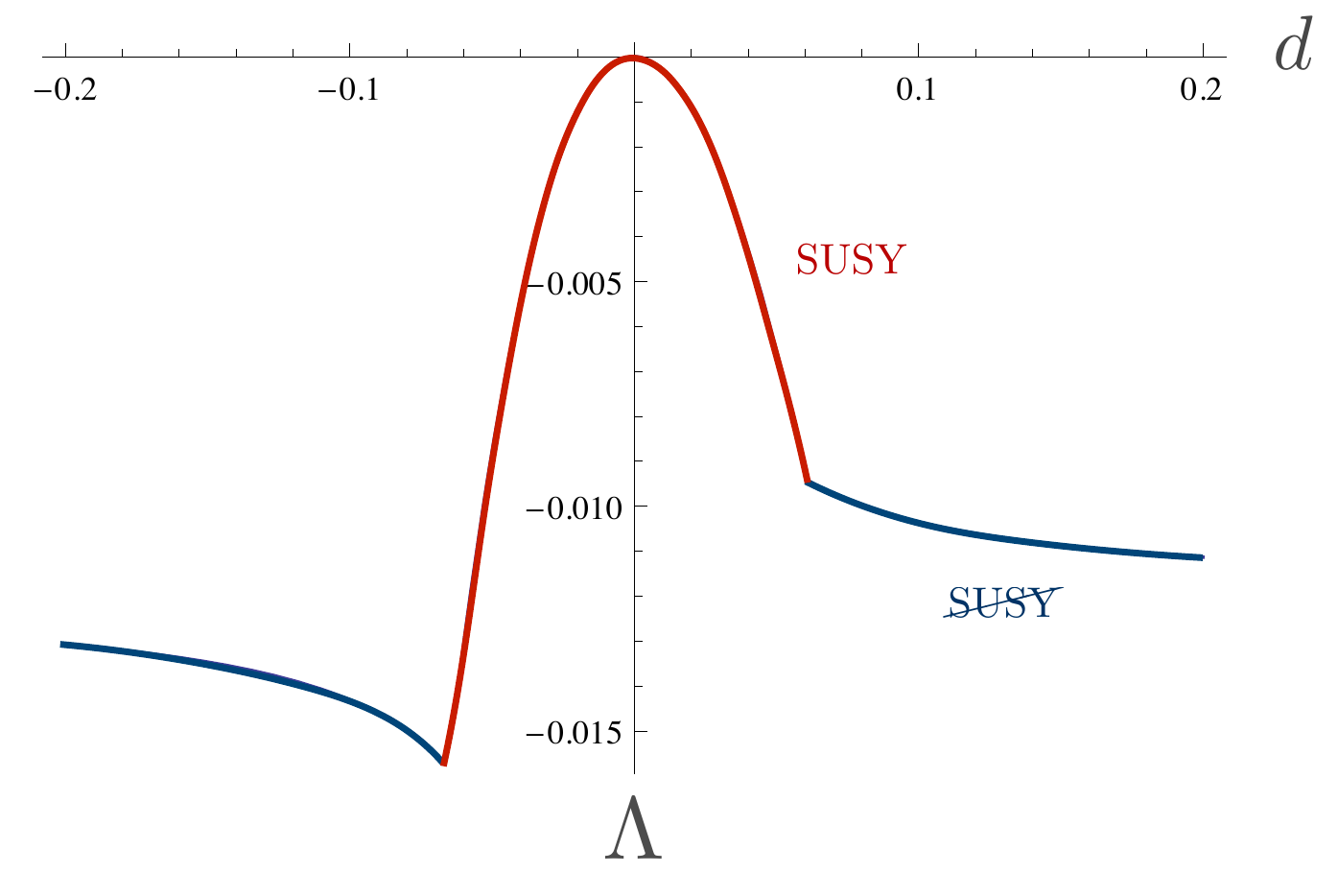}
\includegraphics[width=6cm]{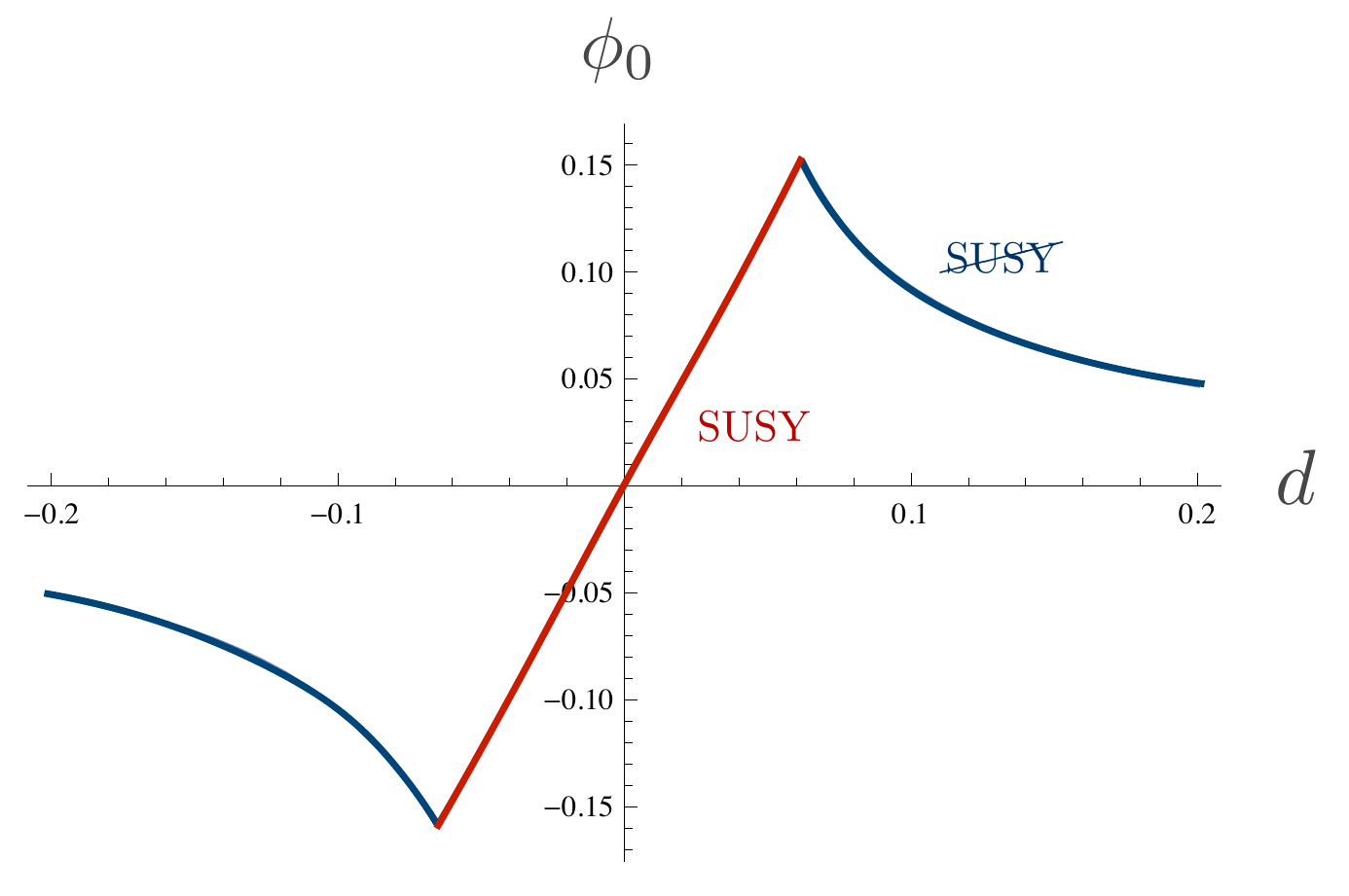}
\vspace*{0.5cm}
\caption{\it The value of the cosmological constant (left panel) in the minimum and its location $\phi_0$ (right panel) as a function of the constant term $d$ in the superpotential \eqref{Wd}. The two branches of solutions (SUSY and non-SUSY), within the fundamental physical domain $|\phi|\lesssim0.15$, are shown in different colors.  At larger (positive or negative) values of the constant, both the CC and the location $\phi_0$ level off to a constant. Plots obtained for $m = \zeta = 1$.}\label{small-const}
\end{center}
\vspace{-0.5cm}
\end{figure}

We will start by perturbing this model via a constant term such as
\be\label{Wd}
W= m \left(\tfrac{1}{2}\Phi^2 +d\right)\,.
\ee
The inflationary regime is unaffected by such correction and the scalar potential still reads $V=\tfrac{1}{2}m^2\chi^2$, at $\phi=0$. However, the vacuum of $V(\phi,\chi)$ will move away from the supersymmetric Minkowski minimum, originally placed at $\Phi=0$, but just in the $\phi$-direction (because the superpotential is symmetric). Then, for small parameter values, the minimum of $\phi$ moves as
\be
\phi_0= \sqrt{6}d-\sqrt{\frac{3}{2}}d^2\,.
\ee
This shift immediately leads to an AdS phase which, at small values of $d$, goes as
\be
\Lambda= -\sqrt{3}m^2d^2\,,
\ee
which is fully in line with the no-go theorem \cite{Kallosh:2014oja} summarized in the Introduction. These solutions do not break supersymmetry and they can be obtained by the equation $\mathcal{D}_{\Phi} W=0$. As $|d|$ increases, such a SUSY vacuum moves further away from the origin and, at one point, it crosses the singular boundary of the moduli space. Then, if we search for numerical solutions within the strip corresponding to the correct sign of the kinetic terms (this means for $|\phi|\lesssim0.15$), we run into a feature which will be common also in other examples: for specific values of $d$, the SUSY-branch of vacuum solutions leaves the fundamental physical domain $|\phi|\lesssim0.15$ and it is replaced by a new branch of vacua with broken supersymmetry. This is shown in Fig.~\ref{small-const}. However, as one keeps increasing the absolute value of $d$, $\phi_0$ approaches a constant value which corresponds to an asymptotic AdS phase. Therefore, perturbing $W$ by means of a constant term does not help to uplift to dS.

As second step, we include a linear correction such that the superpotential reads
\be\label{Wc}
W= m \left(\tfrac{1}{2}\Phi^2 +c\Phi\right)\,,
\ee
where the coefficients are real due to the constraint on\footnote{Perturbing the superpotential by means of a linear term with imaginary coefficient such as $ic\Phi$ is equivalent to adding a positive constant $c^2$. This is a direct consequence of the shift symmetry of the \K\ potential.} $W$.

\begin{figure}[htb]
\begin{center}
\includegraphics[width=6cm]{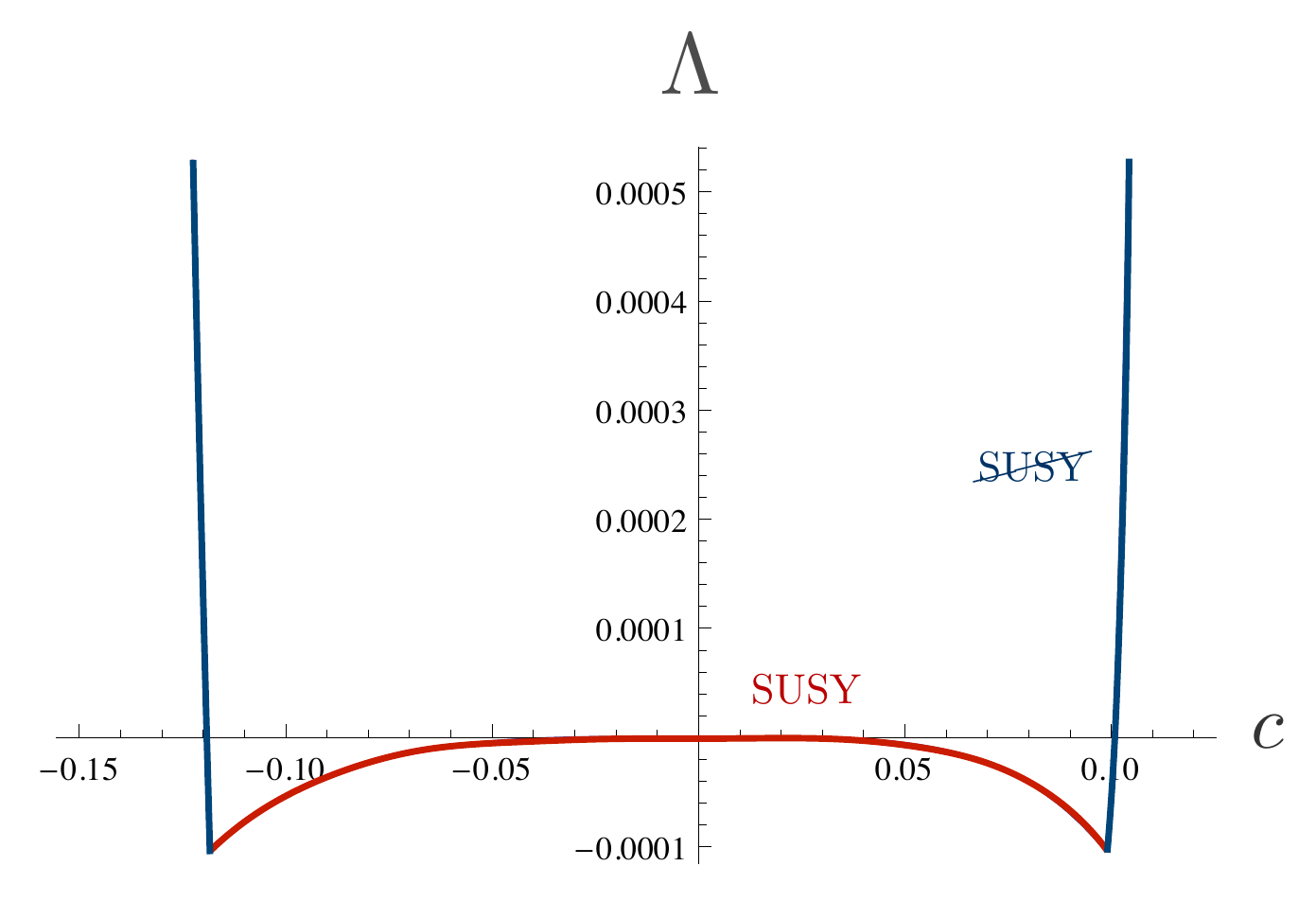}
\includegraphics[width=6cm]{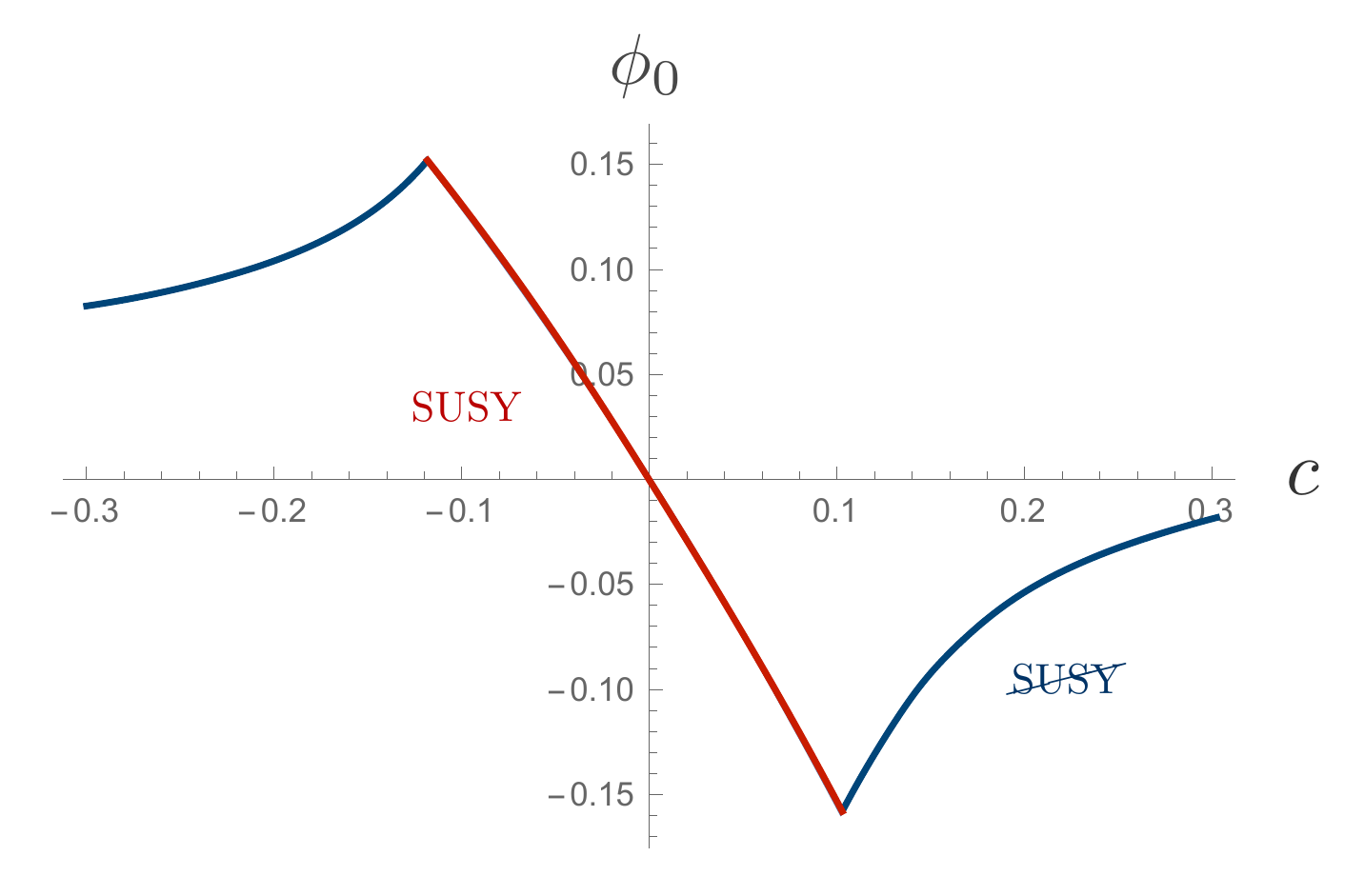}
\vspace*{0.5cm}
\caption{\it The value of the cosmological constant (left panel) in the minimum and its location $\phi_0$ (right panel) as a function of the linear term in the superpotential. The two branches of solutions (SUSY and non-SUSY), within the fundamental physical domain $|\phi|\lesssim0.15$, are shown in different colors. At larger (positive or negative) values of the coefficient $c$, the location $\phi_0$ levels off to a constant while the CC approaches a quadratic shape. Plots obtained for $m = \zeta = 1$.}\label{small-lin}
\end{center}
\vspace{-0.5cm}
\end{figure}

Similarly to the previous case, the SUSY Minkowski vacuum is perturbed by such correction and, at lowest order in $c$, it moves in the $\phi$-direction as
\be
\phi_0=-\sqrt{2}c-\sqrt{\frac{3}{2}}c^2\,,
\ee
leading to a vacuum energy given by
\be
\Lambda= -\sqrt{\frac{3}{4}}m^2c^4\,,
\ee
Then also in this case, as $|c|$ increases, such supersymmetric solutions move towards the boundary $\phi\approx\pm0.15$ and cross it. At the same point in parameter space, a new branch of non-supersymmetric solutions appears and, remarkably, this results into a sharp increase of the scalar potential at the minimum. In fact, this very quickly gives rise to a transition from AdS to dS, as it is shown in Fig.~\ref{small-lin}. 

The exact values for which these transitions happen are as follows. The transition from SUSY to non-SUSY vacua occurs at (calculated for $m= \zeta = 1$)
 \begin{align}
  c =  -0.118162 \,, \qquad c = 0.101918  \,,
 \end{align}
while the CC crosses through Minkowski at
 \begin{align}\label{MinkLin}
  c =  -0.119318 \,, \qquad c = 0.102692 \,.
 \end{align}

Note that, at finite $c$ values, the scalar potential passes through Minkowski. In contrast to the ground state at $c=0$, the new Minkowski vacua are non-supersymmetric, and hence can be deformed into dS without violating the no-go theorem. In fact, these non-supersymmetric Minkowski vacua are exactly the type of  structures that were identified in \cite{Kallosh:2014oja} as promising starting points for uplifts to De Sitter (although there the focus was on a hierarchy of supersymmetry breaking order parameters for different superfields). A minuscule deviation of $c$ from \eqref{MinkLin} will be sufficient to obtain the physical value of cosmological constant $\Lambda\sim10^{-120}$.

It is worthwhile to remark that the order of magnitude of the parameter $c$, for which we get a tiny uplifting to dS, is small with respect to the coefficient of the quadratic term in the superpotential \eqref{Wc}. This translates into the fact that the inflationary predictions will be basically unchanged with respect the simple scenario with a quadratic potential. In fact, the scalar potential in the direction $\phi=0$ reads
\be
V(\phi=0 ,\chi)= \frac{1}{2}\left(1-\sqrt{3}c\right)m^2\chi^2\ + m^2c^2\,.
\ee
At $\chi \lesssim O(1)$, the field $\phi$ no longer vanishes and starts moving towards the minimum of the potential. However, the main stage of inflation happens at $\chi \gg c =O(0.1)$, when $\phi$ nearly vanishes and the inflaton potential is approximately equal to $\frac{1}{2}\left(1-\sqrt{3}c\right)m^2\chi^2$. The main effect of this change of the potential is a slight change of normalization of the amplitude of the perturbations spectrum, which requires a small adjustment for the choice of the parameter $m$: 
\be
m\approx (6+5.2c)\cdot10^{-6}\,.
\ee

However, even though the inflationary regime is essentially unaffected by such a small correction, supersymmetry is strongly broken at the end of inflation, just as in the theory with a simple linear superpotential, discussed in Sec.~\ref{Wlin}. This is a direct consequence of the no-go theorem discussed above and of the impossibility of uplifting the SUSY Minkowski vacuum (corresponding to $c=0$) by an infinitesimal deformation of $W$. In particular, for values of $c$ leading to a realistic dS phase (these values are extremely close to \eqref{MinkLin}, corresponding to non-supersymmetric Minkowski) and  for $\zeta=1$, we obtain the following: for positive $c$, the superpotential at the minimum is $|W|\approx3.4\times 10^{-8}$ and the gravitino mass is $m_{3/2}\sim 4.2 \times 10^{-8}$, in Planck units, i.e. $m_{3/2}\sim 1.0 \times 10^{11}$ GeV; for negative $c$, the superpotential at the minimum is $|W|\approx3.8\times 10^{-8}$ and the gravitino mass is $m_{3/2}\sim 3.2 \times 10^{-8}$, in Planck units, i.e. $m_{3/2}\sim 7.6 \times 10^{10}$ GeV. These values are again well beyond the usual predictions of the low scale of supersymmetry breaking in supergravity phenomenology. 

\subsection{Discussion}
In this Section we have investigated the possibility to realize a model of inflation and dark energy in supergravity. As an example, we considered the class of single chiral superfield models proposed in \cite{Ketov:2014hya}. The models described in  \cite{Ketov:2014hya} share the following feature: The vacuum energy in these models vanishes, and supersymmetry is unbroken. One could expect that this is a wonderful first approximation to describe dS vacua with vanishingly small vacuum energy $\Lambda\sim 10^{{-120}}$ and small supersymmetry breaking with $m_{3/2} \sim 10^{-15}$ or $10^{-13}$ in Planck units. However, we have shown that this is not the case, because of the no-go theorem formulated in \cite{Kallosh:2014oja}. While it is possible to realize an inflationary scenario  that ends in a dS vacuum with $\Lambda\sim 10^{{-120}}$, these vacua cannot be infinitesimally uplifted by making small changes in the \K\ potential and superpotential. One can uplift a stable Minkowski with unbroken SUSY to a dS minimum, but it always requires large uplifting terms, resulting in a strong supersymmetry breaking with $m_{3/2}$ many orders of magnitude higher than the TeV  or even PeV range advocated by many supergravity phenomenologists. 

In our investigation, we also introduced a new model, which contained only linear and constant terms in the superpotential. This superpotential is simpler than those studied in \cite{Ketov:2014hya}, but we have found that this model does describe a consistent inflationary theory with  dS vacuum, which can have $\Lambda\sim 10^{{-120}}$. However, just as in all other cases considered in this Section, we found that supersymmetry is strongly broken after inflation in this model. While we have analyzed only some specific cases in detail, our conclusions apply to a much wider class of models, well beyond the specific models proposed in \cite{Ketov:2014hya}, because of the general nature of the no-go theorem of  \cite{Kallosh:2014oja}. 

Since there is no evidence of low scale supersymmetry at LHC as yet, one could argue that the large scale of supersymmetry breaking is not necessarily a real problem. However, it would be nice to have more flexibility in the model building, which would avoid this issue altogether. One way to get dS uplifting with small supersymmetry breaking, without violating the no-go theorem, is to add other chiral multiplets (e.g. Polonyi fields), and to strongly stabilize them to minimize their influence on the cosmological evolution, see e.g. \cite{Dudas:2012wi}. In certain cases, one can make the Polonyi fields so heavy and strongly stabilized that they do not change much during the cosmological evolution and do not lead to the infamous Polonyi field problem which bothered cosmologists for more than 30 years  \cite{Coughlan:1983ci,Goncharov:1984qm,Banks:1993en,Dvali:1995mj,Dine:1995uk}.  A more radical approach, which allows to have a single scalar field evolution is to use models involving nilpotent chiral superfields \cite{Ferrara:2014kva,Kallosh:2014via,Dall'Agata:2014oka,Kallosh:2014hxa,Linde:2014hfa}, which have an interesting string theory interpretation in terms of D-branes \cite{Kallosh:2014wsa}. This framework will be investigated in the next two sections.


\section{Arbitrary inflation and de Sitter landscape} \label{sectionKLS}

In this Section, we intend to present how the addition of a nilpotent sector allows us to evade the restrictions  presented above in Sec.~\ref{sectionLRS} and yield remarkable simplifications, within a unified cosmological scenario of inflation and dark energy. After reviewing the main properties of the nilpotent superfield $S$, we show how to construct a general class of inflationary models with de Sitter exit and controllable level of SUSY breaking at the minimum. The \Kahler geometry of these scenarios is flat thus allowing for arbitrary inflaton potential, along the line of the general model presented in Sec.~\ref{SECflat}. Finally, we comment on the relation between the supersymmetry breaking directions and the fermionic sector of the supergravity action.

\subsection{The nilpotent superfield}\label{nilpotentSEC}

In the 1970s Volkov and Akulov (VA) \cite{Volkov:1972jx,Volkov:1973ix} proposed to identify the neutrino with the massless Goldstino arising from supersymmetry breaking. They derived the corresponding action which is invariant under non-linear supersymmetry transformations (see the recent investigations \cite{Kallosh:2015pho,Ferrara:2016een,Kallosh:2016hcm}). However, this idea was soon abandoned after the discovery of neutrino oscillations.

Later in \cite{Rocek:1978nb,Ivanov:1978mx,Lindstrom:1979kq,Casalbuoni:1988xh,Komargodski:2009rz}, it was shown that VA Goldstino can be expressed  in the form a constrained superfield (see also the recent works \cite{Dall'Agata:2015lek,Dall'Agata:2016yof}). Specifically, it can be represented by a chiral multiplet $S$ with the nilpotency condition $S^2=0$. We detail this below.

The unconstrained off-shell chiral superfield has the form
\be
S(x, \theta)=s(x)+ \sqrt{2}\, \theta\, \chi^s(x) +\  \theta^2 F^S(x) \ ,
\ee
where $s(x)$ is the scalar part,  $\chi^s(x)$ is a fermion partner and $F^S(x)$ is an auxiliary field. It was shown in \cite{Komargodski:2009rz} that the nilpotent superfield $S^2(x, \theta)=0$ depends only on the fermion $\chi^s$, the VA goldstino,   and an auxiliary field $F^S$. It does not have a fundamental scalar field, that is 
\be
 S(x, \theta)|_{S^2(x, \theta)=0}  =  \ \frac{\chi^s \chi^s}{2 \, F^S} \ + \ \sqrt{2}\, \theta\, \chi^s \ +\  \theta^2 F^S \ , \label{va1}
\ee
since $s(x)$ is replaced by $ \frac{\chi^s \chi^s}{2 \, F^S}$.
For the nilpotent off-shell superfields  the rules for the bosonic action required for cosmology turned out to be very simple. Namely, one has to calculate potentials as functions of all superfields as usual, and then declare that  the scalar part of the nilpotent superfield $s(x)$ vanishes, since it is replaced by a bilinear combination of the fermions. No need to stabilize and study the evolution of the complex   field $s(x)$.

\subsection{Arbitrary inflation, dark energy and SUSY breaking}

Now we turn to the unified cosmological scenario, presented in \cite{Kallosh:2014hxa}, which allows to obtain general inflaton potential and controllable level of dark energy and SUSY breaking.

 The \Kahler potential and superpotential are of the form

\be\label{K&W}
K=-\tfrac{1}{2}\left(\Phi-\bar{\Phi}\right)^2+S\bar{S}\,,\quad W=f(\Phi)+ g(\Phi) S\,,
\ee
where $f$ and $g$ are real holomorphic functions of their arguments and $W$ has the the most general form, provided $S$ is nilpotent. Indeed, due to the nilpotency of $S$ and holomorphicity of the superpotential, $W(\Phi, S)$ in Eq.~\eqref{K&W}  is the most general form of the superpotential depending on $\Phi$ and $S$. This is analogous to the fact that an arbitrary function of a single Grassmann variable $\theta$ can be expanded into a Taylor series which terminates after 2 terms, $F(\theta)= a+b \theta$, since $\theta^2=\theta^3=...=\theta^n...=0$. In our case we have $S^2=S^3=...=S^n...=0$.

Within this class of models, the real part of the field $\Phi$ plays the role of the inflaton, rolling down along $S=0$ and $\Phi=\bar{\Phi}$, and drives a potential which reads
\be\label{pot}
V=g(\Phi)^2+f'(\Phi)^2 -3f(\Phi)^2\,.
\ee
Note that the last two terms are exactly the ones appearing in \eqref{potsingleflat}, that is, for a single superfield model (see Sec.~\ref{subsecsgold}).

After inflation, the journey of $\Re\Phi$ ends into a minimum placed at $\Phi=0$, provided the functions $f$ and $g$ satisfy
\begin{align}\label{derivf&g}
f'(0)=g'(0)=0\,.
\end{align}

The values of $f$ and $g$ at the minimum will allow for a wide spectrum of possibilities in terms of supersymmetry breaking and cosmological constant, along the lines of the string landscape scenario. Supersymmetry is spontaneously broken just in the nilpotent direction\footnote{This allows for a simplification of the fermionic sector of the supergravity action. Specifically, in the unitary gauge, the gravitino interacts just with the fermion of the nilpotent field leading to a simple version of the super-Higgs mechanism \cite{Dall'Agata:2014oka,Kallosh:2014hxa}.}, namely
\begin{align}
D_S W_{min}=g(0)=M\,, \qquad D_\Phi W_{min}=0\,,
\end{align}
where we have introduced $M$ as SUSY breaking parameter. Further, the gravitino mass is given by $m_{3/2}=f(0)$. The value of the cosmological constant is equal to
\be\label{Vminimum}
\Lambda= g^2(0) - 3 f^2(0)=M^2-3m_{3/2}^2\,.
\ee
The vacuum is stable if the masses of both directions, as given by
\be \label{masses}
\begin{aligned}
m^2_{\Re \Phi} (\Phi =0)&= f''(0)^2 + M g''(0) - 3 m_{3/2} f''(0)\,,\\
m^2_{\Im \Phi} (\Phi =0)&= f''(0)^2 - M g''(0) -  m_{3/2} f''(0) + 2 (M^2-m^2_{3/2})\,,
\end{aligned}
\ee
are assured to be positive.

However, the generality of Eq.~\eqref{pot} does not assure always a viable inflationary scenario. The negative term can be dominating at large value of the inflaton field and not give rise to inflation. In the framework defined by Eq.~\eqref{K&W}, a successful choice for the functions  $f$ and $g$ is given by \cite{Dall'Agata:2014oka,Kallosh:2014hxa}
\be\label{g&f}
f(\Phi)=\beta\ g(\Phi)\,,
\ee
with $\beta$ being some constant. The specific relation \eqref{g&f} leads to a situation where the negative contribution in \eqref{pot} is exactly canceled when the minimum \eqref{Vminimum} is Minkowski and, then, by fine-tuning $\beta=1/\sqrt{3}$. Then, the scalar potential turns out to have the simple form
\be
V=\left[f'(\Phi)\right]^2\,.
\ee
Allowing for a small cosmological constant $\Lambda\sim10^{-120}$ (then, having a tiny deviation of $\beta$ from $1/\sqrt{3}$) does not change effectively the inflationary predictions. Other possible choices for $f$ and $g$ are discussed in \cite{Kallosh:2014via,Kallosh:2014hxa}.

This construction is quite flexible in terms of  observational predictions allowing for any possible value of $n_s$ and $r$. Nonetheless, the generality of such construction relies on the relation \eqref{g&f} and turns out to be really sensitive with respect to any other generic coupling between the inflaton and the nilpotent sector.  Moreover, the negative contribution of Eq.~\eqref{pot} is balanced just if one assumes the observational evidence of a negligible cosmological constant. A generic de Sitter landscape would yield important corrections to such construction.

\subsection{Fermionic sector after the exit from inflation}
Now we will describe the fermionic sector of the theory.
The generic mixing term of the  gravitino with the goldstino $v$ can be expressed as a combination of  fermions from chiral multiplets $\chi^i$ such as
\be
 \bar \psi^\mu \gamma_\mu \,  v +h.c= \bar \psi^\mu \gamma_\mu  \sum_i \chi^i e^{K\over 2} D_i W +h.c.
\ee
In case of our two multiplets, we have that the inflatino $\chi^\phi$ as well as the $S$-multiplet fermion $\chi^s$ form a goldstino $v$, which is mixed with the gravitino as
\be
 \bar \psi^\mu \gamma_\mu \,  v= \bar \psi^\mu \gamma_\mu  \left( \chi^\phi e^{K\over 2} D_\phi W +  \chi^s e^{K\over 2} D_S W\right)   \ .
\ee
Therefore, the local supersymmetry gauge-fixing $v=0$ leads to a condition
\be
v= \chi^\phi e^{K\over 2} D_\phi W +  \chi^s e^{K\over 2} D_S W=0 \ .
\ee
This leads to a mixing of the inflatino $\chi^\phi$ with the $S$-multiplet fermion $\chi^s$. The action has many non-linear in $\chi^s$ terms and therefore the fermionic action in terms of a non-vanishing combination of $\chi^\phi$ and $\chi^s$ is extremely complicated. For example,  a non-gravitational part of the action of the fermion of the nilpotent multiplet  is given by
\be
{\cal L}_{VA}= - M^2 +i \partial_\mu \bar  \chi^s \bar \sigma ^\mu  \chi^s + {1\over 4 M^2}  { (\bar \chi^s)}^2 \partial^2  (\chi^s)^2 - {1\over 16 M^6}  (\chi^s)^2   (\bar \chi^s)^2 \partial^2  (\chi^s)^2 \partial^2  ( \bar\chi^s)^2 \ ,
\label{VA}\ee
as shown in \cite{Komargodski:2009rz}. In  supergravity there will be more non-linear couplings of $\chi^s$ with other fields.

In our class of models where the only direction in which supersymmetry is spontaneously broken is the direction of the nilpotent chiral superfield and $D_\Phi W=0$ the coupling is
\be
\bar \psi^\mu \gamma_\mu  \chi^s e^{K\over 2} D_S W|_{\min} +h.c = \bar \psi^\mu \gamma_\mu \,  \chi^s M +h.c.
\ee
and the goldstino is defined only by one spinor
\be
v= \chi^s M  \ .
\ee
The inflatino $\chi^\phi$,  the spinor from the $\Phi$ multiplet does not couple to $\gamma^\mu \Phi_\mu$ since $D_\Phi W|_{\min}=0$. In this case we can make a choice of the unitary gauge  $v=0$, when fixing local supersymmetry.  Since $M\neq 0$  it means that we can remove the spinor from the nilpotent multiplet
\be
\chi^s=0 \ .
\ee
The corresponding gauge is the one where gravitino becomes massive by `eating' a goldstino. The unitary gauge is a gauge where the massive gravitino has both $\pm 3/2$  as well as $\pm1/2$ helicity states.
In our models the fully non-linear fermion action simplifies significantly since it depends only on inflatino. All complicated non-linear terms of the form 
$
{1\over M^2} (\chi^s)^2 \partial^2   (\bar \chi^s)^2
$ and higher power of fermions as well as mixing of the inflatino  $\chi^\phi$ with $\chi^s$ disappear  in this unitary gauge. 

In particular, the fermion masses of the gravitino and the inflatino,  at the minimum, are simply
\be
m_{3/2}=W_{0}=f(0) \ , \qquad m_{\chi^\phi}= e^{K\over 2}\partial_\Phi D_\Phi W = f''(0)- f(0) = f''(0)- m_{3/2} \ .
\ee
Here we have presented the masses of fermions without taking into account the subtleties of the definition of such masses in the de Sitter background. This was explained in details for spin 1/2 and spin 3/2 in  \cite{Kallosh:2000ve}  in case including  $\Lambda>0$. For example, the chiral fermion mass matrix ${\bf m}^{ij}= D^i D^j e^{K\over 2} W$  is replaced by
$
\hat m \equiv {\bf m} + \sqrt {\Lambda/3}\, \gamma_0 
$.


\section{Attractors and de Sitter landscape} \label{sectionScalisi}

In this Section, we provide a unified description of cosmological $\alpha$-attractors and late-time acceleration. As in the case of flat geometry, previously discussed in Sec.~\ref{sectionKLS}, our construction involves two superfields playing distinctive roles: one is the dynamical field and its evolution determines inflation and dark energy, the other is nilpotent and responsible for a landscape of vacua and supersymmetry breaking.

We prove that the attractor nature of the theory is enhanced when combining the two sectors:  cosmological attractors are  very stable with respect to any possible value of the cosmological constant and, interestingly, to any generic coupling of the inflationary sector with the field responsible for uplifting. Finally, as related result,  we show how specific couplings generate an arbitrary inflaton potential in a supergravity framework with varying \Kahler curvature.

\subsection{Uplifting flat $\alpha$-attractors} \label{FlatUp}

In the single superfield framework defined by
\be\label{K&Tsingle}
K=-\tfrac{1}{2}\left(\Phi-\bar{\Phi}\right)^2\,,\qquad W=f(\Phi)\,,
\ee
inflationary models with observational predictions given by \eqref{nsrattractors} and in excellent agreement with Planck were found in \cite{Roest:2015qya}. We have reviewed these models in the previous chapter but we recall here some basics for convenience. These are defined by 
\be\label{fflat}
f(\Phi)= \e^{\sqrt{3} \Phi} - \e^{-\sqrt{3} \Phi} F \left(\e^{- 2 \Phi / \sqrt{3 \alpha}}\right) \,,
\ee
where $F$ is an arbitrary function having an expansion such as $F(x)=\sum_n c_n x^n$ with
\be\label{xflat}
x\equiv\e^{- 2 \Phi / \sqrt{3 \alpha}}\,.
\ee

This class of models, being characterized by exponentials as building blocks of the superpotential, manifestly exhibits its attractor nature through the insensitivity to the structure of $F$. While the constant term $c_0$ would yield a de Sitter plateau $V=12c_0$, the first linear term would define the inflationary fall-off typical of $\alpha$-attractors, such as
\be\label{falloff}
V=V_0+V_1 \e^{- \sqrt{\frac{2}{3\alpha}} \varphi }+...\,,
\ee
at large values of the canonical  scalar field $\varphi= \sqrt{2}\ \Re\Phi$, with $V_0=12c_0$ and $V_1=16c_1$, the latter being negative. Higher order terms would be unimportant for observational predictions. 

This scenario can be naturally embedded in the construction discussed in the previous section. A first step would be simply choosing \eqref{fflat} as function $f$ in Eq.~\eqref{K&W}. In fact, this represents a valid alternative to the specific choice \eqref{g&f}: it yields always a balance of the negative term in \eqref{pot}, independently of the value of the uplifting at the minimum, and, interestingly, it decouples the functional forms of $f$ and $g$.  As second step, one may notice that, given the form of the scalar potential Eq.~\eqref{pot}, any generic expansion such as
\be\label{f&gflat}
f(x)=\sum_n a_n x^n\,,\qquad g(x)=\sum_n b_n x^n\,,
\ee
with $x$ given by Eq.~\eqref{xflat}, would give rise to a fall-off from de Sitter analogous to Eq.~\eqref{falloff} with
\be \label{V0V1}
V_0 =b_0^2-3a_0^2\,,\qquad V_1=2b_0b_1-6a_0a_1\,,
\ee
and, then, yield the universal predictions \eqref{nsrattractors}.

It is remarkable that the attractor structure of the theory is enhanced when combining the inflaton  with the nilpotent sector. The inflationary regime is very stable with respect to any deformation of the superpotential and any value of the uplifting.

Within this construction, the condition \eqref{derivf&g} of a minimum placed at $\Phi=0$ ($x=1$) translates into 
\be\label{condmin}
\sum_{n=1}^{\infty} n\,a_n=0\,, \qquad \sum_{n=1}^{\infty} n\ b_n=0\,.
\ee

Interestingly, the value of the cosmological constant at the minimum is given by 
\be\label{CC}
\Lambda= \left(\sum_{n} b_n\right)^2-3\left(\sum_{n} a_n\right)^2\,,
\ee
and then as a sum of the coefficients of the expansions \eqref{f&gflat} which, separately, determine the gravitino mass and the scale of supersymmetry breaking, such as
\be\label{M32M}
m_{3/2}=\sum_{n} a_n\,, \qquad M=\sum_{n} b_n\,.
\ee

Stability of the inflationary regime in the imaginary direction is always assured, for any value of $\alpha$, as the condition is simply
\be\label{massIm}
|b_0|>|a_0|\,.
\ee
In fact, the mass of $\Im\Phi$ turns out to have a natural expansion at small value of $x$ (large values of $\varphi$) such as
\be\label{MassImaginary}
m_{\Im\Phi}^2= 2 (b_0^2-a_0^2) + \frac{4}{3\alpha} \left[b_0b_1(3\alpha -1) - a_0 a_1 (3\alpha +1) \right] x + ... \,,
\ee
that is an exponential deviation from a constant plateau. Interestingly, this is the typical functional form of the scalar potential of $\alpha$-attractors, where  higher order terms do not play any role. During inflation, the $\Re \Phi$ moves along a valley of constant width. This phenomenon can be appreciated below in Fig.~\ref{Masses}, for a specific example.  Stability at the minimum is model dependent since, generically, the infinite tower of coefficients $a_n$ and $b_n$ contribute to the masses.

\begin{figure}[htb]
\hspace{-3mm}
\begin{center}
\includegraphics[width=7.5cm]{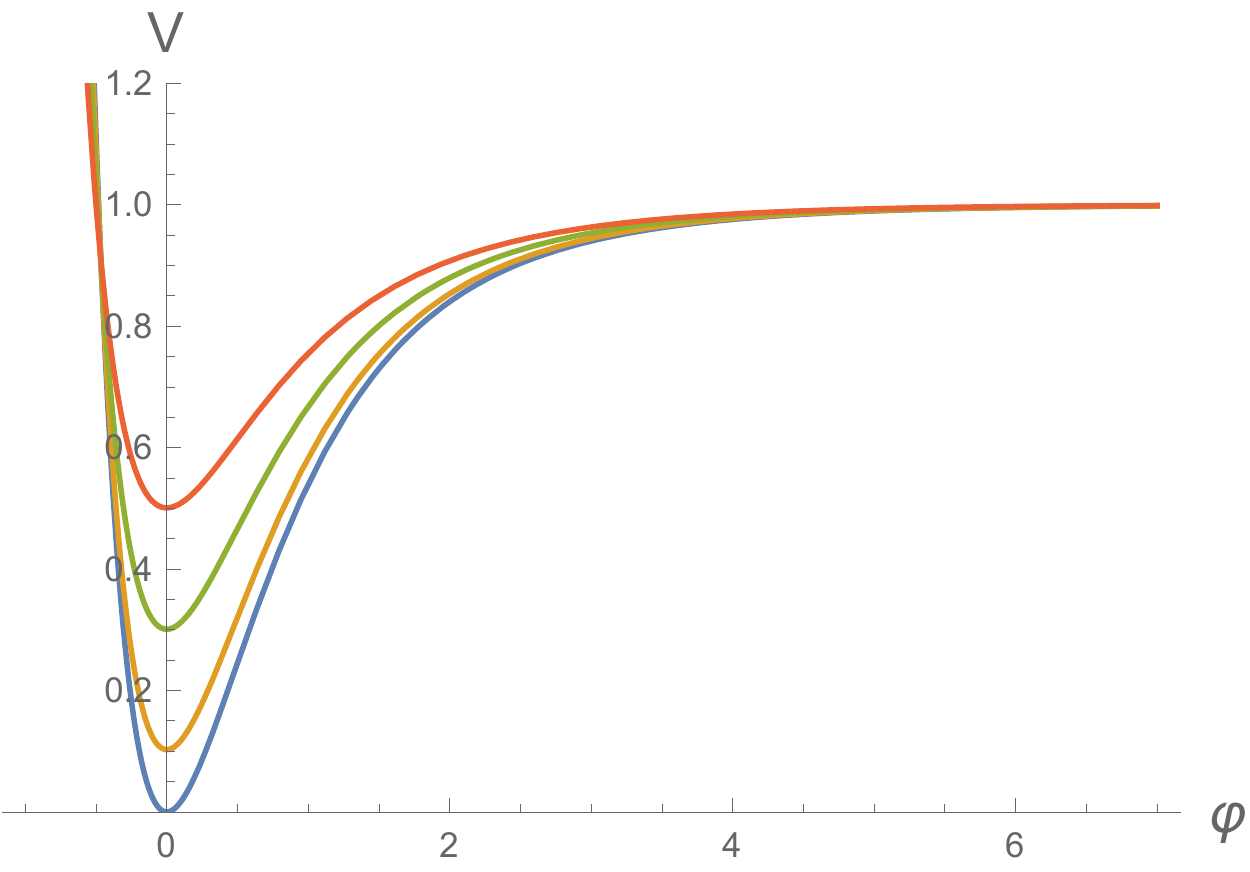}
\vspace*{0.4cm}
\caption{\it Scalar potential of the model defined by Eq.~\eqref{Example} with $\alpha=1$ and uplifting equal to $\Lambda=\{0, 0.1, 0.3, 0.5\}.$}\label{Potential}
\end{center}
\vspace{0cm}
\end{figure}
\vspace{-0.5cm}

The simplest example of such class of models is given by the following choice:
\be\label{Example}
f=a_0+a_1 x+a_2 x^2\,,\qquad g=b_0\,.
\ee

In fact, this is a minimum in order to have a deviation from de Sitter typical of $\alpha$-attractors, which comes from the linear term, and a non-trivial solution of Eq.~\eqref{condmin} to have a minimum placed at the origin, thanks to the quadratic contribution. Higher order terms will not affect neither the inflationary energy  nor the characteristic fall-off, as it is clear from Eq.~\eqref{V0V1}. The scalar potential, for $\alpha=1$ and different amount of uplifting, is shown in Fig.~\ref{Potential}. Stability occurs along the full inflationary trajectory and also at the minimum where both directions of $\Phi$ turn out to be stable, as it is shown in Fig.~\ref{Masses}. Analogous results hold for other values of $\alpha$.

The addition of higher order terms both in $f$ and $g$ would allow for more flexibility in terms of separation of the physical scales. In fact, whereas the inflationary regime would be absolutely insensitive to high order contributions, the coefficients of these terms turn out to be fundamental in determining the scale of SUSY breaking, the gravitino mass and the cosmological constant, as given by Eq.~\eqref{CC} and Eq.~\eqref{M32M}.

\begin{figure}[h!]
\hspace{-3mm}
\begin{center}
\includegraphics[width=7.5cm]{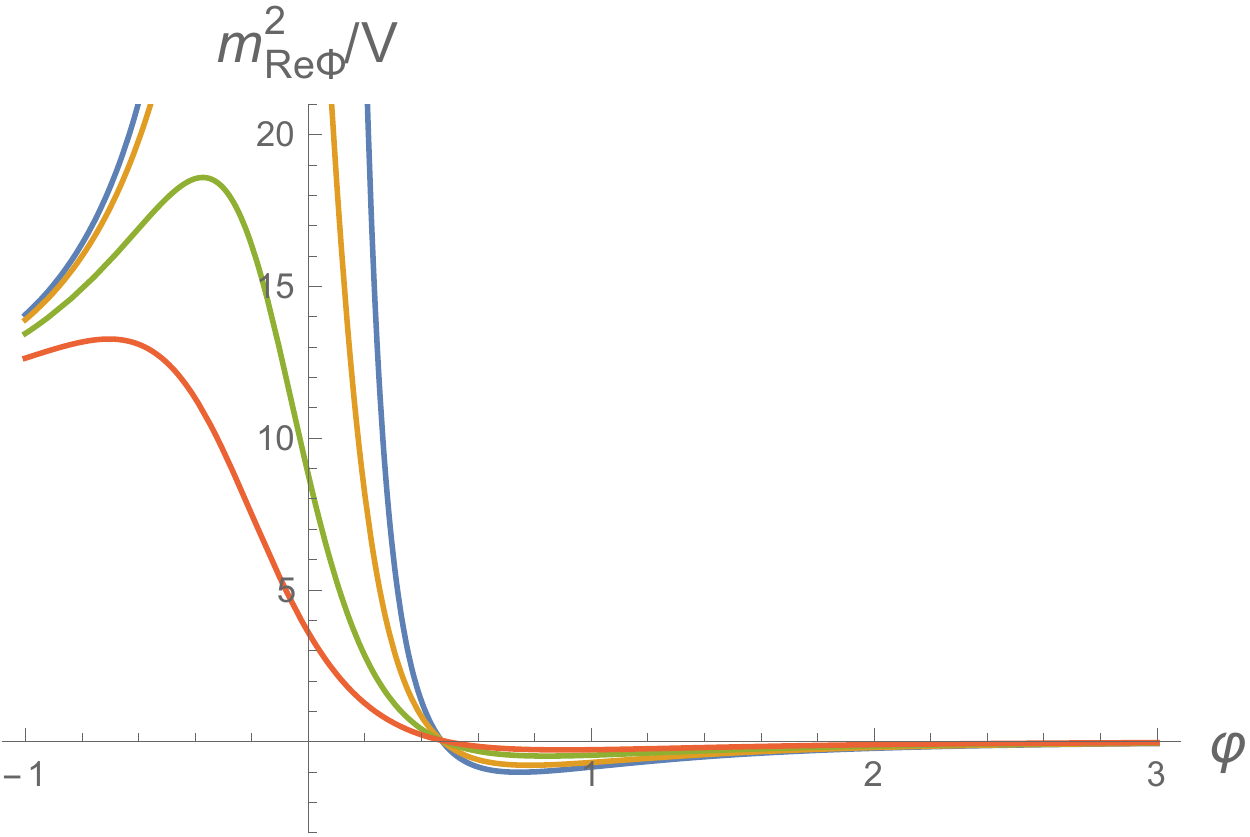}
\includegraphics[width=7.5cm]{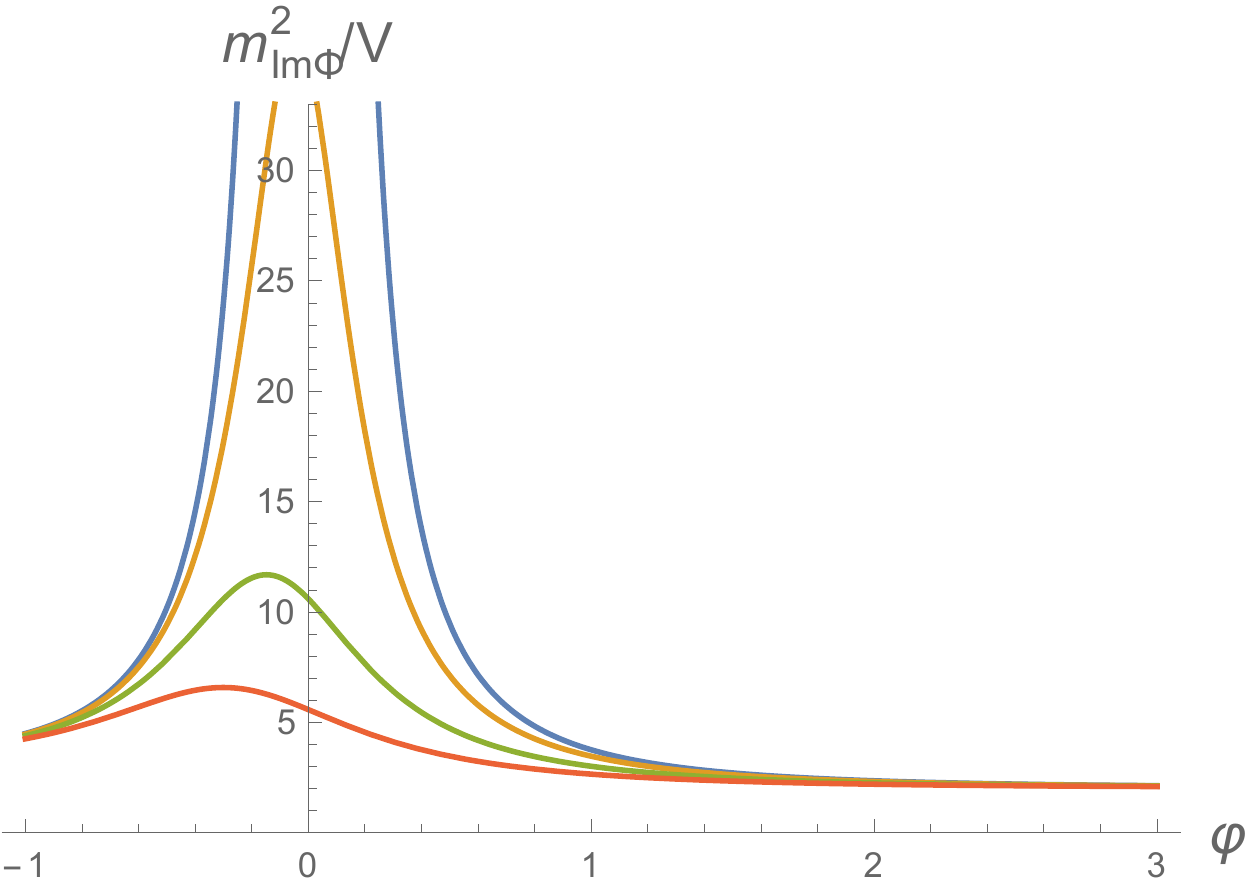}
\vspace*{0.5cm}
\caption{\it Masses of the real and imaginary part of the field $\Phi$ for the model defined by Eq.~\eqref{Example} with $\alpha=1$ and uplifting equal to $\Lambda=\{0, 0.1, 0.3, 0.5\}$. Both scalar parts are massive at the minimum. During inflation, at large values the $\varphi$, the mass of $\Re\Phi$ goes to zero while the mass of $\Im\Phi$ approaches a constant value as defined by Eq.~\eqref{MassImaginary}.}\label{Masses}
\end{center}
\vspace{-0.cm}
\end{figure}

\subsection{Uplifting geometric  $\alpha$-attractors} \label{GeoUp}

The appealing property of the original formulation of $\alpha$-attractors, as discovered in \cite{Kallosh:2013hoa,Kallosh:2013yoa,Kallosh:2014rga}, is the unique relation between the \Kahler geometry and the observational predictions \eqref{nsrattractors}. In particular, the logarithmic \Kahler potential fixes the spectral tilt while its constant curvature
\be\label{Kcurv}
R_K=-\frac{2}{3\alpha}\,,
\ee
determines the amount of primordial gravitational waves. However, these original models require always the presence of a second superfield.

Single superfield geometric formulations have been discovered in \cite{Roest:2015qya,Linde:2015uga}. As shown in \cite{Roest:2015qya}, they originate from a natural deformation of the well-known no-scale constructions\footnote{No scale models, as originally proposed in \cite{Cremmer:1983bf,Ellis:1983sf}, represent a good starting point in order to produce consistent inflationary dynamics (see e.g. \cite{Cecotti:1987sa,Ellis:2013xoa,Ellis:2013nxa,Lahanas:2015jwa,Ellis:2015kqa,Ellis:2015pla}). However, the geometric models of this section emerge from a different construction which naturally leads to stable de Sitter solutions and have scale depending on the parameter $\alpha$ (see \cite{Roest:2015qya} for explicit derivation). The no scale symmetry is intimately related to a specific value of the \Kahler curvature \eqref{Kcurv} and it is restored just in the limit $\alpha\rightarrow 1$.}  and they are defined by 
\be\label{alphascaleKW}
K=-3 \alpha \ln\left(\Phi+\bar\Phi\right) \,, \qquad W=\Phi^{n_-} - \Phi^{n_+} F(\Phi)\,,
\ee
with power coefficients equal to
\be\label{npm}
n_\pm = \frac{3}{2} \left(\alpha \pm \sqrt{\alpha}\right)\,,
\ee
and $F$ having general expansion $F(\Phi)=\sum_n c_n \Phi^n$ which encodes the attractor nature of these scenarios.

This class gives rise to the flat $\alpha$-attractors of the previous section in the limit $\alpha\rightarrow\infty$ and, then, when the curvature becomes flat,  as shown in \cite{Roest:2015qya}. The procedure is the following: one performs a field redefinition such as $\Phi \rightarrow \exp(-2 \Phi / \sqrt{3 \alpha})$, an appropriate \Kahler transformation and, in the singular limit, one obtains canonical and shift-symmetric $K$ and $W$ equal to \eqref{fflat}, with F constant. On top of this, one adds exponential corrections which returns the desired inflationary behavior.

In order to uplift the SUSY Minkoswki minimum of these scenarios, one can add a nilpotent field which breaks supersymmetry and yields a non-zero cosmological constant. The geometric analogous of the flat case, discussed in the previous section, is given by

\be\label{GeoUp}
\begin{aligned}
K=-3 \alpha \ln\left(\Phi+\bar\Phi\right) +S\bar{S}\,, \qquad W=\Phi^{\frac{3}{2}\alpha}\left[f(\Phi)+g(\Phi) S \right]\,.
\end{aligned}
\ee
In fact, along the real axis $\Phi=\bar{\Phi}$ and at $S=0$, this supergravity model yields a scalar potential
\be
V=8^{-\alpha}\left[g(\Phi)^2-3f(\Phi)^2+\frac{4 \Phi^2 f'(\Phi)^2}{3\alpha}\right]\,,
\ee
which, when expressed in terms of the canonical field $\varphi=-\sqrt{3\alpha/2}\ln \Phi$, coincides with the one obtained in the flat case Eq.~\eqref{pot}, up to an overall constant factor. Furthermore, Eq.~\eqref{GeoUp} reduces to Eq.~\eqref{K&W} in the flat singular limit. The \Kahler potential \eqref{GeoUp} parameterizes a manifold $SU(2,1)/U(1)\times U(1)$ and  related analysis with similar settings are performed in \cite{Lahanas:2015jwa,Carrasco:2015pla}.

The correspondence between the scalar potentials of the flat and the geometric construction (for the single superfield case it was proven in \cite{Roest:2015qya}) is remarkable as it allows to identically assume the whole set of results, from Eq.~\eqref{f&gflat} to Eq.~\eqref{MassImaginary}, found and described in the previous section, provided one identifies
\be
x\equiv\Phi\,.
\ee
The functions $f$ and $g$ can be assumed to have generic expansion \eqref{f&gflat} and the inflationary behavior will be of the form \eqref{falloff}. However, in this case, the fall-off will be governed by the curvature of the \Kahler manifold which depends on the parameter $\alpha$. The minimum, placed at $\Phi=1$, provided
\be
f'(1)=g'(1)=0\,,
\ee
will have uplifting equal to \eqref{CC}, gravitino mass and SUSY breaking scale given by \eqref{M32M} and, again, supersymmetry broken just in the $S$ direction, as given by
\be
D_S W_{min} = g(1) = M\,, \qquad D_{\Phi}W_{min}=0,.
\ee

Remarkably, the condition on the stability of the inflationary trajectory turns out to be the same of the previous section. At large value of the canonical field $\varphi$, the mass of $\Im\Phi$ is positive when Eq.~\eqref{massIm} is satisfied, independently of the value of $\alpha$. This represents a considerable improvement  with respect to the single superfield case defined by \eqref{alphascaleKW} which is stable just for $\alpha>1$ \cite{Roest:2015qya}. Furthermore, the mass of $\Im\Phi$ approaches a constant value during inflation as given by \eqref{MassImaginary}, up to an overall constant.

\subsection{General inflaton potential from curved \Kahler geometry}

We have so far developed a general framework in order to obtain inflation together with controllable level of uplifting and SUSY breaking at the minimum when the \Kahler geometry is curved and defined by Eq.~\eqref{GeoUp}. We have proven that generic expansion of $f$ and $g$ gives rise to $\alpha$-attractors with cosmological predictions extremely stable. 

On the other hand, also in this context, it is possible to make the specific choice \eqref{g&f} and consider the geometric analogous of the class of models introduced in \cite{Dall'Agata:2014oka,Kallosh:2014hxa} and reviewed in Sec.~\ref{sectionKLS}. Then, the \Kahler potential and the superpotential read
\be
\begin{aligned}
K=-3 \alpha \ln\left(\Phi+\bar\Phi\right) +S\bar{S}\,, \qquad W=\Phi^{\frac{3}{2}\alpha} f(\Phi)\left(1+\frac{S}{\beta} \right)\,.
\end{aligned}
\ee
The choice $\beta=1/\sqrt{3}$ gives rise to a scalar potential with a Minkowki minimum. Along $\Phi=\bar{\Phi}$ and $S=0$, one has (up to an overall constant factor)
\be
V=\frac{2 }{3\alpha}\Phi^2 f'(\Phi)^2\,,
\ee
which, in terms of the canonical scalar field $\varphi$ reads
\be
V= f'\left(\e^{-\sqrt{\tfrac{2}{3\alpha}}\varphi}\right)^2\,,
\ee
where primes denote derivatives with respect to the variables the function depends on. Then, one can implement an arbitrary inflaton potential, independently of the value of the \Kahler curvature which is parametrised by $\alpha$. Related results for the case $\alpha=1$ were obtained in \cite{Lahanas:2015jwa}. In the case of a flat \Kahler geometry the works \cite{Kallosh:2010xz,Dall'Agata:2014oka,Kallosh:2014hxa} developed analogous constructions.

Within this setup, one can implement even a quadratic potential $V=\tfrac{1}{2}m^2\varphi^2$ by choosing
\be
f(\Phi)=\frac{3\alpha\ m}{4\sqrt{2}}\ln^2(\Phi)\,.
\ee

The properties at the minimum remain the same as in the flat case of Sec.~\ref{sectionKLS}. Then, a small deviation of $\beta$ from the value $1/\sqrt{3}$ yields the desirable tiny uplifting which reproduces the current acceleration of the Universe.

\subsection{Discussion}\label{disc}

In this Section, we have provided evidences for the special role that  $\alpha$-attractors would play in the  cosmological evolution of the Universe. In the simple supergravity framework consisting of two sectors (one containing the inflaton and the other controlling the landscape of possible vacua), any arbitrary expansion of the superpotential would yield automatically such inflationary scenarios. We have obtained these results both in the case of a flat \Kahler geometry, as given by Eq.~\eqref{K&W}, and in the case of the logarithmic \Kahler as defined by Eq.~\eqref{GeoUp} where the geometric properties of the \Kahler manifold determines the observational predictions. In this latter case, the overall factor $\Phi^{\frac{3}{2}\alpha}$ in $W$ can be removed by means of an appropriate \Kahler transformation (this choice makes the shift symmetry of the canonical inflaton $\varphi$ manifest even in the case of a logarithmic \Kahler potential, as pointed out in \cite{Carrasco:2015uma}). However, one would lose immediate contact with string theory scenarios as the form of $K$ would change consequently. In this respect, polynomial contributions to the superpotential, typically arising from flux compactification, would be possible if 
\be
\alpha=\frac{2}{3}n
\ee
with $n$ integer. In particular, the simple choice $n=1$ would give
\be
\begin{aligned}
K&=-2\ln\left(\Phi+\bar\Phi\right) +S\bar{S}\,,\\
 W&=\left(a_0\Phi + a_1\Phi^2 +...\right)+ \left(b_0\Phi + b_1\Phi^2 +...\right) S \,,
\end{aligned}
\ee
where dots stand for higher order terms in $\Phi$ (see \cite{Dudas:2015lga} for a recent analysis of this class of models in the context of supplementary moduli breaking supersymmetry). Then, the minimal addition of a nilpotent sector with canonical $K$ to the class proposed in \cite{Roest:2015qya} leads to a simplification of the original superpotential \eqref{alphascaleKW} and enhancement of stability of the inflationary trajectory, which now occurs for any value of $\alpha$ (see \cite{Carrasco:2015uma} for a discussion on the connection between curvature and stabilization).

We have shown that  cosmological $\alpha$-attractors are absolutely insensitive with respect to any value of the cosmological constant and to the coupling between $\Phi$ and $S$. The plateau and the fall-off turn out to be extremely stable with respect to generic deformations of the superpotential (similar stability can be observed in some examples of \cite{Kallosh:2015lwa}). These scenarios would arise naturally in any possible Universe, independently of the amount of dark energy. In this regard, cosmological attractors seem to be fundamentally compatible with the idea of Multiverse and landscape of vacua.

\clearpage
\thispagestyle{empty}


\chapter{Conclusions}
\label{chapter: conclusions}

\begingroup
\begin{flushright}
\vspace{.5cm}
\end{flushright}
\begin{quote}
{\it We conclude this thesis by discussing the main results obtained and providing an outlook on future perspectives and possible developments.}
\end{quote}

\endgroup

\newpage

\section{Overview of the results}

The main focus of this thesis has been inflation and its intricate connections to UV-physics scenarios. This primordial cosmological phase allows us to effectively probe energies well beyond what any particle accelerator could ever achieve. It provides a great physics arena where to investigate and eventually test our most speculative theories. 
%

We have been particularly interested in extracting generic features of the inflationary mechanism, beyond the specific details of the model at hand. This is important as CMB data allows us to probe just a very limited region of the inflationary trajectory (we have explained this in Ch.~\ref{chapter:Inflation} and in Ch.~\ref{chapter:Universality}). Specifically, we have investigated some of those aspects which are of utmost relevance for a proper embedding of the inflationary paradigm into a complete framework of UV-physics. The results have turned out to be very exciting:

\begin{itemize}

\item On the one hand, in Ch.~\ref{chapter:Universality} we have shown that focusing on {\it universality} properties of inflation can yield surprisingly stringent bounds on its dynamics. First of all, a large set of inflationary models can be organized in universality classes depending on their observational predictions. This is simply a direct consequence of the small sensitivity cosmological experiments have over the whole inflationary trajectory. Secondly, we have investigated the regime where this ignorance becomes of crucial importance for a variable depending on the entire expansion period, such as the inflationary field range $\Delta \phi$. While we have provided explicit examples where a measurement of $n_s$ and $r$ correspond to a wide spectrum of values $\Delta\phi>M_{Pl}$ (see \cite{Garcia-Bellido:2014eva} and Sec.~\ref{SecDegeneracy}), we have also demonstrated that the inflaton range generically exhibits universal behaviour in the sub-Planckian region (see\cite{Garcia-Bellido:2014wfa} and Sec.~\ref{SecUniversalityRange}). In other words, given a point in the $(n_s , r)$ plane, there is a unique estimate for $\Delta\phi<M_{Pl}$. It is then very remarkable that knowledge about a small portion of the inflationary trajectory turns out to be enough in order to infer basic properties of a region not accessible via CMB measurements. The inflaton excursion becomes a function of both the tensor-to-scalar ratio $r$ and the spectral index $n_s$.  This novel and universal dependence has allowed us to strengthen the usual Lyth bound of two order of magnitudes. One has sub-Planckian field ranges and can safely work within an EFT of inflation just for very small values $r\lesssim 2\cdot10^{-5}$.

\item On the other hand, one would like to find a satisfying theoretical underpinning to explaining why the spectral index $n_s$ and the tensor-to-scalar ratio $r$ take the values they do. In Ch.~\ref{chapter:supergravity},  we have proven that non-trivial \Kahler geometries, typically arising in string theory compactifications, unequivocally determine such observables. This is in a nutshell the upshot of $\a$-attractors: the parameter $\a$ controls the curvature of the moduli space and, with it, the amount of primordial gravitational waves; the value of the scalar tilt tends automatically to the ``sweet spot'' of Planck, no matter the details of the superpotential. The \Kahler geometry basically induces an {\it attractor} for observations. We have then taken our investigation in the direction of proving the generality of this attractor mechanism, independently of the other fields involved. This definitively represents an important step towards a consistent string theory realization where many moduli appear naturally. The fundamental nature of this phenomenon appears indeed to be related to the novel $\a$-scale supergravity construction which we originally discovered in \cite{Roest:2015qya} and, then, presented in Sec.~\ref{SECalphascale}. We have thus proved that the attractor mechanism is independent of the specific direction of SUSY breaking but rather related to the \Kahler structure of the inflaton sector. 

\item Finally, the possibility of obtaining a pure de Sitter phase in supergravity by means of a sole nilpotent superfield \cite{Bergshoeff:2015tra,Hasegawa:2015bza,Ferrara:2015gta,Kuzenko:2015yxa,Kallosh:2015tea,Schillo:2015ssx} and its relations to D-brane physics \cite{Kallosh:2014wsa,Bergshoeff:2015jxa,Bandos:2015xnf} acquires particular relevance in the light of constructing a unified framework for inflation and dark energy. We have indeed proved that coupling the inflaton sector with a nilpotent superfield has very striking implications. First of all, it allows to evade the restrictions imposed by the no-go theorem of \cite{Kallosh:2014oja} on the possibility of uplifting a SUSY Minkowski minimum (see \cite{Linde:2014ela} and Sec.~\ref{sectionLRS} for a detailed investigation). Secondly, it greatly simplifies the inflationary dynamics and allows for remarkable phenomenological flexibility (see \cite{Kallosh:2014hxa} and Sec.~\ref{sectionKLS}). However, the greatest surprise actually appears again in the context of $\a$-attractors: in \cite{Scalisi:2015qga} and in Sec.~\ref{sectionScalisi}, we have indeed shown how such a coupling leads to an enhancement of the attractor nature of the theory.

\end{itemize}

\section{Outlook}

The inflationary paradigm has turned to be one of the best and most concrete physical scenarios providing a wonderful testing ground for ideas in quantum gravity. The results of this thesis have proven that there is a remarkable interplay between actual cosmological predictions and deep theoretical aspects characterizing a complete framework of UV-physics.

\begin{figure}[htb]
\hspace{-3mm}
\begin{center}
\includegraphics[width=5.5cm]{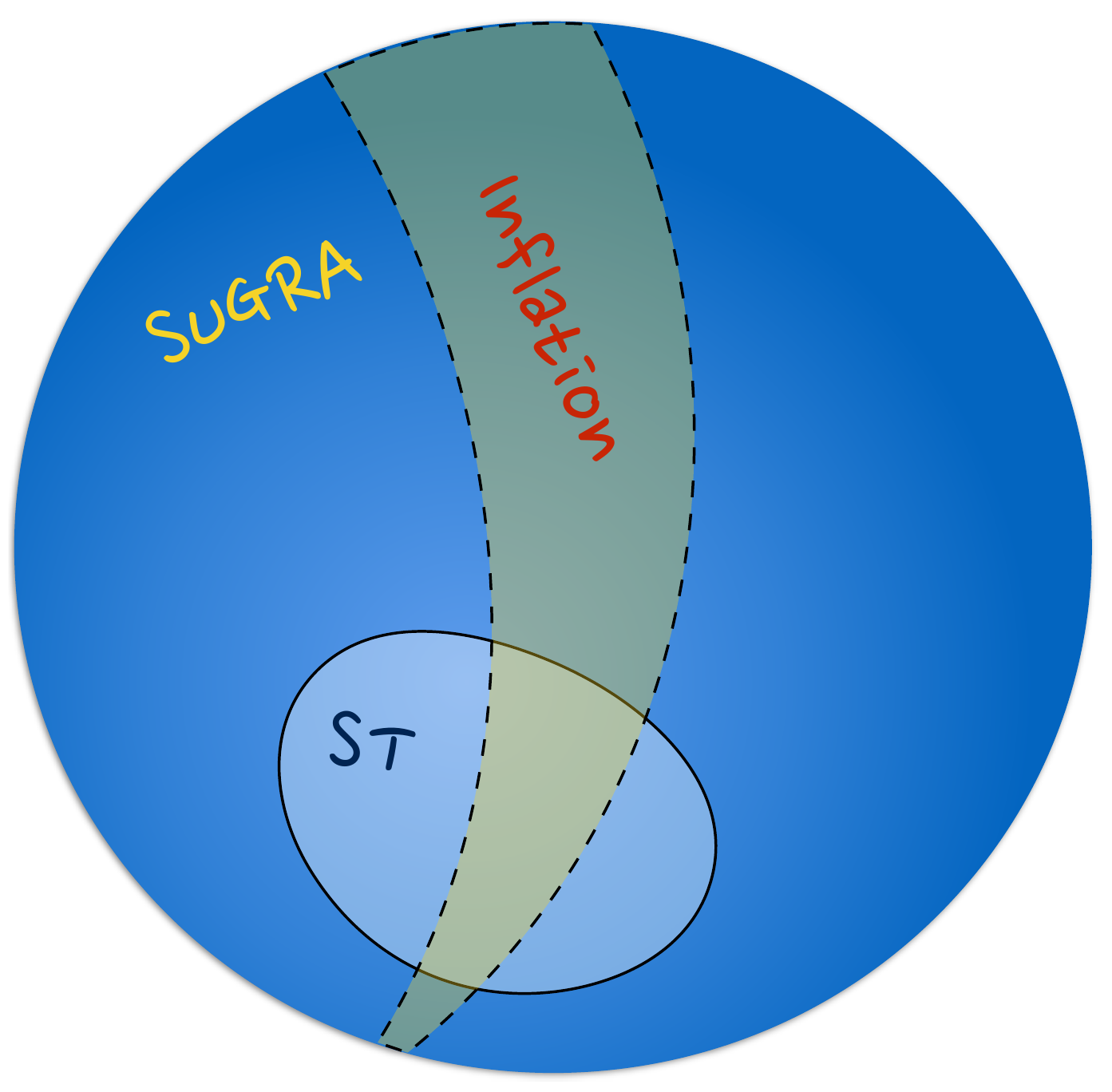}
\vspace*{0.4cm}
\caption{\it The subset of the suitable supergravity constructions for inflation (yellow slice) intersects the subset of the effective supergravities arising from string theory compactifications (labeled by ST). This provides a very useful guidance for the implementation of the inflationary paradigm in string theory.}\label{SUGRAST}
\end{center}
\vspace{-0.cm}
\end{figure}

In particular, our investigation has focused on the generic properties of inflation when this is embedded in a supergravity setting. Then, one may wonder whether this study is enough in order to constrain the physics of inflation in string theory, being this the ultimate goal. We have been taking the following approach. On the one hand, it is certainly true that consistent string theory compactifications will form just a subset of the full spectrum of supergravity possibilites (e.g. when these are expressed in terms of $K$ and $W$). On the other hand, not every supergravity theory will generically allow for a consistent inflationary dynamics. In fact, we have provided compelling evidence of the restrictions the internal \Kahler geometry and SUSY breaking directions may lead on the viability of inflation. Therefore, the subset of supergravity models suitable for inflation will intersect the subset of the proper string theory compactifications, thus yielding important guidance for a consistent cosmological construction within this ultimate theory of Nature. We have graphically condensed this discussion in Fig.~\ref{SUGRAST}.

The discovery that hyperbolic \Kahler geometries make the particular details of the specific model insensitive to the final phenomenological result, thus inducing an attractor, is rather intriguing. Yet more remarkable is the fact that the cosmological predictions are given by
\be\label{nsrattractorsconclusion}
n_s=1-\frac{2}{N}\,, \qquad r=\frac{12 \alpha}{N^2}\,,
\ee
in terms of the number of e-folds $N$, thus being at the center of the Planck dome \cite{Planck:2015xua,Ade:2015lrj}. This belongs to a region in the $(n_s,r)$ plane which seems to be very appealing from the theoretical viewpoint. Higher order terms of the Ricci scalar in a pure gravitational Lagrangian, such as the Starobinsky model \cite{Starobinsky:1980te} and its supergravity realizations \cite{Cecotti:1987sa, Ellis:2013xoa,Kallosh:2013lkr,Buchmuller:2013zfa,Farakos:2013cqa,Ellis:2013nxa}, lead to \eqref{nsrattractors} with $\alpha=1$ (by including an auxiliary vector field one can vary the value of $\alpha$ \cite{Ozkan:2015iva}). Further, models with non-minimal couplings, such as Higgs inflation \cite{Bezrukov:2007ep} and the universal attractor model \cite{Kallosh:2013tua}, yield identical observational predictions. Interestingly, the peculiarity of such a region translates into a common denominator being a pole of order two in the kinetic term of the inflaton \cite{Galante:2014ifa}. Finally, investigations on the excursion of the inflaton field reveal a change of its behavior just around the region defined by \eqref{nsrattractors} (as shown \cite{Garcia-Bellido:2014eva,Garcia-Bellido:2014wfa} and in Ch.~\ref{chapter:Universality}).

A consistent realization of the attractor mechanism within string theory still awaits to be discovered. Several delicate issues must be tackled in order to have full control of the model, once embedded in this rich physical framework. These problems include the divergence of the kinetic term inside the moduli space, the correct identification of the inflationary modulus, the interactions of the latter with the other moduli (see e.g. the recent works \cite{Kallosh:2016ndd,Kallosh:2016gqp}) and others.  Nevertheless, certain approximate stringy incarnations already exist \cite{Cicoli:2008gp,Broy:2015zba,Burgess:2016owb}. These definitively raise very good hopes for successfully reaching this important goal. We look forward to facing this exciting challenge in the very near future.

\clearpage
\thispagestyle{empty}


\renewcommand{\chapterheadstartvskip}{\vspace*{1\baselineskip}}

\chapter*{List of Publications}
\label{chapter:Publications}
\addcontentsline{toc}{chapter}{List of Publications}
\markboth{List of Publications}{List of Publications}
\begin{itemize}

\item[{[}{\sc i}{]}] M.~Arzano, G.~Calcagni, D.~Oriti and M.~Scalisi,\\ ``{\it Fractional and noncommutative spacetimes}',' \\
       \href{http://dx.doi.org/10.1103/PhysRevD.84.125002}{Phys.\ Rev.\ D {\bf 84} (2011) 125002},
       \href{http://arxiv.org/abs/arXiv:1107.5308}{{\ttfamily arXiv:1107.5308}}. \\
       
\item[{[}{\sc ii}{]}] G.~Calcagni, G.~Nardelli and M.~Scalisi,\\ ``{\it Quantum mechanics in fractional and other anomalous spacetimes}',' \\
       \href{http://dx.doi.org/10.1063/1.4757647}{ J.\ Math.\ Phys.\  {\bf 53} (2012) 102110},
       \href{http://arxiv.org/abs/arXiv:1207.4473}{{\ttfamily arXiv:1207.4473}}. \\

\item[{[}{\sc iii}{]}]  D.~Roest, M.~Scalisi and I.~Zavala,\\ ``{\it \Kahler potentials for Planck inflation}'', \\
       \href{http://dx.doi.org/10.1088/1475-7516/2013/11/007}{ JCAP {\bf 1311} (2013) 007},
       \href{http://arxiv.org/abs/arXiv:1307.4343}{{\ttfamily arXiv:1307.4343}}. \\

\item[{[}{\sc iv}{]}]  J.~Garcia-Bellido, D.~Roest, M.~Scalisi and I.~Zavala,\\ ``{\it Can CMB data constrain the inflationary field range?}'', \\
       \href{http://dx.doi.org/10.1088/1475-7516/2014/09/006}{ JCAP {\bf 1409} (2014) 006},
       \href{http://arxiv.org/abs/arXiv:1405.7399}{{\ttfamily arXiv:1405.7399}}. \\

\item[{[}{\sc v}{]}]   J.~Garcia-Bellido, D.~Roest, M.~Scalisi and I.~Zavala,\\ ``{\it Lyth bound of inflation with a tilt}'', \\
       \href{http://dx.doi.org/10.1103/PhysRevD.90.123539}{Phys.\ Rev.\ D {\bf 90} (2014) no.12,  123539},
       \href{http://arxiv.org/abs/arXiv:1408.6839}{{\ttfamily arXiv:1408.6839}}. \\

      \item[{[}{\sc vi}{]}]   R.~Kallosh, A.~Linde and M.~Scalisi,\\ ``{\it Inflation, de Sitter Landscape and Super-Higgs effect}'', \\
       \href{http://dx.doi.org/10.1007/JHEP03(2015)111}{ JHEP {\bf 1503} (2015) 111},
       \href{http://arxiv.org/abs/arXiv:1411.5671}{{\ttfamily arXiv:1411.5671}}. \\

\item[{[}{\sc vii}{]}]   A.~Linde, D.~Roest and M.~Scalisi,\\ ``{\it Inflation and Dark Energy with a Single Superfield}'', \\
       \href{http://dx.doi.org/10.1088/1475-7516/2015/03/017}{JCAP {\bf 1503} (2015) 017},
       \href{http://arxiv.org/abs/arXiv:1412.2790}{{\ttfamily arXiv:1412.2790}}. \\               
  
\item[{[}{\sc viii}{]}]   D.~Roest and M.~Scalisi,\\ ``{\it Cosmological attractors from $\alpha$-scale supergravity}'', \\
       \href{http://dx.doi.org/10.1103/PhysRevD.92.043525}{Phys.\ Rev.\ D {\bf 92} (2015) 043525},
       \href{http://arxiv.org/abs/arXiv:1503.07909}{{\ttfamily arXiv:1503.07909}}. \\            
       
     \item[{[}{\sc ix}{]}]  M.~Scalisi,\\ ``{\it Cosmological $\alpha$-attractors and de Sitter landscape}'', \\
       \href{http://dx.doi.org/10.1007/JHEP12(2015)134}{JHEP {\bf 1512} (2015) 134},
       \href{http://arxiv.org/abs/arXiv:1506.01368}{{\ttfamily arXiv:1506.01368}}. \\   
       
   \item[{[}{\sc x}{]}] D.~Roest and  M.~Scalisi,\\ ``{\it Inflation: observations and attractors}'', \\
  Contribution to Springer Proceedings in Physics (Book 176), ``Theoretical Frontiers in Black Holes and Cosmology'',
      Springer; 1st ed. 2016 edition,  \href{http://www.springer.com/in/book/9783319313511}{ISBN 978-3-319-31351-1},

\end{itemize}

\newpage
\thispagestyle{empty}

{\small
\bibliography{refs}

\providecommand{\href}[2]{#2}\begingroup\raggedright\begin{thebibliography}{100}

\bibitem{Wilson:1971bg}
K.~G. Wilson, ``{Renormalization group and critical phenomena. 1.
  Renormalization group and the Kadanoff scaling picture}'',
\href{http://dx.doi.org/10.1103/PhysRevB.4.3174}{{\em Phys. Rev.} {\bf B4}
  (1971)  3174--3183}.

\bibitem{Wilson:1971dh}
K.~G. Wilson, ``{Renormalization group and critical phenomena. 2. Phase space
  cell analysis of critical behavior}'',
\href{http://dx.doi.org/10.1103/PhysRevB.4.3184}{{\em Phys. Rev.} {\bf B4}
  (1971)  3184--3205}.

\bibitem{Oriti:2009zz}
D.~Oriti, {\em {Approaches to quantum gravity: Toward a new understanding of
  space, time and matter}}.
\newblock Cambridge University Press, 2009.

\bibitem{Kiefer:2012boa}
C.~Kiefer, {\em {Quantum gravity}}.
\newblock Oxford Univ. Pr., Oxford, UK, 2012.

\bibitem{Green:1987sp}
M.~B. Green, J.~H. Schwarz, and E.~Witten, {\em {Superstring Theory. Vol. 1:
  Introduction}}.
\newblock
1988.
\newblock

\bibitem{Green:1987mn}
M.~B. Green, J.~H. Schwarz, and E.~Witten, {\em {Superstring Theory. Vol. 2:
  Loop Amplitudes, Anomalies and Phenomenology}}.
\newblock
1988.
\newblock

\bibitem{Polchinski:1998rq}
J.~Polchinski, {\em {String theory. Vol. 1: An introduction to the bosonic
  string}}.
\newblock Cambridge University Press,
2007.
\newblock

\bibitem{Polchinski:1998rr}
J.~Polchinski, {\em {String theory. Vol. 2: Superstring theory and beyond}}.
\newblock Cambridge University Press,
2007.
\newblock

\bibitem{Guth:1980zm}
A.~H. Guth, ``{The Inflationary Universe: A Possible Solution to the Horizon
  and Flatness Problems}'',
\href{http://dx.doi.org/10.1103/PhysRevD.23.347}{{\em Phys. Rev.} {\bf D23}
  (1981)  347--356}.

\bibitem{Linde:1981mu}
A.~D. Linde, ``{A New Inflationary Universe Scenario: A Possible Solution of
  the Horizon, Flatness, Homogeneity, Isotropy and Primordial Monopole
  Problems}'',
\href{http://dx.doi.org/10.1016/0370-2693(82)91219-9}{{\em Phys. Lett.} {\bf
  B108} (1982)  389--393}.

\bibitem{Albrecht:1982wi}
A.~Albrecht and P.~J. Steinhardt, ``{Cosmology for Grand Unified Theories with
  Radiatively Induced Symmetry Breaking}'',
\href{http://dx.doi.org/10.1103/PhysRevLett.48.1220}{{\em Phys. Rev. Lett.}
  {\bf 48} (1982)  1220--1223}.

\bibitem{Adam:2015rua}
{\bf Planck} Collaboration, R.~Adam {\em et al.}, ``{Planck 2015 results. I.
  Overview of products and scientific results}'',
\href{http://arxiv.org/abs/1502.01582}{{\tt arXiv:1502.01582 [astro-ph.CO]}}.

\bibitem{Planck:2015xua}
{\bf Planck} Collaboration, P.~Ade {\em et al.}, ``{Planck 2015 results. XIII.
  Cosmological parameters}'',
\href{http://arxiv.org/abs/1502.01589}{{\tt arXiv:1502.01589 [astro-ph.CO]}}.

\bibitem{Ade:2015lrj}
{\bf Planck} Collaboration, P.~Ade {\em et al.}, ``{Planck 2015. XX.
  Constraints on inflation}'',
\href{http://arxiv.org/abs/1502.02114}{{\tt arXiv:1502.02114 [astro-ph.CO]}}.

\bibitem{Ahn:2013gms}
{\bf SDSS} Collaboration, C.~P. Ahn {\em et al.}, ``{The Tenth Data Release of
  the Sloan Digital Sky Survey: First Spectroscopic Data from the SDSS-III
  Apache Point Observatory Galactic Evolution Experiment}'',
  \href{http://dx.doi.org/10.1088/0067-0049/211/2/17}{{\em Astrophys. J.
  Suppl.} {\bf 211} (2014)  17},
\href{http://arxiv.org/abs/1307.7735}{{\tt arXiv:1307.7735 [astro-ph.IM]}}.

\bibitem{Betoule:2014frx}
{\bf SDSS} Collaboration, M.~Betoule {\em et al.}, ``{Improved cosmological
  constraints from a joint analysis of the SDSS-II and SNLS supernova
  samples}'', \href{http://dx.doi.org/10.1051/0004-6361/201423413}{{\em Astron.
  Astrophys.} {\bf 568} (2014)  A22},
\href{http://arxiv.org/abs/1401.4064}{{\tt arXiv:1401.4064 [astro-ph.CO]}}.

\bibitem{Keck}
\url{https://www.cfa.harvard.edu/CMB/keckarray}.

\bibitem{Bicep3}
\url{https://www.cfa.harvard.edu/CMB/bicep3}.

\bibitem{Polar}
\url{http://bolo.berkeley.edu/polarbear},.

\bibitem{Abbott:2016blz}
{\bf Virgo, LIGO Scientific} Collaboration, B.~P. Abbott {\em et al.},
  ``{Observation of Gravitational Waves from a Binary Black Hole Merger}'',
  \href{http://dx.doi.org/10.1103/PhysRevLett.116.061102}{{\em Phys. Rev.
  Lett.} {\bf 116} (2016) no.~6, 061102},
\href{http://arxiv.org/abs/1602.03837}{{\tt arXiv:1602.03837 [gr-qc]}}.

\bibitem{Baumann:2014nda}
D.~Baumann and L.~McAllister, {\em {Inflation and String Theory}}.
\newblock Cambridge University Press, 2015.
\newblock
\href{http://arxiv.org/abs/1404.2601}{{\tt arXiv:1404.2601 [hep-th]}}.
\newblock

\bibitem{Dodelson:2003ft}
S.~Dodelson, {\em {Modern Cosmology}}.
\newblock Academic Press, Amsterdam,
2003.
\newblock

\bibitem{Mukhanov:2005sc}
V.~Mukhanov, {\em {Physical Foundations of Cosmology}}.
\newblock Cambridge University Press, Oxford,
2005.
\newblock

\bibitem{Lyth:2009zz}
D.~H. Lyth and A.~R. Liddle, {\em {The primordial density perturbation:
  Cosmology, inflation and the origin of structure}}.
\newblock
2009.
\newblock

\bibitem{Baumann}
D.~Baumann, ``Part {III}: {C}osmology''.
\newblock \url{http://www.damtp.cam.ac.uk/user/db275/Cosmology.pdf}.

\bibitem{Calcagni}
G.~Calcagni, {\em {Classical and Quantum Cosmology}}.
\newblock Springer-Verlag, Berlin, (2016 in press).

\bibitem{Hubble:1929ig}
E.~Hubble, ``{A relation between distance and radial velocity among
  extra-galactic nebulae}'',
\href{http://dx.doi.org/10.1073/pnas.15.3.168}{{\em Proc. Nat. Acad. Sci.} {\bf
  15} (1929)  168--173}.

\bibitem{Wright}
E.~L. Wright, ``Cosmology lecture notes'', 2015.
\newblock \url{http://www.astro.ucla.edu/~wright/A275.pdf}.

\bibitem{Riess:1994nx}
A.~G. Riess, W.~H. Press, and R.~P. Kirshner, ``{Using SN-Ia light curve shapes
  to measure the Hubble constant}'',
  \href{http://dx.doi.org/10.1086/187704}{{\em Astrophys. J.} {\bf 438} (1995)
  L17--20},
\href{http://arxiv.org/abs/astro-ph/9410054}{{\tt arXiv:astro-ph/9410054
  [astro-ph]}}.

\bibitem{Riess:1996pa}
A.~G. Riess, W.~H. Press, and R.~P. Kirshner, ``{A Precise distance indicator:
  Type Ia supernova multicolor light curve shapes}'',
  \href{http://dx.doi.org/10.1086/178129}{{\em Astrophys. J.} {\bf 473} (1996)
  88},
\href{http://arxiv.org/abs/astro-ph/9604143}{{\tt arXiv:astro-ph/9604143
  [astro-ph]}}.

\bibitem{Kirshner}
R.~P. Kirshner, ``{Hubble’s diagram and cosmic expansion}'',
  \href{http://dx.doi.org/10.1073/pnas.2536799100}{{\em Proc. Nat. Acad. Sci.}
  {\bf 101} (2004)  8--13}.

\bibitem{Einstein:1917ce}
A.~Einstein, ``{Cosmological Considerations in the General Theory of
  Relativity}'',
{\em Sitzungsber. Preuss. Akad. Wiss. Berlin (Math. Phys.)} {\bf 1917} (1917)
  142--152.

\bibitem{Cole:2005sx}
{\bf 2dFGRS} Collaboration, S.~Cole {\em et al.}, ``{The 2dF Galaxy Redshift
  Survey: Power-spectrum analysis of the final dataset and cosmological
  implications}'',
  \href{http://dx.doi.org/10.1111/j.1365-2966.2005.09318.x}{{\em Mon. Not. Roy.
  Astron. Soc.} {\bf 362} (2005)  505--534},
\href{http://arxiv.org/abs/astro-ph/0501174}{{\tt arXiv:astro-ph/0501174
  [astro-ph]}}.

\bibitem{Preskill:1979zi}
J.~Preskill, ``{Cosmological Production of Superheavy Magnetic Monopoles}'',
\href{http://dx.doi.org/10.1103/PhysRevLett.43.1365}{{\em Phys. Rev. Lett.}
  {\bf 43} (1979)  1365}.

\bibitem{Guth:2007ng}
A.~H. Guth, ``{Eternal inflation and its implications}'',
  \href{http://dx.doi.org/10.1088/1751-8113/40/25/S25}{{\em J. Phys.} {\bf A40}
  (2007)  6811--6826},
\href{http://arxiv.org/abs/hep-th/0702178}{{\tt arXiv:hep-th/0702178
  [HEP-TH]}}.

\bibitem{Schwarz:2001vv}
D.~J. Schwarz, C.~A. Terrero-Escalante, and A.~A. Garcia, ``{Higher order
  corrections to primordial spectra from cosmological inflation}'',
  \href{http://dx.doi.org/10.1016/S0370-2693(01)01036-X}{{\em Phys. Lett.} {\bf
  B517} (2001)  243--249},
\href{http://arxiv.org/abs/astro-ph/0106020}{{\tt arXiv:astro-ph/0106020
  [astro-ph]}}.

\bibitem{Schwarz:2004tz}
D.~J. Schwarz and C.~A. Terrero-Escalante, ``{Primordial fluctuations and
  cosmological inflation after WMAP 1.0}'',
  \href{http://dx.doi.org/10.1088/1475-7516/2004/08/003}{{\em JCAP} {\bf 0408}
  (2004)  003},
\href{http://arxiv.org/abs/hep-ph/0403129}{{\tt arXiv:hep-ph/0403129
  [hep-ph]}}.

\bibitem{Allahverdi:2010xz}
R.~Allahverdi, R.~Brandenberger, F.-Y. Cyr-Racine, and A.~Mazumdar,
  ``{Reheating in Inflationary Cosmology: Theory and Applications}'',
  \href{http://dx.doi.org/10.1146/annurev.nucl.012809.104511}{{\em Ann. Rev.
  Nucl. Part. Sci.} {\bf 60} (2010)  27--51},
\href{http://arxiv.org/abs/1001.2600}{{\tt arXiv:1001.2600 [hep-th]}}.

\bibitem{Cook:2015vqa}
J.~L. Cook, E.~Dimastrogiovanni, D.~A. Easson, and L.~M. Krauss, ``{Reheating
  predictions in single field inflation}'',
  \href{http://dx.doi.org/10.1088/1475-7516/2015/04/047}{{\em JCAP} {\bf 1504}
  (2015)  047},
\href{http://arxiv.org/abs/1502.04673}{{\tt arXiv:1502.04673 [astro-ph.CO]}}.

\bibitem{Seljak:1996ti}
U.~Seljak, ``{Measuring polarization in cosmic microwave background}'',
  \href{http://dx.doi.org/10.1086/304123}{{\em Astrophys. J.} {\bf 482} (1997)
  6},
\href{http://arxiv.org/abs/astro-ph/9608131}{{\tt arXiv:astro-ph/9608131
  [astro-ph]}}.

\bibitem{Kamionkowski:1996zd}
M.~Kamionkowski, A.~Kosowsky, and A.~Stebbins, ``{A Probe of primordial gravity
  waves and vorticity}'',
  \href{http://dx.doi.org/10.1103/PhysRevLett.78.2058}{{\em Phys. Rev. Lett.}
  {\bf 78} (1997)  2058--2061},
\href{http://arxiv.org/abs/astro-ph/9609132}{{\tt arXiv:astro-ph/9609132
  [astro-ph]}}.

\bibitem{Seljak:1996gy}
U.~Seljak and M.~Zaldarriaga, ``{Signature of gravity waves in polarization of
  the microwave background}'',
  \href{http://dx.doi.org/10.1103/PhysRevLett.78.2054}{{\em Phys. Rev. Lett.}
  {\bf 78} (1997)  2054--2057},
\href{http://arxiv.org/abs/astro-ph/9609169}{{\tt arXiv:astro-ph/9609169
  [astro-ph]}}.

\bibitem{Zaldarriaga:1996xe}
M.~Zaldarriaga and U.~Seljak, ``{An all sky analysis of polarization in the
  microwave background}'',
  \href{http://dx.doi.org/10.1103/PhysRevD.55.1830}{{\em Phys. Rev.} {\bf D55}
  (1997)  1830--1840},
\href{http://arxiv.org/abs/astro-ph/9609170}{{\tt arXiv:astro-ph/9609170
  [astro-ph]}}.

\bibitem{Kamionkowski:1996ks}
M.~Kamionkowski, A.~Kosowsky, and A.~Stebbins, ``{Statistics of cosmic
  microwave background polarization}'',
  \href{http://dx.doi.org/10.1103/PhysRevD.55.7368}{{\em Phys. Rev.} {\bf D55}
  (1997)  7368--7388},
\href{http://arxiv.org/abs/astro-ph/9611125}{{\tt arXiv:astro-ph/9611125
  [astro-ph]}}.

\bibitem{Hu:1997hv}
W.~Hu and M.~J. White, ``{A CMB polarization primer}'',
  \href{http://dx.doi.org/10.1016/S1384-1076(97)00022-5}{{\em New Astron.} {\bf
  2} (1997)  323},
\href{http://arxiv.org/abs/astro-ph/9706147}{{\tt arXiv:astro-ph/9706147
  [astro-ph]}}.

\bibitem{Mukhanov:1990me}
V.~F. Mukhanov, H.~A. Feldman, and R.~H. Brandenberger, ``{Theory of
  cosmological perturbations. Part 1. Classical perturbations. Part 2. Quantum
  theory of perturbations. Part 3. Extensions}'',
\href{http://dx.doi.org/10.1016/0370-1573(92)90044-Z}{{\em Phys. Rept.} {\bf
  215} (1992)  203--333}.

\bibitem{Brandenberger:2003vk}
R.~H. Brandenberger, ``{Lectures on the theory of cosmological
  perturbations}'', {\em Lect. Notes Phys.} {\bf 646} (2004)  127--167,
  \href{http://arxiv.org/abs/hep-th/0306071}{{\tt arXiv:hep-th/0306071
  [hep-th]}}.
[,127(2003)].

\bibitem{Baumann:2009ds}
D.~Baumann,
  \href{http://dx.doi.org/10.1142/9789814327183_0010}{``{Inflation}'',} in {\em
  {Physics of the large and the small, TASI 09, proceedings of the Theoretical
  Advanced Study Institute in Elementary Particle Physics, Boulder, Colorado,
  USA, 1-26 June 2009}}, pp.~523--686.
\newblock 2011.
\newblock
\href{http://arxiv.org/abs/0907.5424}{{\tt arXiv:0907.5424 [hep-th]}}.
\newblock

\bibitem{Bunch:1978yq}
T.~S. Bunch and P.~C.~W. Davies, ``{Quantum Field Theory in de Sitter Space:
  Renormalization by Point Splitting}'',
\href{http://dx.doi.org/10.1098/rspa.1978.0060}{{\em Proc. Roy. Soc. Lond.}
  {\bf A360} (1978)  117--134}.

\bibitem{Guth:1982ec}
A.~H. Guth and S.~Y. Pi, ``{Fluctuations in the New Inflationary Universe}'',
\href{http://dx.doi.org/10.1103/PhysRevLett.49.1110}{{\em Phys. Rev. Lett.}
  {\bf 49} (1982)  1110--1113}.

\bibitem{Penzias:1965wn}
A.~A. Penzias and R.~W. Wilson, ``{A Measurement of excess antenna temperature
  at 4080-Mc/s}'',
\href{http://dx.doi.org/10.1086/148307}{{\em Astrophys. J.} {\bf 142} (1965)
  419--421}.

\bibitem{Hu:2008hd}
W.~Hu, ``{Lecture Notes on CMB Theory: From Nucleosynthesis to
  Recombination}'',
\href{http://arxiv.org/abs/0802.3688}{{\tt arXiv:0802.3688 [astro-ph]}}.

\bibitem{COBE}
G.~F. Smoot, C.~Bennett, A.~Kogut, E.~Wright, J.~Aymon, {\em et al.},
  ``{Structure in the COBE differential microwave radiometer first year
  maps}'',
\href{http://dx.doi.org/10.1086/186504}{{\em Astrophys.J.} {\bf 396} (1992)
  L1--L5}.

\bibitem{Ade:2015tva}
{\bf BICEP2, Planck} Collaboration, P.~Ade {\em et al.}, ``{Joint Analysis of
  BICEP2/$Keck Array$ and $Planck$ Data}'',
  \href{http://dx.doi.org/10.1103/PhysRevLett.114.101301}{{\em Phys. Rev.
  Lett.} {\bf 114} (2015)  101301},
\href{http://arxiv.org/abs/1502.00612}{{\tt arXiv:1502.00612 [astro-ph.CO]}}.

\bibitem{Array:2015xqh}
{\bf BICEP2, Keck Array} Collaboration, P.~A.~R. Ade {\em et al.}, ``{Improved
  Constraints on Cosmology and Foregrounds from BICEP2 and Keck Array Cosmic
  Microwave Background Data with Inclusion of 95 GHz Band}'',
  \href{http://dx.doi.org/10.1103/PhysRevLett.116.031302}{{\em Phys. Rev.
  Lett.} {\bf 116} (2016)  031302},
\href{http://arxiv.org/abs/1510.09217}{{\tt arXiv:1510.09217 [astro-ph.CO]}}.

\bibitem{Salopek:1990jq}
D.~S. Salopek and J.~R. Bond, ``{Nonlinear evolution of long wavelength metric
  fluctuations in inflationary models}'',
\href{http://dx.doi.org/10.1103/PhysRevD.42.3936}{{\em Phys. Rev.} {\bf D42}
  (1990)  3936--3962}.

\bibitem{Muslimov:1990be}
A.~G. Muslimov, ``{On the Scalar Field Dynamics in a Spatially Flat Friedman
  Universe}'',
\href{http://dx.doi.org/10.1088/0264-9381/7/2/015}{{\em Class. Quant. Grav.}
  {\bf 7} (1990)  231--237}.

\bibitem{Garcia-Bellido:2014gna}
J.~Garcia-Bellido and D.~Roest, ``{Large-$N$ running of the spectral index of
  inflation}'', \href{http://dx.doi.org/10.1103/PhysRevD.89.103527}{{\em
  Phys.Rev.} {\bf D89} (2014) no.~10, 103527},
\href{http://arxiv.org/abs/1402.2059}{{\tt arXiv:1402.2059 [astro-ph.CO]}}.

\bibitem{Martin:2013tda}
J.~Martin, C.~Ringeval, and V.~Vennin, ``{Encyclopædia Inflationaris}'',
  \href{http://dx.doi.org/10.1016/j.dark.2014.01.003}{{\em Phys. Dark Univ.}
  {\bf 5-6} (2014)  75--235},
\href{http://arxiv.org/abs/1303.3787}{{\tt arXiv:1303.3787 [astro-ph.CO]}}.

\bibitem{Garcia-Bellido:2014eva}
J.~Garcia-Bellido, D.~Roest, M.~Scalisi, and I.~Zavala, ``{Can CMB data
  constrain the inflationary field range?}'',
  \href{http://dx.doi.org/10.1088/1475-7516/2014/09/006}{{\em JCAP} {\bf 1409}
  (2014)  006},
\href{http://arxiv.org/abs/1405.7399}{{\tt arXiv:1405.7399 [hep-th]}}.

\bibitem{Dai:2014jja}
L.~Dai, M.~Kamionkowski, and J.~Wang, ``{Reheating constraints to inflationary
  models}'', \href{http://dx.doi.org/10.1103/PhysRevLett.113.041302}{{\em Phys.
  Rev. Lett.} {\bf 113} (2014)  041302},
\href{http://arxiv.org/abs/1404.6704}{{\tt arXiv:1404.6704 [astro-ph.CO]}}.

\bibitem{Martin:2010kz}
J.~Martin and C.~Ringeval, ``{First CMB Constraints on the Inflationary
  Reheating Temperature}'',
  \href{http://dx.doi.org/10.1103/PhysRevD.82.023511}{{\em Phys. Rev.} {\bf
  D82} (2010)  023511},
\href{http://arxiv.org/abs/1004.5525}{{\tt arXiv:1004.5525 [astro-ph.CO]}}.

\bibitem{Adshead:2010mc}
P.~Adshead, R.~Easther, J.~Pritchard, and A.~Loeb, ``{Inflation and the Scale
  Dependent Spectral Index: Prospects and Strategies}'',
  \href{http://dx.doi.org/10.1088/1475-7516/2011/02/021}{{\em JCAP} {\bf 1102}
  (2011)  021},
\href{http://arxiv.org/abs/1007.3748}{{\tt arXiv:1007.3748 [astro-ph.CO]}}.

\bibitem{Mielczarek:2010ag}
J.~Mielczarek, ``{Reheating temperature from the CMB}'',
  \href{http://dx.doi.org/10.1103/PhysRevD.83.023502}{{\em Phys. Rev.} {\bf
  D83} (2011)  023502},
\href{http://arxiv.org/abs/1009.2359}{{\tt arXiv:1009.2359 [astro-ph.CO]}}.

\bibitem{Dodelson:2003vq}
S.~Dodelson and L.~Hui, ``{A Horizon ratio bound for inflationary
  fluctuations}'', \href{http://dx.doi.org/10.1103/PhysRevLett.91.131301}{{\em
  Phys. Rev. Lett.} {\bf 91} (2003)  131301},
\href{http://arxiv.org/abs/astro-ph/0305113}{{\tt arXiv:astro-ph/0305113
  [astro-ph]}}.

\bibitem{Easther:2011yq}
R.~Easther and H.~V. Peiris, ``{Bayesian Analysis of Inflation II: Model
  Selection and Constraints on Reheating}'',
  \href{http://dx.doi.org/10.1103/PhysRevD.85.103533}{{\em Phys. Rev.} {\bf
  D85} (2012)  103533},
\href{http://arxiv.org/abs/1112.0326}{{\tt arXiv:1112.0326 [astro-ph.CO]}}.

\bibitem{Remmen:2014mia}
G.~N. Remmen and S.~M. Carroll, ``{How Many $e$-Folds Should We Expect from
  High-Scale Inflation?}'',
  \href{http://dx.doi.org/10.1103/PhysRevD.90.063517}{{\em Phys. Rev.} {\bf
  D90} (2014) no.~6, 063517},
\href{http://arxiv.org/abs/1405.5538}{{\tt arXiv:1405.5538 [hep-th]}}.

\bibitem{Mukhanov:2013tua}
V.~Mukhanov, ``{Quantum Cosmological Perturbations: Predictions and
  Observations}'', \href{http://dx.doi.org/10.1140/epjc/s10052-013-2486-7}{{\em
  Eur.Phys.J.} {\bf C73} (2013)  2486},
\href{http://arxiv.org/abs/1303.3925}{{\tt arXiv:1303.3925 [astro-ph.CO]}}.

\bibitem{Roest:2013fha}
D.~Roest, ``{Universality classes of inflation}'',
  \href{http://dx.doi.org/10.1088/1475-7516/2014/01/007}{{\em JCAP} {\bf 01}
  (2014)  007},
\href{http://arxiv.org/abs/1309.1285}{{\tt arXiv:1309.1285 [hep-th]}}.

\bibitem{Boyanovsky:2005pw}
D.~Boyanovsky, H.~J. de~Vega, and N.~G. Sanchez, ``{Clarifying Inflation
  Models: Slow-roll as an expansion in $1/N_{efolds}$}'',
  \href{http://dx.doi.org/10.1103/PhysRevD.73.023008}{{\em Phys.Rev.} {\bf D73}
  (2006)  023008},
\href{http://arxiv.org/abs/astro-ph/0507595}{{\tt arXiv:astro-ph/0507595
  [astro-ph]}}.

\bibitem{Binetruy:2014zya}
P.~Binetruy, E.~Kiritsis, J.~Mabillard, M.~Pieroni, and C.~Rosset,
  ``{Universality classes for models of inflation}'',
  \href{http://dx.doi.org/10.1088/1475-7516/2015/04/033}{{\em JCAP} {\bf 1504}
  (2015) no.~04, 033},
\href{http://arxiv.org/abs/1407.0820}{{\tt arXiv:1407.0820 [astro-ph.CO]}}.

\bibitem{Creminelli:2014nqa}
P.~Creminelli, S.~Dubovsky, D.~López~Nacir, M.~Simonović, G.~Trevisan,
  G.~Villadoro, and M.~Zaldarriaga, ``{Implications of the scalar tilt for the
  tensor-to-scalar ratio}'',
  \href{http://dx.doi.org/10.1103/PhysRevD.92.123528}{{\em Phys. Rev.} {\bf
  D92} (2015) no.~12, 123528},
\href{http://arxiv.org/abs/1412.0678}{{\tt arXiv:1412.0678 [astro-ph.CO]}}.

\bibitem{Cheung:2007st}
C.~Cheung, P.~Creminelli, A.~L. Fitzpatrick, J.~Kaplan, and L.~Senatore, ``{The
  Effective Field Theory of Inflation}'',
  \href{http://dx.doi.org/10.1088/1126-6708/2008/03/014}{{\em JHEP} {\bf 03}
  (2008)  014},
\href{http://arxiv.org/abs/0709.0293}{{\tt arXiv:0709.0293 [hep-th]}}.

\bibitem{Weinberg:2008hq}
S.~Weinberg, ``{Effective Field Theory for Inflation}'',
  \href{http://dx.doi.org/10.1103/PhysRevD.77.123541}{{\em Phys. Rev.} {\bf
  D77} (2008)  123541},
\href{http://arxiv.org/abs/0804.4291}{{\tt arXiv:0804.4291 [hep-th]}}.

\bibitem{Senatore:2010wk}
L.~Senatore and M.~Zaldarriaga, ``{The Effective Field Theory of Multifield
  Inflation}'', \href{http://dx.doi.org/10.1007/JHEP04(2012)024}{{\em JHEP}
  {\bf 04} (2012)  024},
\href{http://arxiv.org/abs/1009.2093}{{\tt arXiv:1009.2093 [hep-th]}}.

\bibitem{Lyth:1996im}
D.~H. Lyth, ``{What would we learn by detecting a gravitational wave signal in
  the cosmic microwave background anisotropy?}'',
  \href{http://dx.doi.org/10.1103/PhysRevLett.78.1861}{{\em Phys. Rev. Lett.}
  {\bf 78} (1997)  1861--1863},
\href{http://arxiv.org/abs/hep-ph/9606387}{{\tt arXiv:hep-ph/9606387
  [hep-ph]}}.

\bibitem{Boubekeur:2005zm}
L.~Boubekeur and D.~Lyth, ``{Hilltop inflation}'',
  \href{http://dx.doi.org/10.1088/1475-7516/2005/07/010}{{\em JCAP} {\bf 0507}
  (2005)  010},
\href{http://arxiv.org/abs/hep-ph/0502047}{{\tt arXiv:hep-ph/0502047
  [hep-ph]}}.

\bibitem{Boubekeur:2012xn}
L.~Boubekeur, ``{Theoretical bounds on the tensor-to-scalar ratio in the cosmic
  microwave background}'',
  \href{http://dx.doi.org/10.1103/PhysRevD.87.061301}{{\em Phys. Rev.} {\bf
  D87} (2013) no.~6, 061301},
\href{http://arxiv.org/abs/1208.0210}{{\tt arXiv:1208.0210 [astro-ph.CO]}}.

\bibitem{Garcia-Bellido:2014wfa}
J.~Garcia-Bellido, D.~Roest, M.~Scalisi, and I.~Zavala, ``{Lyth bound of
  inflation with a tilt}'',
  \href{http://dx.doi.org/10.1103/PhysRevD.90.123539}{{\em Phys.Rev.} {\bf D90}
  (2014) no.~12, 123539},
\href{http://arxiv.org/abs/1408.6839}{{\tt arXiv:1408.6839 [hep-th]}}.

\bibitem{Liddle:2003as}
A.~R. Liddle and S.~M. Leach, ``{How long before the end of inflation were
  observable perturbations produced?}'',
  \href{http://dx.doi.org/10.1103/PhysRevD.68.103503}{{\em Phys. Rev.} {\bf
  D68} (2003)  103503},
\href{http://arxiv.org/abs/astro-ph/0305263}{{\tt arXiv:astro-ph/0305263
  [astro-ph]}}.

\bibitem{Freese:1990rb}
K.~Freese, J.~A. Frieman, and A.~V. Olinto, ``{Natural inflation with pseudo -
  Nambu-Goldstone bosons}'',
\href{http://dx.doi.org/10.1103/PhysRevLett.65.3233}{{\em Phys. Rev. Lett.}
  {\bf 65} (1990)  3233--3236}.

\bibitem{Linde:1993cn}
A.~D. Linde, ``{Hybrid inflation}'',
  \href{http://dx.doi.org/10.1103/PhysRevD.49.748}{{\em Phys. Rev.} {\bf D49}
  (1994)  748--754},
\href{http://arxiv.org/abs/astro-ph/9307002}{{\tt arXiv:astro-ph/9307002
  [astro-ph]}}.

\bibitem{Hotchkiss:2011gz}
S.~Hotchkiss, A.~Mazumdar, and S.~Nadathur, ``{Observable gravitational waves
  from inflation with small field excursions}'',
  \href{http://dx.doi.org/10.1088/1475-7516/2012/02/008}{{\em JCAP} {\bf 1202}
  (2012)  008},
\href{http://arxiv.org/abs/1110.5389}{{\tt arXiv:1110.5389 [astro-ph.CO]}}.

\bibitem{German:2014qza}
G.~German, ``{On the Lyth bound and single field slow-roll inflation}'',
\href{http://arxiv.org/abs/1405.3246}{{\tt arXiv:1405.3246 [astro-ph.CO]}}.

\bibitem{Gao:2014pca}
Q.~Gao, Y.~Gong, and T.~Li, ``{Modified Lyth bound and implications of BICEP2
  results}'', \href{http://dx.doi.org/10.1103/PhysRevD.91.063509}{{\em Phys.
  Rev.} {\bf D91} (2015)  063509},
\href{http://arxiv.org/abs/1405.6451}{{\tt arXiv:1405.6451 [gr-qc]}}.

\bibitem{Bramante:2014rva}
J.~Bramante, S.~Downes, L.~Lehman, and A.~Martin, ``{Last stand of single small
  field inflation}'', \href{http://dx.doi.org/10.1103/PhysRevD.90.023530}{{\em
  Phys. Rev.} {\bf D90} (2014) no.~2, 023530},
\href{http://arxiv.org/abs/1405.7563}{{\tt arXiv:1405.7563 [astro-ph.CO]}}.

\bibitem{Antusch:2014cpa}
S.~Antusch and D.~Nolde, ``{BICEP2 implications for single-field slow-roll
  inflation revisited}'',
  \href{http://dx.doi.org/10.1088/1475-7516/2014/05/035}{{\em JCAP} {\bf 1405}
  (2014)  035},
\href{http://arxiv.org/abs/1404.1821}{{\tt arXiv:1404.1821 [hep-ph]}}.

\bibitem{Hebecker:2013zda}
A.~Hebecker, S.~C. Kraus, and A.~Westphal, ``{Evading the Lyth Bound in Hybrid
  Natural Inflation}'',
  \href{http://dx.doi.org/10.1103/PhysRevD.88.123506}{{\em Phys. Rev.} {\bf
  D88} (2013)  123506},
\href{http://arxiv.org/abs/1305.1947}{{\tt arXiv:1305.1947 [hep-th]}}.

\bibitem{McDonald:2014oza}
J.~McDonald, ``{Sub-Planckian Two-Field Inflation Consistent with the Lyth
  Bound}'', \href{http://dx.doi.org/10.1088/1475-7516/2014/09/027}{{\em JCAP}
  {\bf 1409} (2014) no.~09, 027},
\href{http://arxiv.org/abs/1404.4620}{{\tt arXiv:1404.4620 [hep-ph]}}.

\bibitem{Easther:2006qu}
R.~Easther, W.~H. Kinney, and B.~A. Powell, ``{The Lyth bound and the end of
  inflation}'', \href{http://dx.doi.org/10.1088/1475-7516/2006/08/004}{{\em
  JCAP} {\bf 0608} (2006)  004},
\href{http://arxiv.org/abs/astro-ph/0601276}{{\tt arXiv:astro-ph/0601276
  [astro-ph]}}.

\bibitem{Kallosh:2013tua}
R.~Kallosh, A.~Linde, and D.~Roest, ``{Universal Attractor for Inflation at
  Strong Coupling}'',
  \href{http://dx.doi.org/10.1103/PhysRevLett.112.011303}{{\em Phys. Rev.
  Lett.} {\bf 112} (2014) no.~1, 011303},
\href{http://arxiv.org/abs/1310.3950}{{\tt arXiv:1310.3950 [hep-th]}}.

\bibitem{Kallosh:2013yoa}
R.~Kallosh, A.~Linde, and D.~Roest, ``{Superconformal Inflationary
  $\alpha$-Attractors}'', \href{http://dx.doi.org/10.1007/JHEP11(2013)198}{{\em
  JHEP} {\bf 11} (2013)  198},
\href{http://arxiv.org/abs/1311.0472}{{\tt arXiv:1311.0472 [hep-th]}}.

\bibitem{Galante:2014ifa}
M.~Galante, R.~Kallosh, A.~Linde, and D.~Roest, ``{Unity of Cosmological
  Inflation Attractors}'',
  \href{http://dx.doi.org/10.1103/PhysRevLett.114.141302}{{\em Phys. Rev.
  Lett.} {\bf 114} (2015) no.~14, 141302},
\href{http://arxiv.org/abs/1412.3797}{{\tt arXiv:1412.3797 [hep-th]}}.

\bibitem{Roest:2015qya}
D.~Roest and M.~Scalisi, ``{Cosmological attractors from $\alpha$-scale
  supergravity}'', \href{http://dx.doi.org/10.1103/PhysRevD.92.043525}{{\em
  Phys. Rev.} {\bf D92} (2015)  043525},
\href{http://arxiv.org/abs/1503.07909}{{\tt arXiv:1503.07909 [hep-th]}}.

\bibitem{Scalisi:2015qga}
M.~Scalisi, ``{Cosmological $\alpha$-attractors and de Sitter landscape}'',
  \href{http://dx.doi.org/10.1007/JHEP12(2015)134}{{\em JHEP} {\bf 12} (2015)
  134},
\href{http://arxiv.org/abs/1506.01368}{{\tt arXiv:1506.01368 [hep-th]}}.

\bibitem{Burgess:2014tja}
C.~P. Burgess, M.~Cicoli, F.~Quevedo, and M.~Williams, ``{Inflating with Large
  Effective Fields}'',
  \href{http://dx.doi.org/10.1088/1475-7516/2014/11/045}{{\em JCAP} {\bf 1411}
  (2014)  045},
\href{http://arxiv.org/abs/1404.6236}{{\tt arXiv:1404.6236 [hep-th]}}.

\bibitem{Maleknejad:2011jw}
A.~Maleknejad and M.~M. Sheikh-Jabbari, ``{Gauge-flation: Inflation From
  Non-Abelian Gauge Fields}'',
  \href{http://dx.doi.org/10.1016/j.physletb.2013.05.001}{{\em Phys. Lett.}
  {\bf B723} (2013)  224--228},
\href{http://arxiv.org/abs/1102.1513}{{\tt arXiv:1102.1513 [hep-ph]}}.

\bibitem{Baumann:2011ws}
D.~Baumann and D.~Green, ``{A Field Range Bound for General Single-Field
  Inflation}'', \href{http://dx.doi.org/10.1088/1475-7516/2012/05/017}{{\em
  JCAP} {\bf 1205} (2012)  017},
\href{http://arxiv.org/abs/1111.3040}{{\tt arXiv:1111.3040 [hep-th]}}.

\bibitem{BenDayan:2009kv}
I.~Ben-Dayan and R.~Brustein, ``{Cosmic Microwave Background Observables of
  Small Field Models of Inflation}'',
  \href{http://dx.doi.org/10.1088/1475-7516/2010/09/007}{{\em JCAP} {\bf 1009}
  (2010)  007},
\href{http://arxiv.org/abs/0907.2384}{{\tt arXiv:0907.2384 [astro-ph.CO]}}.

\bibitem{Bouchet:2011ck}
{\bf COrE} Collaboration, F.~R. Bouchet {\em et al.}, ``{COrE (Cosmic Origins
  Explorer) A White Paper}'',
\href{http://arxiv.org/abs/1102.2181}{{\tt arXiv:1102.2181 [astro-ph.CO]}}.

\bibitem{Core}
\url{http://www.core-mission.org}.

\bibitem{Andre:2013afa}
{\bf PRISM} Collaboration, P.~Andre {\em et al.}, ``{PRISM (Polarized Radiation
  Imaging and Spectroscopy Mission): A White Paper on the Ultimate Polarimetric
  Spectro-Imaging of the Microwave and Far-Infrared Sky}'',
\href{http://arxiv.org/abs/1306.2259}{{\tt arXiv:1306.2259 [astro-ph.CO]}}.

\bibitem{Andre:2013nfa}
{\bf PRISM} Collaboration, P.~André {\em et al.}, ``{PRISM (Polarized
  Radiation Imaging and Spectroscopy Mission): An Extended White Paper}'',
  \href{http://dx.doi.org/10.1088/1475-7516/2014/02/006}{{\em JCAP} {\bf 1402}
  (2014)  006},
\href{http://arxiv.org/abs/1310.1554}{{\tt arXiv:1310.1554 [astro-ph.CO]}}.

\bibitem{Prism}
\url{http://www.prism-mission.org}.

\bibitem{Grana:2005jc}
M.~Grana, ``{Flux compactifications in string theory: A Comprehensive
  review}'', \href{http://dx.doi.org/10.1016/j.physrep.2005.10.008}{{\em Phys.
  Rept.} {\bf 423} (2006)  91--158},
\href{http://arxiv.org/abs/hep-th/0509003}{{\tt arXiv:hep-th/0509003
  [hep-th]}}.

\bibitem{Douglas:2006es}
M.~R. Douglas and S.~Kachru, ``{Flux compactification}'',
  \href{http://dx.doi.org/10.1103/RevModPhys.79.733}{{\em Rev. Mod. Phys.} {\bf
  79} (2007)  733--796},
\href{http://arxiv.org/abs/hep-th/0610102}{{\tt arXiv:hep-th/0610102
  [hep-th]}}.

\bibitem{Giddings:2001yu}
S.~B. Giddings, S.~Kachru, and J.~Polchinski, ``{Hierarchies from fluxes in
  string compactifications}'',
  \href{http://dx.doi.org/10.1103/PhysRevD.66.106006}{{\em Phys. Rev.} {\bf
  D66} (2002)  106006},
\href{http://arxiv.org/abs/hep-th/0105097}{{\tt arXiv:hep-th/0105097
  [hep-th]}}.

\bibitem{Kachru:2003aw}
S.~Kachru, R.~Kallosh, A.~D. Linde, and S.~P. Trivedi, ``{De Sitter vacua in
  string theory}'', \href{http://dx.doi.org/10.1103/PhysRevD.68.046005}{{\em
  Phys. Rev.} {\bf D68} (2003)  046005},
\href{http://arxiv.org/abs/hep-th/0301240}{{\tt arXiv:hep-th/0301240
  [hep-th]}}.

\bibitem{Baumann:2009ni}
D.~Baumann and L.~McAllister, ``{Advances in Inflation in String Theory}'',
  \href{http://dx.doi.org/10.1146/annurev.nucl.010909.083524}{{\em Ann. Rev.
  Nucl. Part. Sci.} {\bf 59} (2009)  67--94},
\href{http://arxiv.org/abs/0901.0265}{{\tt arXiv:0901.0265 [hep-th]}}.

\bibitem{Silverstein:2013wua}
E.~Silverstein, ``{Les Houches lectures on inflationary observables and string
  theory}'',
\href{http://arxiv.org/abs/1311.2312}{{\tt arXiv:1311.2312 [hep-th]}}.

\bibitem{Westphal:2014ana}
A.~Westphal, ``{String cosmology — Large-field inflation in string theory}'',
  \href{http://dx.doi.org/10.1142/S0217751X15300240}{{\em Int. J. Mod. Phys.}
  {\bf A30} (2015) no.~09, 1530024},
\href{http://arxiv.org/abs/1409.5350}{{\tt arXiv:1409.5350 [hep-th]}}.

\bibitem{Cicoli:2016ygh}
M.~Cicoli, ``{Recent developments in string model-building and cosmology}'',
\newblock 2016.
\newblock
\href{http://arxiv.org/abs/1604.00904}{{\tt arXiv:1604.00904 [hep-th]}}.
\newblock

\bibitem{Dvali:1998pa}
G.~R. Dvali and S.~H.~H. Tye, ``{Brane inflation}'',
  \href{http://dx.doi.org/10.1016/S0370-2693(99)00132-X}{{\em Phys. Lett.} {\bf
  B450} (1999)  72--82},
\href{http://arxiv.org/abs/hep-ph/9812483}{{\tt arXiv:hep-ph/9812483
  [hep-ph]}}.

\bibitem{Burgess:2001fx}
C.~P. Burgess, M.~Majumdar, D.~Nolte, F.~Quevedo, G.~Rajesh, and R.-J. Zhang,
  ``{The Inflationary brane anti-brane universe}'',
  \href{http://dx.doi.org/10.1088/1126-6708/2001/07/047}{{\em JHEP} {\bf 07}
  (2001)  047},
\href{http://arxiv.org/abs/hep-th/0105204}{{\tt arXiv:hep-th/0105204
  [hep-th]}}.

\bibitem{Kachru:2003sx}
S.~Kachru, R.~Kallosh, A.~D. Linde, J.~M. Maldacena, L.~P. McAllister, and
  S.~P. Trivedi, ``{Towards inflation in string theory}'',
  \href{http://dx.doi.org/10.1088/1475-7516/2003/10/013}{{\em JCAP} {\bf 0310}
  (2003)  013},
\href{http://arxiv.org/abs/hep-th/0308055}{{\tt arXiv:hep-th/0308055
  [hep-th]}}.

\bibitem{Kallosh:2007ig}
R.~Kallosh, ``{On inflation in string theory}'',
  \href{http://dx.doi.org/10.1007/978-3-540-74353-8_4}{{\em Lect. Notes Phys.}
  {\bf 738} (2008)  119--156},
\href{http://arxiv.org/abs/hep-th/0702059}{{\tt arXiv:hep-th/0702059
  [HEP-TH]}}.

\bibitem{Cicoli:2008gp}
M.~Cicoli, C.~P. Burgess, and F.~Quevedo, ``{Fibre Inflation: Observable
  Gravity Waves from IIB String Compactifications}'',
  \href{http://dx.doi.org/10.1088/1475-7516/2009/03/013}{{\em JCAP} {\bf 0903}
  (2009)  013},
\href{http://arxiv.org/abs/0808.0691}{{\tt arXiv:0808.0691 [hep-th]}}.

\bibitem{Hebecker:2011hk}
A.~Hebecker, S.~C. Kraus, D.~Lust, S.~Steinfurt, and T.~Weigand, ``{Fluxbrane
  Inflation}'', \href{http://dx.doi.org/10.1016/j.nuclphysb.2011.08.025}{{\em
  Nucl. Phys.} {\bf B854} (2012)  509--551},
\href{http://arxiv.org/abs/1104.5016}{{\tt arXiv:1104.5016 [hep-th]}}.

\bibitem{Hebecker:2012aw}
A.~Hebecker, S.~C. Kraus, M.~Kuntzler, D.~Lust, and T.~Weigand, ``{Fluxbranes:
  Moduli Stabilisation and Inflation}'',
  \href{http://dx.doi.org/10.1007/JHEP01(2013)095}{{\em JHEP} {\bf 01} (2013)
  095},
\href{http://arxiv.org/abs/1207.2766}{{\tt arXiv:1207.2766 [hep-th]}}.

\bibitem{Burgess:2013sla}
C.~P. Burgess, M.~Cicoli, and F.~Quevedo, ``{String Inflation After Planck
  2013}'', \href{http://dx.doi.org/10.1088/1475-7516/2013/11/003}{{\em JCAP}
  {\bf 1311} (2013)  003},
\href{http://arxiv.org/abs/1306.3512}{{\tt arXiv:1306.3512 [hep-th]}}.

\bibitem{Hebecker:2014eua}
A.~Hebecker, S.~C. Kraus, and L.~T. Witkowski, ``{D7-Brane Chaotic
  Inflation}'', \href{http://dx.doi.org/10.1016/j.physletb.2014.08.028}{{\em
  Phys. Lett.} {\bf B737} (2014)  16--22},
\href{http://arxiv.org/abs/1404.3711}{{\tt arXiv:1404.3711 [hep-th]}}.

\bibitem{Achucarro:2008sy}
A.~Achucarro, S.~Hardeman, and K.~Sousa, ``{Consistent Decoupling of Heavy
  Scalars and Moduli in N=1 Supergravity}'',
  \href{http://dx.doi.org/10.1103/PhysRevD.78.101901}{{\em Phys. Rev.} {\bf
  D78} (2008)  101901},
\href{http://arxiv.org/abs/0806.4364}{{\tt arXiv:0806.4364 [hep-th]}}.

\bibitem{Achucarro:2008fk}
A.~Achucarro, S.~Hardeman, and K.~Sousa, ``{F-term uplifting and the
  supersymmetric integration of heavy moduli}'',
  \href{http://dx.doi.org/10.1088/1126-6708/2008/11/003}{{\em JHEP} {\bf 11}
  (2008)  003},
\href{http://arxiv.org/abs/0809.1441}{{\tt arXiv:0809.1441 [hep-th]}}.

\bibitem{Achucarro:2010jv}
A.~Achucarro, J.-O. Gong, S.~Hardeman, G.~A. Palma, and S.~P. Patil, ``{Mass
  hierarchies and non-decoupling in multi-scalar field dynamics}'',
  \href{http://dx.doi.org/10.1103/PhysRevD.84.043502}{{\em Phys. Rev.} {\bf
  D84} (2011)  043502},
\href{http://arxiv.org/abs/1005.3848}{{\tt arXiv:1005.3848 [hep-th]}}.

\bibitem{Achucarro:2010da}
A.~Achucarro, J.-O. Gong, S.~Hardeman, G.~A. Palma, and S.~P. Patil,
  ``{Features of heavy physics in the CMB power spectrum}'',
  \href{http://dx.doi.org/10.1088/1475-7516/2011/01/030}{{\em JCAP} {\bf 1101}
  (2011)  030},
\href{http://arxiv.org/abs/1010.3693}{{\tt arXiv:1010.3693 [hep-ph]}}.

\bibitem{Achucarro:2012sm}
A.~Achucarro, J.-O. Gong, S.~Hardeman, G.~A. Palma, and S.~P. Patil,
  ``{Effective theories of single field inflation when heavy fields matter}'',
  \href{http://dx.doi.org/10.1007/JHEP05(2012)066}{{\em JHEP} {\bf 05} (2012)
  066},
\href{http://arxiv.org/abs/1201.6342}{{\tt arXiv:1201.6342 [hep-th]}}.

\bibitem{Achucarro:2012yr}
A.~Achucarro, V.~Atal, S.~Cespedes, J.-O. Gong, G.~A. Palma, and S.~P. Patil,
  ``{Heavy fields, reduced speeds of sound and decoupling during inflation}'',
  \href{http://dx.doi.org/10.1103/PhysRevD.86.121301}{{\em Phys. Rev.} {\bf
  D86} (2012)  121301},
\href{http://arxiv.org/abs/1205.0710}{{\tt arXiv:1205.0710 [hep-th]}}.

\bibitem{Achucarro:2012fd}
A.~Achúcarro, J.-O. Gong, G.~A. Palma, and S.~P. Patil, ``{Correlating
  features in the primordial spectra}'',
  \href{http://dx.doi.org/10.1103/PhysRevD.87.121301}{{\em Phys. Rev.} {\bf
  D87} (2013) no.~12, 121301},
\href{http://arxiv.org/abs/1211.5619}{{\tt arXiv:1211.5619 [astro-ph.CO]}}.

\bibitem{Achucarro:2013cva}
A.~Achúcarro, V.~Atal, P.~Ortiz, and J.~Torrado, ``{Localized correlated
  features in the CMB power spectrum and primordial bispectrum from a transient
  reduction in the speed of sound}'',
  \href{http://dx.doi.org/10.1103/PhysRevD.89.103006}{{\em Phys. Rev.} {\bf
  D89} (2014) no.~10, 103006},
\href{http://arxiv.org/abs/1311.2552}{{\tt arXiv:1311.2552 [astro-ph.CO]}}.

\bibitem{Achucarro:2014msa}
A.~Achucarro, V.~Atal, B.~Hu, P.~Ortiz, and J.~Torrado, ``{Inflation with
  moderately sharp features in the speed of sound: Generalized slow roll and
  in-in formalism for power spectrum and bispectrum}'',
  \href{http://dx.doi.org/10.1103/PhysRevD.90.023511}{{\em Phys. Rev.} {\bf
  D90} (2014) no.~2, 023511},
\href{http://arxiv.org/abs/1404.7522}{{\tt arXiv:1404.7522 [astro-ph.CO]}}.

\bibitem{Buchmuller:2014pla}
W.~Buchmuller, E.~Dudas, L.~Heurtier, and C.~Wieck, ``{Large-Field Inflation
  and Supersymmetry Breaking}'',
  \href{http://dx.doi.org/10.1007/JHEP09(2014)053}{{\em JHEP} {\bf 09} (2014)
  053},
\href{http://arxiv.org/abs/1407.0253}{{\tt arXiv:1407.0253 [hep-th]}}.

\bibitem{Buchmuller:2015oma}
W.~Buchmuller, E.~Dudas, L.~Heurtier, A.~Westphal, C.~Wieck, and M.~W. Winkler,
  ``{Challenges for Large-Field Inflation and Moduli Stabilization}'',
  \href{http://dx.doi.org/10.1007/JHEP04(2015)058}{{\em JHEP} {\bf 04} (2015)
  058},
\href{http://arxiv.org/abs/1501.05812}{{\tt arXiv:1501.05812 [hep-th]}}.

\bibitem{Dudas:2015lga}
E.~Dudas and C.~Wieck, ``{Moduli backreaction and supersymmetry breaking in
  string-inspired inflation models}'',
  \href{http://dx.doi.org/10.1007/JHEP10(2015)062}{{\em JHEP} {\bf 10} (2015)
  062},
\href{http://arxiv.org/abs/1506.01253}{{\tt arXiv:1506.01253 [hep-th]}}.

\bibitem{Nath:1975nj}
P.~Nath and R.~L. Arnowitt, ``{Generalized Supergauge Symmetry as a New
  Framework for Unified Gauge Theories}'',
\href{http://dx.doi.org/10.1016/0370-2693(75)90297-X}{{\em Phys. Lett.} {\bf
  B56} (1975)  177}.

\bibitem{Freedman:1976xh}
D.~Z. Freedman, P.~van Nieuwenhuizen, and S.~Ferrara, ``{Progress Toward a
  Theory of Supergravity}'',
\href{http://dx.doi.org/10.1103/PhysRevD.13.3214}{{\em Phys. Rev.} {\bf D13}
  (1976)  3214--3218}.

\bibitem{Freedman:2012zz}
D.~Z. Freedman and A.~Van~Proeyen, {\em {Supergravity}}.
\newblock Cambridge Univ. Press, Cambridge, UK,
2012.
\newblock

\bibitem{Gervais:1971ji}
J.-L. Gervais and B.~Sakita, ``{Field Theory Interpretation of Supergauges in
  Dual Models}'',
\href{http://dx.doi.org/10.1016/0550-3213(71)90351-8}{{\em Nucl. Phys.} {\bf
  B34} (1971)  632--639}.

\bibitem{Volkov:1973ix}
D.~V. Volkov and V.~P. Akulov, ``{Is the Neutrino a Goldstone Particle?}'',
\href{http://dx.doi.org/10.1016/0370-2693(73)90490-5}{{\em Phys. Lett.} {\bf
  B46} (1973)  109--110}.

\bibitem{Ramond:1971gb}
P.~Ramond, ``{Dual Theory for Free Fermions}'',
\href{http://dx.doi.org/10.1103/PhysRevD.3.2415}{{\em Phys. Rev.} {\bf D3}
  (1971)  2415--2418}.

\bibitem{Wess:1974tw}
J.~Wess and B.~Zumino, ``{Supergauge Transformations in Four-Dimensions}'',
\href{http://dx.doi.org/10.1016/0550-3213(74)90355-1}{{\em Nucl. Phys.} {\bf
  B70} (1974)  39--50}.

\bibitem{Cremmer:1979up}
E.~Cremmer and B.~Julia, ``{The SO(8) Supergravity}'',
\href{http://dx.doi.org/10.1016/0550-3213(79)90331-6}{{\em Nucl. Phys.} {\bf
  B159} (1979)  141}.

\bibitem{Cremmer:1982en}
E.~Cremmer, S.~Ferrara, L.~Girardello, and A.~Van~Proeyen, ``{Yang-Mills
  Theories with Local Supersymmetry: Lagrangian, Transformation Laws and
  SuperHiggs Effect}'',
\href{http://dx.doi.org/10.1016/0550-3213(83)90679-X}{{\em Nucl. Phys.} {\bf
  B212} (1983)  413}.

\bibitem{Wands:2007bd}
D.~Wands, ``{Multiple field inflation}'',
  \href{http://dx.doi.org/10.1007/978-3-540-74353-8_8}{{\em Lect. Notes Phys.}
  {\bf 738} (2008)  275--304},
\href{http://arxiv.org/abs/astro-ph/0702187}{{\tt arXiv:astro-ph/0702187
  [ASTRO-PH]}}.

\bibitem{Copeland:1994vg}
E.~J. Copeland, A.~R. Liddle, D.~H. Lyth, E.~D. Stewart, and D.~Wands, ``{False
  vacuum inflation with Einstein gravity}'',
  \href{http://dx.doi.org/10.1103/PhysRevD.49.6410}{{\em Phys.Rev.} {\bf D49}
  (1994)  6410--6433},
\href{http://arxiv.org/abs/astro-ph/9401011}{{\tt arXiv:astro-ph/9401011
  [astro-ph]}}.

\bibitem{Roest:2013aoa}
D.~Roest, M.~Scalisi, and I.~Zavala, ``{K\"ahler potentials for Planck
  inflation}'', \href{http://dx.doi.org/10.1088/1475-7516/2013/11/007}{{\em
  JCAP} {\bf 1311} (2013)  007},
\href{http://arxiv.org/abs/1307.4343}{{\tt arXiv:1307.4343}}.

\bibitem{Kawasaki:2000yn}
M.~Kawasaki, M.~Yamaguchi, and T.~Yanagida, ``{Natural chaotic inflation in
  supergravity}'', \href{http://dx.doi.org/10.1103/PhysRevLett.85.3572}{{\em
  Phys.Rev.Lett.} {\bf 85} (2000)  3572--3575},
\href{http://arxiv.org/abs/hep-ph/0004243}{{\tt arXiv:hep-ph/0004243
  [hep-ph]}}.

\bibitem{Chen:2009we}
X.~Chen and Y.~Wang, ``{Large non-Gaussianities with Intermediate Shapes from
  Quasi-Single Field Inflation}'',
  \href{http://dx.doi.org/10.1103/PhysRevD.81.063511}{{\em Phys. Rev.} {\bf
  D81} (2010)  063511},
\href{http://arxiv.org/abs/0909.0496}{{\tt arXiv:0909.0496 [astro-ph.CO]}}.

\bibitem{Baumann:2011nk}
D.~Baumann and D.~Green, ``{Signatures of Supersymmetry from the Early
  Universe}'', \href{http://dx.doi.org/10.1103/PhysRevD.85.103520}{{\em Phys.
  Rev.} {\bf D85} (2012)  103520},
\href{http://arxiv.org/abs/1109.0292}{{\tt arXiv:1109.0292 [hep-th]}}.

\bibitem{GomezReino:2006dk}
M.~Gomez-Reino and C.~A. Scrucca, ``{Locally stable non-supersymmetric
  Minkowski vacua in supergravity}'',
  \href{http://dx.doi.org/10.1088/1126-6708/2006/05/015}{{\em JHEP} {\bf 05}
  (2006)  015},
\href{http://arxiv.org/abs/hep-th/0602246}{{\tt arXiv:hep-th/0602246
  [hep-th]}}.

\bibitem{GomezReino:2006wv}
M.~Gomez-Reino and C.~A. Scrucca, ``{Constraints for the existence of flat and
  stable non-supersymmetric vacua in supergravity}'',
  \href{http://dx.doi.org/10.1088/1126-6708/2006/09/008}{{\em JHEP} {\bf 09}
  (2006)  008},
\href{http://arxiv.org/abs/hep-th/0606273}{{\tt arXiv:hep-th/0606273
  [hep-th]}}.

\bibitem{Covi:2008cn}
L.~Covi, M.~Gomez-Reino, C.~Gross, J.~Louis, G.~A. Palma, and C.~A. Scrucca,
  ``{Constraints on modular inflation in supergravity and string theory}'',
  \href{http://dx.doi.org/10.1088/1126-6708/2008/08/055}{{\em JHEP} {\bf 08}
  (2008)  055},
\href{http://arxiv.org/abs/0805.3290}{{\tt arXiv:0805.3290 [hep-th]}}.

\bibitem{Borghese:2012yu}
A.~Borghese, D.~Roest, and I.~Zavala, ``{A Geometric bound on F-term
  inflation}'', \href{http://dx.doi.org/10.1007/JHEP09(2012)021}{{\em JHEP}
  {\bf 09} (2012)  021},
\href{http://arxiv.org/abs/1203.2909}{{\tt arXiv:1203.2909 [hep-th]}}.

\bibitem{Goncharov:1983mw}
A.~B. Goncharov and A.~D. Linde, ``{Chaotic Inflation in Supergravity}'',
\href{http://dx.doi.org/10.1016/0370-2693(84)90027-3}{{\em Phys. Lett.} {\bf
  B139} (1984)  27}.

\bibitem{Goncharov:1985yu}
A.~S. Goncharov and A.~D. Linde, ``{CHAOTIC INFLATION OF THE UNIVERSE IN
  SUPERGRAVITY}'', {\em Sov. Phys. JETP} {\bf 59} (1984)  930--933.
[Zh. Eksp. Teor. Fiz.86,1594(1984)].

\bibitem{AlvarezGaume:2010rt}
L.~Alvarez-Gaume, C.~Gomez, and R.~Jimenez, ``{Minimal Inflation}'',
  \href{http://dx.doi.org/10.1016/j.physletb.2010.04.069}{{\em Phys. Lett.}
  {\bf B690} (2010)  68--72},
\href{http://arxiv.org/abs/1001.0010}{{\tt arXiv:1001.0010 [hep-th]}}.

\bibitem{AlvarezGaume:2011xv}
L.~Alvarez-Gaume, C.~Gomez, and R.~Jimenez, ``{A Minimal Inflation Scenario}'',
  \href{http://dx.doi.org/10.1088/1475-7516/2011/03/027}{{\em JCAP} {\bf 1103}
  (2011)  027},
\href{http://arxiv.org/abs/1101.4948}{{\tt arXiv:1101.4948 [hep-th]}}.

\bibitem{Achucarro:2012hg}
A.~Achucarro, S.~Mooij, P.~Ortiz, and M.~Postma, ``{Sgoldstino inflation}'',
  \href{http://dx.doi.org/10.1088/1475-7516/2012/08/013}{{\em JCAP} {\bf 1208}
  (2012)  013},
\href{http://arxiv.org/abs/1203.1907}{{\tt arXiv:1203.1907 [hep-th]}}.

\bibitem{Ketov:2014qha}
S.~V. Ketov and T.~Terada, ``{Inflation in supergravity with a single chiral
  superfield}'', \href{http://dx.doi.org/10.1016/j.physletb.2014.07.036}{{\em
  Phys. Lett.} {\bf B736} (2014)  272--277},
\href{http://arxiv.org/abs/1406.0252}{{\tt arXiv:1406.0252 [hep-th]}}.

\bibitem{Ketov:2014hya}
S.~V. Ketov and T.~Terada, ``{Generic Scalar Potentials for Inflation in
  Supergravity with a Single Chiral Superfield}'',
  \href{http://dx.doi.org/10.1007/JHEP12(2014)062}{{\em JHEP} {\bf 12} (2014)
  062},
\href{http://arxiv.org/abs/1408.6524}{{\tt arXiv:1408.6524 [hep-th]}}.

\bibitem{Linde:2014ela}
A.~Linde, D.~Roest, and M.~Scalisi, ``{Inflation and Dark Energy with a Single
  Superfield}'', \href{http://dx.doi.org/10.1088/1475-7516/2015/03/017}{{\em
  JCAP} {\bf 1503} (2015)  017},
\href{http://arxiv.org/abs/1412.2790}{{\tt arXiv:1412.2790 [hep-th]}}.

\bibitem{Linde:2014hfa}
A.~Linde, ``{Does the first chaotic inflation model in supergravity provide the
  best fit to the Planck data?}'',
  \href{http://dx.doi.org/10.1088/1475-7516/2015/02/030}{{\em JCAP} {\bf 1502}
  (2015)  030},
\href{http://arxiv.org/abs/1412.7111}{{\tt arXiv:1412.7111 [hep-th]}}.

\bibitem{Linde:2015uga}
A.~Linde, ``{Single-field $\alpha$-attractors}'',
  \href{http://dx.doi.org/10.1088/1475-7516/2015/05/003}{{\em JCAP} {\bf 1505}
  (2015) no.~05, 003},
\href{http://arxiv.org/abs/1504.00663}{{\tt arXiv:1504.00663 [hep-th]}}.

\bibitem{Terada:2015sna}
T.~Terada, {\em {Inflation in Supergravity with a Single Superfield}}.
\newblock PhD thesis, Tokyo U., 2015.
\newblock
\href{http://arxiv.org/abs/1508.05335}{{\tt arXiv:1508.05335 [hep-th]}}.
\newblock

\bibitem{Ketov:2015tpa}
S.~V. Ketov and T.~Terada, ``{Single-Superfield Helical-Phase Inflation}'',
  \href{http://dx.doi.org/10.1016/j.physletb.2015.11.039}{{\em Phys. Lett.}
  {\bf B752} (2016)  108--112},
\href{http://arxiv.org/abs/1509.00953}{{\tt arXiv:1509.00953 [hep-th]}}.

\bibitem{Cremmer:1983bf}
E.~Cremmer, S.~Ferrara, C.~Kounnas, and D.~V. Nanopoulos, ``{Naturally
  Vanishing Cosmological Constant in N=1 Supergravity}'',
\href{http://dx.doi.org/10.1016/0370-2693(83)90106-5}{{\em Phys.Lett.} {\bf
  B133} (1983)  61}.

\bibitem{Ellis:1983sf}
J.~R. Ellis, A.~Lahanas, D.~V. Nanopoulos, and K.~Tamvakis, ``{No-Scale
  Supersymmetric Standard Model}'',
\href{http://dx.doi.org/10.1016/0370-2693(84)91378-9}{{\em Phys.Lett.} {\bf
  B134} (1984)  429}.

\bibitem{Lahanas:1986uc}
A.~Lahanas and D.~V. Nanopoulos, ``{The Road to No Scale Supergravity}'',
\href{http://dx.doi.org/10.1016/0370-1573(87)90034-2}{{\em Phys.Rept.} {\bf
  145} (1987)  1}.

\bibitem{Kallosh:2010ug}
R.~Kallosh and A.~Linde, ``{New models of chaotic inflation in supergravity}'',
  \href{http://dx.doi.org/10.1088/1475-7516/2010/11/011}{{\em JCAP} {\bf 1011}
  (2010)  011},
\href{http://arxiv.org/abs/1008.3375}{{\tt arXiv:1008.3375 [hep-th]}}.

\bibitem{Kallosh:2010xz}
R.~Kallosh, A.~Linde, and T.~Rube, ``{General inflaton potentials in
  supergravity}'', \href{http://dx.doi.org/10.1103/PhysRevD.83.043507}{{\em
  Phys.Rev.} {\bf D83} (2011)  043507},
\href{http://arxiv.org/abs/1011.5945}{{\tt arXiv:1011.5945 [hep-th]}}.

\bibitem{Cecotti:1987sa}
S.~Cecotti, ``{HIGHER DERIVATIVE SUPERGRAVITY IS EQUIVALENT TO STANDARD
  SUPERGRAVITY COUPLED TO MATTER. 1.}'',
\href{http://dx.doi.org/10.1016/0370-2693(87)90844-6}{{\em Phys. Lett.} {\bf
  B190} (1987)  86}.

\bibitem{Kallosh:2013lkr}
R.~Kallosh and A.~Linde, ``{Superconformal generalizations of the Starobinsky
  model}'', \href{http://dx.doi.org/10.1088/1475-7516/2013/06/028}{{\em JCAP}
  {\bf 1306} (2013)  028},
\href{http://arxiv.org/abs/1306.3214}{{\tt arXiv:1306.3214 [hep-th]}}.

\bibitem{Ellis:2013nxa}
J.~Ellis, D.~V. Nanopoulos, and K.~A. Olive, ``{Starobinsky-like Inflationary
  Models as Avatars of No-Scale Supergravity}'',
  \href{http://dx.doi.org/10.1088/1475-7516/2013/10/009}{{\em JCAP} {\bf 1310}
  (2013)  009},
\href{http://arxiv.org/abs/1307.3537}{{\tt arXiv:1307.3537}}.

\bibitem{Kallosh:2014rga}
R.~Kallosh, A.~Linde, and D.~Roest, ``{Large field inflation and double
  $\alpha$-attractors}'', \href{http://dx.doi.org/10.1007/JHEP08(2014)052}{{\em
  JHEP} {\bf 1408} (2014)  052},
\href{http://arxiv.org/abs/1405.3646}{{\tt arXiv:1405.3646 [hep-th]}}.

\bibitem{Carrasco:2015rva}
J.~J.~M. Carrasco, R.~Kallosh, and A.~Linde, ``{Cosmological Attractors and
  Initial Conditions for Inflation}'',
  \href{http://dx.doi.org/10.1103/PhysRevD.92.063519}{{\em Phys. Rev.} {\bf
  D92} (2015) no.~6, 063519},
\href{http://arxiv.org/abs/1506.00936}{{\tt arXiv:1506.00936 [hep-th]}}.

\bibitem{Carrasco:2015pla}
J.~J.~M. Carrasco, R.~Kallosh, and A.~Linde, ``{$\alpha $-Attractors: Planck,
  LHC and Dark Energy}'', \href{http://dx.doi.org/10.1007/JHEP10(2015)147}{{\em
  JHEP} {\bf 10} (2015)  147},
\href{http://arxiv.org/abs/1506.01708}{{\tt arXiv:1506.01708 [hep-th]}}.

\bibitem{Lahanas:2015jwa}
A.~Lahanas and K.~Tamvakis, ``{Inflation in no-scale supergravity}'',
  \href{http://dx.doi.org/10.1103/PhysRevD.91.085001}{{\em Phys.Rev.} {\bf D91}
  (2015) no.~8, 085001},
\href{http://arxiv.org/abs/1501.06547}{{\tt arXiv:1501.06547 [hep-th]}}.

\bibitem{Hardeman:2010fh}
S.~Hardeman, J.~M. Oberreuter, G.~A. Palma, K.~Schalm, and T.~van~der Aalst,
  ``{The everpresent eta-problem: knowledge of all hidden sectors required}'',
  \href{http://dx.doi.org/10.1007/JHEP04(2011)009}{{\em JHEP} {\bf 04} (2011)
  009},
\href{http://arxiv.org/abs/1012.5966}{{\tt arXiv:1012.5966 [hep-ph]}}.

\bibitem{LopesCardoso:1994is}
G.~Lopes~Cardoso, D.~Lust, and T.~Mohaupt, ``{Moduli spaces and target space
  duality symmetries in (0,2) Z(N) orbifold theories with continuous Wilson
  lines}'', \href{http://dx.doi.org/10.1016/0550-3213(94)90594-0}{{\em Nucl.
  Phys.} {\bf B432} (1994)  68--108},
\href{http://arxiv.org/abs/hep-th/9405002}{{\tt arXiv:hep-th/9405002
  [hep-th]}}.

\bibitem{Antoniadis:1994hg}
I.~Antoniadis, E.~Gava, K.~S. Narain, and T.~R. Taylor, ``{Effective mu term in
  superstring theory}'',
  \href{http://dx.doi.org/10.1016/0550-3213(94)90599-1}{{\em Nucl. Phys.} {\bf
  B432} (1994)  187--204},
\href{http://arxiv.org/abs/hep-th/9405024}{{\tt arXiv:hep-th/9405024
  [hep-th]}}.

\bibitem{Kallosh:2011qk}
R.~Kallosh, A.~Linde, K.~A. Olive, and T.~Rube, ``{Chaotic inflation and
  supersymmetry breaking}'',
  \href{http://dx.doi.org/10.1103/PhysRevD.84.083519}{{\em Phys. Rev.} {\bf
  D84} (2011)  083519},
\href{http://arxiv.org/abs/1106.6025}{{\tt arXiv:1106.6025 [hep-th]}}.

\bibitem{Kallosh:2015zsa}
R.~Kallosh and A.~Linde, ``{Escher in the Sky}'',
  \href{http://dx.doi.org/10.1016/j.crhy.2015.07.004}{{\em Comptes Rendus
  Physique} {\bf 16} (2015)  914--927},
\href{http://arxiv.org/abs/1503.06785}{{\tt arXiv:1503.06785 [hep-th]}}.

\bibitem{Kallosh:2013hoa}
R.~Kallosh and A.~Linde, ``{Universality Class in Conformal Inflation}'',
  \href{http://dx.doi.org/10.1088/1475-7516/2013/07/002}{{\em JCAP} {\bf 1307}
  (2013)  002},
\href{http://arxiv.org/abs/1306.5220}{{\tt arXiv:1306.5220 [hep-th]}}.

\bibitem{Ozkan:2015kma}
M.~Ozkan and D.~Roest, ``{Universality Classes of Scale Invariant Inflation}'',
\href{http://arxiv.org/abs/1507.03603}{{\tt arXiv:1507.03603 [hep-th]}}.

\bibitem{Starobinsky:1980te}
A.~A. Starobinsky, ``{A New Type of Isotropic Cosmological Models Without
  Singularity}'',
\href{http://dx.doi.org/10.1016/0370-2693(80)90670-X}{{\em Phys.Lett.} {\bf
  B91} (1980)  99--102}.

\bibitem{Ellis:2013xoa}
J.~Ellis, D.~V. Nanopoulos, and K.~A. Olive, ``{No-Scale Supergravity
  Realization of the Starobinsky Model of Inflation}'',
  \href{http://dx.doi.org/10.1103/PhysRevLett.111.129902,
  10.1103/PhysRevLett.111.111301}{{\em Phys.Rev.Lett.} {\bf 111} (2013) no.~12,
  111301},
\href{http://arxiv.org/abs/1305.1247}{{\tt arXiv:1305.1247 [hep-th]}}.

\bibitem{Buchmuller:2013zfa}
W.~Buchmuller, V.~Domcke, and K.~Kamada, ``{The Starobinsky Model from
  Superconformal D-Term Inflation}'',
  \href{http://dx.doi.org/10.1016/j.physletb.2013.08.042}{{\em Phys.Lett.} {\bf
  B726} (2013)  467--470},
\href{http://arxiv.org/abs/1306.3471}{{\tt arXiv:1306.3471 [hep-th]}}.

\bibitem{Farakos:2013cqa}
F.~Farakos, A.~Kehagias, and A.~Riotto, ``{On the Starobinsky Model of
  Inflation from Supergravity}'',
  \href{http://dx.doi.org/10.1016/j.nuclphysb.2013.08.005}{{\em Nucl.Phys.}
  {\bf B876} (2013)  187--200},
\href{http://arxiv.org/abs/1307.1137}{{\tt arXiv:1307.1137}}.

\bibitem{Bezrukov:2007ep}
F.~L. Bezrukov and M.~Shaposhnikov, ``{The Standard Model Higgs boson as the
  inflaton}'', \href{http://dx.doi.org/10.1016/j.physletb.2007.11.072}{{\em
  Phys.Lett.} {\bf B659} (2008)  703--706},
\href{http://arxiv.org/abs/0710.3755}{{\tt arXiv:0710.3755 [hep-th]}}.

\bibitem{Ferrara:2013rsa}
S.~Ferrara, R.~Kallosh, A.~Linde, and M.~Porrati, ``{Minimal Supergravity
  Models of Inflation}'',
  \href{http://dx.doi.org/10.1103/PhysRevD.88.085038}{{\em Phys.Rev.} {\bf D88}
  (2013) no.~8, 085038},
\href{http://arxiv.org/abs/1307.7696}{{\tt arXiv:1307.7696 [hep-th]}}.

\bibitem{Mosk:2014cba}
B.~Mosk and J.~P. van~der Schaar, ``{Chaotic inflation limits for non-minimal
  models with a Starobinsky attractor}'',
  \href{http://dx.doi.org/10.1088/1475-7516/2014/12/022}{{\em JCAP} {\bf 1412}
  (2014) no.~12, 022},
\href{http://arxiv.org/abs/1407.4686}{{\tt arXiv:1407.4686 [hep-th]}}.

\bibitem{Broy:2015qna}
B.~J. Broy, M.~Galante, D.~Roest, and A.~Westphal, ``{Pole inflation — Shift
  symmetry and universal corrections}'',
  \href{http://dx.doi.org/10.1007/JHEP12(2015)149}{{\em JHEP} {\bf 12} (2015)
  149},
\href{http://arxiv.org/abs/1507.02277}{{\tt arXiv:1507.02277 [hep-th]}}.

\bibitem{Terada:2016nqg}
T.~Terada, ``{Generalized Pole Inflation: Hilltop, Natural, and Chaotic
  Inflationary Attractors}'',
\href{http://arxiv.org/abs/1602.07867}{{\tt arXiv:1602.07867 [hep-th]}}.

\bibitem{Antusch:2008pn}
S.~Antusch, M.~Bastero-Gil, K.~Dutta, S.~F. King, and P.~M. Kostka, ``{Solving
  the eta-Problem in Hybrid Inflation with Heisenberg Symmetry and Stabilized
  Modulus}'', \href{http://dx.doi.org/10.1088/1475-7516/2009/01/040}{{\em JCAP}
  {\bf 0901} (2009)  040},
\href{http://arxiv.org/abs/0808.2425}{{\tt arXiv:0808.2425 [hep-ph]}}.

\bibitem{Antusch:2009ty}
S.~Antusch, M.~Bastero-Gil, K.~Dutta, S.~F. King, and P.~M. Kostka, ``{Chaotic
  Inflation in Supergravity with Heisenberg Symmetry}'',
  \href{http://dx.doi.org/10.1016/j.physletb.2009.08.022}{{\em Phys. Lett.}
  {\bf B679} (2009)  428--432},
\href{http://arxiv.org/abs/0905.0905}{{\tt arXiv:0905.0905 [hep-th]}}.

\bibitem{Cecotti:2014ipa}
S.~Cecotti and R.~Kallosh, ``{Cosmological Attractor Models and Higher
  Curvature Supergravity}'',
  \href{http://dx.doi.org/10.1007/JHEP05(2014)114}{{\em JHEP} {\bf 05} (2014)
  114},
\href{http://arxiv.org/abs/1403.2932}{{\tt arXiv:1403.2932 [hep-th]}}.

\bibitem{Carrasco:2015uma}
J.~J.~M. Carrasco, R.~Kallosh, A.~Linde, and D.~Roest, ``{Hyperbolic geometry
  of cosmological attractors}'',
  \href{http://dx.doi.org/10.1103/PhysRevD.92.041301}{{\em Phys. Rev.} {\bf
  D92} (2015) no.~4, 041301},
\href{http://arxiv.org/abs/1504.05557}{{\tt arXiv:1504.05557 [hep-th]}}.

\bibitem{Covi:2008ea}
L.~Covi, M.~Gomez-Reino, C.~Gross, J.~Louis, G.~A. Palma, and C.~A. Scrucca,
  ``{de Sitter vacua in no-scale supergravities and Calabi-Yau string
  models}'', \href{http://dx.doi.org/10.1088/1126-6708/2008/06/057}{{\em JHEP}
  {\bf 06} (2008)  057},
\href{http://arxiv.org/abs/0804.1073}{{\tt arXiv:0804.1073 [hep-th]}}.

\bibitem{Kallosh:2015lwa}
R.~Kallosh and A.~Linde, ``{Planck, LHC, and $\alpha$-attractors}'',
\href{http://arxiv.org/abs/1502.07733}{{\tt arXiv:1502.07733 [astro-ph.CO]}}.

\bibitem{GomezReino}
M.~Gomez-Reino and C.~A. Scrucca, ``{Locally stable non-supersymmetric
  Minkowski vacua in supergravity}'',
  \href{http://dx.doi.org/10.1088/1126-6708/2006/05/015}{{\em JHEP} {\bf 0605}
  (2006)  015},
\href{http://arxiv.org/abs/hep-th/0602246}{{\tt arXiv:hep-th/0602246
  [hep-th]}}.

\bibitem{Kallosh:2014oja}
R.~Kallosh, A.~Linde, B.~Vercnocke, and T.~Wrase, ``{Analytic Classes of
  Metastable de Sitter Vacua}'',
  \href{http://dx.doi.org/10.1007/JHEP10(2014)011}{{\em JHEP} {\bf 1410} (2014)
   11},
\href{http://arxiv.org/abs/1406.4866}{{\tt arXiv:1406.4866 [hep-th]}}.

\bibitem{Kallosh:2014via}
R.~Kallosh and A.~Linde, ``{Inflation and Uplifting with Nilpotent
  Superfields}'', \href{http://dx.doi.org/10.1088/1475-7516/2015/01/025}{{\em
  JCAP} {\bf 1501} (2015) no.~01, 025},
\href{http://arxiv.org/abs/1408.5950}{{\tt arXiv:1408.5950 [hep-th]}}.

\bibitem{Dall'Agata:2014oka}
G.~Dall'Agata and F.~Zwirner, ``{On sgoldstino-less supergravity models of
  inflation}'', \href{http://dx.doi.org/10.1007/JHEP12(2014)172}{{\em JHEP}
  {\bf 1412} (2014)  172},
\href{http://arxiv.org/abs/1411.2605}{{\tt arXiv:1411.2605 [hep-th]}}.

\bibitem{Kallosh:2014hxa}
R.~Kallosh, A.~Linde, and M.~Scalisi, ``{Inflation, de Sitter Landscape and
  Super-Higgs effect}'', \href{http://dx.doi.org/10.1007/JHEP03(2015)111}{{\em
  JHEP} {\bf 1503} (2015)  111},
\href{http://arxiv.org/abs/1411.5671}{{\tt arXiv:1411.5671 [hep-th]}}.

\bibitem{Ellis:2014gxa}
J.~Ellis, M.~A.~G. Garcia, D.~V. Nanopoulos, and K.~A. Olive, ``{A No-Scale
  Inflationary Model to Fit Them All}'',
  \href{http://dx.doi.org/10.1088/1475-7516/2014/08/044}{{\em JCAP} {\bf 1408}
  (2014)  044},
\href{http://arxiv.org/abs/1405.0271}{{\tt arXiv:1405.0271 [hep-ph]}}.

\bibitem{Ellis:2014opa}
J.~Ellis, M.~A.~G. Garcia, D.~V. Nanopoulos, and K.~A. Olive, ``{Two-Field
  Analysis of No-Scale Supergravity Inflation}'',
  \href{http://dx.doi.org/10.1088/1475-7516/2015/01/010}{{\em JCAP} {\bf 1501}
  (2015) no.~01, 010},
\href{http://arxiv.org/abs/1409.8197}{{\tt arXiv:1409.8197 [hep-ph]}}.

\bibitem{Riess:1998cb}
{\bf Supernova Search Team} Collaboration, A.~G. Riess {\em et al.},
  ``{Observational evidence from supernovae for an accelerating universe and a
  cosmological constant}'', \href{http://dx.doi.org/10.1086/300499}{{\em
  Astron.J.} {\bf 116} (1998)  1009--1038},
\href{http://arxiv.org/abs/astro-ph/9805201}{{\tt arXiv:astro-ph/9805201
  [astro-ph]}}.

\bibitem{Perlmutter:1998np}
{\bf Supernova Cosmology Project} Collaboration, S.~Perlmutter {\em et al.},
  ``{Measurements of Omega and Lambda from 42 high redshift supernovae}'',
  \href{http://dx.doi.org/10.1086/307221}{{\em Astrophys.J.} {\bf 517} (1999)
  565--586},
\href{http://arxiv.org/abs/astro-ph/9812133}{{\tt arXiv:astro-ph/9812133
  [astro-ph]}}.

\bibitem{Hinshaw:2012aka}
{\bf WMAP} Collaboration, G.~Hinshaw {\em et al.}, ``{Nine-Year Wilkinson
  Microwave Anisotropy Probe (WMAP) Observations: Cosmological Parameter
  Results}'', \href{http://dx.doi.org/10.1088/0067-0049/208/2/19}{{\em
  Astrophys.J.Suppl.} {\bf 208} (2013)  19},
\href{http://arxiv.org/abs/1212.5226}{{\tt arXiv:1212.5226 [astro-ph.CO]}}.

\bibitem{Linde:1986fd}
A.~D. Linde, ``{Eternally Existing Selfreproducing Chaotic Inflationary
  Universe}'',
\href{http://dx.doi.org/10.1016/0370-2693(86)90611-8}{{\em Phys.Lett.} {\bf
  B175} (1986)  395--400}.

\bibitem{Weinberg:1987dv}
S.~Weinberg, ``{Anthropic Bound on the Cosmological Constant}'',
\href{http://dx.doi.org/10.1103/PhysRevLett.59.2607}{{\em Phys.Rev.Lett.} {\bf
  59} (1987)  2607}.

\bibitem{Bousso:2000xa}
R.~Bousso and J.~Polchinski, ``{Quantization of four form fluxes and dynamical
  neutralization of the cosmological constant}'',
  \href{http://dx.doi.org/10.1088/1126-6708/2000/06/006}{{\em JHEP} {\bf 0006}
  (2000)  006},
\href{http://arxiv.org/abs/hep-th/0004134}{{\tt arXiv:hep-th/0004134
  [hep-th]}}.

\bibitem{Douglas:2003um}
M.~R. Douglas, ``{The Statistics of string / M theory vacua}'',
  \href{http://dx.doi.org/10.1088/1126-6708/2003/05/046}{{\em JHEP} {\bf 0305}
  (2003)  046},
\href{http://arxiv.org/abs/hep-th/0303194}{{\tt arXiv:hep-th/0303194
  [hep-th]}}.

\bibitem{Susskind:2003kw}
L.~Susskind, ``{The Anthropic landscape of string theory}'',
\href{http://arxiv.org/abs/hep-th/0302219}{{\tt arXiv:hep-th/0302219
  [hep-th]}}.

\bibitem{Volkov:1972jx}
D.~Volkov and V.~Akulov, ``{Possible universal neutrino interaction}'',
{\em JETP Lett.} {\bf 16} (1972)  438--440.

\bibitem{Rocek:1978nb}
M.~Rocek, ``{Linearizing the Volkov-Akulov Model}'',
\href{http://dx.doi.org/10.1103/PhysRevLett.41.451}{{\em Phys.Rev.Lett.} {\bf
  41} (1978)  451--453}.

\bibitem{Ivanov:1978mx}
E.~Ivanov and A.~Kapustnikov, ``{General Relationship Between Linear and
  Nonlinear Realizations of Supersymmetry}'',
\href{http://dx.doi.org/10.1088/0305-4470/11/12/005}{{\em J.Phys.} {\bf A11}
  (1978)  2375--2384}.

\bibitem{Lindstrom:1979kq}
U.~Lindstrom and M.~Rocek, ``{Constrained Local Superfields}'',
\href{http://dx.doi.org/10.1103/PhysRevD.19.2300}{{\em Phys.Rev.} {\bf D19}
  (1979)  2300--2303}.

\bibitem{Casalbuoni:1988xh}
R.~Casalbuoni, S.~De~Curtis, D.~Dominici, F.~Feruglio, and R.~Gatto,
  ``{Nonlinear Realization of Supersymmetry Algebra From Supersymmetric
  Constraint}'',
\href{http://dx.doi.org/10.1016/0370-2693(89)90788-0}{{\em Phys.Lett.} {\bf
  B220} (1989)  569}.

\bibitem{Komargodski:2009rz}
Z.~Komargodski and N.~Seiberg, ``{From Linear SUSY to Constrained
  Superfields}'', \href{http://dx.doi.org/10.1088/1126-6708/2009/09/066}{{\em
  JHEP} {\bf 0909} (2009)  066},
\href{http://arxiv.org/abs/0907.2441}{{\tt arXiv:0907.2441 [hep-th]}}.

\bibitem{Dudas:2015eha}
E.~Dudas, S.~Ferrara, A.~Kehagias, and A.~Sagnotti, ``{Properties of Nilpotent
  Supergravity}'', \href{http://dx.doi.org/10.1007/JHEP09(2015)217}{{\em JHEP}
  {\bf 09} (2015)  217},
\href{http://arxiv.org/abs/1507.07842}{{\tt arXiv:1507.07842 [hep-th]}}.

\bibitem{Bergshoeff:2015tra}
E.~A. Bergshoeff, D.~Z. Freedman, R.~Kallosh, and A.~Van~Proeyen, ``{Pure de
  Sitter Supergravity}'',
  \href{http://dx.doi.org/10.1103/PhysRevD.92.085040}{{\em Phys. Rev.} {\bf
  D92} (2015) no.~8, 085040},
\href{http://arxiv.org/abs/1507.08264}{{\tt arXiv:1507.08264 [hep-th]}}.

\bibitem{Hasegawa:2015bza}
F.~Hasegawa and Y.~Yamada, ``{Component action of nilpotent multiplet coupled
  to matter in 4 dimensional $ \mathcal{N}=1 $ supergravity}'',
  \href{http://dx.doi.org/10.1007/JHEP10(2015)106}{{\em JHEP} {\bf 10} (2015)
  106},
\href{http://arxiv.org/abs/1507.08619}{{\tt arXiv:1507.08619 [hep-th]}}.

\bibitem{Ferrara:2015gta}
S.~Ferrara, M.~Porrati, and A.~Sagnotti, ``{Scale invariant Volkov–Akulov
  supergravity}'', \href{http://dx.doi.org/10.1016/j.physletb.2015.08.066}{{\em
  Phys. Lett.} {\bf B749} (2015)  589--591},
\href{http://arxiv.org/abs/1508.02939}{{\tt arXiv:1508.02939 [hep-th]}}.

\bibitem{Kuzenko:2015yxa}
S.~M. Kuzenko, ``{Complex linear Goldstino superfield and supergravity}'',
  \href{http://dx.doi.org/10.1007/JHEP10(2015)006}{{\em JHEP} {\bf 10} (2015)
  006},
\href{http://arxiv.org/abs/1508.03190}{{\tt arXiv:1508.03190 [hep-th]}}.

\bibitem{Kallosh:2015tea}
R.~Kallosh and T.~Wrase, ``{De Sitter Supergravity Model Building}'',
  \href{http://dx.doi.org/10.1103/PhysRevD.92.105010}{{\em Phys. Rev.} {\bf
  D92} (2015) no.~10, 105010},
\href{http://arxiv.org/abs/1509.02137}{{\tt arXiv:1509.02137 [hep-th]}}.

\bibitem{Schillo:2015ssx}
M.~Schillo, E.~van~der Woerd, and T.~Wrase, ``{The general de Sitter
  supergravity component action}'', in {\em {21st European String Workshop: The
  String Theory Universe Leuven, Belgium, September 7-11, 2015}}.
\newblock 2015.
\newblock
\href{http://arxiv.org/abs/1511.01542}{{\tt arXiv:1511.01542 [hep-th]}}.
\newblock

\bibitem{Antoniadis:2014oya}
I.~Antoniadis, E.~Dudas, S.~Ferrara, and A.~Sagnotti, ``{The
  Volkov-Akulov-Starobinsky supergravity}'',
  \href{http://dx.doi.org/10.1016/j.physletb.2014.04.015}{{\em Phys.Lett.} {\bf
  B733} (2014)  32--35},
\href{http://arxiv.org/abs/1403.3269}{{\tt arXiv:1403.3269 [hep-th]}}.

\bibitem{Ferrara:2014kva}
S.~Ferrara, R.~Kallosh, and A.~Linde, ``{Cosmology with Nilpotent
  Superfields}'', \href{http://dx.doi.org/10.1007/JHEP10(2014)143}{{\em JHEP}
  {\bf 1410} (2014)  143},
\href{http://arxiv.org/abs/1408.4096}{{\tt arXiv:1408.4096 [hep-th]}}.

\bibitem{KLnil}
R.~Kallosh and A.~Linde, ``{Inflation and Uplifting with Nilpotent
  Superfields}'', \href{http://dx.doi.org/10.1088/1475-7516/2015/01/025}{{\em
  JCAP} {\bf 1501} (2015) no.~01, 025},
\href{http://arxiv.org/abs/1408.5950}{{\tt arXiv:1408.5950 [hep-th]}}.

\bibitem{Kallosh:2014wsa}
R.~Kallosh and T.~Wrase, ``{Emergence of Spontaneously Broken Supersymmetry on
  an Anti-D3-Brane in KKLT dS Vacua}'',
  \href{http://dx.doi.org/10.1007/JHEP12(2014)117}{{\em JHEP} {\bf 1412} (2014)
   117},
\href{http://arxiv.org/abs/1411.1121}{{\tt arXiv:1411.1121 [hep-th]}}.

\bibitem{Dudas:2012wi}
E.~Dudas, A.~Linde, Y.~Mambrini, A.~Mustafayev, and K.~A. Olive, ``{Strong
  moduli stabilization and phenomenology}'',
  \href{http://dx.doi.org/10.1140/epjc/s10052-012-2268-7}{{\em Eur. Phys. J.}
  {\bf C73} (2013) no.~1, 2268},
\href{http://arxiv.org/abs/1209.0499}{{\tt arXiv:1209.0499 [hep-ph]}}.

\bibitem{Kallosh:2014qta}
R.~Kallosh, A.~Linde, B.~Vercnocke, and W.~Chemissany, ``{Is Imaginary
  Starobinsky Model Real?}'',
  \href{http://dx.doi.org/10.1088/1475-7516/2014/07/053}{{\em JCAP} {\bf 1407}
  (2014)  053},
\href{http://arxiv.org/abs/1403.7189}{{\tt arXiv:1403.7189 [hep-th]}}.

\bibitem{Coughlan:1983ci}
G.~D. Coughlan, W.~Fischler, E.~W. Kolb, S.~Raby, and G.~G. Ross,
  ``{Cosmological Problems for the Polonyi Potential}'',
\href{http://dx.doi.org/10.1016/0370-2693(83)91091-2}{{\em Phys. Lett.} {\bf
  B131} (1983)  59}.

\bibitem{Goncharov:1984qm}
A.~S. Goncharov, A.~D. Linde, and M.~I. Vysotsky, ``{COSMOLOGICAL PROBLEMS FOR
  SPONTANEOUSLY BROKEN SUPERGRAVITY}'',
\href{http://dx.doi.org/10.1016/0370-2693(84)90116-3}{{\em Phys. Lett.} {\bf
  B147} (1984)  279}.

\bibitem{Banks:1993en}
T.~Banks, D.~B. Kaplan, and A.~E. Nelson, ``{Cosmological implications of
  dynamical supersymmetry breaking}'',
  \href{http://dx.doi.org/10.1103/PhysRevD.49.779}{{\em Phys. Rev.} {\bf D49}
  (1994)  779--787},
\href{http://arxiv.org/abs/hep-ph/9308292}{{\tt arXiv:hep-ph/9308292
  [hep-ph]}}.

\bibitem{Dvali:1995mj}
G.~R. Dvali, ``{Inflation versus the cosmological moduli problem}'',
\href{http://arxiv.org/abs/hep-ph/9503259}{{\tt arXiv:hep-ph/9503259
  [hep-ph]}}.

\bibitem{Dine:1995uk}
M.~Dine, L.~Randall, and S.~D. Thomas, ``{Supersymmetry breaking in the early
  universe}'', \href{http://dx.doi.org/10.1103/PhysRevLett.75.398}{{\em Phys.
  Rev. Lett.} {\bf 75} (1995)  398--401},
\href{http://arxiv.org/abs/hep-ph/9503303}{{\tt arXiv:hep-ph/9503303
  [hep-ph]}}.

\bibitem{Kallosh:2015pho}
R.~Kallosh, A.~Karlsson, and D.~Murli, ``{From linear to nonlinear
  supersymmetry via functional integration}'',
  \href{http://dx.doi.org/10.1103/PhysRevD.93.025012}{{\em Phys. Rev.} {\bf
  D93} (2016) no.~2, 025012},
\href{http://arxiv.org/abs/1511.07547}{{\tt arXiv:1511.07547 [hep-th]}}.

\bibitem{Ferrara:2016een}
S.~Ferrara, R.~Kallosh, A.~Van~Proeyen, and T.~Wrase, ``{Linear Versus
  Non-linear Supersymmetry, in General}'',
  \href{http://dx.doi.org/10.1007/JHEP04(2016)065}{{\em JHEP} {\bf 04} (2016)
  065},
\href{http://arxiv.org/abs/1603.02653}{{\tt arXiv:1603.02653 [hep-th]}}.

\bibitem{Kallosh:2016hcm}
R.~Kallosh, A.~Karlsson, B.~Mosk, and D.~Murli, ``{Orthogonal Nilpotent
  Superfields from Linear Models}'',
  \href{http://dx.doi.org/10.1007/JHEP05(2016)082}{{\em JHEP} {\bf 05} (2016)
  082},
\href{http://arxiv.org/abs/1603.02661}{{\tt arXiv:1603.02661 [hep-th]}}.

\bibitem{Dall'Agata:2015lek}
G.~Dall'Agata and F.~Farakos, ``{Constrained superfields in Supergravity}'',
  \href{http://dx.doi.org/10.1007/JHEP02(2016)101}{{\em JHEP} {\bf 02} (2016)
  101},
\href{http://arxiv.org/abs/1512.02158}{{\tt arXiv:1512.02158 [hep-th]}}.

\bibitem{Dall'Agata:2016yof}
G.~Dall’Agata, E.~Dudas, and F.~Farakos, ``{On the origin of constrained
  superfields}'', \href{http://dx.doi.org/10.1007/JHEP05(2016)041}{{\em JHEP}
  {\bf 05} (2016)  041},
\href{http://arxiv.org/abs/1603.03416}{{\tt arXiv:1603.03416 [hep-th]}}.

\bibitem{Kallosh:2000ve}
R.~Kallosh, L.~Kofman, A.~D. Linde, and A.~Van~Proeyen, ``{Superconformal
  symmetry, supergravity and cosmology}'',
  \href{http://dx.doi.org/10.1088/0264-9381/17/20/308}{{\em Class. Quant.
  Grav.} {\bf 17} (2000)  4269--4338},
  \href{http://arxiv.org/abs/hep-th/0006179}{{\tt arXiv:hep-th/0006179
  [hep-th]}}.
[Erratum: Class. Quant. Grav.21,5017(2004)].

\bibitem{Ellis:2015kqa}
J.~Ellis, M.~A.~G. Garcia, D.~V. Nanopoulos, and K.~A. Olive,
  ``{Phenomenological Aspects of No-Scale Inflation Models}'',
  \href{http://dx.doi.org/10.1088/1475-7516/2015/10/003}{{\em JCAP} {\bf 1510}
  (2015) no.~10, 003},
\href{http://arxiv.org/abs/1503.08867}{{\tt arXiv:1503.08867 [hep-ph]}}.

\bibitem{Ellis:2015pla}
J.~Ellis, M.~A.~G. Garcia, D.~V. Nanopoulos, and K.~A. Olive, ``{Calculations
  of Inflaton Decays and Reheating: with Applications to No-Scale Inflation
  Models}'', \href{http://dx.doi.org/10.1088/1475-7516/2015/07/050}{{\em JCAP}
  {\bf 1507} (2015) no.~07, 050},
\href{http://arxiv.org/abs/1505.06986}{{\tt arXiv:1505.06986 [hep-ph]}}.

\bibitem{Bergshoeff:2015jxa}
E.~A. Bergshoeff, K.~Dasgupta, R.~Kallosh, A.~Van~Proeyen, and T.~Wrase, ``{$
  \overline{\mathrm{D}3} $ and dS}'',
  \href{http://dx.doi.org/10.1007/JHEP05(2015)058}{{\em JHEP} {\bf 05} (2015)
  058},
\href{http://arxiv.org/abs/1502.07627}{{\tt arXiv:1502.07627 [hep-th]}}.

\bibitem{Bandos:2015xnf}
I.~Bandos, L.~Martucci, D.~Sorokin, and M.~Tonin, ``{Brane induced
  supersymmetry breaking and de Sitter supergravity}'',
  \href{http://dx.doi.org/10.1007/JHEP02(2016)080}{{\em JHEP} {\bf 02} (2016)
  080},
\href{http://arxiv.org/abs/1511.03024}{{\tt arXiv:1511.03024 [hep-th]}}.

\bibitem{Ozkan:2015iva}
M.~Ozkan, Y.~Pang, and S.~Tsujikawa, ``{Planck constraints on inflation in
  auxiliary vector modified $f(R)$ theories}'',
  \href{http://dx.doi.org/10.1103/PhysRevD.92.023530}{{\em Phys. Rev.} {\bf
  D92} (2015) no.~2, 023530},
\href{http://arxiv.org/abs/1502.06341}{{\tt arXiv:1502.06341 [astro-ph.CO]}}.

\bibitem{Kallosh:2016ndd}
R.~Kallosh, A.~Linde, and T.~Wrase, ``{Coupling the Inflationary Sector to
  Matter}'', \href{http://dx.doi.org/10.1007/JHEP04(2016)027}{{\em JHEP} {\bf
  04} (2016)  027},
\href{http://arxiv.org/abs/1602.07818}{{\tt arXiv:1602.07818 [hep-th]}}.

\bibitem{Kallosh:2016gqp}
R.~Kallosh and A.~Linde, ``{Cosmological Attractors and Asymptotic Freedom of
  the Inflaton Field}'',
  \href{http://dx.doi.org/10.1088/1475-7516/2016/06/047}{{\em JCAP} {\bf 1606}
  (2016) no.~06, 047},
\href{http://arxiv.org/abs/1604.00444}{{\tt arXiv:1604.00444 [hep-th]}}.

\bibitem{Broy:2015zba}
B.~J. Broy, D.~Ciupke, F.~G. Pedro, and A.~Westphal, ``{Starobinsky-Type
  Inflation from $\alpha'$-Corrections}'',
  \href{http://arxiv.org/abs/1509.00024}{{\tt arXiv:1509.00024 [hep-th]}}.
[JCAP1601,001(2016)].

\bibitem{Burgess:2016owb}
C.~P. Burgess, M.~Cicoli, S.~de~Alwis, and F.~Quevedo, ``{Robust Inflation from
  Fibrous Strings}'',
  \href{http://dx.doi.org/10.1088/1475-7516/2016/05/032}{{\em JCAP} {\bf 1605}
  (2016) no.~05, 032},
\href{http://arxiv.org/abs/1603.06789}{{\tt arXiv:1603.06789 [hep-th]}}.

\end{thebibliography}\endgroup
\bibliographystyle{utphys}
}

\newpage
\thispagestyle{empty}

\ 



\renewcommand{\chapterheadstartvskip}{\vspace*{1\baselineskip}}

\chapter*{Acknowledgments}
\label{Chap:acknowledgments}
\addcontentsline{toc}{chapter}{Acknowledgments}
\markboth{Acknowledgments}{Acknowledgments}

This thesis is probably the best condensate of the scientific results of my research of the last four years. However, the people who made all this happen and (directly or indirectly) contributed to this piece of work are numerous. They belong to a huge variety of contexts and I would like to thank them all here below. 

Without any doubt, I owe very much to my supervisor {\it Diederik Roest}. His great enthusiasm and extreme keenness in doing research have been a fantastic example for my personal and academic development.  I have got many valuable lessons from him, probably one of the most important being how to turn a project into fun and always be optimistic when facing a challenge. Thank you very much for all this!

Next to my supervisor, I would like to express my deepest gratitude to the people who co-authored the  publications of my PhD period. Their collaboration, together with the many enlightening discussions and comments, has been precious in terms of critically developing and successfully finalizing each project. This thesis would have certainly been of a different weight without the contributions of {\it Juan Garcia-Bellido}, {\it Andrei Linde}, {\it Renata Kallosh}, and {\it Ivonne Zavala}. Thank you so much again. 

My sincere thanks also goes to the members of the assessment committee,  {\it Ana Ach\'ucarro}, {\it Arthur Hebecker} and {\it Rien van de Weijgaert}, for carefully reading and positively assessing my thesis. Their in-depth comments and remarks undoubtedly helped me a lot to improve my final work.

Being a PhD student in the Netherlands has given me the great opportunity to travel a lot and often be in contact with a very nice and active community of theoretical physicists and cosmologists. The numerous discussions, questions, debates, remarks and criticisms have simply enriched my scientific background in a wonderful way. Besides, my research work has definitively benefited from all of this. I would like then to thank, first, the synergetic Dutch cosmology community (THC$@$NL), whose regular meetings have effectively reduced the distances between Groningen and the others. Specifically, I would like to thank {\it Ana Ach\'ucarro} (again), {\it Vicente Atal}, {\it Daniel Baumann}, {\it Dra\v{z}en Glavan}, {\it Roberto Gobetti}, {\it Ben Freivogel}, {\it Enrico Pajer}, {\it Marieke Postma}, {\it Tomislav Prokopec}, {\it Pablo Ortiz}, {\it Wessel Valkenburg}, {\it Jan Pieter van der Schaar}, {\it Drian van der Woude} and {\it Yvette Welling}. Crossing the borders of this small country, my thanks goes also to {\it \mbox{Benedict} Broy}, {\it Wissam Chemissany}, {\it Michele Cicoli}, {\it David Ciupke}, {\it Paolo \mbox{Creminelli}}, {\it Ulf Danielsson}, {\it Keshav Dasgupta}, {\it Mafalda Dias}, {\it Giuseppe Dibitetto}, {\it Emilian Dudas}, {\it Jonathan Frazer}, {\it Adolfo Guarino}, {\it Thomas Hertog}, {\it Sugumi Kanno}, {\it David Lyth}, {\it M. C. David Marsh}, {\it Anupam Mazumdar}, {\it Evan McDonough}, {\it Sander Mooji}, {\it Carlos N\'u\~nez}, {\it Gonzalo Palma}, {\it Francisco Pedro}, {\it Fernando Quevedo}, {\it Fabrizio Rompineve}, {\it Augusto Sagnotti}, {\it Leonardo Senatore}, {\it Dmitri Sorokin}, {\it Alexei Starobinsky}, {\it Gianmassimo Tasinato}, {\it Takahiro Terada}, {\it Alessandro Tomasiello}, {\it Gabriele Trevisan}, {\it Antoine van Proeyen}, {\it Thomas van Riet}, {\it Vincent Vennin}, {\it Bert Vercnocke}, {\it David Wands}, {\it Alexander Westphal}, {\it Clemens Wieck}, {\it Timm Wrase} and {\it Aleksandr Zheltukhin}.

Next to them, I would like to thank my former supervisors and teachers, whose (scientific and non-scientific) support has always been so important to me, during the last years. Thanks to {\it Giuseppe Angilella}, {\it Alfio Bonanno}, {\it Gianluca Calcagni}, {\it Daniele Oriti} and {\it Andrea Rapisarda}.

I absolutely do not regret the time and the energies I spent in order to organize my three-months stay at Stanford in 2014. Having no financial support from my research group in Groningen, my determination and personal search for external funding possibilities turned out to be essential. Among the several places I visited during my PhD, Stanford  provided undoubtedly one of the most exciting experiences. The incredibly stimulating atmosphere of the SITP, the passionate corridor-discussions, the clear sky and warm sun of California simply created fantastic conditions to fruitfully work and develop ideas. Then, I  am very grateful to {\it Andrei Linde} and {\it Renata Kallosh} for giving me the great chance to join their group. Besides them, I would like to deeply thank the people who made my stay there so enjoyable. Thanks to my office-mates {\it Hideki Perrier} and {\it Matteo Cataneo} for sharing their daily stories, jokes and their so diverse personalities in that small room. Further, a very genuine thanks goes to {\it Emanuela Dimastrogiovanni} and {\it Matteo Fasiello}, whose authentic smiles and sympathy let me feel always ``a casa''. A special thanks is reserved to {\it Edward Mazenc}, a person you rarely happen to come across nowadays. His superb cuisine, his wide spectrum of interests and expertise, his profound sense of loyalty and friendship have always contributed to that unique atmosphere I have always enjoyed any time I have been with him. Grazie davvero Edward! Finally, I would like to stress that this visit would have been not possible without the financial support coming from two different sources. I gratefully acknowledge a grant awarded by the COST Action MP1210 and the ``Marco Polo Fund'' provided by the University of Groningen. Then, I take the opportunity to thank the people who helped me and dealt with the administration of the relevant documents. I would like to sincerely thank {\it Fiorella Brustolin}, {\it Anna Ceresole} and {\it Silvia Penati}.

Of course, the stimulating and pleasant atmosphere provided by my closest colleagues has been an essential ingredient for successfully managing my research work. I would like then to thank all the people of the Van Swinderen Institute and of the Zernike Institute in Groningen. Specifically, I would like to address my ``thanks'' to {\it Aditya}, {\it Adolfo}, {\it Ana}, {\it Anna}, {\it Andrea}, {\it Andr\'es}, {\it Arsalan}, {\it Blaise} (thanks for being my paranymph!), {{\it Dennis}, {\it Dries}, {\it Giuseppe}, {\it G\"okhan}, {\it Hamid}, {\it Jan}, {\it Jan-Willem}, {\it Julian}, {\it Keri}, {\it Lorena}, {\it Luca}, {\it Luca}, {\it Marija}, {\it Mario}, {\it Mark}, {\it Marwa}, {\it Mehmet}, {\it Pelle}, {\it Pulastya}, {\it Remko}, {\it Roel}, {\it Shanka}, {\it Sjoerd}, {\it Souvik}, {\it Thomas}, {\it Tiago}, {\it Victor}, {\it Wout}, {\it Wouter} and {\it Yihao}. Among the staff, let me thank {\it Daniel}, {\it Elisabetta}, {\it Eric}, {\it Kyriakos} and {\it Rob}. My sincere gratitude also goes to {\it Annelien} and {\it Iris}: thank you very much for dealing with all my requests (!) and for running the institute so well. Finally, I wish to deliver my thanks to those colleagues, outside Groningen, who have been very good companions of this academic adventure. Specifically, let me thank {\it Adolfo}, {\it Blaza}, {\it Ellen}, {\it Gabriele}, {\it Jules}, {\it Marco} and {\it Teresa}.

Even not being directly involved in the scientific part of this venture, many people have contributed to my personal cheerfulness and peace of mind of these last four years. I would like to thank {\it Alessio}, {\it Felipe}, {\it Luciana}, {\it Mariella}, {\it Martina}, le piccole {\it Margherita} e {\it Mariasole}, {\it Salvo}, {\it Sandra}, {\it Sara} and {\it Stefano} for the very nice time spent together. Vorrei inoltre ringraziare {\it Filippo}, {\it Francesco} ed {\it Elena} per avermi sempre regalato dei bei momenti insieme, ogni volta che son tornato a Berlino in questi anni.

Grazie ai miei amici di sempre {\it Adriana}, {\it Alessio}, {\it Andrea}, {\it Azzurra}, {\it Caterina}, {\it Emilia}, {\it Francesca}, {\it Gabriele}, {\it Giada}, {\it Giovanni}, {\it Gnappi}, {\it Marta}, {\it Mario}, {\it Paolo}, {\it Rita}, {\it Simone}, {\it Valentina}. Rimanete sempre una delle mie maggiori motivazioni per scendere in Sicilia e stare più a lungo possibile (ovviamente dopo la granita al pistacchio!). 

Un grazie immenso va anche a {\it mamma}, {\it pap\'a} e {\it Valeria} per il grandissimo supporto quotidiano che sanno darmi in qualunque occasione, anche a migliaia di chilometri di distanza.

Schlie{\ss}lich m\"ochte ich {\it Anika} danken, der Person, der ich diese Arbeit gewidmet habe. Tutto, senza di te, avrebbe brillato di una luce più fioca. Wir haben dieses sehr sch\"one Kapitel unserer Geschichte zusammen abgeschlossen. Non vedo l'ora di scrivere il seguito insieme a te.

\begin{flushright}
Marco Scalisi\\
May 2016
\end{flushright}

\clearpage
\thispagestyle{empty}

\vspace*{3.5cm}
\begin{center}
{\it ...still...}
\end{center}
\vspace{0.5cm}
\begin{figure}[htb]
\begin{center}
\includegraphics[width=6.5cm,keepaspectratio]{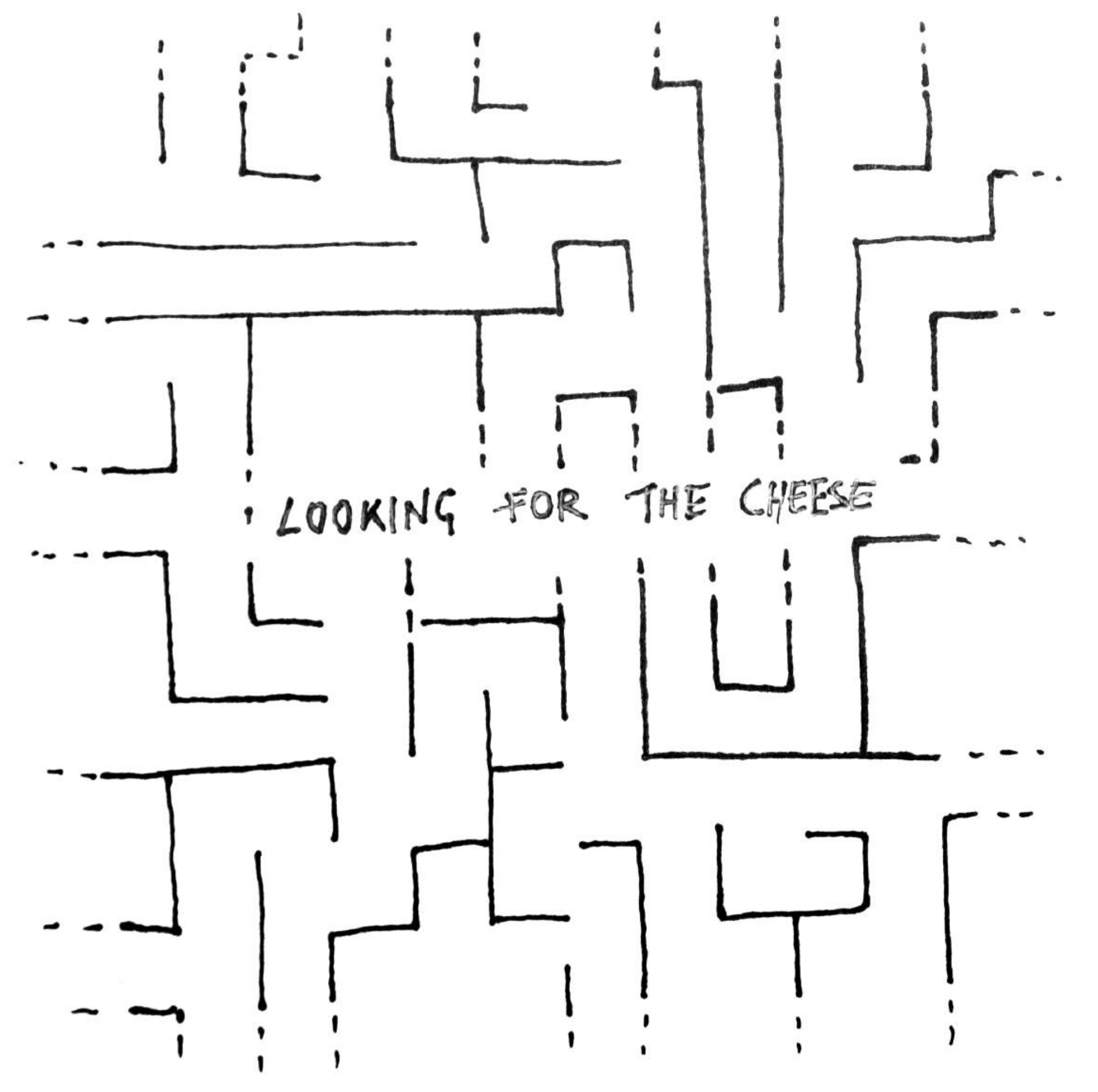}
\end{center}
\end{figure}
\vspace*{\fill}

\end{document}